%% file: tesi.tex
\documentclass[11pt,a4paper,twoside]{book} 

\usepackage{amsmath}
\usepackage{amscd}
\usepackage{amssymb}
\usepackage{amsfonts}
\usepackage{eufrak}
\usepackage[mathscr]{euscript}
\usepackage{pstricks}
\usepackage{pst-node}
\usepackage{cite}
\usepackage[dvips]{pstcol}\textheight=8.5truein
\usepackage[dvips]{graphicx}
\usepackage{fancyheadings}

\newcommand{\s}{\section}
\newcommand{\la}{\label}

\newcommand{\bi}{\begin{itemize}}
\newcommand{\ei}{\end{itemize}}
\newcommand{\ben}{\begin{enumerate}}
\newcommand{\een}{\end{enumerate}}
\newcommand{\be}{\begin{equation}}
\newcommand{\ee}{\end{equation}}
\newcommand{\noi}{\noindent}

\newcommand{\Tr}{\operatorname{Tr}}
\newcommand{\Det}{\operatorname{Det}}
\newcommand{\re}{\operatorname{Re}}
\newcommand{\im}{\operatorname{Im}}

\newcommand{\bz}{\vert {\rm {\bf B}^0} \rangle}
\newcommand{\abz}{\vert {\rm \overline{\bf B}^0} \rangle}
\newcommand{\bo}{\vert {\rm {\bf B}}_1 \rangle}
\newcommand{\bt}{\vert {\rm {\bf B}}_2 \rangle}
\newcommand{\bl}{\vert {\rm {\bf B}}_L \rangle}
\newcommand{\bh}{\vert {\rm {\bf B}}_H \rangle}
\newcommand{\nel}{\vert \nu_e \rangle}
\newcommand{\nm}{\vert \nu_{\mu} \rangle}
\newcommand{\no}{\vert \nu_1 \rangle}
\newcommand{\nt}{\vert \nu_2 \rangle}

\begin{document}

\newtheorem{Def}{Definition}[chapter]
\newtheorem{The}{Theorem}[chapter]
\newtheorem{Cor}{Corollary}[chapter]
\newtheorem{Lem}{Lemma}[chapter]
\newtheorem{Rem}{Remark}[chapter]
\newtheorem{Pro}{Proposition}[chapter]

\mainmatter
\begin{titlepage}
\include{title}

\end{titlepage}
\thispagestyle{empty}

\begin{titlepage}
\include{title3}
\end{titlepage}
\thispagestyle{empty}

\frontmatter
\tableofcontents

\mainmatter
\thispagestyle{empty}

\begin{titlepage}
\include{title2}
\end{titlepage}
\thispagestyle{empty}

\include{introduction} 
\newpage \thispagestyle{plain}
 
\include{chapter1} 
\newpage \thispagestyle{plain}

\include{chapter2} 
\newpage \thispagestyle{plain}

\include{chapter3} 
\newpage \thispagestyle{plain}

\include{chapter4} 
\newpage \thispagestyle{plain}

\include{chapter5} 
\newpage \thispagestyle{plain}

\begin{appendix}
\setcounter{chapter}{0}

\include{appendixA} %
\newpage \thispagestyle{plain}
\setcounter{chapter}{1}

\include{appendixB} %
\newpage \thispagestyle{plain}
\setcounter{chapter}{2}

\include{appendixC} %
\newpage \thispagestyle{plain}
\setcounter{chapter}{3}

\include{appendixD} %
\newpage \thispagestyle{plain}
\setcounter{chapter}{4}

\include{appendixE} %
\newpage \thispagestyle{plain}

\end{appendix}

\end{document}

%% file: title.tex
\begin{center}
{\sc \LARGE UNIVERSIT\`A DEGLI STUDI DI TRIESTE}



\vspace{0.5 cm}

{\sc Dottorato di Ricerca in Fisica - XV ciclo}

\vspace{0.5 cm}

{\sc Anno Accademico 2001-2002}

\vspace{2.0 cm}

{\Huge \bf Dissipative Dynamics}

\vspace{0.5cm}

{\Huge \bf in Particle Physics}

\vspace{1.0cm}

{\huge \bf Raffaele Romano}

\vspace{1.0cm}

{\large Ph.D. Thesis}

\end{center}

\vskip 1cm

\begin{figure}[h]

\begin{center}
\includegraphics[angle=0,width=0.3\textwidth]{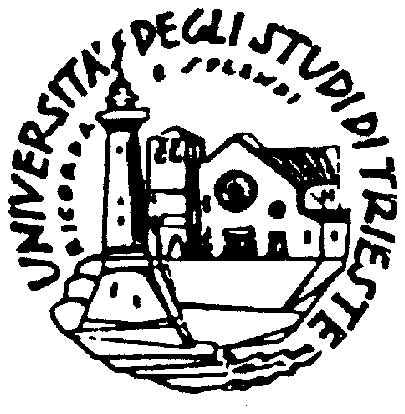}
\end{center}

\end{figure}

\vskip 1cm

\begin{center}

{\sc Tutore}: Dott. Fabio Benatti (Universit\`a di Trieste)

\vspace{0.5cm} 

{\sc Coordinatore}: Prof.ssa Francesca Matteucci (Universit\`a di Trieste)

\end{center}


%% file: title3.tex
\begin{center}

\vspace{7.0 cm}

{\Huge  \sc{DISSIPATIVE DYNAMICS}}

\vspace{1.0cm}

{\Huge \sc{IN PARTICLE PHYSICS}}

\vspace{4.0cm}

{\Large Raffaele Romano}

\end{center}

%% file: title2.tex
{\footnotesize

12 - {\it Vom Ziele der Wissenschaft\,} - Wie? Das letzte Ziel der Wissenschaft sei, 
dem Menschen m{\" o}glichst viel Lust und m{\" o}glichst wenig Unlust zu schaffen? 
Wie, wenn nun Lust und Unlust so mit einem Stricke zusammengekn{\" u}pft w{\" a}ren, 
dass, wer m{\" o}glichst viel von der einen haben will, auch m{\" o}glichst viel 
von der andern haben muss,-dass, wer das ``Himmelhoch-Jauchzen'' lernen will, sich 
auch f{\" u}r das ``zum-Todebetr{\" u}bt'' bereit halten muss? Und so steht es 
vielleicht! Die Stoiker glaubten wenigstens, dass es so stehe, und waren consequent, 
als sie nach m{\" o}glichst wenig Lust begehrten, um m{\" o}glichst wenig
Unlust vom Leben zu haben (wenn man den Spruch im Munde f{\" u}hrte ``Der Tugendhafte 
ist der Gl{\" u}cklichste'', so hatte man in ihm sowohl ein Aush{\" a}ngeschild der 
Schule f{\" u}r die grosse Masse, als auch eine casuistische Feinheit f{\" u}r die 
Feinen). Auch heute noch habt ihr die Wahl: entweder m{\" o}glichst wenig Unlust, 
kurz Schmerzlosigkeit-und im Grunde d{\" u}rften Socialisten und Politiker aller 
Parteien ihren Leuten ehrlicher Weise nicht mehr verheissen-oder m{\" o}glichst 
viel Unlust als Preis f{\" u}r das Wachstum einer F{\" u}lle von feinen und bisher 
selten gekosteten L{\" u}sten und Freuden! Entschliesst ihr euch f{\" u}r das Erstere, 
wollt ihr also die Schmerzhaftigkeit der Menschen herabdr{\" u}cken und vermindern, 
nun, so m{\" u}sst ihr auch ihre F{\" a}higkeit zur Freude herabdr{\" u}cken und 
vermindern. In der Tat kann man mit der Wissenschaft das eine wie das andere Ziel 
f{\" o}rdern! Vielleicht ist sie jetzt noch bekannter wegen ihrer Kraft, den 
Menschen um seine Freuden zu bringen, und ihn k{\" a}lter, statuenhafter, stoischer 
zu machen. Aber sie k{\" o}nnte auch noch als die grosse Schmerzbringerin entdeckt 
werden!-Und dann w{\" u}rde vielleicht zugleich ihre Gegenkraft entdeckt sein, ihr 
ungeheures Verm{\" o}gen, neue Sternenwelten der Freude aufleuchten zu lassen!

\vspace{1.0 cm}

\begin{flushright}
F. Nietzsche, Die fr{\" o}hliche Wissenschaft (``la gaya scienza''), 1882
\end{flushright}
}

%% file: introduction.tex
\chapter*{Introduction}
\addcontentsline{toc}{chapter}{\numberline{}Introduction}
\markboth{\small{\rm{INTRODUCTION}}}{\small{\rm{INTRODUCTION}}}

The subject of this thesis is the study of dissipative dynamics
and their properties in particle physics.

Physical systems can be considered closed when they are isolated
from their environment; otherwise, due to interactions between them,
there occur exchanges of energy, entropy and matter from system to
environment and vice versa. Physical systems are termed open in such a
case.

Typically, the physical systems we will be concerned with have 
few degrees of freedom in comparison
to the very (infinitely) many of the environment and matter exchanges
are neglected. Closed quantum systems evolve reversibly in time according to
the unitary dynamics generated by the Schr{\" o}dinger equation; in
particular, there cannot be decoherence since Hilbert space vectors
evolve in time into Hilbert space vectors.

On the contrary, the time-evolution of open quantum systems is quite
complicated due to the coupling with the environment. It can be formally
derived from the unitary dynamics of the closed system formed by the open
system plus the environment, by eliminating (or summing over) the degrees 
of freedom of the latter.

The result is a so-called reduced dynamics for the open system alone
that does not conserve its energy and entropy; the dynamics is thus
dissipative and, moreover, does not transform pure states into themselves,
rather it destroys quantum coherence. 

The reduced dynamics generally contains memory
effects, that is it depends not only on the initial conditions and the
observation time, but on the whole history of the system in between.
However, if the coupling with 
the environment is weak, it is possible to approximate this dynamics with
semigroups of linear transformations in which the memory effects have
disappeared, namely by Markovian semigroups which constitute the
class of time-evolutions studied in this thesis~\cite{alic1,breu,davi1}.

Open quantum systems are met in many branches of physics; in particular,
the approach
based on semigroups of linear transformations has been applied in
several contexts, ranging from phenomena of quantum relaxation in
magnetic resonance~\cite{slic}, 
to quantum optics~\cite{loui,puri,gard,scul}, to foundational aspects of 
quantum mechanics~\cite{fond,nakaz}, to the study of non-standard dynamics
in elementary particle physics~\cite{benak,benab,benan,benan2} 
and of interference phenomena with photons or 
neutrons~\cite{benai1,benai2,benai3}.

Since reduced dynamics are decohering,
the states of open systems must be described by density matrices;
their eigenvalues represent probabilities, so that they have to be positive.
In order to have a consistent statistical interpretation of the formalism,
the dynamical maps describing the time-evolution must preserve such
positivity; technically speaking, they need to be positivity preserving
or positive for short.
However, when there is entanglement between the open system and its 
surroundings, positivity is no longer sufficient to guarantee the 
consistency of the physical description. A stronger request 
need to be enforced: the so-called complete positivity. Formally, complete 
positivity assures the positivity of states of composite 
systems in which there is a statistical but not a dynamical coupling
between the system under study and a generic $n$-level system.

Due to its rather abstract definition, complete positivity
is sometime considered as a mere mathematical property rather than a 
physically motivated request. Indeed, the literature about open quantum
systems abounds in reduced dynamics that are not completely positive
(and some examples will be discussed below)~\cite{budi,lair,suar}. 

The physical systems and the discussions presented in this thesis
provide some arguments
in favor of complete positivity by showing explicitly the unavoidable
problems that arise by considering dynamics that are not completely positive.
Moreover, since the models that will be presented represent concrete 
systems amenable to actual experiments, they offer the possibility of direct 
tests of this property.

In Chapter~\ref{ch1}, after introducing the basics of positivity and
complete positivity for generic time-evolutions, we develop the formalism 
leading to the reduced dynamics of open quantum systems; in particular 
the weak coupling limit and the resulting general structure of semigroups.

In Chapter~\ref{ch2} we restrict our attention to open quantum systems
described by two-dimensional Hilbert spaces, for the concrete physical systems
we shall consider can be described in this framework. We introduce the 
coefficients that characterize general irreversible reduced 
dynamics fulfilling the Markov property; in order to guarantee either 
the positivity or the complete positivity request, they must 
satisfy some constraints, expressed by sets of inequalities.

In the next three chapters this formalism is applied to three
specific contexts: the study of irreversible dynamics of neutral {\bf B} mesons
in a gravitational environment (Chapter~\ref{ch3}), 
of a beam of neutrinos crossing a layer of matter (Chapter~\ref{ch4}), and
finally of neutrons in interferometric experiments in presence of stochastic 
magnetic fields (Chapter~\ref{ch5}).

The analysis of a possible dissipative time-evolution for the neutral 
{\bf B} mesons is suggested by general considerations about the behavior
of quantum gravity (or, from a more fundamental level, string theory)
at the Planck scale. The approach is
phenomenological, that is the microscopic details of the interaction between
system and gravitational bath are unspecified; in this sense,
the adopted dissipative time-evolution for the {\bf B} mesons
is completely general. We have explicitly
evaluated the expressions of the decay probabilities of neutral 
mesons for some selected decay channels and, by means of them, some 
asymmetries, that is quantities usually employed to evaluate the violation
of the discrete symmetries (parity, time reversal and charge conjugation). 
These asymmetries can be directly used to fit experimental data. In this 
respect, entangled {\bf B} mesons produced in the decay of spin-$1$
resonances are particularly useful: in such a case, the dissipative
effects induced by the non-standard dynamics produce characteristic
signatures, that, at least in principle, can be experimentally probed.
Indeed, the experiments involving entangled neutral {\bf B} mesons at the
so-called {\bf B}-factories have reached a high level of sensitivity,
very close to the one required for detecting the non-standard, 
dissipative phenomena. This is a further motivation for the
investigations presented here. Moreover, since complete positivity 
is tightly related to the quantum correlations expressed by  
entanglement, the {\bf B}-factories offer the possibility of performing
experimental tests of this fundamental property.

The analysis of neutrino interactions and oscillations in matter
described in Chapter~\ref{ch4} is not purely phenomenological in character,
rather it is based on a microscopic model. The formalism is nevertheless 
very general
and, provided the Markovian approximation is valid, it can be applied to study 
dissipative phenomena in neutrino physics induced by various matter
effects, e.g. the one induced by the thermal fluctuations inside
the Sun. We have computed the modified form of
the transition probability between two different flavors; for thermal
fluctuations in the Sun we find no meaningful changements to the standard
expression, in agreement with previous results coming from different 
approaches.

Finally, in the last chapter, we devote our attention to neutron
interferometers and discuss how these apparata can be used to study
dissipative dynamics, in a controlled environment. The basic idea is to 
insert the interferometer in an external stochastic magnetic field.
By suitably adjusting the parameters characterizing the fluctuating
external field, one can then reproduce the physical situations adopted to
justify widely used dynamics (e.g. of the Redfield-Bloch type).
It turns out that such time-evolutions are not completely positive
and sometimes not even positive; neutron interferometry would then
allow direct experimental tests of the inconsistencies that such dynamics
necessarily produce.

Neutron interferometers have been proposed as one of the possible set-ups 
to study non-contextuality in quantum mechanics, by testing suitable Bell's 
inequalities. In presence of dissipative effects leading to loss of quantum 
coherence, one expects a weakening of the effectiveness of these Bell's
inequality tests. As a byproduct of the previous investigations, we have
studied in detail how the expressions of these inequalities change under
the action of dissipative dynamics.

In this thesis, elementary particle phenomenology and quantum interferometry 
are shown to provide, under specific conditions, very good laboratories
to investigate the properties of reduced dynamics of open systems, in 
particular the notion of complete positivity. The motivations about relevance 
and necessity of complete positivity result strengthened by the arguments 
developed in what follows. 

%% file: chapter1.tex
\chapter{Open Quantum Systems}
\la{ch1}
\markboth{\small{\rm{\thechapter. OPEN QUANTUM SYSTEMS}}}{}

In this chapter we introduce the concept of quantum open systems and 
their reduced time-evolutions.
After some basic definitions, we focus our attention
on the properties of positivity and complete positivity of reduced
time-evolutions, stressing their relevance and the problems that
arise if they are not positive or not completely positive.
Complete positivity shall be the leit-motiv of this thesis, 
since it is usually 
believed that any dynamical map describing the time-evolution of
an open quantum system should enjoy this property in order to avoid any fallacy
in the interpretation of the formalism. However, in the literature about
open quantum systems many examples of not completely positive reduced
dynamics are encountered; we will present some of them to show
the problems they suffer.

In Section~\ref{s1.3} we explicitly show how, by means of the projector
formalism, it is possible to extract the dynamics of a subsystem of
a larger system whose time-evolution is known. The reduced dynamics so
obtained is in general rather complex because it contains memory
effects negligible only when some particular conditions are fulfilled,
typically involving the weakness of the coupling between system and
environment. In view of this, in Section~\ref{s1.4} we
introduce the so-called Markovian dynamics, that is the class of 
time-evolutions in which memory terms are made to 
disappear; they are described 
by linear maps satisfying the semigroup composition law. This is the class of
dynamics we shall consider in the following chapters.

A different approach to the time-evolution of an open quantum
system is presented in Section~\ref{s1.7}, based on a Langevin-type
equation.

In the final part of this chapter some considerations about the long-time
behavior of states evolving under a Markovian dynamics are included.


\s{States, observables and dynamical maps}
\la{s1.1}
\markright{\small{\rm{\thesection. STATES, OBSERVABLES AND DYNAMICAL MAPS}}}

In this section we introduce the basic elements to describe a quantum system:
the states, the observables and their time-evolutions.

\noi
Given a Hilbert space $\mathfrak{H}$, we recall the following definitions:

\begin{Def}
\la{def1.1.1}
{\rm A linear operator $A$ acting on $\mathfrak{H}$ is said to be} bounded 
{\rm iff $\| A \|_{\infty} < \infty$, where 
\begin{equation*}
\| A \|_{\infty} \equiv
\sup_{\| \psi \| \leqslant 1} \| A \psi \|, \quad \psi \in \mathfrak{H}.
\end{equation*}}
\end{Def}

\begin{Def}
\la{def1.1.2}
{\rm A linear operator $A$ acting on $\mathfrak{H}$ is said to be} trace class 
{\rm iff $A \in \cal{B}(\mathfrak{H})$ and $\| A \|_{1} \equiv 
\Tr(A^{\dagger}A)^{1/2} < + \infty$.}
\end{Def}

\noi
We introduce two complex linear spaces of operators acting on 
$\mathfrak{H}$~\cite{take}:

\bi
\item $\cal{B}(\mathfrak{H})$: the complex
linear space of bounded operators, which 
is complete in the operator norm $\| \cdot \|_{\infty}$;
\item $\cal{T}(\mathfrak{H})$: the complex
linear space of trace-class operators, 
which is complete in the trace norm $\| \cdot \|_{1}$.
\ei

\begin{Rem}
\la{rem1.1.1}
{\rm If $\mathfrak{H}$ is finite-dimensional, say
${\rm dim}(\mathfrak{H}) = n$, then ${\cal B}(\mathfrak{H}) = 
{\cal T}(\mathfrak{H}) = M_n({\bf C})$, the algebra of $n \times n$
matrices. If ${\rm dim}(\mathfrak{H}) = \infty$, ${\cal T}(\mathfrak{H})
\subset {\cal B}(\mathfrak{H})$; in particular ${\cal B}(\mathfrak{H})$
is an algebra whereas ${\cal T}(\mathfrak{H})$ is not.}
\end{Rem}

The physical states of quantum systems with finitely many degrees of
freedom can always be represented by positive (then 
Hermitian) trace-class operators with trace one\footnote{Or, as we 
shall shortly see, less than one if the state represents an unstable 
system}, called density matrices or statistical operators:

\be
\la{eq1.1.1}
\rho \in {\cal T}({\mathfrak H}),\quad \rho \geqslant 0,\quad
(\rho^{\dagger} = \rho), \quad \Tr(\rho) = 1.
\ee

They constitute a convex set in ${\cal T}({\mathfrak H})$, denoted by
${\cal P}({\mathfrak H})$, whose extremal 
points are projectors on one-dimensional subspaces; in spectral 
representation, they read

\be
\la{eq1.1.2}
\rho = \sum_{i} \lambda_i \vert e_i \rangle \langle e_i \vert;\quad
\langle e_i \vert e_j \rangle = \delta_{ij},\quad \lambda_i \geqslant 0,\quad
\sum_i \lambda_i = 1.
\ee

\begin{Def}
\la{def1.1.3}
{\rm A state $\rho$ is said to be} pure {\rm if the sum in~(\ref{eq1.1.2}) 
can be reduced to only one term (after a suitable choice of the basis vectors
$\vert e_i \rangle$), otherwise it is said to be} mixed {\rm and it 
represents a statistical convex superposition of pure states. In such a case 
the eigenvalues are the statistical weights of the corresponding 
eigenprojectors in the mixture.}
\end{Def}

\begin{Rem}
\la{rem1.1.0.add}
{\rm Density matrices are not uniquely expressible as convex linear 
combinations of other states. For instance any mixture $\sum_i \mu_i
\sigma_i$ with generic density matrices $\sigma_i$ and $\mu_i > 0$, 
$\sum_i \mu_i = 1$, can always be spectralized and written also
in the form~(\ref{eq1.1.2})}
\end{Rem}

\noi
Pure states are fully characterized as follows.

\begin{Rem}
\la{rem1.1.0}
{\rm A state $\rho \in {\cal T}(\mathfrak{H})$ is pure iff
$\rho^2 = \rho$.}
\end{Rem}

The system observables are represented by Hermitian operators $A = A^{\dagger}$
belonging to ${\cal B}({\mathfrak H})$. The expectation value of  
physical observables $A$ with respect to a state $\rho$, in this formalism, 
is given by the trace operation:

\be
\la{eq1.1.3} 
\langle A \rangle_{\rho} = \Tr(\rho A) = \sum_i \lambda_i \langle e_i
\vert A \vert e_i \rangle.
\ee

In order to describe the time-evolution of quantum systems, two approaches
are possible. In the Schr{\"o}dinger picture a state evolves
as $\rho(0) \rightarrow \rho(t) = \Lambda_t[\rho(0)]$ where $\Lambda_t$
is a linear operator, $\Lambda_t: {\cal P}({\mathfrak H})\rightarrow
{\cal P}({\mathfrak H})$ and $t$ is time; the one-parameter
family of linear maps \{$\Lambda_t, t\in \mathfrak{R}$\} is called 
time-evolution of the system; $\mathfrak{R}$ is the
range over which $t$ can vary and it depends on the characteristics of 
the considered dynamics.


In the Heisenberg picture, we introduce the 
one-parameter family of linear maps \{$\Lambda_t^*, t \in \mathfrak{R}$\},
producing the time-evolution of the observables: $A(0) \rightarrow
A(t) = \Lambda^*_t[A(0)]$. $\Lambda_t^*$ is called the dual map 
to $\Lambda_t$, defined by the duality relation: 

\be
\la{eq1.1.4}
\Tr[\rho(t)A] = \Tr[\rho A(t)].
\ee

\noi
We distinguish the following classes of dynamical maps.

\begin{Def}
\la{def1.1.4}
{\rm A dynamical map is called} reversible {\rm iff
$\Lambda_t$ admits an inverse $\Lambda_t^{-1}$, $\forall t \in \mathfrak{R}$. 
Otherwise it is called} irreversible.
\end{Def}

\begin{Def}
\la{def1.1.5}
{\rm A dynamical map is called} trace-preserving {\rm iff
$\Tr [\rho(t)] = \Tr [\rho(0)]$, $\forall t \in \mathfrak{R}$.}
\end{Def}

For the physical consistency of the formalism, any dynamical map must be 
Hermiticity-preserving, that is $\Lambda_t[\rho]^{\dagger} = \Lambda_t[\rho]$,
$\forall \rho \in {\cal P}({\cal H})$, $\forall t \in \mathfrak{R}$.
Instead, reversibility and trace preservation may be satisfyed or not, 
depending on the physical systems we are dealing with.

In particular, trace-preserving maps describe stable systems
in which the total probability is conserved; on the other hand,
the time-evolution of unstable systems (that is decaying systems) is 
modeled by maps that do not preserve the trace of states: the total
probability has indeed to decrease in time.

The standard quantum dynamics is given by the time-evolution of 
stable closed systems, that in the Schr{\"o}dinger picture reads 
(putting $\hbar = 1$):

\be
\la{eq1.1.5}
\rho(t) = \Lambda_t[\rho(0)] = e^{-i H t}\rho(0) e^{i H t}
\ee

\noi
or, in differential form,

\be
\la{eq1.1.6}
\dot{\rho}(t) = -i [H, \rho(t)],
\ee

\noi
where $H = H^{\dagger}$ is the Hamiltonian describing the system. 
In the following we denote by $L_H[\,\cdot\,] = -i[H, \,\cdot\,]$ the Liouville
operator which generates of the time-evolution maps (or dynamical maps):

\be
\la{eq1.1.7}
\Lambda_t = e^{L_H t} \quad \forall t\in{\bf R}.
\ee

\noi
The same evolution in the Heisenberg picture reads

\be
\la{eq1.1.8}
A(t) = \Lambda_t^*[A(0)] = e^{i H t} A(0) e^{-i H t},
\ee

\noi
which is generated by

\be
\la{eq1.1.9}
\dot{A}(t) = i [H, A(t)],
\ee

\noi
where $A$ is any observable of the system. 

In such a case, the set of 
linear operators $\{\Lambda_t, t \in {\bf R}\}$ (and the set of their 
dual maps $\{\Lambda_t^*, t \in {\bf R}\}$) forms a group of 
transformations. The group structure implies that for each $\Lambda_t$
there is an inverse map $\Lambda_t^{-1} = \Lambda_{-t}$, that is the
dynamics is reversible; moreover the composition law
$\Lambda_{t_1 + t_2} = \Lambda_{t_1}\circ \Lambda_{t_2}$,
$\forall t_1, t_2 \in {\bf R}$ is satisfied, as can be seen 
from~(\ref{eq1.1.7}). The invariance of the trace under
cyclic permutations produces the trace-preserving
property for the dynamical maps. Moreover, it is Hermiticity-preserving
and it maintains the coherence of the states over which it acts:
indeed, from~(\ref{eq1.1.5}), $\rho^2(t) = \rho(t)$. This is another
expression of the fact that Hilbert space vectors $\vert \psi \rangle$ are 
mapped into themselves: $\vert \psi \rangle \rightarrow \vert \psi_t \rangle
 = e^{-i H t} \vert \psi \rangle$.

Another example of dynamical map is the one describing a closed unstable
system. The description of decaying systems is performed introducing
a non-Hermitian Hamiltonian $H$ in the Weisskopf-Wigner form~\cite{nach}, 
that is composed by a Hermitian part and an anti-Hermitian one:

\be
\la{eq1.1.1.add}
H = M - \frac{i}{2} \it{\Gamma}; \quad M = M^{\dagger}, \quad 
\it{\Gamma} = \it{\Gamma}^{\dagger}
\ee

\noi
with positive $M$ and $\it{\Gamma}$ characterizing masses and natural 
widths of decaying states. In this case, in the Schr{\" o}dinger picture,
the states evolution reads 

\be
\la{eq1.1.2.add}
\rho(t) = \Lambda_t[\rho(0)] = e^{-i H t}\rho(0) e^{i H^{\dagger} t}
\ee

\noi
or, in differential form,

\be
\la{eq1.1.3.add}
\dot{\rho}(t) = -i H \rho(t) + i \rho(t) H^{\dagger} = -i[M, \rho(t)] 
- \frac{1}{2}\{\it{\Gamma}, \rho(t)\}
\ee

\noi
leading to the property

\be
\la{eq1.1.4.add}
\frac{d}{dt} \Tr[\rho(t)] = - \Tr[\rho(t) {\it \Gamma}] < 0.
\ee

In the Heisenberg picture the time-evolution of the observables is
described by equations analogous to~(\ref{eq1.1.2.add}) and 
(\ref{eq1.1.3.add}).

In this case the range $\mathfrak{R}$ over which the dynamics makes
sense is not the real axis ${\bf R}$ but rather the positive semi-axis
${\bf R}^+$. Indeed for $t < 0$ equation~(\ref{eq1.1.4.add}) should
imply $\Tr [\rho(t)] > 1$ since $\Tr [\rho(0)] = 1$, and thus the statistical 
interpretation of the eigenvalues of $\rho(t)$ would be compromised.
Since this time-evolution is defined only for positive time, it is 
irreversible; moreover it is
not trace-preserving, as can be seen from~(\ref{eq1.1.4.add}), and it
maintains the coherence of the states, i.e. Hilbert space vectors
transform into Hilbert space vectors, though not normalized, since
$H \ne H^{\dagger}$.

In the study of open quantum systems we will meet irreversible 
time-evolutions either trace-preserving or 
not, and decohering, that is causing the transition from pure to 
mixed states.

We have to discuss the fundamental property of the preservation of
positivity of the states subject to a dynamical map. For its
relevance and its prominent interest we reserve to this issue the next
section.


\s{Positivity and Complete Positivity}
\la{s1.2}
\markright{\small{\rm{\thesection. POSITIVITY AND COMPLETE POSITIVITY}}}

Let us start with the definitions of positivity and complete positivity:

\begin{Def}
\la{def1.2.1}
{\rm A linear transformation 
$\Gamma: {\cal B}({\mathfrak H}) \rightarrow {\cal B}
({\mathfrak H})$ 
is said to be} positive {\rm (or} positivity-preserving{\rm ) iff 
$A \geqslant 0 \Rightarrow \Gamma(A) \geqslant 0$, $\forall A 
\in {\cal B}({\mathfrak H})$.} 
\end{Def}

\begin{Def}
\la{def1.2.2}
{\rm A linear transformation 
$\Gamma: {\cal B}({\mathfrak H}) \rightarrow {\cal B}
({\mathfrak H})$ 
is said to be} $n$-positive {\rm iff $\Gamma \otimes {\bf I}_n: 
{\cal B}({\mathfrak H}\otimes {\mathfrak H}_n) \rightarrow {\cal B}
({\mathfrak H}\otimes {\mathfrak H}_n)$ is positive, with $n \in {\bf N}$ 
and ${\bf I}_n$ the identity on the $n$-dimensional Hilbert space
${\mathfrak H}_n$.}
\end{Def}

\begin{Def}
\la{def1.2.3}
{\rm A linear transformation 
$\Gamma: {\cal B}({\mathfrak H}) \rightarrow {\cal B}
({\mathfrak H})$ 
is said to be} completely positive (CP) {\rm iff it is $n$-positive 
$\forall n \in {\bf N}$.}
\end{Def}

\begin{Rem}
\la{rem1.2.1.add}
{\rm Complete positivity is a stronger property than positivity 
which amounts to $1$-positivity.}
\end{Rem}

\begin{Rem}
\la{rem1.2.1}
{\rm Historically, the notion of complete positivity was introduced for
linear transformations acting on bounded operators, that is, in a
dynamical context, not for the time-evolution of states, $\Lambda_t$,
but for its dual $\Lambda_t^*$, the time-evolution of operators.
In this thesis, the notion of complete positivity will be referred to
dynamical maps on states in the sense that a map 
$\Lambda_t: {\cal P}(\mathfrak{H}) \rightarrow 
{\cal P}(\mathfrak{H})$ is CP if and only if its dual map 
$\Lambda_t^*: {\cal B}(\mathfrak{H}) \rightarrow 
{\cal B}(\mathfrak{H})$ is CP.}
\end{Rem}

In the context of dynamical maps describing generic time-evolutions
of states, the property of positivity is required in order to preserve at any
time the positivity of the eigenvalues of the statistical operator
representing the system. As previously stated, these eigenvalues
represent the weights of a statistical superposition; then they must 
be positive in order to guarantee the correct probabilistic interpretation
of the formalism. For any non-positive time-evolution $\Lambda_t$
there necessarily exists a positive initial state $\rho(0)$ mapped
by $\Lambda_t$ in a non-positive one, $\rho(t) = \Lambda_t[\rho(0)]$,
leading to physical inconsistency of the formalism.

It is less intuitive to understand the relevance and necessity
of complete positivity for time-evolutions of states. Let $\Lambda_t$
be a generic dynamical map describing the time-evolution of a state
$\rho_S$; following the definition of CP, we have to consider the map
$\Lambda_t \otimes {\bf I}_n$, where ${\bf I}_n$ is the identity
in a $n$-dimensional Hilbert space and $n$ is any natural number.
This map represents the factorized time-evolution of a composite
system $T = S + S_n$, where $S$ is the system described by $\rho_S$
and evolving under $\Lambda_t$ while $S_n$ is a generic $n$-level 
system characterized by a trivial Hamiltonian $H_n = 0$. 
Since the dynamics of the composite
system has the factorized form $\Lambda_t \otimes {\bf I}_n$ there
is no dynamical coupling between $S$ and $S_n$. Now, the positivity
of any state $\rho_T$, describing the composite system $T$, must
be preserved under the action of $\Lambda_t \otimes {\bf I}_n$, but
this is not guaranteed by the positivity of $\Lambda_t$: a completely
positive $\Lambda_t$ is needed.

To clarify this point, we define a fundamental property of the state
describing the composite system $T$, tightly connected to the notion of
complete positivity.

\begin{Def}
\la{def1.2.4}
{\rm Given a state $\rho_T$ describing a composite system $T = S_1 + S_2$, 
it is called} separable {\rm if it can be written as $\rho_T = \sum_{i,j}
c_{ij} \rho_{S_1}^i \otimes \rho_{S_2}^j$ with $c_{ij} > 0$,
$\forall i, j$; otherwise it is called} entangled.
\end{Def}

It turns out that the request of complete positivity is necessary
when the evolution of entangled states $\rho_T$ is taken into account.
Indeed, any factorized statistical operator $\rho_T$ of the bipartite system
$T = S + S_n$, evolving under the time-evolution $\Lambda_t \otimes 
{\bf I}_n$, where $\Lambda_t$ is positive-definite, is mapped into a positive
operator. In fact, 

\be
\la{eq1.2.rem2.add}
\rho_T(0) = \sum_{i,j} c_{ij} \rho^i_S(0) \otimes
\rho^j_{S_n}(0)
\ee

\noi
with $c_{ij} > 0$, so that, since $\Lambda_t \geqslant 0$,

\be
\la{eq1.2.rem2}
\rho_T(t) = (\Lambda_t \otimes {\bf I}_n)[\rho_T(0)] = \sum_{i,j}
c_{ij}\Lambda_t[\rho^i_S(0)] \otimes \rho^j_{S_n}(0) \geqslant 0.
\ee

Then it is the presence of quantum entanglement that asks for complete 
positivity, that is only entangled states of $S + S_n$ can lose their
positivity under the dynamical map $\Lambda_t \otimes {\bf I}_n$ when
$\Lambda_t$ is not completely positive. We shall shortly come back to
this point after discussing the structural characterization
of completely positive maps.

The following theorem is due to Kraus~\cite{krau} who used a result 
by Stinespring~\cite{stin}, obtained in the general framework of 
completely positive maps on $C^*$-algebras.

\begin{The}
\la{the1.2.1}
{\rm (Kraus, Stinespring) A linear transformation 
$\Gamma: {\cal B}({\mathfrak H}) \rightarrow 
{\cal B}({\mathfrak H})$ is CP iff $\Gamma (B) = \sum_{k \in K} 
A^{\dagger}_k B A_k$, $\forall B \in {\cal B}({\mathfrak H})$, 
where $K$ is a finite or countably infinite set; the operators 
$A_k \in {\cal B}({\mathfrak H})$ satisfy 
$\sum_{k \in K} A^{\dagger}_k A_k \leqslant {\bf I}$. The set 
$\{A_k, k \in K\}$ is not uniquely determined by $\Gamma$.}
\end{The}

\begin{Cor}
\la{cor1.2.1}
{\rm A dynamical map $\Lambda_t: {\cal P}({\mathfrak H}) \rightarrow
{\cal P}({\mathfrak H})$ is CP iff 
$\Lambda_t [\rho] = \sum_{k \in K} A_k \rho A^{\dagger}_k$,
$\forall \rho \in {\cal P}({\mathfrak H})$, with
$\sum_{k \in K} A^{\dagger}_k A_k \leqslant {\bf I}$.}
\end{Cor}

\begin{Cor}
\la{cor1.2.2}
{\rm For a trace-preserving completely positive map the operators
$A_k$ in Theorem~\ref{the1.2.1} must satisfy $\sum_{k \in K} A^{\dagger}_k 
A_k = {\bf I}$.}
\end{Cor}

When $\mathfrak{H}_S$ is finite-dimensional, say ${\rm dim}(\mathfrak{H}_S) 
= d$, one doesn't need to check all the maps $\Lambda_t \otimes {\bf I}_n$, 
but can use the following result.

\begin{The}
\la{the1.2.2}
{\rm (Choi~\cite{choi,choi2}) 
Given a finite $d$-dimensional Hilbert space ${\mathfrak H}_d$, a linear
transformation $\Gamma :{\cal B}({\mathfrak H}_d) \rightarrow 
{\cal B}({\mathfrak H}_d)$ is CP iff it is $d$-positive.}
\end{The}

\begin{Rem}
\la{rem.add.1}
{\rm Note that the standard evolutions for systems, 
either stable or unstable (equations~(\ref{eq1.1.5}) and (\ref{eq1.1.2.add})), 
are in the form of Corollary~\ref{cor1.2.1}, thus they are completely positive.}
\end{Rem}

\begin{Rem}
\la{rem.add.2}
{\rm A simple example of a positive but not completely positive map is given 
by the transposition in a two-dimensional Hilbert space,
$\tau: M_2({\bf C}) \rightarrow M_2({\bf C})$. Positivity holds for
$\tau [\rho] \geqslant 0$, $\forall \rho \in {\cal P}({\cal H}_2)$;
to see that complete positivity does not hold, consider the singlet state
$\rho = \vert \psi \rangle \langle \psi \vert$ where
$\vert \psi \rangle = (\vert e_1 \rangle \otimes \vert e_2 \rangle -
\vert e_2 \rangle \otimes \vert e_1 \rangle)/ \sqrt{2}$. $\rho$ is 
represented by the $4\times 4$ positive matrix:
\be
\la{eq.rem.add.1}
\rho = \frac{1}{2}\left( \begin{array}{cccc}
0 & 0 & 0 & 0 \\
0 & 1 & -1 & 0 \\
0 & -1 & 1 & 0 \\
0 & 0 & 0 & 0
\end{array} \right).
\ee
After explicit computation, we find that $\rho_{\tau} \equiv {\tau}
\otimes{\bf I}_2$ is given by
\be
\la{eq.rem.add.2}
\rho_{\tau} = \frac{1}{2}\left( \begin{array}{cccc}
0 & 0 & 0 & -1 \\
0 & 1 & 0 & 0 \\
0 & 0 & 1 & 0 \\
-1 & 0 & 0 & 0
\end{array} \right),
\ee
that is no longer positive, since one of its eigenvalues is $-1/2$. 
Consequently, $\tau$ is not completely positive.}
\end{Rem}

Using Theorem~\ref{the1.2.2}, we stress the tight connection between 
complete positivity and quantum entanglement.

\begin{Pro}
\la{pro1.2.1}
{\rm If $\Lambda_t$ is not completely positive, there exists an entangled
state $\rho_T$ in $S + S_d$ whose positivity is not preserved by 
$\Lambda_t \otimes {\bf I}_d$.

\noi
{\it Proof}: If $\Lambda_t$ is not completely 
positive, the same holds for its dual $\Lambda_t^*$. Then,
because of Theorem~\ref{the1.2.2}, there must be a positive
bounded operator $A \in M_d({\bf C}) \otimes M_d({\bf C})$ such that 
$(\Lambda^*_t \otimes {\bf I}_d)[A]$ has a negative eigenvalue. Let
$\rho_T$ be the corresponding eigenprojection, then
\be
\la{eq1.2.rem4}
\Tr \left[ \rho_T (\Lambda_t^* \otimes {\bf I}_d)[A] \right] = 
\Tr \left[ (\Lambda_t \otimes {\bf I}_d)[\rho_T] A \right] \leqslant 0.
\ee
Thus, $(\Lambda_t \otimes {\bf I}_d)[\rho_T]$ must have negative 
eigenvalues since $A \geqslant 0$.}
\end{Pro}

To summarize, positivity of a dynamical map $\Lambda_t$ guarantees that
the eigenvalues of any state $\rho_S$ evolving under $\Lambda_t$ remain 
positive at any time. Complete positivity of the dynamical map $\Lambda_t$
guarantees that the eigenvalues of any entangled state $\rho_T$ of $S + S_n$
remain positive at any time, where $S_n$ is an arbitrary $n$-level system,
possibly very far from $S$, and the coupling between $S$ and $S_n$ 
is statistical, not dynamical.

Because of the abstractness of the definition of CP, the necessity of
positivity of maps of the form $\Lambda_t \otimes {\bf I}_n$, and thus
of complete positivity of $\Lambda_t$, is often refused as a technical 
artifact rather than a physical constraint. 

In fact, the motivations for considering the system of interest $S$
jointly with an arbitrary $n$-level system $S_n$, 
not evolving and without interaction with $S$, are not obvious. All the more
so, since this coupling should motivate the special form that, according
to Corollary~\ref{cor1.2.1}, the time-evolution would inherit.

In Section~\ref{s1.6} we will show that complete positivity
of $\Lambda_t$ is a necessary and sufficient condition to preserve the 
positivity of factorized maps in the form $\Lambda_t \otimes \Lambda_t$. 
This gives the opportunity to study the notion of CP in the treatment of 
the dissipative evolution of entangled pairs of neutral mesons, given in 
Chapter~\ref{ch2}. More in general, time-evolutions of the form
$\Lambda_t \otimes \Lambda_t$ are definitely physically plausible and less
abstract than $\Lambda_t \otimes {\bf I}_n$.

Nevertheless, in 
Chapter~\ref{ch5} we shall deal with a physical setup reproducing
a factorized dynamics in the form $\Lambda_t \otimes {\bf I}_n$ and
involving entanglement between the degrees of freedom of two 
subsystems: then, it will be possible to show explicitly, even with
$\Lambda_t \otimes {\bf I}_n$, that lack of complete positivity of $\Lambda_t$
produces necessarily physically concrete inconsistencies.


\s{Reduced dynamics: the Generalized Master Equation}
\la{s1.3}
\markright{\small{\rm{\thesection. REDUCED DYNAMICS: THE GENERALIZED MASTER EQUATION}}}

In the previous sections we have described the general features 
characterizing the time-evolutions of quantum systems $\rho \in {\cal P}
({\cal H})$. 
Now, we proceed to describe the time-evolution equations
for open systems.

We denote the system of interest by $S$, the environment by $B$ and
assume that the Hamiltonian $H_T$ of the compound system $T = S + B$ is given.
Such a Hamiltonian consists of contributions $H_S$ and $H_B$ from system
and environment alone, plus an interaction term $H_I$ : 

\be
\la{eq1.3.1.add}
H_T = H_S + H_B + H_I.
\ee

\noi
$H_I$ is usually assumed to be of the linear form

\be
\la{eq1.3.1.addd}
H_I = \sum_j V_S^j \otimes V_B^j,
\ee

\noi
with $V_S^j$, $V_B^j$ Hermitian operators of $S$, respectively $B$.

We want to describe states and observables belonging to $S$ only (because
of lack of control over all the degrees of freedom of $B$). 
The formalism we follow in order to obtain the 
dynamics of $S$ when the dynamics of $T$ is known, consists in the 
projector technique developed independently by Nakajima~\cite{naka} and 
Zwanzig~\cite{zwan} (see~\cite{haak} for a review). There are also different 
approaches~\cite{pech,roye,rome,suar,gnut,yu,gasp} 
to describe the time-evolution of open systems; in 
Section~\ref{s1.7} we shall present a stochastic approach~\cite{budi}.

The Hilbert spaces characterizing $S$, $B$ and $T$ are such that
${\mathfrak H}_T = {\mathfrak H}_S \otimes {\mathfrak H}_B$. We are 
interested in observables $A \in {\cal B}({\mathfrak H}_S)$, whose
expectation values are:

\be
\la{eq1.3.1}
\langle A \rangle _S = \Tr (A \rho_T) = \Tr_S \Tr_B 
[(A\otimes {\bf I}_B) \rho_T] = \Tr_S (A \rho_S),
\ee

\noi
where the statistical operator of $S$ is obtained from the one of $T$ 
by the partial trace over the degrees of freedom of $B$:

\be
\la{eq1.3.1.bis}
\rho_S = \Tr_B (\rho_T).
\ee 

We are interested in reduced dynamics of the system $S$ alone, that is 
in linear maps $\Lambda_t$ describing the time-evolution $\rho_S(0) 
\rightarrow \Lambda_t [\rho_S(0)]$. In full generality, the state of $S$
at time $t$, denoted by $\rho_S(t)$, is obtained from the unitary evolution
of $T = S + B$, $U_t = e^{-iH_T t}$, as follows:

\be
\la{finita!}
\rho_S(t) = \Tr_B[U_t \rho_T(0) U_t^{\dagger}].
\ee

However, $\rho_S(t)$ is not always expressible as $\Lambda_t [\rho_S(0)]$,
because there is no general prescription for extracting $\rho_S(0)$ from
the initial global state $\rho_T(0)$. Indeed, there can be two different 
initial global states $\rho_T^a(0) \ne \rho_T^b(0)$ that lead to the same 
reduced state of $S$, $\rho_S^a(0) = \rho_S^b(0)$, and neverthless at time $t$
$\rho_S^a(t) \ne \rho_S^b(t)$. In this case, the dynamics $\Lambda_t$ 
is not properly defined because a state of $S$ at time $t = 0$ would evolve 
in two distinct states at time $t$. It is thus necessary to define an 
assignment map 
$\Phi:{\cal P}(\mathfrak{H}_S) \rightarrow {\cal P}(\mathfrak{H}_T)$
giving, for each initial state $\rho_S(0)$, a state of the total system 
$T$ such that
\be
\la{eq1.3.8.add}
\rho_S(t) = \Lambda_t[\rho_S(0)] = 
\Tr_B(U_t \Phi[\rho_S(0)] U_t^{\dagger}),
\ee
and the diagram in Figure~\ref{fig1.3.1} commute. Following this 
procedure, it is possible to define the reduced dynamics $\Lambda_t$.
The conditions usually imposed on the map $\Phi$ are: 1. $\Phi$ preserves 
mixtures: $\Phi \Bigl[ \sum_i \lambda_i \rho_S^i \Bigr] = \sum_i 
\lambda_i \Phi [\rho_S^i]$; 2. $\Phi$ is consistent: $\Tr_B(\Phi[\rho_S]) 
= \rho_S$; 3. $\Phi[\rho_S]$
is positive for any positive $\rho_S$. It has been shown that, for a
two-dimensional Hilbert space $\mathfrak{H}_S$, the only assignment 
maps satisfying these three conditions for all initial density matrices 
are the product maps $\Phi[\rho_S] = \rho_S \otimes \rho_B$, where 
$\rho_B$  is a fixed reference state in the bath~\cite{pech}. This
factorized state is usually justified in the context of a weak coupling 
between $S$ and $B$; however there are physical situations in which such 
a factorized initial state cannot be used~\cite{roye}.

\begin{Rem}
\la{rem1.3.3}
{\rm Since $\Lambda_t$ in~(\ref{eq1.3.8.add}) is obtained by 
the composition of three maps, its properties depends on those 
of the trace over the environment degrees of freedom,
of the standard evolution and of the assignment map $\Phi$.
If the choice $\Phi[\rho_S] = \rho_S \otimes \rho_B$ is assumed,
the reduced dynamics $\Lambda_t$ is completely positive since it is
the composition of maps that are CP. However, in some cases the initial 
state has not the form of a product state, namely, when there exist 
initial correlations between $S$ and $B$~\cite{pech,roye}. 
In~\cite{pech} an argument against the complete
positivity of a reduced dynamics has been discussed. It has been argued 
that, with correlated initial conditions, dynamical maps $\Lambda_t$ can be 
properly defined only on a subset of initial system states and the linear 
extension of these maps on all possible initial states may not be physically 
realizable and may not even be positive. In this case the condition 3 
is not fulfilled; an example of this procedure is
given by the assignment map proposed in~\cite{rome}. In literature many
examples of not completely positive dynamical maps are 
encountered~\cite{suar,gnut,yu,gasp,1gasp};
however, in~\cite{alic2} it has been shown that it is possible in any case 
to preserve the complete positivity of $\Lambda_t$, giving up either the 
condition 1 or 2; it has also been stressed that,
beyond the weak coupling regime, there exists no unique definition of the
reduced dynamics $\Lambda_t$ for the system $S$.}
\end{Rem}

\begin{figure}
\la{fig1.3.1}
\begin{equation*}
\begin{CD}
\rho_S(0) \in {\cal P}({\cal H}_S) @>{\Lambda_t}>> \rho_S(t)  
\in {\cal P}({\cal H}_S) \\
@VV{\Phi}V @AA{\Tr_B}A \\
\Phi[\rho_S(0)] \in {\cal P}({\cal H}_T) @>>> U_t\Phi[\rho_S(0)]U_t^{\dagger}
\in {\cal P}({\cal H}_T) 
\end{CD}
\end{equation*}
\caption{\footnotesize The procedure required in order to define consistently
the reduced dynamics $\Lambda_t$ acting on the states of the system $S$
using the projector technique.}
\end{figure}

Let us now present the Nakajima-Zwanzig procedure. We assume
the dynamics of the total system $T$ to be given; furthermore, to be as 
general as possible, we do not consider the latter to be necessarily closed.
Namely, we consider, for the states $\rho_T$ of $T$, time-evolution equations
of the form

\be
\la{eq1.3.2}
\dot{\rho}_T(t) = L_T[\rho_T(t)],
\ee

\noi
where $L_T = L_S + L_B + L_I$ is the
generator of the time-evolution of $\rho_T$; only when $T$ is a closed system, 
$L_T$ is proportional to the commutator with the total Hamiltonian 
$H_T = H_S + H_B + H_I$, see equation~(\ref{eq1.1.6}). 

We define the two projectors $P, Q: {\cal P}({\mathfrak H}_T)\rightarrow 
{\cal P}({\mathfrak H}_T)$ that map states of $T$ into themselves as
follows:

\be
\la{eq1.3.3}
P[\rho_T]\equiv \Tr_B (\rho_T) \otimes \rho_B = \rho_S \otimes \rho_B; 
\quad Q = {\bf I} - P.
\ee

They depend on a normalized reference state $\rho_B$ of the bath, 
$\Tr_B({\rho_B}) = 1$, that one usually choose to be invariant under the 
dynamics of $B$ : $L_B [\rho_B] = 0$ . The following manipulations are 
independent from the choice of this state. Inserting in 
equation~(\ref{eq1.3.2}) $\rho_T(t) = P[\rho_T(t)] + Q[\rho_T(t)]$, we obtain

\be
\la{eq1.3.4}
\left\{
\begin{array}{l}
P [\dot{\rho}_T(t)] = P L_T P [\rho_T(t)] + 
P L_T Q [\rho_T(t)], \\ \\
Q [\dot{\rho}_T(t)] = Q L_T P [\rho_T(t)] + 
Q L_T Q [\rho_T(t)].
\end{array}
\right.
\ee

\noi
Formally, solving the second of these equations, one gets

\be
\la{eq1.3.5}
Q [\rho_T(t)] = e^{Q L_T t}Q [\rho_T(0)] + \int_0^t dt^{\prime}
e^{Q L_T t^{\prime}}Q L_T P [\rho_T(t - t^{\prime})].
\ee

\noi
Then upon inserting~(\ref{eq1.3.5}) in the first equation 
in~(\ref{eq1.3.4}), we find a closed
equation for the relevant part $P [\rho_T(t)]$; after a partial trace over
$B$ we are left with the equation of motion of $S$ alone, a so-called
generalized master equation (GME):

\be
\la{eq1.3.6}
\dot{\rho}_S(t) = L^{eff}_S [\rho_S(t)] + \int_0^t dt^{\prime}
{\cal K}(t^{\prime}) [\rho_S(t - t^{\prime})] + {\cal I}(t)
\ee

\noi
with

\be
\la{eq1.3.7}
\begin{split}
&L^{eff}_S \equiv L_S + \Tr_B (L_I [\rho_B]); \\
&{\cal K}(t)[\rho_S] \equiv \Tr_B \Bigl(L_I e^{Q L_T t}Q L_T 
[\rho_S \otimes \rho_B]\Bigr); \\
&{\cal I}(t) \equiv \Tr_B \Bigl(L_I e^{Q L_T t}Q [\rho_T(0)]\Bigr).
\end{split}
\ee

\noi
These expressions have been obtained using that
$P L_{S,B}Q = 0$.

The GME is the equation we were looking for; it describes in general 
the behavior of an open system $S$ because no assumptions have been made
about the nature of the system and of the bath, and the interaction is
completely arbitrary; we have only assumed the existence of a reference
bath-equilibrium state $\rho_B$ in ${\cal P}({\mathfrak H}_B)$.
Notice that in absence of interaction
between $S$ and $B$ (i.e. if $L_I = 0$) this equation reduces to the 
time-evolution for $S$ isolated from $B$, that is 
$\dot{\rho}_s(t) = L_S [\rho_S(t)]$; in fact
$L^{eff}_S = L_S$, ${\cal K}(t) = 0$ and ${\cal I}(t) = 0$.

Let us now describe the various terms appearing in the GME. 
$L^{eff}_S$ generates the effective time-evolution 
in $\rho_S$, redefined by $\Tr_B(L_I[\rho_B])$ which
represents a mean field term. ${\cal K}(t)$ is the kernel of the
convolution term, that is the memory term depending on the whole history
of $\rho_S$. 
The inhomogeneity ${\cal I}(t)$ is related to initial conditions: 
if they are factorized, $\rho_T(0) = \rho_S(0)\otimes\rho_B$, then 
$Q\rho_T(0) = 0$ and consequently ${\cal I}(t) = 0$. In the following we 
will assume this condition to be fulfilled. 

\noi
The reduced dynamics 
$\Lambda_t$ is well defined and it is completely positive.

If $T$ is a closed system it is possible to simplify~(\ref{eq1.3.6})
and (\ref{eq1.3.7}). In such a case the operators $L$ are 
proportional to the
commutators with the Hamiltonians and the following relations hold:
$[P, L_S] = 0$; $P L_B = L_B P = 0$. Then the GME becomes:

\be
\la{eq1.3.8}
\begin{split}
&\dot{\rho}_S(t) = L^{eff}_S [\rho_S(t)] + \int_0^t
dt^{\prime} {\cal K}(t^{\prime})[\rho_S(t - t^{\prime})]; \\
&L^{eff}_S[\rho_S] = -i [H_S + \Tr_B(H_I \rho_B),\rho_S], \\
&{\cal K}(t)[\rho_S] = \Tr_B \Bigl( L_I e^{QL_TQ t}QL_I 
[\rho_S \otimes \rho_B] \Bigr).
\end{split}
\ee

Let us focus our attention on cases in which the interaction between
$S$ and $B$ is weak. We introduce a small dimensionless parameter
$\varepsilon$ in the interaction Hamiltonian; consequently
$H_T = H_S + H_B + \varepsilon H_I$. In this way we can write a 
perturbative expansion in $\varepsilon$ for the kernel ${\cal K}(t)$
in equation~(\ref{eq1.3.8})~\cite{frig}:

\be
\la{eq1.3.9}
{\cal K}(t) = e^{L_S t}\Bigl[ {\cal K}_0(t) + \sum_{n = 1}^{\infty}
\varepsilon^n \int dt_1 \ldots dt_n {\cal K}_n(t \vert t_1 \ldots t_n)\Bigr],
\ee

\noi
with $0\leqslant t_n \leqslant \ldots \leqslant t_1 \leqslant t$, and

\be
\la{eq1.3.10}
\begin{split}
&{\cal K}_0(t)[\rho_S] = \Tr_B (L_I^i(t) Q L_I^i[\rho_S \otimes \rho_B]), \\
&{\cal K}_n(t\vert t_1 \ldots t_n)[\rho_S] = \Tr_B(L_I^i(t) Q L_I^i(t_1) Q 
\ldots Q L_I^i(t_n)Q L_I^i [\rho_S \otimes \rho_B])
\end{split}
\ee

\noi
where 

\be
\la{eq1.3.10.bis}
L_I^i(t_j)[\,\cdot\,] = -i[H_I^i(t_j), \,\cdot\,] = 
-i[e^{i(H_S + H_B)t_j}H_Ie^{-i(H_S + H_B)t_j},\,\cdot\,] 
\ee

\noi
is the Liouville 
generator in the interaction picture. ${\cal K}_n(t\vert t_1 \ldots t_n)$
contains the multi-time correlation function of the bath

\be
\la{eq1.3.10.tris}
h_{j,j_1 \ldots j_{n+1}}(t, t_1, \ldots t_n) = \Tr_B(e^{iH_B t}V^{j}_B 
e^{-iH_B t} e^{iH_B t_1}V^{j_1}_B e^{-iH_B t_1}\ldots V^{j_{n+1}}_B \rho_B),
\ee

\noi
where, as alredy mentioned in~(\ref{eq1.3.1.addd}), we have assumed 
the interaction Hamiltonian to be $H_I = \sum_j V^j_S \otimes V^j_B$.

Usually, one proceeds with a
Born approximation of the GME, namely with neglecting in the kernel
expansion~(\ref{eq1.3.9}) all the terms except ${\cal K}_0(t)$; in this
case only two-time correlation functions appear.


\s{The Markovian approximation. The Weak Coupling Limit}
\la{s1.4}
\markright{\small{\rm{\thesection. THE MARKOVIAN APPROXIMATION. THE WEAK COUPLING LIMIT}}}

As we have seen, the GME obtained in the previous section, 
eq.~(\ref{eq1.3.8}) (or, more in general, eq.~(\ref{eq1.3.6}))
contains a memory term that depends on the kernel ${\cal K}(t)$.
If ${\cal K}(t)$ decays rather fastly, it is possible to
extract the density matrix from the convolution integral, then obtaining
an equation without memory term~\cite{alic1,spoh,gori1}; in such a case it is 
possible to introduce a time-independent generator of the reduced dynamics.

\begin{Def}
\la{def1.4.1}
{\rm The dynamical map \{$\Lambda_t, t \in \mathfrak{R}$\} 
is said to be} Markovian {\rm iff there exists a linear bounded operator $L$
on the states space ${\cal P}({\cal H})$ which is time-independent, 
such that $\Lambda_t = e^{Lt} \; \; \forall t 
\in \mathfrak{R}$; $L$ is the generator of the dynamical map.}
\end{Def}

We want to extract a Markovian time-evolution from the GME presented
in the previous section: in this way we obtain a more manageable
time-evolution for the system considered.

\begin{Rem}
\la{rem1.4.1}
{\rm A Markovian time-evolution \{$\Lambda_t, t\in \mathfrak{R}$\} 
satisfies the semigroup 
composition law $\Lambda_{t_1 + t_2} = \Lambda_{t_1} \circ \Lambda_{t_2}$,
$\forall t_1,t_2 \in \mathfrak{R}$, as can be seen from 
Definition~\ref{def1.4.1}.} 
\end{Rem}

As alredy stated, in order to perform a Markovian approximation of the 
dynamics~(\ref{eq1.3.6}) or (\ref{eq1.3.8}) we need the decay time of the 
kernel ${\cal K}(t)$ to be much shorter than the typical evolution time
of the system $S$. We find this situation when the interaction Hamiltonian
can be parametrized by a suitable coefficient $\varepsilon$ which is then
let to vanish: $H_I \rightarrow \varepsilon H_I$. 
The total Hamiltonian is thus scaled as $H_T \rightarrow H_T^{\varepsilon}$:

\be
\la{eq1.4.5}
H_T^{\varepsilon} = H_S + H_B + \varepsilon H_I,
\ee

\noi 
and the corresponding dynamical map is denoted by $\Lambda_t^{\varepsilon}$:

\be
\la{eq1.4.1}
\Lambda_t^{\varepsilon}[\rho_S(0)] = \Tr_B(U_t^{\varepsilon}
\rho_S(0)\otimes\rho_B U_t^{\varepsilon \dagger}),
\ee

\noi
where $U_t^{\varepsilon} = e^{-i H_T^{\varepsilon} t}$. 

As we shall shortly see, for small $\varepsilon$ the reduced dynamics 
$\Lambda^{\varepsilon}_t$ becomes Markovian~\cite{dumc} over times scaling
as $\varepsilon^{-2}$: this procedure is called Weak Coupling Limit.

Since the memory term is of order $\varepsilon^2$ (being quadratic in 
the interaction Hamiltonian), the evolution has to be studied over a 
time-scale of order $\varepsilon^{-2}$. It is then appropriate to
rescale the time as $t^{\varepsilon} = \varepsilon^{-2} t$; in other words, 
the evolution is observed over a time long enough so that the memory 
effects can be neglected. On this time scale the convolution term in the 
GME heuristically behaves as:

\be
\la{eq1.4.2}
\int_0^{t^{\varepsilon}} {\cal K}(t^{\prime})
[\rho_S(t^{\varepsilon} - t^{\prime})] dt^{\prime} 
\approx \left( \int_0^{+\infty} {\cal K}
(t^{\prime}) dt^{\prime} \right) [\rho_S(t^{\varepsilon})] 
\equiv K [\rho_S(t^{\varepsilon})]
\ee

\noi
and thus the GME becomes

\be
\la{eq1.4.3}
\dot{\rho}_S(t) = L^{eff}_S [\rho_S(t)] + \varepsilon^2
\int_0^t {\cal K}(t^{\prime}) [\rho_S(t - t^{\prime})] dt^{\prime}
\approx (L^{eff}_S + \varepsilon^2 K)[\rho_S(t)],
\ee

\noi
whose solution is

\be
\la{eq1.4.4}
\rho_S(t) = \Lambda^{\varepsilon}_t[\rho_S(0)] =
e^{(L^{eff}_S + \varepsilon^2 K)t}[\rho_S(0)].
\ee

Thus $L = L^{eff}_S + \varepsilon^2 K$ is the time-independent generator,
and the time-evolution $\Lambda^{\varepsilon}_t$ is Markovian. 
This is the basic idea
underlying any Markovian approximation of the generalized master equation;
there are however many ways to perform the Markov limit, leading to
Markovian reduced dynamics with different properties~\cite{dumc}.

\begin{Rem}
\la{rem1.4.1.add}
{\rm The prescription~(\ref{eq1.4.2}) generally breaks the complete positivity 
and even positivity of the reduced dynamics~\cite{dumc}}
\end{Rem} 


The convergence condition of the weak coupling limit depends
on the time decay of the correlations of the bath. If we suppose
$H_I = \sum_j V^j_S \otimes V^j_B$ this condition reads

\be
\la{eq1.4.7}
\int_0^{\infty} dt \vert h_{jk}(t) \vert (1 + t)^{\delta} < \infty, \;\;
\forall \delta > 0,
\ee

\noi
where 

\be
\la{eq1.4.7.ad}
h_{jk}(t) = \Tr_B \Bigl(e^{iH_B t}V^j_B e^{-iH_B t} V^k_B \rho_B\Bigr).
\ee

\noi
The condition~(\ref{eq1.4.7}) can be fulfilled only if the reservoir $B$ is 
infinite~\cite{spoh,gori1}.

The Markovian approximations of the GME usually does not preserve 
the positivity of the states over which they act~\cite{dumc}.
The following theorem defines Markovian approximations that do not
suffer from lack of positivity. 
Noticeably, they turn out to be not only positive but also completely 
positive.

\begin{The}
\la{the1.4.1}
{\rm (Davies~\cite{davi2})
Suppose the condition~(\ref{eq1.4.7}) is valid and the spectrum of $H_S$ 
is discrete, then

\be
\la{eq1.4.8}
\lim_{\varepsilon \rightarrow 0} \sup_{0 \leqslant \varepsilon^2 t 
\leqslant \tau} \| \Lambda_t^{\varepsilon} [\rho_S(0)] -
e^{(L^{eff}_S + \varepsilon^2 K^{\natural})t}[\rho_S(0)]\|_1 
= 0,
\ee

\noi
where $\| \cdot \|_1$ is the norm introduced in Definition~\ref{def1.1.2} and

\be
\la{eq1.4.9}
K^{\natural}[\rho_S] \equiv \lim_{T\rightarrow \infty}
\frac{1}{2T} \int_{-T}^T dt e^{-L_S t} K e^{L_S t}[\rho_S]
\ee

\noi
with

\be
\la{eq1.4.10}
 K[\rho_S] \equiv \int_0^{\infty} dt^{\prime} e^{-L_S t^{\prime}}
\Tr_B(L_I e^{(L_S + L_B)t^{\prime}}L_I[\rho_B])[\rho_S].
\ee

\noi
The Markovian evolution defined by~(\ref{eq1.4.8}) is completely positive.}
\end{The}

In~(\ref{eq1.4.8}) we have to consider the averaged evolutor $K^{\natural}$ 
rather than $K$ since the free motion and the dissipation have different time 
scales. 





Another kind of Markovian approximation is the so called Singular Coupling 
Limit (or Singular Reservoir Limit), introduced in~\cite{hepp};
in this case we do not rescale the Hamiltonian $H_T$ but the two-time
correlation functions of the environment: 
$h_{jk}(t) \rightarrow h_{jk}^{\varepsilon}(t)$, 
with $h_{jk}^{\varepsilon}\rightarrow c_{jk} \delta (t)$ as 
$\varepsilon \rightarrow 0$. It has been shown in~\cite{palm}
that such a case is equivalent to a weak coupling limit
with a rescaled Hamiltonian in the form

\be
\la{eq1.4.6}
H_T^{\varepsilon} = \varepsilon^2 H_S + H_B + \varepsilon H_I,
\ee

\noi
where the free Hamiltonian $H_S$ is scaled by a factor $\varepsilon^2$
and then the uncoupled and dissipative evolution are of the same order.

\begin{Rem}
\la{rem1.4.2}
{\rm The singular coupling limit is particularly relevant in the study
of Markovian quantum open systems. Indeed, it is possible to show that 
any family of completely positive maps satisfying the semigroup composition 
law, describing the dynamics of a $n$-level system, can be derived by a 
singular coupling~\cite{gori2}.}
\end{Rem}

\begin{Rem}
\la{rem1.4.3}
{\rm Another relevant property of the singular coupling
is that, in this limit, the correlation functions of order higher than 2 
vanish, that is the generalized master equation reduces to its Born 
approximation, and it is Markovian~\cite{frig}.}
\end{Rem}

Historically the Markovian approximations have been studied considering
two typical times, $\tau_S$ and $\tau_B$, characterizing the system
and the bath respectively; the density matrix $\rho_S$ can be extracted
from the integral kernel ${\cal K}(t)$ under the hypothesis $\tau_S/ \tau_B
\rightarrow \infty$. There are two possibilities to realize this
condition: either $\tau_S\rightarrow \infty$ and $\tau_B$ remains constant,
or $\tau_B\rightarrow 0$ while $\tau_S$ remains constant~\cite{gori1}. 
The two alternatives correspond to weak coupling and singular coupling 
respectively.




\s{Quantum Dynamical Semigroups}
\la{s1.5}
\markright{\small{\rm{\thesection. QUANTUM DYNAMICAL SEMIGROUPS}}}

In the previous section we have introduced the set of dynamical maps
we are interested in, namely, the set of Markovian time-evolutions 
describing the irreversible evolution of an open system under the
influence of an external environment. It turns out that the structure
of the generators of these dynamical maps is fully characterized by two
theorems that we shall present after introducing the notion of 
quantum dynamical semigroups.

\begin{Def}
\la{def1.5.1}
{\rm A set of time-evolutions \{$\Lambda_t, t\geqslant 0$\} forms a} 
quantum dynamical semigroup {\rm \cite{alic1,spoh} iff the following 
properties are fulfilled:
\renewcommand{\labelenumi}{\theenumi.i)}
\ben
\item $\Lambda_t$ is completely positive $\forall t \geqslant 0$;
\item $\Lambda_t$ is trace-preserving: $\Tr (\Lambda_t[\rho]) = \Tr (\rho)
\;\; \forall t \geqslant 0$, $\forall \rho \in {\cal P}({\mathfrak H})$;
\item $\Lambda_t$ satisfies the forward in time composition law:
$\Lambda_{t_1 + t_2}[\rho] = \Lambda_{t_1} \circ \Lambda_{t_2}[\rho]
\;\; \forall t_1, t_2 \geqslant 0$, $\forall \rho \in {\cal P}({\mathfrak H})$;
\item the $\Lambda_t$ are strongly continuous: 
$\lim_{t\rightarrow 0} \| \Lambda_t [\rho] 
- \rho \|_1 = 0 \;\; \forall \rho \in {\cal P}({\mathfrak H})$.
\een
In the Heisenberg picture, defined by~(\ref{eq1.1.4}), the corresponding
properties of the dual map $\Lambda_t^*$ are:
\renewcommand{\labelenumi}{\theenumi.ii)}
\ben
\item $\Lambda_t^*$ is completely positive $\forall t \geqslant 0$;
\item $\Lambda_t^* {\bf I} = {\bf I}$, where ${\bf I}$ is the identity
on ${\cal B}({\mathfrak H})$;
\item $\Lambda_t^*$ satisfies the forward in time composition law:
$\Lambda_{t_1 + t_2}^*[A] = \Lambda_{t_1}^* \circ \Lambda_{t_2}^*[A]
\;\; \forall t_1, t_2 \geqslant 0$, $\forall A \in {\cal B}({\mathfrak H})$;
\item $\Lambda^*_t$ ultraweakly continuous: 
$\lim_{t\rightarrow 0} \Tr [\rho 
(\Lambda_t^*[A] - A)] = 0\;\; \forall \rho \in {\cal P}({\mathfrak H})$, 
$\forall A \in {\cal B}({\mathfrak H})$;
\item $\Lambda_t^*$ is normal.
\een
}
\end{Def}

\begin{Rem}
\la{rem1.5.1}
{\rm The properties 3.i) and 4.i) imply that $\Lambda_t = e^{Lt}$ , where 
$L$ is the generator of the semigroup (and $\dot\rho(t) = L [\rho(t)]$); then
the maps \{$\Lambda_t, t\geqslant 0$\} are Markovian. Analogously, 
$\Lambda_t^* = e^{L^* t}$ (i.e. $\dot A(t) = L^* [A(t)]$) and $L^*$ is the 
generator of the Markovian semigroup \{$\Lambda^*, t\geqslant 0$\}.}
\end{Rem}

\begin{Rem}
\la{rem1.5.2}
{\rm The property 5.ii) guarantees that $\Lambda_t^*$ is the dual map of
$\Lambda_t$~\cite{spoh}.}
\end{Rem}


The form of the generator of a quantum dynamical semigroup, if it is bounded, 
as a linear operator on the space of states ${\cal P}({\cal H})$, 
is completely determined as follows.

\begin{The}
\la{the1.5.1}
{\rm (Lindblad~\cite{lind}) 
Given a generic separable Hilbert space ${\mathfrak H}$, a bounded linear
transformation $L: {\cal P}({\mathfrak H})\rightarrow {\cal P}({\mathfrak H})$
is the generator of a quantum dynamical semigroup on ${\cal P}(\mathfrak{H})$ 
iff
\be
\la{eq1.5.1}
L[\rho] = -i[H,\rho] + \frac{1}{2}\sum_{j}\left( 
[V_j\rho,V_j^{\dagger}] + [V_j,\rho V_j^{\dagger}]\right),
\ee
\noi
where $H = H^{\dagger}$ and the $V_j \in {\cal B}({\cal H})$ satisfy
$\sum_j V_j^{\dagger}V_j \in {\cal B}({\mathfrak H})$. The dual generator 
$L^*$, in the Heisenberg picture, then reads
\be
\la{eq1.5.2}
L^*[A] = i[H,A] + \frac{1}{2}\sum_{j}\left( 
V_j^{\dagger}[A,V_j] + [V_j^{\dagger},A] V_j\right).
\ee}
\end{The}

\noi
Independently, Gorini {\it et al.} arrived at~(\ref{eq1.5.1}) in the case 
of finite dimensional system $S$.

\begin{The}
\la{the1.5.2}
{\rm (Gorini, Kossakowski, Sudarshan\cite{gori3}) 
Given a finite $n$-dimensional Hilbert space,
${\mathfrak H}_n$, a bounded operator $L: {\cal P}
({\mathfrak H}_n)\rightarrow {\cal P}({\mathfrak H}_n)$ is the generator 
of a quantum dynamical semigroup on ${\cal P}({\mathfrak H}_n)$ iff 
\be
\la{eq1.5.3}
L[\rho] = -i[H,\rho] + \frac{1}{2}\sum_{k,l = 1}^{n^2 - 1} c_{kl}
\left( [F_k\rho,F_l^{\dagger}] + [F_k,\rho F_l^{\dagger}] \right),
\ee
\noi
with $H = H^{\dagger}$, \{$F_k, k=0,\ldots n^2-1$\}
a basis in $M_n({\bf C})$ (the space of $n\times n$ matrices with 
complex entries), $F_0 = {\bf I}_n / \sqrt{n}$, $\Tr(F_k) = 0$
$\forall k \neq 0$, $\Tr(F_i^{\dagger}F_j) = \delta_{ij}$ and 
$C = [c_{kl}]$ a positive matrix (whose entries depend on the choice 
of the basis \{$F_k$\}). The Heisenberg generator $L^*: {\cal B}
({\mathfrak H}_n) \rightarrow {\cal B}({\mathfrak H}_n)$ is written as
\be
\la{eq1.5.4}
L^*[A] = i[H,A] + \frac{1}{2} \sum_{k,l = 1}^{n^2 -1} c_{kl}
\left( F_l^{\dagger}[A,F_k] + [F_l^{\dagger},A]F_k \right),\quad
[c_{kl}]\geqslant 0.
\ee
}
\end{The}

\begin{Cor}
\la{cor1.5.1}
{\rm Given a $n$-dimensional Hilbert space,
${\mathfrak H}_n$, with $n$ finite, a bounded operator $L: {\cal P}
({\mathfrak H}_n)\rightarrow {\cal P}({\mathfrak H}_n)$ is the generator of a 
trace and Hermiticity-preserving semigroup of maps iff it has the form 
in~(\ref{eq1.5.3}), with $H = H^{\dagger}$ the effective Hamiltonian 
and $C = [c_{kl}]$ a Hermitian matrix.}
\end{Cor}

\begin{Rem}
\la{rem.eno}
{\rm The form of the generator of a generic semigroup does not change if we 
relax the request of complete positivity, but the positivity of $C$ is
lost.}
\end{Rem}

The entries of the matrix $C$ are not all independent because of the
Hermiticity (or positivity) condition. A relation between these entries
we shall use in the following is expressed by Lemma~\ref{lem1.5.1}.

\begin{Lem}
\la{lem1.5.1}
{\rm The entries of the matrix $C$ in the generator $L$ of a trace and 
Hermiticity-preserving semigroup of maps acting on ${\cal P}(\mathfrak{H})$, 
eq.~(\ref{eq1.5.3}), satisfy the constraints
\be
\la{eq1.5.5}
\vert c_{kl} \vert \leqslant \frac{1}{2}(c_{kk} + c_{ll}),
\;\;  k,l = 1 , \ldots  n^2 - 1
\ee
{\it Proof:}
Expressing the operators in~(\ref{eq1.5.1}) as $V_j = \sum_{k = 0}^{n^2 - 1}
v^j_k F_k$ we have $c_{kl} = \sum_j v^j_k v^{j*}_l$. Then, applying Schwartz
inequality,
\begin{equation*}
\vert c_{kl} \vert \leqslant \sqrt{c_{kk} c_{ll}} \leqslant \frac{1}{2}
(c_{kk} + c_{ll}).
\end{equation*}}
\end{Lem}

\begin{Rem}
\la{rem.arecontro}
{\rm By rearranging equations~(\ref{eq1.5.1}) and (\ref{eq1.5.3}),
the generator $L$ can also be written as
\begin{subequations}
\la{eq1.5.6}
\begin{gather}
\la{eq1.5.6.a}
L[\rho] = -i[H,\rho] + \sum_j \left( V_j \rho V_j^{\dagger}
- \frac{1}{2} \{ \rho,V_j^{\dagger} V_j \} \right) \\
\la{eq1.5.6.b}
L[\rho] = -i[H,\rho] + \sum_{k,l = 1}^{n^2 - 1} c_{kl} \left(
F_k \rho F_l^{\dagger} - \frac{1}{2} \{ \rho,F_l^{\dagger}F_k \} \right);
\end{gather}
\end{subequations}
all these expressions for $L$ are generically called Lindblad forms.}
\end{Rem}

\begin{Rem}
\la{rem1.5.3}
{\rm The Hamiltonian $H$ is not the free Hamiltonian $H_S$ of the system 
of interest; indeed, in general
it contains dissipative terms produced by the interaction with
the environment: contributions of first order in this interaction, 
the mean field term $\Tr_B(L_I[\rho_B])$ in~(\ref{eq1.3.7}), 
and of second order, as we shall see explicitely in Chapter~\ref{ch4}.}
\end{Rem}

\noi
Let us now analyze separately the contributions
that appear in~(\ref{eq1.5.6.a}).

The commutator with the Hamiltonian $H$ generates a time-evolution 
trace and coherence preserving. The anticommutator between
$\rho$ and the positive operator $R\equiv \sum_j V_j^{\dagger} V_j$ may
be interpreted as a friction term; indeed, let us consider the 
time-evolution produced only by the anticommutator and the commutator:

\be
\la{eq1.5.7}
\begin{split}
\dot \rho (t) & = -i[H,\rho] - \frac{1}{2}\{R,\rho \} = -iK\rho(t) +i\rho(t)
K^{\dagger}, \\
\rho(t) & = e^{-iKt} \rho(0) e^{iK^{\dagger}t},
\end{split}
\ee

\noi
where $K = H - \frac{i}{2} R$. This evolution has the Weisskopf-Wigner form
and it is typical of an unstable system, as previously remarked; 
in fact it causes a loss of probability, as we have seen in Section~\ref{s1.1}:

\begin{equation*}
\frac{d}{dt}\Tr [\rho(t)] = \Tr [\dot \rho (t)] =
-\Tr [\rho(t)R] \leqslant 0 \quad 
\forall t\geqslant 0.
\end{equation*}

Notice that, despite loss of probability, the time-evolution~(\ref{eq1.5.7}) 
preserves the coherence of states. In fact, in order to get decoherence, the 
term $\sum_j V_j \rho V_j^{\dagger}$ in~(\ref{eq1.5.6.a}), called noise, 
is needed. Starting with a pure state, $\rho = \vert \psi \rangle \langle 
\psi \vert$, 

\be
\la{eq1.5.9}
\begin{split}
\frac{d}{dt} \Tr[\rho (t)^2]\vert_{t = 0} =
2 \Tr [\dot \rho (0) \rho(0)] & = 2 \sum_j \Tr [V_j\rho(0) V_j^{\dagger} -
V_j^{\dagger} V_j \rho(0)] \\
& = -2 \sum_j(\langle \psi \vert 
V_j^{\dagger} V_j\vert \psi\rangle - \langle \psi \vert
V_j \vert \psi \rangle \langle \psi \vert V_j^{\dagger} 
\vert \psi \rangle) \leqslant 0
\end{split}
\ee

Then, in general $\rho(t)^2 \ne \rho(t)$, so that pure states 
become mixtures. Moreover, the noise term balances the probability loss due 
to pure friction and the overall evolution is probability preserving:

\be
\la{eq1.5.10}
\frac{d}{dt} \Tr[\rho(t)] = \sum_j \Tr\Bigl[V_j \rho(t)V_j^{\dagger} -
\frac{1}{2}\{ V_j^{\dagger}V_j, \rho(t) \}\Bigr] = 0, 
\quad \forall t \geqslant 0.
\ee


\s{Factorized dynamics of quantum dynamical semigroups}
\la{s1.6}
\markright{\small{\rm{\thesection. FACTORIZED DYNAMICS OF QUANTUM DYNAMICAL SEMIGROUPS}}}

In this section we consider two systems of the same kind, $S_1$ and $S_2$,
in weak interaction with a same environment. We further suppose 
that the composite system $S_1 + S_2$ evolves 
in time according to a reduced dynamics of factorized 
form $\Lambda_t \otimes\Lambda_t$, where $\Lambda_t$ is the dynamics of 
either systems $S_1$ or $S_2$ and \{$\Lambda_t, t \geqslant 0$\} 
is assumed to form a Markovian semigroup so that 
the same is true for the set of maps
\{$\Lambda_t \otimes\Lambda_t, t\geqslant 0$\}. We focus 
our attention on the positivity properties of these maps; in particular,
in Theorem~\ref{the1.6.1} we shall prove that, in the case of $S_1$ and 
$S_2$ $n$-dimensional systems, the positivity of $\Lambda_t \otimes\Lambda_t$
necessarily requires the complete positivity of $\Lambda_t$~\cite{benfac}. 

Given a semigroup $\{\Lambda_t, t\geqslant 0\}$ generated 
by~(\ref{eq1.5.3}), the justification why $\Lambda_t$ should be completely 
positive and thus the matrix $C = [c_{kl}]$ positive, is based on the fact 
that, otherwise, $\Lambda_t \otimes {\bf I}_n$ would fail to preserve the
positivity of entangled states in ${\cal P}({\mathfrak H}_n\otimes 
{\mathfrak H}_n)$ (see Proposition~\ref{pro1.2.1}).

However, while the first factor in ${\mathfrak H}_n\otimes {\mathfrak H}_n$
refers to a concrete open quantum system evolving in time according
to $\Lambda_t$, because of the possible interaction with a certain
environment, the second factor represents a mere possibility of
entanglement with anything described by a $n$-dimensional system and
generically out of physical control. 

Instead, we argue that complete positivity is necessary to
avoid physical inconsistencies in compound systems consisting of two
$n$-dimensional systems that interact with a same environment, but not
among themselves, neither directly, nor indirectly, that is
through the environment itself.
In such a case, the two systems are expected to evolve according to
semigroups of linear maps 
\{$\Lambda_t \otimes
\Lambda_t, t\geqslant 0$\}, where $\Lambda_t$ is the single
open system dynamics obtained when only one of them is present in the
environment. 

A necessary request for the physical consistency of such dynamics 
is that the $\Lambda_t \otimes \Lambda_t$'s preserve 
the positivity of all separable and entangled states of the compound system,
which now describe physically concrete and controllable settings.

Then for Markovian dynamics of open quantum systems the argument in 
favour of the necessity of complete positivity results strengthened 
with respect to the argument based on $\Lambda_t \otimes {\bf I}_n$
(see the definition of complete positivity in Section~\ref{s1.2}).

\begin{The}
\la{the1.6.1}
{\rm If \{$\Lambda_t, t\geqslant 0$\} is a semigroup of linear maps 
on ${\cal P}({\mathfrak H}_n)$ generated as in~(\ref{eq1.5.3}), 
then the semigroup
\{$\Lambda_t \otimes \Lambda_t, t\geqslant 0$\} of linear
maps over ${\cal P}({\mathfrak H}_n \otimes {\mathfrak H}_n)$ is
positivity-preserving if and only if \{$\Lambda_t, t\geqslant 0$\} consists 
of completely positive maps~\cite{benfac}.}
\end{The}

The proof of Theorem~\ref{the1.6.1} is divided into several steps. We need just
show the only if part; indeed, if $\Lambda_t$  is completely positive,
$\Lambda_t \otimes {\bf I}_n$ and ${\bf I}_n \otimes \Lambda_t$ are both
positive and  such is the composite map
$\Lambda_t \otimes \Lambda_t = (\Lambda_t \otimes
{\bf I}_n) \circ ({\bf I}_n \otimes \Lambda_t)$.
Actually, it is completely positive (see e.g.~\cite{take}).

Before starting with the proof of this theorem, some remarks are needed.

\begin{Rem}
\la{rem1.6.1}
{\rm If the $\Lambda_t$'s preserve the positivity of states of
${\mathfrak H}_n$, $\Lambda_t \otimes \Lambda_t$ preserves the positivity
of separable states of ${\mathfrak H}_n \otimes {\mathfrak H}_n$: this
follows by the definition of separability (Def.~\ref{def1.2.4}).}
\end{Rem}

\begin{Rem}
\la{rem1.6.2}
{\rm For generic positive linear maps $\Lambda_t$ on the states of
${\mathfrak H}_n$, it does not follow that, if $\Lambda_t \otimes \Lambda_t$
is positivity-preserving, then $\Lambda_t$ is completely positive.
A counter example is the transposition $\tau$ over $M_2({\bf C})$,
presented in Remark~\ref{rem.add.2}:
$\tau \otimes \tau$ is positivity-preserving, but $\tau$ is not
completely positive. 
We notice, however, that $\tau$ cannot be among the $\Lambda_t$ of a
continuous semigroup over the states of $M_2({\bf C})$ since it is
not connected to the identity map.}
\end{Rem}

\begin{Rem}
\la{rem1.6.3}
{\rm Actually, there are experimental situations that are describable by
semigroups \{$\Lambda_t \otimes \Lambda_t, t\geqslant 0$\}.
For instance, as we shall see in Chapter~\ref{ch3}, 
neutral mesons may be imagined to suffer from
dissipative effects due to a noisy background determined by Planck's
scale physics. As decay products of spin $1$ resonances, these mesons 
are produced in maximally entangled states and, while
independently flying apart back to back, they arguably evolve
according to semigroups \{$\Lambda_t \otimes
\Lambda_t, t\geqslant 0$\}. 
In such a context, whether $\Lambda_t \otimes \Lambda_t$ is 
positivity-preserving is crucial for concrete physical 
consistency~\cite{benak,1benak,2benak,3benak,4benak,benab}.}
\end{Rem}

The proof of Theorem~\ref{the1.6.1}
proceeds as follows.

\begin{Lem}
\la{lem1.6.1}
{\rm If \{$\Lambda_t, t\geqslant 0$\} is a semigroup
generated by~(\ref{eq1.5.3}), the set
\{$\Lambda_t \otimes \Lambda_t, t\geqslant 0$\} consists of
positivity-preserving maps only if

\be
\la{eq1.6.1}
{\cal L}_{\phi, \psi} \equiv \langle \phi \vert (L\otimes {\bf I}_n +
{\bf I}_n \otimes L)[\vert 
\psi \rangle \langle \psi \vert] \vert \phi \rangle \geqslant 0
\ee

\noi
for all orthogonal vector states $\vert \phi \rangle$, $\vert \psi
\rangle$ in ${\mathfrak H}_n$, where $L$ is the generator on the right hand
side of~(\ref{eq1.5.3}) and ${\bf I}_n$ is the identity operation on
$M_n({\bf C})$.

\noi
{\it Proof:}
The request of positivity preservation demands that

\be
\la{eq1.6.2}
{\cal G}_{\phi, \psi}(t)\equiv
\langle \phi \vert (\Lambda_t \otimes \Lambda_t) [\vert \psi \rangle
\langle \psi \vert ] \vert \phi \rangle \geqslant 0, 
\ee

\noi
for all  $\vert \phi\rangle$ and $\vert \psi \rangle\, \in 
{\mathfrak H}_n \otimes {\mathfrak H}_n$. 
Choosing $\langle \phi \vert \psi \rangle = 0$, if 
${\rm d}{\cal G}_{\phi, \psi}(t)/{\rm d}t\vert_{t=0} < 0$, then
${\cal G}_{\phi, \psi}(t)\geqslant 0$ is violated in a neighborhood of $t=0$. 
Thus~(\ref{eq1.6.1}) follows.}
\end{Lem}

\begin{Lem}
\la{lem1.6.2}
{\rm In the hypothesis of Lemma~\ref{lem1.6.1}, let
\{$\vert j \rangle, j = 1, n$\} be an orthonormal basis of ${\mathfrak H}_n$, 
and $\Phi$, $\Psi$ the $n \times n$ matrices $\Phi = [\varphi_{ij}]$, 
$\Psi = [\psi_{ij}]$ consisting of the coefficients of the expansion 
of $\vert \phi \rangle$ and $\vert \psi \rangle$ with respect to the 
basis \{$\vert j \rangle \otimes \vert k \rangle; j,k = 1, n$\} 
of ${\mathfrak H}_n \otimes {\mathfrak H}_n$. Then

\be
\la{eq1.6.3}
\begin{split}
{\cal L}_{\phi, \psi} = 
& \sum_{a,b = 1}^{n^2 - 1} c_{ab} \left[\Tr(\Psi
\Phi^{\dagger}F_a)\Tr(\Phi \Psi^{\dagger}F_b^{\dagger}) + 
\Tr((\Phi^{\dagger}\Psi)^T F_a)\Tr((\Psi^{\dagger} \Phi)^T
F_b^{\dagger}) \right]\ ,
\end{split}
\ee

\noi
where $C=[c_{ab}]$ is the matrix of coefficients and
$F_a, F_b$ the traceless matrices appearing
in~(\ref{eq1.5.3}), while $X^T$ denotes
transposition of $X$ with respect to the chosen basis.

\noi
{\it Proof:}
Let $\vert \phi \rangle = \sum_{j,k = 1}^{n}\varphi_{jk} \vert j
\rangle \otimes \vert k \rangle$, 
$\vert \psi \rangle = \sum_{j,k =
1}^{n} \psi_{jk} \vert j \rangle \otimes \vert k \rangle$;
then, one calculates

\be
\la{eq1.6.4}
\begin{split}
{\cal L}_{\phi, \psi} & =
\sum_{ij} \sum_{kl} \sum_{pr} (
\varphi^*_{ij} \varphi_{kl} \psi_{pj} \psi^*_{rl} + 
\varphi^*_{ji} \varphi_{lk} \psi_{jp} \psi^*_{lr}
) 
\langle
i \vert L[\vert p \rangle \langle r \vert] \vert k \rangle  
\\ 
& =
\sum_{ik} \sum_{pr} [
(\Psi \Phi^{\dagger})_{pi}(\Phi
\Psi^{\dagger})_{kr} + 
(\Phi^{\dagger} \Psi)_{ip}(\Psi^{\dagger}
\Phi)_{rk} ] 
\langle i \vert L[\vert p \rangle \langle r \vert]
\vert k \rangle.
\end{split}
\ee

The commutator and the anticommutator in
the generator $L$ drop from this equation since, 
given any $K \in M_n({\bf C})$,
$\langle i \vert (K \vert p \rangle\langle r \vert)\vert k \rangle =
K_{ip}\delta_{rk}$. In~(\ref{eq1.6.4}), we can further sum over
either $r = k$ or $i = p$; in either cases, as $\langle \phi
\vert \psi \rangle = 0$, we find $\Tr \Psi
\Phi^{\dagger} = (\Tr \Phi \Psi^{\dagger})^* = 0$ 
and the result follows.}
\end{Lem}

\begin{Lem}
\la{lem1.6.3}
{\rm In the hypothesis of Lemma~\ref{lem1.6.1}, the matrix $C=[c_{ab}]$
in~(\ref{eq1.6.3}) must be positive definite.

\noi
{\it Proof:}
With any $\vec{w} = \{ w_a, a = 1, n^2 - 1\} \in {\bf C}^{n^2 -1}$, 
we consider $W = \frac{1}{2} \sum_{a = 1}^{n^2 - 1} w_a^* F_a$, 
which is a traceless $n\times n$ matrix.
If matrices $\Psi$ and $\Phi$ exist such that $\Phi \Psi^{\dagger} = W$
and $\Psi^{\dagger} \Phi = W^T$, then, from Lemmas~\ref{lem1.6.1}
and~\ref{lem1.6.2} and from the orthogonality of the matrices $F_a$ 
it follows

\be
\la{eq1.6.5}
{\cal L}_{\phi, \psi} = \sum_{a,b = 1}^{n^2 - 1} c_{ab} w^*_a w_b
\geqslant 0\ ,
\ee

\noi
whence the positivity of $C = [c_{ab}]$ and the proof of 
Theorem~\ref{the1.6.1}.
Any matrix $W$ and its transposed with respect to the given basis, $W^T$,
have the same elementary divisors; therefore, they are similar to the same
canonical Jordan form and thus similar to each other~\cite{gelf}.
Let $\Phi$ such that $\Phi^{-1} W \Phi = W^T$, that is we take as
vector $\vert \phi \rangle \in {\mathfrak H}_n \otimes {\mathfrak H}_n$ the one
whose components $\phi_{ij}$ are the elements of the similarity matrix
transforming the given $W$ into its transposed $W^T$.
It then follows that $\Psi^{\dagger} = \Phi^{-1} W$ and moreover
$\Psi^{\dagger} \Phi = \Phi^{-1} W \Phi = W^T$, which is what we
need.}
\end{Lem}

\s{The stochastic approach}
\la{s1.7}
\markright{\small{\rm{\thesection. THE STOCHASTIC APPROACH}}}

The description of an open quantum system can be performed employing 
an approach different from the projector technique presented in 
Section~(\ref{s1.3}), that is using a stochastic equation~\cite{budi}. 
The point of view is similar to that developed by Langevin to describe the 
Brownian motion: we introduce a time dependent term in the system 
Hamiltonian: $H_S(t) = H_0 + H_I(t)$, where $H_I(t)$ characterizes the
interaction between $S$ and reservoir. The statistical operator that
describes $S$ is obtained by an averaging procedure over the noise:
$\rho_S(t) = \langle \Sigma_S(t) \rangle$, and $\Sigma_S(t)$ is the 
stochastic density matrix of $S$:

\be
\la{eq1.7.1}
\dot \Sigma_S(t) = L_S(t)[\Sigma_S(t)] = (L_0 + L_I(t))[\Sigma_S(t)]
\ee

\noi
with Liouville operators $L_0[\Sigma_S(t)] = -i[H_0,\Sigma_S(t)]$ and
$L_I(t)[\Sigma_S(t)] = -i[H_I(t),\Sigma_S(t)]$. In the interaction 
picture~(\ref{eq1.7.1}) reads

\be
\la{eq1.7.2}
\dot \Sigma^i_S(t) = L_I^i(t)[\Sigma_S^i(t)],
\ee

\noi
where

\be
\la{eq1.7.2.bis}
\Sigma_S^i(t) = e^{itH_0} \Sigma_S(t) e^{-itH_0}, \quad 
L_I^i(t) = e^{-L_0t}L_I(t)e^{L_0t}.
\ee

\noi
The solution of~(\ref{eq1.7.2}) is

\be
\la{eq1.7.3}
\Sigma^i_S(t) = {\bf T}e^{\int_0^t L_I^i(t^{\prime}) dt^{\prime}}
\Sigma_S^i(0),
\ee

\noi
where ${\bf T}$ means time-ordering. Since $\rho_S^i(t) = 
\langle \Sigma_S^i(t) \rangle$, it follows that

\be
\la{eq1.7.4}
\rho_S^i(t) = \sum_{k = 0}^{\infty} M_k(t)[\rho_S^i(0)]
\ee

\noi 
with

\be
\la{eq1.7.5}
M_k(t) \equiv \int_0^t ds_1 \int_0^{s_1} ds_2 \ldots \int_0^{s_{k-1}}ds_k
\langle L_I^i(s_1)L_I^i(s_2) \ldots L_I^i(s_k)\rangle.
\ee

We assume that the initial state $\Sigma_S(0)$ is decoupled from the noisy
environment so that $\rho_S(0) = \langle \Sigma_S(0) \rangle = 
\Sigma_S(0)$ (and then $\rho_S^i(0) = \Sigma_S^i(0)$). Eq.~(\ref{eq1.7.4})
is a series expansion in the interaction, containing the correlation
functions of the fields defining $H_I(t)$. If we suppose these stochastic
fields to be Gaussian, only the even terms in~(\ref{eq1.7.4}) survive.

Setting $M(t) \equiv \sum_k M_k(t) = \sum_k M_{2k}(t)$, we rewrite the 
differential form of equation~(\ref{eq1.7.4}) as follows:

\be
\la{eq1.7.6}
\dot \rho_S^i(t) = \dot M(t) M^{-1}(t)[\rho_S^i(t)] =
\left( \dot M_2(t) + \dot M_4(t) - \dot M_2(t) M_2(t) + \ldots \right)
[\rho_S^i(t)]
\ee

\noi
using the formal expressions

\be
\la{eq1.7.7}
M(t) = {\bf I} + M_2(t) + M_4(t) + \ldots, \quad
M^{-1}(t) = {\bf I} - M_2(t) + [M_2(t)]^2 - M_4(t) + \ldots. 
\ee

Assuming the system $S$ to be a $n$-level system,
it is possible to express the interaction Hamiltonian as $H_I(t) = 
\sum_{k =0}^{n^2 - 1}V_k(t) F_k$, where $F_k$ is the basis of 
$M_n({\bf C})$ introduced in~(\ref{eq1.5.3}) and $V_k(t)$ are Gaussian
stochastic fields with correlation functions $h_{kl}(t - s) = 
\langle V_k(t) V_l(s) \rangle$. Then, up to the second order in the 
interaction, eq.~(\ref{eq1.7.6}) assumes the form

\be
\la{eq1.7.8}
\dot \rho_S^i(t) = - \sum_{k,l = 1}^{n^2 - 1} \int_0^t
ds h_{kl}(s)[F_k(t),[F_l(t - s),\rho_S^i(t)]],
\ee

\noi
with $F_k(t) \equiv e^{L_0 t}[F_k] = e^{itH_0} F_k e^{-itH_0}$.

Turning back to the Schr{\" o}dinger picture, $\rho_S(t) = 
e^{-itH_0} \rho_S^i(t) e^{itH_0}$ and we find:

\be
\la{eq1.7.9}
\dot \rho_S(t) = -i[H_0,\rho_S(t)] + \sum_{k,j = 1}^{n^2 - 1}
d_{kj}(t)[F_k,[F_j,\rho_S(t)]],
\ee

\noi
with 

\be
\la{eq1.7.10}
d_{kj}(t) = - \sum_{l = 1}^{n^2 - 1}\int_0^t h_{kl}(s) u_{lj}(-s) ds
\ee

\noi
and $u_{lj}(-s)$ defined by $F_l(-s) = \sum_{j = 0}^{n^2 - 1}
u_{lj}(-s)F_j$.

The Markovian approximation of the reduced dynamics~(\ref{eq1.7.9}) is
performed with the hypothesis that the correlation functions $h_{kl}(t)$
decay so fast that we can substitute $+ \infty$ for the parameter $t$
in~(\ref{eq1.7.10}); in this way the coefficients $d_{kj}(t)$ become 
time independent.
This assumption, as we have already seen, is the same as considering 
the system evolution for times long enough, so that the transient terms 
(the memory contributions, non Markovian) disappear.

The reduced dynamics so obtained, although Markovian, in general is not
completely positive nor positive (in fact the
generator of the dissipative part in general has a different form than that
expressed in Theorems~\ref{the1.5.1} and \ref{the1.5.2}): the stochastic
approach suffers from this failure. Equations of motion in the 
form~(\ref{eq1.7.9}), in the two-dimensional case, after the Markovian limit, 
are called Redfield-Bloch equations~\cite{suar,budi,lair}; 
in Chapter~\ref{ch5} we shall explicitly show instances where lack of 
positivity occurs.

It has been argued that it is possible to consistently use these 
time-evolutions, even if they are not positive, 
provided they act only on a suitable subset 
of ${\cal P}({\mathfrak H}_n)$, constituted by statistical operators whose 
positivity is preserved. Such a restriction is usually explained in terms
of a slippage of initial conditions~\cite{gasp,1gasp}: the initial transient, 
neglected in the Markov limit, is supposed to select from 
${\cal P}({\mathfrak H}_n)$ exactly the set of states which remain positive 
under the dynamical map considered. 
In such a perspective, the long-time Markov approximation 
of the dynamics, possibly not positive, should be coupled to a redefinition 
of initial conditions, generated by the action, on ${\cal P}
({\mathfrak H}_n)$, of the short-time non-Markovian dynamics. 
In such a way the positivity of $\rho_S(t)$ should be saved. 
However, it can be shown that the prescription needed to guarantee the
consistency of the action of a Markovian dynamical map $\Lambda_t$ in 
general does not hold when two systems are immersed in the same environment 
and evolve with the factorized dynamics $\Lambda_t \otimes \Lambda_t$, starting
from an entangled state: a new prescription has to be enforced~\cite{pian}. 
This argument can be repeated  when three or more systems, living in the same
environment, are taken into account: there 
does not exist a general prescription,
valid for any number of systems interacting with the same environment.
In this sense, complete positivity appears at least as a cheaper tool
to avoid physical inconsistencies.


\s{Entropy and long time behavior in open quantum systems}
\la{s1.8}
\markright{\small{\rm{\thesection. ENTROPY AND LONG TIME BEHAVIOR IN OPEN QUANTUM SYSTEMS}}}

Typically, an open system exchanges energy and entropy (also matter
in general) with its environment. Because of the vastly different
number of their degrees of freedom, one expects the noise coming from
the environment to always increase the randomness of the system and thus 
its entropy.

For later use, in this section we shortly review some results about the
long time behavior of a state evolving according to quantum dynamical 
semigroups, and of its entropy.

\begin{Def}
\la{def2.3.1}
{\rm Given a statistical operator $\rho \in {\cal P}({\mathfrak H})$, its 
von Neumann entropy $S(\rho)$ is defined by 
$S(\rho) \equiv -\Tr (\rho \log \rho)$.}
\end{Def}

\noi
For a review about entropy and its properties, see~\cite{wehr}; here
it is sufficient to remark that the von Neumann entropy is always 
non-negative, $S(\rho) \geqslant 0$, $\forall \rho \in {\cal P}
({\mathfrak H})$, and $S(\rho) = 0$ if and only $\rho$ is a pure state. 
Furthermore, if $\mathfrak{H}$ is finite dimensional, say ${\rm dim}
(\mathfrak{H}) = n$, the entropy is greatest for the maximally mixed
state $\rho_n \equiv {\bf I}_n / n$: $S(\rho_n) = \log n$.

The time-evolution of entropy is related to the long time behavior of 
states evolving under quantum dynamical semigroups, to the
existence of invariant states, and to the approach to equilibrium of 
the considered open systems.

\begin{Def}
\la{def1.8.1}
{\rm A state $\rho_0\in {\cal P}({\cal H})$ is said} invariant {\rm (or
} stationary {\rm ) under the action of a quantum dynamical semigroup 
$\{\Lambda_t, t\geqslant 0\}$ iff $\Lambda_t \rho_0 = \rho_0$, $\forall t 
\geqslant 0$ or, equivalently, $L[\rho_0] = 0$, where $L$ is the
generator of the semigroup, expressed in~(\ref{eq1.5.1}).}
\end{Def}

\begin{Def}
\la{def1.8.2}
{\rm The dynamical semigroup $\{ \Lambda_t, t \geqslant 0\}$ is called}
relaxing {\rm iff there exists an invariant state $\rho_0 \in {\cal P}
({\cal H})$ such that $\forall \rho \in {\cal P}({\cal H})$,
$\lim_{t\rightarrow \infty} \Lambda_t \rho = \rho_0$.}
\end{Def}

A complete classification of all types of stationary states associated
to relaxing semigroups is given in~\cite{lend}; see also~\cite{spo,fri}.
For later reference we are only interested in the following considerations.

As we have seen in Section~\ref{s1.5}, the dual time-evolution $\Lambda_t^*$
always preserves the identity ${\bf I}$, and then its generator, denoted 
by $L^*$, maps ${\bf I}$ into the null operator: $L^*[{\bf I}] = 0$. In 
general, the same does not hold for the evolution map $\Lambda_t$ and
$L[{\bf I}] \ne 0$. However, as we shall shortly see, there is a tight
connection between the entropy time-evolution of the system and the 
invariance of ${\bf I}$, technically called unitality. 

\begin{Def}
\la{def1.8.3}
{\rm A quantum dynamical semigroup \{$\Lambda_t, t\geqslant0$\} is called}
unital {\rm iff the identity operator ${\bf I}\in {\cal P}({\cal H})$,
is invariant under its action.}
\end{Def}

\begin{Rem}
\la{rem1.8.1}
{\rm If $\Lambda_t$ is unital, the maximally mixed state $\rho_n$ is
invariant.}
\end{Rem}

\begin{Rem}
\la{rem1.8.2}
{\rm If $\Lambda_t$ is unital, the operators $A_k$ defined in 
Corollary~\ref{cor1.2.1} must satisfy $\sum_k A_k A_k^{\dagger} = {\bf I}$.}
\end{Rem}

\begin{Rem}
\la{rem1.8.3}
{\rm If $\Lambda_t$ is unital, the operators $V_k$ defined in 
Theorem~\ref{the1.5.1} must satisfy $\sum_k [V_k, V_k^{\dagger}] = 0$.}
\end{Rem}

Now, it turns out that, for an $n$-level system, the unitality of the 
dynamics is a necessary condition for the non-decreasing behavior of
its entropy. Indeed, if unitality is not fulfilled, the maximally mixed state 
$\rho_n$ is not invariant under the reduced dynamics considered:
$\dot{\rho}_n = L[\rho_n] \ne 0$; therefore
$\rho_n(t) = \Lambda_t[\rho_n] \ne \rho_n$ and then, necessarily, 
the entropy decreases: $S[\rho_n(t)] < S(\rho_n)$. 

In the next chapter we shall prove that, for $n = 2$, the unitality 
of quantum dynamical semigroups is also a sufficient
condition in order to have a non-decreasing entropy.

%% file: chapter2.tex
\chapter{Bidimensional Systems}
\la{ch2}
\markboth{\small{\rm{\thechapter. BIDIMENSIONAL SYSTEMS}}}{}

In view of later applications, in this chapter we 
restrict our attention to quantum open systems
described by two-dimensional Hilbert spaces. The parameters describing 
general Markovian dynamics are introduced; positivity and 
complete positivity of the time-evolution are expressed by sets of 
inequalities among them.

Although the equations of motion can be integrated exactly,
in many cases it is preferable to express their solutions as 
perturbative expansions
in the dissipative parameters, provided they are small with respect to the
parameters describing the dynamics of the system alone, in absence of the
environment. This computation, to leading order in the dissipative 
parameters, is performed in Section~\ref{s2.4}.



\s{Evolution of states}
\la{s2.1}
\markright{\small{\rm{\thesection. EVOLUTION OF STATES}}}

The physical systems we shall consider in the next chapters will be 
described by means of two-dimensional Hilbert spaces. Therefore,
their states will be represented by density matrices corresponding to
positive elements of $M_2({\bf C})$ with unit trace. 
We will use two different basis for the algebra $M_2({\bf C})$, namely:

\begin{itemize}

\item the set $\{P_i, i = 1,\ldots 4\}$, where

\be
\la{eq2.1.1}
P_1 = \left( 
\begin{array}{cc}
1 & 0 \\
0 & 0 \\
\end{array}
\right), \quad
P_2 = \left( 
\begin{array}{cc}
0 & 0 \\
0 & 1 \\
\end{array}
\right), \quad
P_3 = \left( 
\begin{array}{cc}
0 & 1 \\
0 & 0 \\
\end{array}
\right), \quad
P_4 = \left( 
\begin{array}{cc}
0 & 0 \\
1 & 0 \\
\end{array}
\right),
\ee

\noi
with $P_{1,2}^{\dagger} = P_{1,2}$ and $P_{3,4}^{\dagger} = P_{4,3}$;

\item the set $\{\sigma_i, i = 0,\ldots 3\}$, that is the set of Pauli
matrices plus the identity:

\be
\la{eq2.1.2}
\sigma_0 = {\bf I}_2 = 
\left(
\begin{array}{cc}
1 & 0 \\
0 & 1 \\
\end{array}
\right), \quad
\sigma_1 = 
\left(
\begin{array}{cc}
0 & 1 \\
1 & 0 \\
\end{array}
\right), \quad
\sigma_2 =  
\left(
\begin{array}{cc}
0 & -i \\
i & 0 \\
\end{array}
\right), \quad
\sigma_3 =  
\left(
\begin{array}{cc}
1 & 0 \\
0 & -1 \\
\end{array}
\right),
\ee

\noi
satisfying $\sigma_i^{\dagger} = \sigma_i$, $i = 0,\ldots 3$;
$\Tr(\sigma_i) = 0$, $i = 1, \ldots 3$; $\Tr(\sigma_i^{\dagger}
\sigma_j) = 2\delta_{ij}$, $i,j = 0,\ldots 3$.
\end{itemize}

\noi
Accordingly, states $\rho$ will be written, with respect to these
two basis, as

\be
\la{eq2.1.3}
\begin{split}
\rho & = \sum_{i = 1}^4 \rho_i P_i =
\left(
\begin{array}{cc}
\rho_1 & \rho_3 \\
\rho_4 & \rho_2 \\
\end{array}
\right) \\
& = \sum_{i = 0}^3 \rho^i \sigma_i = 
\left(
\begin{array}{cc}
\rho^0 + \rho^3 & \rho^1 - i \rho^2 \\
\rho^1 + i \rho^2 & \rho^0 - \rho^3 \\
\end{array}
\right).
\end{split}
\ee

We use upper and lower indices to distinguish the coefficients in the 
two different expansions. We have $\rho_i \in {\bf C}$ and 
$\rho_4 = \rho_3^{\dagger}$, while $\rho^i \in {\bf R}$. 

We are interested in Markovian dynamics for the states $\rho$; they are
fully characterized by Theorem~\ref{the1.5.2} and Corollary~\ref{cor1.5.1}.
The basis $\{F_i \equiv \sigma_i/\sqrt{2},i = 0, \ldots 3\}$ fulfills
the properties required in Theorem~\ref{the1.5.2}: $F_0 = {\bf I}_2/ 
\sqrt{2}$, $\Tr(F_i) = 0$ $\forall i \ne 0$, $\Tr(F_i^{\dagger} F_j)
= \delta_{ij}$; then, the more general form for the generator of a 
Markovian dynamics is

\be
\la{eq2.1.4.add}
L[\rho] = (L_H + L_D) [\rho],
\ee

\noi
where $L_H[\cdot]$ and $L_D[\cdot]$ are the Hamiltonian, 
respectively dissipative part, of the generator. 
$L_H$, for a stable system, is proportional to the commutator with 
a Hamiltonian $H = H^{\dagger}$ that can be chosen traceless without loss
of generality, then $H = \sum_{i = 1}^3 h_i \sigma_i$, 
$h_i \in {\bf R}$; $L_D$ can be written in the Kossakowski form:

\be
\la{eq2.1.4}
L_H[\rho] \equiv -i[H,\rho], \quad 
L_D[\rho] \equiv \frac{1}{2}\sum_{k,l = 1}^{3} c_{kl} \left(
\sigma_k \rho \sigma_l - \frac{1}{2}\{\rho,\sigma_l \sigma_k\}\right)
\ee 

\noi
with $C = [c_{kl}]$ a Hermitian matrix.

For unstable systems $H$ is non-Hermitian (i.e. $h_i \in 
{\bf C}$) of Weisskopf-Wigner type (see the discussion in 
Section~\ref{s1.1}) and the time-evolution does not preserve the trace of 
the states $\rho$; the generator has thus the form~(\ref{eq2.1.4.add}) with
a modified Hamiltonian part:

\be
\la{eq2.1.4.bis}
L_H[\rho] \equiv -iH \rho + i \rho H^{\dagger} 
\ee 

\noi 
and will be used when dealing with neutral mesons systems; at present
we restrict our attention to trace preserving dynamics with $L_H$ in
form~(\ref{eq2.1.4}).

It is convenient to introduce a vector notation and write the density matrix
$\rho$ as a $4$-component vector $\vert \rho \rangle \equiv 
(\rho^0, \rho^1, \rho^2, \rho^3)$ whose entries are the coefficients 
of the expansion of $\rho$ in the basis~(\ref{eq2.1.2}); in fact, since 
equation~(\ref{eq2.1.4.add}) is linear in $\rho$, it can be rewritten in
vectorial form as a kind of Schr{\" o}dinger-like equation:

\be
\la{eq2.1.5}
\frac{d}{dt}\vert \rho(t) \rangle =
{\cal L}\vert \rho(t) \rangle = ({\cal H} + {\cal D})\vert
\rho(t)\rangle,
\ee 

\noi 
where ${\cal H}$ and ${\cal D}$ are $4\times 4$ real matrices, representing
the Hamiltonian, respectively the dissipative part of~(\ref{eq2.1.4.add}). 
These matrices have the form

\be
\la{eq2.1.6}
{\cal H} = -2 
\left(
\begin{array}{cccc}
0 & 0 & 0 & 0 \\
0 & 0 & h_3 & -h_2 \\
0 & -h_3 & 0 & h_1 \\
0 & h_2 & -h_1 & 0 \\
\end{array}
\right),
\; \;
{\cal D} = - 2 
\left(
\begin{array}{cccc}
0 & 0 & 0 & 0 \\
u & a & b & c \\
v & b & \alpha & \beta \\
w & c & \beta & \gamma \\
\end{array}
\right).
\ee

In~(\ref{eq2.1.6}) we have introduced nine real coefficients to express the
dissipative part ${\cal D}$. They depend on the entries of the matrix $C$
in~(\ref{eq2.1.4}); explicitly~\cite{alic1,benat}:

\be
\la{eq2.1.7}
{\cal D} = 
\left(
\begin{array}{cccc}
0 & 0 & 0 & 0 \\
2 \im c_{32} & -c_{22} - c_{33} & \re c_{12} & \re c_{13} \\
2 \im c_{13} & \re c_{12} & -c_{11} - c_{33} & \re c_{23} \\
2 \im c_{21} & \re c_{13} & \re c_{23} & -c_{11} - c_{22} \\
\end{array}
\right);
\ee

\noi
conversely, the matrix $C$ can be written in terms of these dissipative
parameters:

\be
\la{eq2.1.8}
C = 
\left(
\begin{array}{ccc}
\alpha + \gamma - a & -2b + iw & -2c - iv \\
-2b - iw & a + \gamma - \alpha & -2 \beta + iu \\
-2c + iv & -2 \beta - iu & a + \alpha - \gamma \\
\end{array}
\right).
\ee

\begin{Rem}
\la{rem2.1.1}
{\rm The Hamiltonian contribution ${\cal H}$ is skew-symmetric since it
represent a commutator.}
\end{Rem}

\begin{Rem}
\la{rem2.1.2}
{\rm The entries of the first row of both ${\cal H}$ and ${\cal D}$
vanish since we have required the dynamics to be trace preserving.
Indeed, from~(\ref{eq2.1.6}) it follows that $\dot{\rho}^0(t) = 0$, as
it should be since
$\Tr[\rho(t)] = 2 \rho^0(t)$ and then $d \Tr[\rho(t)] / dt = 0$.}
\end{Rem}

Using the basis~(\ref{eq2.1.1}) $\rho$ gets represented by the $4$-vector
$\vert \hat{\rho} \rangle \equiv (\rho_1, \rho_2, \rho_3, \rho_4)$, 
which evolves in time according to

\be
\la{eq2.1.9}
\frac{d}{dt}\vert \hat{\rho}(t) \rangle = \hat{\cal L}\vert \rho(t) 
\rangle = (\hat{\cal H} + \hat{\cal D})\vert
\hat{\rho}(t)\rangle,
\ee 

\noi
where $\hat{\cal H}$ and $\hat{\cal D}$ are respectively the Hamiltonian 
and the dissipative contribution in this representation\footnote{We use
hats to characterize states as well as operators in this representation}. 
From the relations in~(\ref{eq2.1.3}) it follows that the two vectors 
representing the state $\rho$ are related by

\be
\la{eq2.1.10}
\vert \hat{\rho}(t) \rangle = {\cal V} \vert \rho(t) \rangle, \quad
{\cal V} =
\left(
\begin{array}{cccc}
1 & 0 & 0 & 1 \\
1 & 0 & 0 & -1 \\
0 & 1 & -i & 0 \\
0 & 1 & i & 0 \\
\end{array}
\right).
\ee

\noi
Thus,

\be
\la{eq2.1.11}
\hat{\cal H} = {\cal V} {\cal H} {\cal V}^{-1} = 
\left(
\begin{array}{cccc}
0 & 0 & -h_2 + i h_1 & -h_2 - i h_1 \\
0 & 0 & h_2 - i h_1 & h_2 + i h_1 \\
h_2 + i h_1 & -h_2 - i h_1 & -2i h_3 & 0 \\
h_2 - i h_1 & -h_2 + i h_1 & 0 & 2i h_3 \\
\end{array}
\right)
\ee

\noi
and

\be
\la{eq2.1.12}
\hat{\cal D} = {\cal V} {\cal D} {\cal V}^{-1} = 
\left(
\begin{array}{cccc}
-\gamma - w & \gamma - w & -C & -C^* \\
\gamma + w & w - \gamma & C & C^* \\
-C^* - U^* & C^* - U^* & -A & B \\
-C - U & C - U & B^* & -A \\
\end{array}
\right)
\ee

\noi
with

\be
\la{eq2.1.13}
\begin{array}{l}
A \equiv a + \alpha, \quad
B \equiv \alpha - a + 2ib, \quad
C \equiv c + i \beta, \quad
U \equiv u + iv.
\end{array}
\ee

Before discussing solutions of the equations of motion we discuss 
the positivity of the dynamics.


\s{Positivity and Complete Positivity}
\la{s2.2}
\markright{\small{\rm{\thesection. POSITIVITY AND COMPLETE POSITIVITY}}}

The requests of positivity or complete positivity for the time-evolutions
described in the previous section are expressed by different constraints 
on the dissipative parameters appearing in ${\cal D}$.

\begin{Rem}
\la{rem2.2.1}
{\rm The positivity of $A \in M_2({\bf C})$ is equivalent to 
$\Tr(A) \geqslant 0$ and $\Det(A) \geqslant 0$, since we are in two 
dimensions. Since states $\rho$ have $\Tr(\rho) = 1$, 
so that $\rho^0 = 1/2$, they are positive
if and only if $\Det(\rho) = 1/4 - \sum_{j = 1}^{3}(\rho^j)^2 \geqslant 0$.
A state $\rho$ is pure (i.e. a one-dimensional projection) if and
only if $\Det(\rho) = 0$. A mixed state has $\sum_{j = 1}^3 (\rho^j)^2 
< 1/4$.}
\end{Rem}

\begin{The}
\la{the2.2.1}
{\rm If the trace and Hermiticity preserving
time-evolution~(\ref{eq2.1.5}) with ${\cal H}$ and ${\cal D}$ 
in~(\ref{eq2.1.6}) is positive, then the real entries of ${\cal D}$
satisfy the inequalities:
\be
\la{eq2.2.1}
\left\{
\begin{array}{l}
a \geqslant 0 \\
\alpha \geqslant 0 \\
\gamma \geqslant 0 \\
\end{array}
\right. ; \quad
\left\{
\begin{array}{l}
a \alpha \geqslant b^2 \\
a \gamma \geqslant c^2 \\
\alpha \gamma \geqslant \beta^2 \\
\end{array};
\right . \quad
a \alpha \gamma + 2bc \beta - a \beta^2 - \alpha c^2 - \gamma b^2
\geqslant 0. 
\ee
{\it Proof:}
$\rho(t)\geqslant 0$ if and only if $\Det[\rho(t)] \geqslant 0$ 
(Remark~\ref{rem2.2.1}). Choosing the initial
state $\vert \rho_a(0) \rangle = (1/2, \rho^1, \rho^2, \rho^3)$ to be pure,
that is $\sum_{j = 1}^3 (\rho^j)^2 = 1/4$,
then $d \Det[\rho_a(t)]/ dt \vert_{t = 0}
\geqslant 0$ must hold if $\rho_a(t)$ has to remain a state. Explicitly,
\be
\la{eq2.2.2}
\begin{split}
\frac{d}{dt}\Det[\rho_a(t)]\vert_{t = 0} 
&= - 2 \sum_{j = 1}^3 \rho^j(t) \dot{\rho}^j(t)\vert _{t = 0} \\
&= - 2 \sum_{j,k = 1}^3 ({\cal H} + {\cal D})_{jk}
\rho^j(0) \rho^k(0) - \sum_{j = 1}^3 {\cal D}_{j0}\rho^j(0) 
\geqslant 0. \\
\end{split}
\ee
Analogously, starting with the pure state $\vert \rho_b(0) \rangle = (1/2,
-\rho^1, -\rho^2, -\rho^3)$ we find 
\be
\la{eq2.2.3}
\frac{d}{dt}\Det[\rho_b(t)]\vert_{t = 0} = - 2 \sum_{j,k = 1}^3 ({\cal H} 
+ {\cal D})_{jk}\rho^j(0) \rho^k(0) + \sum_{j = 1}^3 {\cal D}_{j0}\rho^j(0) 
\geqslant 0;
\ee
adding equations~(\ref{eq2.2.2}) and (\ref{eq2.2.3}) and considering that 
${\cal H}$ is a skew symmetric matrix, we are left with
\be
\la{eq2.2.4}
\sum_{j,k = 1}^3 {\cal D}_{jk} \rho^k(0)\rho^j(0) \leqslant 0.
\ee
Introducing the sub-matrix $\tilde{\cal D}$ as 
\be
\la{eq2.2.5}
{\cal D} = -2 \left(
\begin{array}{cccc}
0 & 0 & 0 & 0 \\
u \\
v & & \tilde{\cal D} \\
w \\
\end{array}
\right), \quad
\tilde{\cal D} \equiv
\left(
\begin{array}{ccc}
a & b & c \\
b & \alpha & \beta \\
c & \beta & \gamma \\
\end{array}
\right),
\ee
the condition~(\ref{eq2.2.4}) becomes $\tilde{\cal D} \geqslant 0$;
which is equivalent to~(\ref{eq2.2.1}) by considering the minors of 
$\tilde{\cal D}$.}
\end{The}

\begin{The}
\la{the2.2.2}
{\rm \cite{alic1} The trace and Hermiticity preserving 
time-evolution~(\ref{eq2.1.5}) is completely positive iff the entries of
${\cal D}$ satisfy the inequalities:
\be
\la{eq2.2.6}
\begin{array}{c}
\left\{
\begin{array}{l}
2R \equiv \alpha + \gamma - a \geqslant 0 \\
2S \equiv a + \gamma - \alpha \geqslant 0 \\
2T \equiv a + \alpha - \gamma \geqslant 0 \\
\end{array} 
\right. ; \quad
\left\{
\begin{array}{l}
RS \geqslant b^2 + w^2/4 \\
RT \geqslant c^2 + v^2/4 \\
ST \geqslant \beta^2 + u^2/4 \\
\end{array}
\right. ; \\ \\
RST - 2bc\beta + \frac{1}{2} buv + \frac{1}{2} \beta vw + \frac{1}{2} cuw -
S \left( c^2 + \frac{1}{4} v^2 \right) - 
R \left( \beta^2 + \frac{1}{4} u^2 \right) -
T \left( b^2 + \frac{1}{4} w^2 \right) \geqslant 0.
\end{array}
\ee
{\it Proof:}
The thesis follows from the fact that
the time-evolution is completely positive if and only if the matrix $C$ 
in~(\ref{eq2.1.8}) is positive (Theorem~\ref{the1.5.2}).}
\end{The}

\begin{Rem}
\la{rem2.2.3}
{\rm Note that the constraints expressed in these two theorems are not all
independent. Also, the constraints~(\ref{eq2.2.6}) imply those 
in~(\ref{eq2.2.1}). Indeed, CP is stronger than positivity.}
\end{Rem}


\s{Positivity and CP for entropy increasing evolutions}
\la{s2.3}
\markright{\small{\rm{\thesection. POSITIVITY AND CP FOR ENTROPY INCREASING EVOLUTIONS}}}


There are physical situations in which the reduced time-evolution of 
states does necessarily increase the von Neumann entropy
(see Definition~\ref{def2.3.1}) and 
states becomes less ordered as the time passes. For instance,
in Chapters~\ref{ch3} and \ref{ch5} we shall describe irreversible
time-evolutions that are entropy increasing.

As anticipated in Section~\ref{s1.8},
any entropy non decreasing time-evolution can be fully characterized as
follows.

\begin{The}
\la{the2.3.1.bis}
{\rm The trace and Hermiticity preserving time-evolution~(\ref{eq2.1.5})
with ${\cal H}$ and ${\cal D}$ in~(\ref{eq2.1.6})
has a non decreasing von Neumann entropy 
iff $u = v = w = 0$ and $\tilde{\cal D} \geqslant 0$ (i.e. the time 
evolution~(\ref{eq2.1.5}) is unital),
where $u$, $v$ and $w$ are the parameters defined in 
equations~(\ref{eq2.1.6}) and (\ref{eq2.1.7}) and $\tilde{\cal D}$ is the
matrix in eq.~(\ref{eq2.2.5}). Moreover, the matrix $C$ 
in~(\ref{eq2.1.8}) is symmetric~\cite{gori2,benat}.

\noi
{\it Proof:}
The eigenvalues, 
respectively eigenvectors, of a generic initial state represented by the 
vector $\vert \rho(0) \rangle = (1/2,\rho^1,
\rho^2,\rho^3)$ with $\vert \rho \vert ^2 \equiv
\sum_{j = 1}^3 (\rho^j)^2 \leqslant 1/4$, are 
\be
\la{eq2.3.1}
\rho_{\pm} = \frac{1}{2}(1 \pm 2 \vert \rho \vert); \; \;
\vert \rho_{\pm} \rangle = \frac{1}{2 \vert \rho \vert} (\vert
\rho \vert , \pm \rho^1, \pm \rho^2, \pm \rho^3).
\ee
Consider now
\be
\la{eq2.3.2}
\frac{d}{dt}S[\rho(t)]\vert_{t = 0} 
= -\frac{d}{dt} \Tr[\rho(t) \log \rho(t)]\vert_{t = 0}
= - \Tr[\dot \rho (t) \vert _{t = 0} \log \rho (0)]
\ee
\noi
where we have made use of the trace preserving property of the time-evolution.
Evaluating the trace over the states $\vert \rho _+ \rangle$ and 
$\vert \rho _- \rangle$ we find
\be
\la{eq2.3.3}
\begin{split}
\frac{d}{dt}S[\rho(t)]\vert_{t = 0} 
&= - \log \rho_+\Tr[\dot \rho (t) \vert _{t = 0} \vert \rho_+
\rangle \langle \rho_+ \vert] - 
\log \rho_-\Tr[\dot \rho (t) \vert _{t = 0} \vert \rho_- 
\rangle \langle \rho_- \vert] =  \\
&= - \frac{1}{\vert \rho \vert}\left[ \frac{1}{2}\sum_{i = 1}^3
{\cal D}_{i0} \rho^i(0) + \sum_{i,j = 1}^3 {\cal D}_{ij} 
\rho^i(0) \rho^j(0) \right]\log \frac{\rho_+}{\rho_-},
\end{split}
\ee
\noi
having taken into account the skew-symmetry of ${\cal H}$. 

\noi
Suppose $u = v = w = 0$ (that is ${\cal D}_{i0} = 0 \; \forall i$) and
$\tilde{\cal D} \geqslant 0$. Then 
\be
\la{eq2.3.4}
\frac{d}{dt}S[\rho(t)]\vert_{t = 0} = \frac{2}{\vert \rho \vert}
\sum_{i,j = 1}^3 \tilde{\cal D}_{ij} 
\rho^i(0) \rho^j(0)\log \frac{\rho_+}{\rho_-}
\ee
\noi
and, as $\rho_+ \geqslant \rho_-$, the von Neumann entropy is not decreasing.

\noi
Vice versa, suppose $\frac{d}{dt} S[\rho(t)] \vert _{t = 0} \geqslant 0$
and choose $\rho^1 = \varepsilon$, $\rho^2 = \rho^3 = 0$, with $\vert
\varepsilon \vert \leqslant 1/2$. Then
\be
\la{eq2.3.5}
\frac{d}{dt}S[\rho(t)]\vert_{t = 0} = \frac{2}{\vert \varepsilon \vert}\left(
\frac{1}{2} u \varepsilon + a \varepsilon ^2\right) \log 
\frac{\rho_+}{\rho_-}.
\ee
Since $\vert u \vert \leqslant a$ (see Lemma~\ref{lem1.5.1}), 
the right hand side of~(\ref{eq2.3.5}) could be made negative by 
choosing ${\rm sign} (\varepsilon) = - {\rm sign} (u)$
and $\vert \varepsilon \vert \leqslant \vert u \vert / 2a$, unless
$u = 0$. The same argument implies $v = 0$ and $w = 0$ choosing $\rho^2 = 
\varepsilon$, $\rho^1 = \rho^3 = 0$ and respectively $\rho^3 = \varepsilon$, 
$\rho^1 = \rho^2 = 0$. Then eq.~(\ref{eq2.3.4}) holds, whose positivity imply
$\tilde{\cal D}_{ij} \geqslant 0$.}
\end{The}

\begin{The}
\la{the2.3.2}
{\rm The trace and Hermiticity preserving, entropy non decreasing, time 
evolution~(\ref{eq2.1.5}) is positive iff the real parameters of ${\cal D}$ 
satisfy the inequalities~(\ref{eq2.2.1}).

\noi
{\it Proof:}
The if part is a direct consequence of Theorem~\ref{the2.2.1}. 
Vice versa, considering a generic initial state $\vert \rho(0) \rangle
= (1/2, \rho^1, \rho^2, \rho^3)$ with $\Det [\rho(0)]\geqslant 0$,
we have 
\be
\la{eq2.3.6}
\frac{d}{dt} \Det [\rho(t)]\vert_{t = 0} = \sum_{i,j = 1}^{3}
\tilde{\cal D}_{ij} \rho^i(0)\rho^j(0) \geqslant 0
\ee
\noi
since $\tilde{\cal D}\geqslant 0$. Consequently $\Det [\rho(t)] \geqslant 0$,
$\forall t$; thus the evolution is positive (Remark~\ref{rem2.2.1}).
Indeed, suppose there is a time $\overline{t}$ for which 
$\Det [\rho(\overline{t})] < 0$. By continuity, there must be a time
$t^{\prime}$, with $0 < t^{\prime} < \overline{t}$ such that
$\Det [\rho(t^{\prime})] = 0$ and $d \Det [\rho(t^{\prime})]/dt < 0$.
But this can never happen, if $\tilde{\cal D} \geqslant 0$.}
\end{The}

\begin{The}
\la{the2.3.3}
{\rm \cite{alic1,benak,1benak,2benak,3benak,4benak,benab,benat} 
The trace and Hermiticity preserving, 
entropy non decreasing, time-evolution~(\ref{eq2.1.5}) is completely 
positive iff the real parameters of ${\cal D}$ satisfy the inequalities
\be
\la{eq2.3.7}
\left\{
\begin{array}{l}
2R \equiv \alpha + \gamma - a \geqslant 0 \\
2S \equiv a + \gamma - \alpha \geqslant 0 \\
2T \equiv a + \alpha - \gamma \geqslant 0 \\
\end{array} 
\right. ; \;
\left\{
\begin{array}{l}
RS \geqslant b^2 \\
RT \geqslant c^2 \\
ST \geqslant \beta^2 \\
\end{array}
\right. ; \;
RST - 2bc\beta -
S c^2 - 
R \beta^2 -
T b^2 \geqslant 0.
\ee
{\it Proof:}
The thesis follows from Theorem~\ref{the2.2.2} putting $u = v = w = 0$.}
\end{The}


\s{Perturbative expansion}
\la{s2.4}
\markright{\small{\rm{\thesection. PERTURBATIVE EXPANSION}}}

We want to integrate the equation of motion for $\rho$, i.e. we want to
solve either~(\ref{eq2.1.5}) or~(\ref{eq2.1.9}). In vectorial notation
the solutions will be denoted by

\be
\la{eq2.4.1}
\vert \rho(t) \rangle = {\cal G}_t \vert \rho(0) \rangle, \quad
{\cal G}_t \equiv e^{{\cal L}t} = e^{({\cal H}+{\cal D})t}
\ee

\noi
and

\be
\la{eq2.4.2}
\vert \hat{\rho}(t) \rangle = \hat{\cal G}_t \vert \hat{\rho}(0) \rangle, \quad
\hat{\cal G}_t \equiv e^{\hat{\cal L}t} = e^{(\hat{\cal H}+\hat{\cal D})t}
\ee

\noi 
respectively. Both ${\cal G}_t$ and $\hat{\cal G}_t$ are $4 \times 4$
matrices and $\hat{\cal G}_t = {\cal V} {\cal G}_t {\cal V}^{-1}$,
with ${\cal V}$ expressed in~(\ref{eq2.1.10}).

Although it is always possible, at least in principle, to compute the form of 
${\cal G}_t$ and $\hat{\cal G}_t$ without approximations, often these 
explicit expressions are rather involved and not particularly
useful in order to capture the essence of the dissipative processes.

Usually, however, the parameters appearing in the dissipative part 
of either~(\ref{eq2.1.5}) or~(\ref{eq2.1.9})
can be considered small with respect to those characterizing the Hamiltonian
contribution. It is thus possible to express both ${\cal G}_t$ and 
$\hat{\cal G}_t$ as series expansions in these parameters~\cite{benak,1benak,
2benak,3benak,4benak,benab},

\be
\la{eq2.4.3}
{\cal G}_t = {\cal G}_t^0 + {\cal G}_t^1 + {\cal G}_t^2 + \ldots
\ee

\noi
with

\be
\la{eq2.4.3.bis}
\begin{split}
&{\cal G}_t^0 \equiv e^{{\cal H}t}, \quad
{\cal G}_t^1 \equiv \int_0^t ds_1 e^{{\cal H}(t - s_1)}
{\cal D}e^{{\cal H}s_1}, \\
&{\cal G}_t^2 \equiv \int_0^t ds_1 \int_0^{s_1} ds_2 
e^{{\cal H}(t - s_1)}{\cal D}e^{{\cal H}(s_1 - s_2)}{\cal D}e^{{\cal H}s_2}, 
\quad \ldots
\end{split}
\ee

\noi
Equivalent expressions hold for $\hat{\cal G}_t$.

\begin{Rem}
\la{rem2.4.1}
{\rm Note that $\hat{\cal G}_t^j = {\cal V} {\cal G}_t^j {\cal V}^{-1}$ 
for any $j$, that is the similarity relation holds at any order in the 
perturbative expansion.}
\end{Rem}

In this section we write the time-evolutor to leading order 
in the dissipative coefficients.  Indeed, in the coming chapters, we
will encounter physical situations in which the dissipative parameters
are small with respect to the parameters describing 
the free time-evolution (i.e. without environment) and the formalism 
here developed will be applied. We perform 
the computation in the representation~(\ref{eq2.1.9}).

We suppose $H$ as general 
as possible, so that we allow for a non Hermitian part. Hence its 
eigenvalues $\epsilon_1$ and $\epsilon_2$ of $H$ can be complex and its 
eigenstates non orthogonal:

\be
\la{eq2.4.4}
\left\{
\begin{array}{l}
H \vert \epsilon_1 \rangle = \epsilon_1 \vert \epsilon_1 \rangle \\ \\
H \vert \epsilon_2 \rangle = \epsilon_2 \vert \epsilon_2 \rangle 
\end{array}
\right.
\quad
\epsilon_i = m_i - \frac{i}{2}\gamma_i \quad
m_i, \gamma_i \in {\bf R}, \quad
\langle \epsilon_1 \vert \epsilon_2 \rangle \neq 0;
\ee

\noi
$m_i$ and $\gamma_i$ are the energies, respectively the widths, of the 
eigenstates.

Since the underlying Hilbert space is two-dimensional, in any given basis
the two eigenvectors
can be written as $\vert \epsilon_1 \rangle = (p_1, q_1)$ and 
$\vert \epsilon_2 \rangle = (p_2, -q_2)$, with the normalization 
conditions

\be
\la{eq2.4.5}
\vert p_1 \vert^2 + \vert q_1 \vert^2 = 1, \;\;
\vert p_2 \vert^2 + \vert q_2 \vert^2 = 1.
\ee

\begin{Rem}
\la{rem2.4.2}
{\rm If $H$ is Hermitian then $\langle \epsilon_1 \vert \epsilon_2 \rangle = 0$
and $\gamma_i = 0$, that is $\epsilon_i \in {\bf R}$.
The condition of orthogonality of the two eigenstates reads
\be
\la{eq2.4.5.bis}
p_1 p_2^* - q_1 q_2^* = 0.
\ee }
\end{Rem}

It is convenient to work in the basis that diagonalizes the matrix 
$\hat{\cal H}$, hence $\hat{\cal D}$ will be transformed in a more 
complex matrix\footnote{Notice that the choice of the representation 
of the state and the choice of the basis of the Hilbert space are completely
independent}; the whole computation will result simpler~\cite{benab}. 
Accordingly we 
choose as basis for our bidimensional Hilbert space ${\mathfrak H}_2$ the 
pair of eigenstates of the Hamiltonian, given in~(\ref{eq2.4.4}). 
$H$ is diagonalized by means of an invertible transformation $U$
(unitary if $H$ is Hermitian); its diagonal form will be denoted by an 
upper index $d$, as any operator represented in the basis~(\ref{eq2.4.4}):

\be
\la{eq2.4.6}
H^d = U^{-1} H U; \quad
H^d = \left(
\begin{array}{cc}
\epsilon_1 & 0 \\
0 & \epsilon_2 \\
\end{array}
\right),
\quad
U = \left(
\begin{array}{cc}
p_1 & p_2 \\
q_1 & -q_2 \\
\end{array}
\right).
\ee

\noi
It is useful to introduce the two ratios

\be
\la{eq2.4.7}
r_1 \equiv \frac{p_1}{q_1}, 
\quad
r_2 \equiv \frac{p_2}{q_2};
\ee

\noi
this allows to write $U = R Q$, with

\be
\la{eq2.4.8.add}
R = \left(
\begin{array}{cc}
r_1 & r_2 \\
1 & -1 \\
\end{array}
\right),
\quad
Q = \left(
\begin{array}{cc}
q_1 & 0 \\
0 & q_2 \\
\end{array}
\right),
\ee

\noi
so that the similarity transformation 
connecting $H$ and $H^d$ depends solely on the parameters~(\ref{eq2.4.7}):

\be
\la{eq2.4.8}
H^d = R^{-1} H R.
\ee

\begin{Rem}
\la{rem2.4.3}
{\rm Note that $R$ is not unitary even if $H$ is Hermitian.
In this case $U$ is unitary and then $r_1 r_2^* = 1$.}
\end{Rem}

The evolution of a state $\rho$, expressed in~(\ref{eq2.1.4.add}) 
with Hamiltonian and dissipative contributions as in~(\ref{eq2.1.4})
or~(\ref{eq2.1.4.bis}), in the new basis takes the form

\be
\la{eq2.4.10}
\dot{\rho}^d(t) = (L_H^d + L_D^d)
[\rho^d(t)],
\ee

\noi
where

\be
\la{eq2.4.11}
\rho^d(t) = R^{-1} \rho(t) R^{-1 \dagger}, \;\;
L_{H,D}^d[\cdot] = R^{-1} L_{H,D}[R \cdot R^{\dagger}]R^{-1 \dagger}.
\ee

\noi
Using the vector representation of $\rho^d$ we have

\be
\la{eq2.4.12}
\frac{d}{dt}\vert \hat{\rho}^d(t) \rangle = \hat{\cal L}^d 
\vert \rho(t) \rangle = (\hat{\cal H}^d + 
\hat{\cal D}^d)\vert \hat{\rho}^d(t)\rangle,
\ee

\noi
where $\vert \hat{\rho}^d(t) \rangle$ is the vector whose components 
are the coefficients of the expansion of $\rho^d(t)$ in the  
basis~(\ref{eq2.1.1}). It is obtained from $\vert \hat{\rho}(t) \rangle$ 
by the linear transformation

\be
\la{eq2.4.13}
\vert \hat{\rho}^d(t) \rangle = {\cal R} \vert \hat{\rho}(t) \rangle,
\ee

\noi
whose explicit expressions can be obtained from~(\ref{eq2.4.11}):

\be
\la{eq2.4.14}
{\cal R} = \frac{1}{\vert r_1 + r_2 \vert ^2}
\left(
\begin{array}{cccc}
1 & \vert r_2 \vert ^2 & r_2^* & r_2 \\
1 & \vert r_1 \vert ^2 & -r_1^* & -r_1 \\
1 & -r_2 r_1^* & -r_1^* & r_2 \\
1 & -r_1 r_2^* & r_2^* & -r_1 \\
\end{array}
\right).
\ee

\noi
The $4 \times 4$ matrices in~(\ref{eq2.4.12}) can now be explicitly obtained:

\be
\la{eq2.4.15}
\hat{\cal H}^d = {\cal R} \hat{\cal H} {\cal R}^{-1} =
\left(
\begin{array}{cccc}
-\gamma_1 & 0 & 0 & 0 \\
0 & -\gamma_2 & 0 & 0 \\
0 & 0 & -\Gamma_- & 0 \\
0 & 0 & 0 & -\Gamma_+ \\
\end{array}
\right),
\ee

\noi
with $\Gamma_{\pm} \equiv \Gamma \pm i \Delta m$, $\Gamma \equiv
(\gamma_1 + \gamma_2)/2$ and $\Delta m \equiv m_2 - m_1$; and

\be
\la{eq2.4.16}
\hat{\cal D}^d = {\cal R} \hat{\cal D} {\cal R}^{-1} =
\left(
\begin{array}{cccc}
\Lambda + \lambda_1 & \Sigma + \sigma & 
\Delta + \delta_1 & \Delta^* + \delta_1^* \\ \\
\Xi + \xi & \Lambda + \lambda_2 & 
\Phi + \phi_1 & \Phi^* + \phi_1^* \\ \\
\phi_2 - \Phi^* & \delta_2 - \Delta^* & 
\Omega + \omega & \Theta + \theta \\ \\
\phi^*_2 - \Phi & \delta^*_2 - \Delta & 
\Theta^* + \theta^* & \Omega + \omega^* 
\end{array}
\right),
\ee

\noi
where small and capital Greek letters refer respectively to contributions 
related to the entropy behavior (i.e. functions of $u$, $v$ and $w$ 
in~(\ref{eq2.1.6})) and remaining parameters (i.e. those that are functions 
of $a$, $b$, $c$, $\alpha$, $\beta$ and $\gamma$ in~(\ref{eq2.1.6})). 
Their expressions are listed in Appendix~\ref{a1}, both for stable and unstable
systems.

\begin{Rem}
\la{rem2.4.4}
{\rm For later convenience we observe that $\lambda_1$, $\lambda_2$, $\sigma$, 
$\xi$ $\in {\bf R}$ and $\delta_1$, $\delta_2$, $\phi_1$, $\phi_2$, 
$\omega$, $\theta$ $\in {\bf C}$; on the other hand $\Lambda$, $\Xi$, 
$\Sigma$, $\Omega$ $\in {\bf R}$ and $\Delta$, $\Phi$, $\Theta$ $\in {\bf C}$.}
\end{Rem}

In the basis of the eigenvectors of $H$ and in the chosen vectorial 
representation, the evolution operator is denoted by $\hat{\cal G}^d_t$. 
Up to first order 
contributions in the dissipative parameters, $\hat{\cal G}^d_t = 
\hat{\cal G}^{d0}_t + \hat{\cal G}^{d1}_t$; from~(\ref{eq2.4.15}) we find

\be
\la{eq2.4.19}
\hat{\cal G}_t^{d0} = e^{\hat{\cal H}^d t} = 
\left(
\begin{array}{cccc}
e^{-\gamma_1 t} & 0 & 0 & 0 \\
0 & e^{-\gamma_2 t} & 0 & 0 \\
0 & 0 & e^{-\Gamma_- t} & 0 \\
0 & 0 & 0 & e^{-\Gamma_+ t} \\
\end{array}
\right).
\ee

The next order contribution is obtained, accordingly with~(\ref{eq2.4.3.bis}),
as

\begin{equation*}
\hat{\cal G}_t^{d1} = \int_0^t e^{\hat{\cal H}^d (t-s)}
\hat{\cal D}^d e^{\hat{\cal H}^d s}.
\end{equation*}

The operator $\hat{\cal G}_t^d$ so obtained contains terms which are 
linear in $t$; they disappear if we reconstruct the exponential
dependences out of them. For example, summing the contributions from 
$\hat{\cal G}^{d0}_t$ and $\hat{\cal G}^{d1}_t$, the entry in first 
column and row of $\hat{\cal G}^{d}_t$ for an unstable system reads 

\begin{equation*}
(\hat{\cal G}_t^d)_{11} = [1 + (\Lambda + \lambda_1)t] e^{-\gamma_1 t}
\approx e^{(\Lambda + \lambda_1)t} e^{-\gamma_1 t};
\end{equation*}

\noi
by looking at this expression we observe that, 
at the first order in the dissipative
contributions, what we physically mean by width of the state
$\vert \epsilon_1 \rangle$ is not the $\gamma_1$ appearing in the 
Hamiltonian $H$, but rather $\gamma_1 - \Lambda - \lambda_1$. 
Accordingly, we redefine $\gamma_1$ so that

\begin{equation*}
(\hat{\cal G}_t^d)_{11} = e^{-(\gamma_1 - \Lambda - \lambda_1)t}
\rightarrow e^{- \gamma_1 t}.
\end{equation*}

In the same way also $\gamma_2$ and $\Delta \Gamma_{\pm} \equiv \Delta 
\Gamma \pm 2i \Delta m$ get redefined.
The final, explicit expressions of the components of $\hat{\cal G}_t^d$,
both for stable or unstable systems, are collected in Appendix~\ref{a2}.

\begin{Rem}
\la{rem2.4.5}
{\rm In general, the $n$-th term in the expansion~(\ref{eq2.4.3}) contains
polynomials in $t$ of order $n$; these terms represent truncated exponential
behaviors and it is possible to consistently re-absorb them at any order 
in the perturbative expansion by reconstructing the exponentials they
represent. A truncated perturbative expansions does not
hold for a too long time since the neglected terms become more
and more relevant as the time passes, then the reabsorbtion of the 
contributions polynomial in $t$ could seem useless rather then physically 
justified. However, this procedure allows to
redefine the parameters $m_i$ and $\gamma_i$ by dissipative terms; 
in such a way they coincide with physical quantities (see Chapter~\ref{ch3}
for further details).}
\end{Rem}

\begin{Rem}
\la{rem2.4.6}
{\rm The time-evolutors in the physical basis are obtained as $\hat{\cal G}_t
= {\cal R}^{-1}\hat{\cal G}^d_t {\cal R}$ and ${\cal G}_t = {\cal V}^{-1}
\hat{\cal G}_t {\cal V}$ in the two different representations adopted.
However their explicit expressions are unnecessary for the later
developments; indeed in the formalism of density matrices the mean value 
of any observable, represented by a Hermitian operator 
$A$, is obtained by a trace operation over the evolving state
$\rho(t)$; this operation is invariant under any change of basis in the 
underlying Hilbert space:
\be
\la{eq2.4.20}
\langle A \rangle_{\rho} \equiv \Tr[A \rho(t)] = \Tr[A^d \rho^d(t)].
\ee
It turns out to be more convenient to represent the operator $A$ in 
the basis of the eigenstates of $H$ and to compute $\langle A \rangle_{\rho}$
via~(\ref{eq2.4.20}), rather than re-expressing $\rho^d(t)$
in the physical basis.}
\end{Rem}

\begin{Rem}
\la{rem2.4.7}
{\rm In the case of a stable system, the reabsorption of the terms 
linear in time is more difficult because it involves also the off diagonal 
entries of $\hat{\cal G}_t^d$. The final expressions are reported in 
Appendix~\ref{a2}.}
\end{Rem}

These results will be used in the coming chapters for the description
of the phenomenology of stable and unstable particles immersed
in various specific environments.

%% file: chapter3.tex
\chapter{Neutral {\bf B} Mesons}
\la{ch3}
\markboth{\small{\rm{\thechapter. NEUTRAL {\bf B} MESONS}}}{}

In this chapter we concentrate on the study of the neutral {\bf B} mesons
system treated as an open system.

This system is presently under both theoretical and experimental 
investigation since it enables the study of some specific properties
of the Standard Model, e.g. the topic of the violation of the CP 
symmetry~\cite{bran,bigi} (charge conjugation plus space reversal), but 
also of the CPT symmetry (CP plus time reversal) and the flavors 
oscillations. In particular single {\bf B} mesons or entangled 
pairs of {\bf B} mesons physics are intensively studied at the colliders 
or at the so-called {\bf B} mesons factories~\cite{baba}.

Usually, the neutral {\bf B} mesons system is considered as a closed system. 
Instead, in the following, we assume this 
system to be immersed in a thermal bath of gravitational origin. 
We shall adopt a phenomenological approach, that does not make any hypothesis 
about the microscopic nature of the interaction between 
{\bf B} mesons and environment; we suppose solely the coupling to be weak
and the entropy non-decreasing. It turns out that the resulting 
irreversible dynamics for the {\bf B} mesons can be parameterized by 
six phenomenological coefficients that affect in a very specific way the
relevant physical quantities (decays rates and asymmetries)~\cite{benab}. 
In principle, an experimental study of these quantities enable us to 
evaluate the magnitude of these dissipative parameters. 

As usual, we shall describe our system by a bidimensional Hilbert space,
using the formalism developed in Chapter~\ref{ch2} and in particular the 
time evolution operator calculated in Section~\ref{s2.4}.

This dissipative approach has been also applied to describe the neutral 
{\bf K} mesons system~\cite{mari,1mari,2mari,elli2,benak,1benak,2benak,
3benak,4benak}. 
For another approach
to decoherence in the neutral {\bf B} mesons system, characterized by
only one dissipative parameter, see~\cite{bert}.


\s{The physics of {\bf B} mesons}
\la{s3.1}
\markright{\small{\rm{\thesection. THE PHYSICS OF {\bf B} MESONS}}}

In the standard phenomenological approach
the neutral {\bf B} meson system can be described by means of a 
state or, more in general, a statistical operator, in a
two-dimensional Hilbert space ${\mathfrak H}_2$. The effective Hamiltonian 
$H$ of this system is a non Hermitian operator of the Wigner-Weisskopf 
type~\cite{bran,bigi}, 

\be
\la{eq3.1.0}
H = M - \frac{i}{2} \it{\Gamma}; \quad \quad
H = 
\left(
\begin{array}{cc}
H_1 & H_3 \\
H_4 & H_2 \\
\end{array}
\right)
\ee

\noi
Several basis in ${\mathfrak H}_2$ are of interest for the discussion 
that follows:

\bi
\item The set \{$\bz$, $\abz$\} of the physical states (called also flavor 
states); 
\item The set \{$\bl$, $\bh$\} of the eigenstates of the Hamiltonian 
of the system:

\be
\la{eq3.1.1}
\left\{
\begin{array}{l}
H \bl = \epsilon_L \bl, \quad \epsilon_L = m_L - \frac{i}{2}\gamma_L \\
H \bh = \epsilon_H \bh, \quad \epsilon_H = m_H - \frac{i}{2}\gamma_H \\
\end{array}
\right.
\ee

\noi
where $m$ and $\gamma$ denote respectively the masses and the widths 
of the two eigenstates and the subscripts $L$, $H$ means ``light'', 
``heavy'' respectively, since it turns out that 
$m_L \ll m_H$~\cite{bran,bigi}.
These states are not ortogonal: $\langle {\rm {\bf B}}_L\bh \ne 0$;

\item The set \{$\bo$, $\bt$\} of the eigenstates of $CP$, the operator
performing the CP transformation (charge conjugation plus parity):

\be
\la{eq3.1.2}
\begin{split}
CP \bo &= \bo  \\
CP \bt &= - \bt \\
\end{split}
\ee

The eigenvalues of $CP$ are $1$ and $-1$ since $(CP)^2 = \sigma_0$. 
The $CP$ operator exchange the two physical states:

\be
\la{eq3.1.2.bis}
\begin{split}
CP \bz &= e^{i \varphi} \abz  \\
CP \abz &= e^{-i \varphi} \bz
\end{split}
\ee

\noi
and $\varphi$ is a phase depending on the convention adopted in defining
the physical states.
 
\ei

\begin{Rem}
\la{rem3.1.1}
{\rm In the treatement of kaon phenomenology
the distinction between the two eigenstates of $H$ is based on the
large difference between the two widths and the terms ``short'' and
``long'' are used.}
\end{Rem} 

\noi
The relations between the basis~(\ref{eq3.1.1}) and (\ref{eq3.1.2}) are:

\be
\la{eq3.1.3}
\begin{split}
\bl &= p_L \bz + q_L \abz , \quad \vert p_L \vert^2 + \vert q_L \vert^2 = 1 \\
\bh &= p_H \bz - q_H \abz , \quad \vert p_H \vert^2 + \vert q_H \vert^2 = 1 \\
\end{split}
\ee

\noi
and

\be
\la{eq3.1.3.bis}
\begin{split}
\bo &= \frac{1}{\sqrt{2}} (\bz + \abz) \\
\bt &= \frac{1}{\sqrt{2}} (\bz - \abz) \\
\end{split}
\ee

\noi
The distinction between these three basis is one of the reasons of the 
physical relevance of the {\bf B} mesons system. 

Since the physical states, 
produced in the interactions, are linear superpositions of the eigenstates
of $H$, an initially pure state, e.g. $\bz$, would develope a 
non-vanishing component on the other state, e.g. $\abz$, as time passes. 
We have then transitions between the
states $\bz$ and $\abz$, called flavor oscillations, as the probability of
transition between the two flavors behaves like a periodic function.

On the other side the distinction between eigenstates of $CP$ and $H$ is a 
consequence of CP violation: it is not possible to diagonalise at the same 
time both these operators. In the standard Model, CP violation is due to the 
presence, in the Cabibbo Kobayashi Maskawa matrix\footnote{This is the matrix 
giving the mixing in the hadronic sector of the Standard Model}, of a 
complex phase that 
cannot be reabsorbed by a redefinition of the fields describing the 
quarks~\cite{bran,bigi}; this violation is
expected to be greater in the {\bf B} mesons sector with respect to the {\bf K}
mesons case. CP invariance is equivalent to $\vert p_H \vert = 
\vert q_H \vert = \vert p_L \vert = \vert q_L \vert$.

The adopted phenomenological approach can further account for violations
of other symmetries that are preserved by the Standard Model, like CPT 
invariance and the phenomenological rule $\Delta B = \Delta Q$. In this relation, $B$ 
is the beauty quantum number and $Q$ the charge of the decaying quarks.

Further, it is possible
to describe time-evolutions different from the standard ones, taking
into account a possible interaction with an external
environment. As discussed below, the modifications 
of the relevant quantities produced by the irreversible dynamics turn out 
to have a characteristic signature: the open system dynamics cannot 
be confused with the others possible symmetries violations. To be as 
general as possible in our discussion, we shall consider all of these 
different possibilities together.

\noi
After the definition of the two ratios

\be
\la{eq3.1.5}
r_L = \frac{p_L}{q_L}, \quad
r_H = \frac{p_H}{q_H}; 
\ee

\noi
the Hamiltonian~(\ref{eq3.1.0}) can be rewritten as

\be
\la{eq3.1.6}
H = \frac{1}{r_L + r_H}
\left(
\begin{array}{cc}
r_L \epsilon_L + r_H \epsilon_H & r_L r_H (\epsilon_L - \epsilon_H) \\ \\
\epsilon_L - \epsilon_H & r_L \epsilon_H + r_H \epsilon_L \\
\end{array}
\right).
\ee

For future reference we find useful to define three complex parameters,
$\sigma$, $\xi$ and $\theta$~\cite{bran,bigi}:

\be
\la{eq3.1.7}
\sigma = r_H r_L = \frac{H_3}{H_4}, \quad 
\xi = \frac{\vert \sigma \vert - 1}{\vert \sigma \vert + 1}, \quad
\theta = \frac{r_L - r_H}{r_L + r_H} = \frac{H_1 - H_2}{\epsilon_L
- \epsilon_H};
\ee

\noi 
as we shall see in the coming sections,
by means of them it is possible to parametrize in a convenient way
violations of the discrete symmetries.


\s{The gravitational bath}
\la{s3.2}
\markright{\small{\rm{\thesection. THE GRAVITATIONAL BATH}}}

The physical motivation of an open quantum system treatement of the neutral
{\bf B} mesons system (or a generic neutral mesons system) comes from
quantum gravity. A satisfactory and complete theory of quantum gravity 
has not been found yet. Nevertheless, general physical arguments involving 
quantum fluctuations of the gravitational field lead to the conclusion
that space-time could lose its continuum character at the Planck's scale. 
Several model have been proposed to account for this striking conclusion, 
for example the existence, at the Planck's scale, of virtual black
holes~\cite{hawk,1hawk,2hawk,3hawk}; 
also the dynamics of string and branes could cause the appearance
of a foamy nature of the space-time at microscopic 
scales~\cite{elli1,1elli1,bena1}. In all these cases
the space-time could act as an effective environment and all the physical
systems would necessarily interact with it. Since the degrees of freedom 
characterizing such an environment are not accessible, any physical system
should exhibit an open system dynamics, leading to loss of quantum
coherence (decoherence)~\cite{mari,1mari,2mari,hawk,elli2,cole,gidd,sred,unru,
gara,1gara}.

The microscopic details of the interaction between physical systems and
gravitational bath are unnecessary to parametrize this dissipative evolution:
we adopt a phenomenological approach, that is independent on the 
underlying fundamental dynamics. Indeed, the form of the reduced dynamics
can be uniquely fixed by the requirements of being a completely positive, 
Markovian, entropy increasing evolution.

The Markovian hypothesis is justified by the fact that the action of the
gravitational thermal bath on the neutral mesons system is very weak.
The magnitude of the dissipative parameters may be estimeted to be 
of order $E_S^2/ E_B$, 
where $E_S$ is the characteristic energy of the system and $E_B$ is the 
fundamental scale for the bath. In our case  
$E_S = m_{\bf B} \approx 5~{\rm GeV}$, the neutral {\bf B} mesons mass, 
and $E_B = M_P \approx 10^{19}~{\rm GeV}$, the Planck mass (the scale
at which the space-time is expected to exhibit a kind of granularity). The
magnitude of the dissipative parameters has thus an upper bound of
about $10^{-18}~{\rm GeV}$. The Markovian dynamics is justified, since
this scale is very small respect to the scale characterizing the usual
Hamiltonian dynamics of the {\bf B} mesons system.

\begin{Rem}
\la{rem3.2.1}
{\rm The gravitational thermal bath should affect any physical system.
However, its effects on the majority of these systems would be completely
negligible, so the systems characterized by energies in the sub-nuclear range 
or higher are the only that could feel it. For example, for atomic phenomena,
characterized by energies of about $100~{\rm KeV}$, the dissipative 
parameters would be as small as $10^{-27}~{\rm GeV}$.}
\end{Rem}

The hypothesis of entropy non decreasing evolution is based on the 
choice of time-evolutions leading to states less-ordered in time.
The correlations contained in the coherence of states are lost as time
passes.

The accuracy of present experiments on neutral {\bf B} mesons is very
high, so that one expects the dissipative effects induced by gravitational 
phenomena to be in the reach of present or planned experiments.
This is certainly one of the motivations that justify the coming
considerations.


\s{The dissipative evolution}
\la{s3.3}
\markright{\small{\rm{\thesection. THE DISSIPATIVE EVOLUTION}}}

The equation of motion for the state $\rho$ representing the {\bf B} mesons
is assumed to have the form of a quantum dynamical semigroup, with a
non standard contribution $L_D$ in the generator, due to the gravitational
bath:

\be
\la{eq3.3.1}
\dot{\rho}(t) = -i H \rho(t) + i \rho(t) H^{\dagger} + L_D[\rho(t)].
\ee

\noi
with $H$ as in~(\ref{eq3.1.0}) and $L_D$ expressed in~(\ref{eq2.1.4.bis}).

Working in the basis of physical states \{$\bz, \abz$\} and using
the vector notation introduced in the previous chapter, we have 

\be
\la{eq3.3.2}
\frac{d}{dt}\vert \hat{\rho}(t) \rangle = (\hat{\cal H} + 
\hat{\cal D}^{\prime}) \vert \hat{\rho}(t) \rangle
\ee

\noi
with

\be
\la{eq3.3.3}
\hat{\cal H} = 
\left(
\begin{array}{cccc}
2 \im H_1 & 0 & i H_3^* & -i H_3 \\
0 & 2 \im H_2 & -i H_4 & i H_4^* \\
i H_4^* & -i H_3 & i (H_2^* - H_1) & 0 \\
-i H_4 & i H_3^* & 0 & i (H_1^* - H_2) \\ 
\end{array}
\right)
\ee

\noi
the matrix containing the standard Hamiltonian terms, the masses being 
redefined to take into account possible Hamiltonian dissipative 
contributions, while the matrix containing the non Hamiltonian contribution, 
causing decoherence of states and entropy increase, is

\be
\la{eq3.3.4}
\hat{\cal D}^{\prime} = 
\left(
\begin{array}{cccc}
-a & a & -C^{\prime} & -C^{\prime *} \\
a & -a & C^{\prime} & C^{\prime *} \\
-C^{\prime *} & C^{\prime *} & -A^{\prime} & B^{\prime} \\
-C^{\prime} & C^{\prime} & B^{\prime *} & -A^{\prime} \\
\end{array}
\right)
\ee

\noi
where

\be
\la{eq3.3.5}
A^{\prime} \equiv \gamma + \alpha, \quad
B^{\prime} \equiv \alpha - \gamma - 2i\beta, \quad
C^{\prime} \equiv c - i b.
\ee

The matrix in~(\ref{eq3.3.4}) differs from that appearing
in~(\ref{eq2.1.12}) because here we assume the dissipative contribution
to have the form ${\cal D}$ in~(\ref{eq2.1.6}), with $u = v = w = 0$; 
$\hat{\cal D}^{\prime}$ is obtained from $\hat{\cal D}$ in~(\ref{eq2.1.12})
by a change of basis, from the eigenstates of $CP$ to 
the basis of physical states.

The dissipative parameters $a$, $b$, $c$, $\alpha$, $\beta$ and $\gamma$
are small with respect to the parameters characterizing the 
Hamiltonian contribution; then it is possible to apply the formalism 
developed in Section~\ref{s2.4}. The only difference is the use of 
$\hat{\cal D}^{\prime}$ instead of $\hat{\cal D}$.
The parameters $a$, $A^{\prime}$, 
$B^{\prime}$ and $C^{\prime}$ take the place of $\gamma$, $A$, $B$ and 
$C$ respectively. 

\noi
In the basis \{$\bl, \bh$\}, that diagonalizes $H$, the statistical 
operator becomes

\be
\la{eq3.3.ciccio}
\rho^d = R^{-1} \rho R^{-1 \dagger}, \quad
R = \left(
\begin{array}{cc}
r_L & r_H \\
1 & -1 \\
\end{array}
\right)
\ee

\noi
while~(\ref{eq3.3.2}) takes the form

\be
\la{eq3.3.2.bis}
\frac{d}{dt}\vert \hat{\rho}^d(t) \rangle = (\hat{\cal H}^d + 
\hat{\cal D}^{\prime d}) \vert \hat{\rho}^d(t) \rangle
\ee

\noi
where $\hat{\cal H}^d$ is a diagonal Hamiltonian part,
see~(\ref{eq2.4.15}), and $\hat{\cal D}^{\prime d}$ is a more involved
dissipative part:

\be
\la{eq3.3.6}
\hat{\cal D}^{\prime d} = 
\left(
\begin{array}{cccc}
\Lambda^{\prime} & \Sigma^{\prime} & 
\Delta^{\prime} & \Delta^{\prime *} \\
\Xi^{\prime} & \Lambda^{\prime} & 
\Phi^{\prime} & \Phi^{\prime *} \\
- \Phi^{\prime *} & - \Delta^{\prime *} & 
\Omega^{\prime} & \Theta^{\prime} \\
- \Phi^{\prime} & - \Delta^{\prime} & 
\Theta^{\prime *} & \Omega^{\prime} \\
\end{array}
\right);
\ee

\noi
the coefficients $\Lambda^{\prime}$, $\Sigma^{\prime}$, $\Xi^{\prime}$,
$\Delta^{\prime}$, $\Phi^{\prime}$, $\Theta^{\prime}$ and $\Omega^{\prime}$
are expressed as the corresponding
parameters in Appendix~\ref{a1} with the substitutions $\gamma 
\rightarrow a$, $A \rightarrow A^{\prime}$, $B \rightarrow B^{\prime}$ and
$C \rightarrow C^{\prime}$. They are linear superpositions of
$a$, $b$, $c$, $\alpha$, $\beta$ and $\gamma$ with coefficients that depend 
on $r_L$ and $r_H$, or, equivalently, on $\theta$ and $\sigma$. In 
the next section we will show that $\theta$ and $\sigma$ are related to
violations of the discrete symmetries, that are expected to be small;
it is then justified to neglect the terms where these coefficients
multiply the dissipative parameters. We finally find

\begin{eqnarray}
\la{eq3.3.6.bis}
\Lambda^{\prime} &=& \frac{1}{2}\Bigl\{ \re[(\alpha - \gamma + 2i 
\beta)\sigma] - (\alpha + \gamma) \Bigr\}, \nonumber\\
\Sigma^{\prime} &=& - \Lambda^{\prime}, \nonumber\\
\Delta^{\prime} &=& - \frac{1}{2}\Bigl\{ 2 \re [(c - ib)\sqrt{\sigma}] + 
i \im [(\alpha - \gamma + 2i \beta) \sigma] \Bigr\}, \nonumber\\
\Xi^{\prime} &=& - \Lambda^{\prime}, \\
\Phi^{\prime} &=& - \Delta^{\prime}, \nonumber\\
\Omega^{\prime} &=& - \frac{1}{2}\Bigl\{ (\alpha + \gamma) + 2a +
\re [(\alpha -\gamma + 2i \beta) \sigma] \Bigr\}, \nonumber\\
\Theta^{\prime} &=& - \frac{1}{2}\Bigl\{ 2a - (\alpha + \gamma) -
\re [(\alpha - \gamma + 2i \beta)\sigma] + 4i 
\im [(c - ib) \sqrt{\sigma}] \Bigr\}. \nonumber
\end{eqnarray}

Since we have request that entropy be non decreasing, the remaining 
coefficients in Appendix~\ref{a1} vanish. (see Section~\ref{s2.3})

\noi
By means of the coefficients~(\ref{eq3.3.6.bis})
we can compute the time evolutor $\hat{\cal G}_t^d$ acting on our
states; see Appendix~\ref{a2}:

\be
\la{eq3.3.7}
\vert \hat{\rho}_{\rm{\bf B}^0}^d (t) \rangle = 
\hat{\cal G}_t^{d} \vert \hat{\rho}_{\rm{\bf B}^0}^d (0) \rangle,
\quad
\vert \hat{\rho}_{\overline{\rm {\bf B}}^0}^d (t) \rangle = 
\hat{\cal G}_t^{d} \vert \hat{\rho}_{\overline{\rm {\bf B}}^0}^d (0)
\rangle
\ee

\noi
with the initial conditions

\be
\la{eq3.3.8}
\rho _{\rm{\bf B}^0} (0) = 
\bz \langle {\rm{\bf B}^0} \vert, \quad
\rho _{\overline{\rm{\bf B}}^0} (0) = 
\abz \langle {\overline{\rm{\bf B}}^0} \vert.
\ee

The operators $\rho_{\rm{\bf B}^0}^d(t)$ and 
$\rho_{\overline{\rm {\bf B}}^0}^d(t)$ so obtained will be used to compute 
the decay rates of the neutral {\bf B} mesons.

As alredy observed in Remark~\ref{rem2.4.6}, it is convenient to perform 
all the computations in the basis of the eigenstates of $H$.


\s{Invariance properties}
\la{s3.4}
\markright{\small{\rm{\thesection. INVARIANCE PROPERTIES}}}

Any symmetry transformation on our system can be modeled by a unitary
or antiunitary transformation acting on the basis vectors $\bz$ and
$\abz$. In our formalism, we have to consider the corresponding 
transformation on the vector $\vert \hat{\rho} \rangle$, realized
by a $4\times 4$ matrix $\hat{\cal U}$:

\be
\la{eq3.4.1}
\vert \hat{\rho} \rangle \rightarrow \hat{\cal U} \vert \hat{\rho} \rangle .
\ee

\noi
The request of invariance of the formalism under the transformation
expressed by $\hat{\cal U}$ has the form

\be
\la{eq3.4.2} 
\hat{\cal U}(\hat{\cal H} + \hat{\cal D}^{\prime})\hat{\cal U}^{-1} = 
(\hat{\cal H} + \hat{\cal D}^{\prime}), \quad
\hat{\cal U}(\hat{\cal H} + \hat{\cal D}^{\prime})\hat{\cal U}^{-1} = 
(\hat{\cal H} + \hat{\cal D}^{\prime})^{\dagger}
\ee

\noi
for unitary, respectively antiunitary, transformations.

\noi
For the discrete CPT, T and CP transformations the form of the
matrix $\hat{\cal U}$ is

\be
\la{eq3.4.3}
\hat{\cal U}_{CPT} =
\left(
\begin{array}{cc}
\sigma_1 & 0 \\
0 & \sigma_0 \\
\end{array}
\right), \quad
\hat{\cal U}_{T} =
\left(
\begin{array}{cc}
\sigma_0 & 0 \\
0 & \sigma_{\varphi} \\
\end{array}
\right), \quad
\hat{\cal U}_{CP} =
\left(
\begin{array}{cc}
\sigma_1 & 0 \\
0 & \sigma_{\varphi} \\
\end{array}
\right),
\ee

\noi
where the $\sigma_i$ are the Pauli matrices and the $2\times 2$
identity and

\be
\la{eq3.4.4}
\sigma_{\varphi} = 
\left(
\begin{array}{cc}
0 & e^{-2i \varphi} \\
e^{2i \varphi} & 0 \\
\end{array}
\right)
\ee

\noi
contains the phase appearing in the CP transformation of the physical
states, eq.~(\ref{eq3.1.2.bis}).

\noi
In the standard case $\hat{\cal D}^{\prime} = 0$;
the conditions of invariance under these discrete transformations reads

\be
\la{eq3.4.5}
r_L = r_H \Rightarrow \theta = 0
\ee

\noi
for the CPT symmetry,

\be
\la{eq3.4.6}
\vert r_L r_H \vert = 1 \Rightarrow \vert \sigma \vert = 1, \; 
\xi = 0
\ee

\noi
for the T symmetry; the CP invariance is obtained imposing both CPT
and T invariance, 

\be
\la{eq3.4.7}
r_L = r_H = \sqrt{\sigma}, \; \sigma = e^{-2i \varphi}.
\ee

\begin{Rem}
\la{rem3.4.1}
{\rm The parameters $\sigma$, $\xi$ and $\theta$ characterize 
possible violations of the discrete symmetries during the time evolution of 
our system; they are called violations in mixing to distinguish them from
violations of the same symmetries that can occour during the decay of
the {\bf B} mesons and which are due to the decay mechanism 
(see Section~\ref{s3.5}).}
\end{Rem}
 
Also the dissipative part $\hat{\cal D}^{\prime}$ can be resposable for 
violations of the discrete symmetries; choosing $\hat{\cal H} = 0$ the 
conditions of invariance are

\begin{subequations}
\la{eq3.4.8} 
\begin{gather}
\la{eq3.4.8.a}
c = b = 0 \\
\la{eq3.4.8.b}
C^{\prime} = e^{2i \varphi} C^{\prime *},\quad 
B^{\prime *} = e^{4i \varphi} B^{\prime} \\
\la{eq3.4.8.c}
C^{\prime} = - e^{2i \varphi} C^{\prime *}, \quad B^{\prime *} = 
e^{4i \varphi} B^{\prime} 
\end{gather}
\end{subequations}

\noi
holding for CPT, T and CP transformations respectively.

We consider now the invariance properties of our formalism under
a redefinition of the phases of the two physical states.
If the states $\bz$ and $\abz$ undergo an independent
change of phase,

\be
\la{eq3.4.9}
\bz \rightarrow e^{i \phi} \bz, \quad 
\abz \rightarrow e^{i \overline{\phi}} \abz,  
\ee

\noi
correspondingly the $4$-vector $\vert \hat{\rho} \rangle$ undergoes
the transformation described by

\be
\la{eq3.4.10} 
{\cal U}_{\phi} = 
\left(
\begin{array}{cccc}
1 & 0 & 0 & 0 \\
0 & 1 & 0 & 0 \\
0 & 0 & e^{i(\phi - \overline{\phi})} & 0 \\
0 & 0 & 0 & e^{-i(\phi - \overline{\phi})} \\
\end{array}
\right),
\ee

\noi
that causes a phase shift of the off-diagonal entries of the density matrix 
$\rho$:

\be
\la{eq3.4.11}
\rho_3 \rightarrow e^{i(\phi - \overline{\phi})} \rho_3, \quad
\rho_4 \rightarrow e^{-i(\phi - \overline{\phi})} \rho_4.
\ee

This transformation is not an invariance for $\hat{\cal H}$; indeed, the
off-diagonal elements $H_3$ and $H_4$ of the effective Hamiltonian $H$ do
change. In particular $r_H$, $r_L$ are found to transform as

\be
\la{eq3.4.12}
r_L \rightarrow e^{i(\phi - \overline{\phi})} r_L, \quad
r_H \rightarrow e^{i(\phi - \overline{\phi})} r_H;
\ee

\noi
consequentely,

\be
\la{eq3.4.13}
\sigma \rightarrow e^{2i(\phi - \overline{\phi})} \sigma
\ee

\noi
whereas $\xi$ and $\theta$ are invariant. The phase redifinition
is not an invariance also for $\hat{\cal D}^{\prime}$: the
dissipative coefficients are found to transform as

\be
\la{eq3.4.14}
C^{\prime} \rightarrow e^{-i(\phi - \overline{\phi})} C^{\prime}, \quad
B^{\prime} \rightarrow e^{2i(\phi - \overline{\phi})} B^{\prime}
\ee

\noi
whereas $a$ and $A^{\prime}$ are invariant. Then equation~(\ref{eq3.3.2}) 
is not invariant under ${\cal U}_{\phi}$, although the physically
meaningful quantities (i.e. mean values of observables) are. 
Some problems can arise when this
equation is solved by perturbative methods: the physically meaningful 
quantities could exhibit an explicit dependence on the phases $\phi$ and 
$\overline{\phi}$.

However, combining the results (\ref{eq3.4.12}) and (\ref{eq3.4.14}) it is 
possible to show that the entries of the matrix $\hat{\cal D}^{\prime d}$ 
are rephasing invariant. The same is true for the matrix $\hat{\cal H}^d$
since it is diagonal. These results ensure the invariance 
of our formalism under any phase redefinition, also for a truncated 
perturbative expansion of the time evolutor. This is another advantage in
the use of the eigenstates of the effective Hamiltonian as basis
of the two-dimensional Hilbert space.


\s{Observables}
\la{s3.5}
\markright{\small{\rm{\thesection. OBSERVABLES}}}

In this section we want to introduce the observables usually used to
describe the physics of the neutral {\bf B} mesons system. In the formalism
of density matrices, these are represented by Hermitian operators 
$O_f$ related to the decay of the {\bf B} mesons into a generic final 
state $f$. In the $\bz$, $\abz$ basis, these operators are represented 
by $2 \times 2$ matrices:

\be
\la{eq3.5.1}
O_f = \left(
\begin{array}{cc}
O_1 & O_3 \\
O_4 & O_2 \\
\end{array}
\right)
\ee

\noi
whose entries, by construction, are related to the two independent decay
amplitudes ${\cal A}({\bf B}^0 \rightarrow f)$ and 
${\cal A}(\overline{\bf B}^0 \rightarrow f)$: 

\be
\la{eq3.5.2}
\begin{split}
O_1 = \vert {\cal A}({\bf B}^0 \rightarrow f) \vert^2, \quad
O_3 = [{\cal A}({\bf B}^0 \rightarrow f)]^* 
{\cal A}(\overline{\bf B}^0 \rightarrow f), \\
O_2 = \vert {\cal A}(\overline{\bf B}^0 \rightarrow f) \vert^2, \quad
O_4 = [{\cal A}(\overline{\bf B}^0 \rightarrow f)]^* 
{\cal A}({\bf B}^0 \rightarrow f).
\end{split}
\ee

\noi
with

\be
\la{eq3.5.2.bis}
{\cal A}({\bf B}^0 \rightarrow f) \equiv 
\langle f \vert T \vert {\bf B}^0 \rangle, \quad
{\cal A}(\overline{\bf B}^0 \rightarrow f) \equiv 
\langle f \vert T \vert \overline{\bf B}^0 \rangle,
\ee

\noi
$T$ being the scattering operator.

Accordingly to~(\ref{eq1.1.3}), the mean value of $O_f$ is obtained 
by a trace operation over the state describing the system at time $t$, 
regardless of the basis of the Hilbert space considered:

\be
\la{eq3.5.3}
\langle O_f(t) \rangle = \Tr [O_f \rho(t)] = \Tr [O_f^d \rho^d(t)],
\ee

\noi
where $O_f^d$ is the matrix representation of $O_f$ in the $\bl$, $\bh$ 
basis:
$O_f^d = R^{\dagger} O_f R$ and $R$ as in~(\ref{eq3.3.ciccio}). 
This mean value gives the probability rate for the decay of the
{\bf B} meson into the final state $f$.

In the analysis of the decays of neutral {\bf B} mesons, it is 
customary to introduce the following rephasing invariant 
phenomenological parameters~\cite{bran,bigi}:

\be
\la{eq3.5.4}
\begin{split}
\lambda^f_L \equiv \frac{q_L}{p_L} \frac{{\cal A}
(\overline{\bf B}^0 \rightarrow f)} 
{{\cal A}({\bf B}^0 \rightarrow f)}, \quad 
\mu^f_L \equiv \frac{1}{\lambda^f_L}; \\
\lambda^f_H \equiv \frac{q_H}{p_H} \frac{{\cal A}
(\overline{\bf B}^0 \rightarrow f)} 
{{\cal A}({\bf B}^0 \rightarrow f)}, \quad 
\mu^f_H \equiv \frac{1}{\lambda^f_H}. 
\end{split}
\ee

\noi
By means of them the matrix $O_f^d$ can be written as

\be
\la{eq3.5.5}
\begin{split}
O_f^d &= \vert {\cal A}({\bf B}^0 \rightarrow f) \vert^2
\left(
\begin{array}{cc}
\vert r_L \vert^2 \vert 1 + \lambda^f_L\vert^2 &
r_H r_L^* (1 - \lambda^f_H)(1 + \lambda^f_L)^* \\ \\
r_L r_H^* (1 + \lambda^f_L)(1 - \lambda^f_H)^* &
\vert r_H \vert^2 \vert 1 - \lambda^f_H\vert^2 \\
\end{array}
\right) \\ \\
&= \vert {\cal A}(\overline{\bf B}^0 \rightarrow f) \vert^2
\left(
\begin{array}{cc}
\vert 1 + \mu_L^f \vert^2 &
- (1 - \mu_H^f)(1 + \mu_L^f)^* \\ \\
- (1 + \mu_L^f)(1 - \mu_H^f)^* &
\vert 1 - \mu_H^f \vert^2 \\
\end{array}
\right).
\end{split}
\ee

A class of decays widely studied is the set of the so-called
semileptonic decays, namely the decays where the final state $f$
is constituted by a charged hadron $h$, a charged lepton $l$ and
its associated neutrino $\nu$: ${\bf B}^0$, $\overline{\bf B}^0 \rightarrow 
h^- l^+ \nu$, $h^+ l^- \overline{\nu}$. In this case, it is convenient
to parametrize the decay amplitudes as follows:

\begin{subequations}
\la{eq3.5.6}
\begin{gather}
\la{eq3.5.6.a}
{\cal A}({\bf B}^0 \rightarrow h^- l^+ \nu) = {\cal M}_h (1 - y_h), \\
\la{eq3.5.6.b}
{\cal A}(\overline{\bf B}^0 \rightarrow h^+ l^- \overline{\nu})
= {\cal M}^*_h (1 + y_h^*), \\
\la{eq3.5.6.c}
{\cal A}({\bf B}^0 \rightarrow h^+ l^- \overline{\nu}) = 
z_h {\cal A}(\overline{\bf B}^0 \rightarrow h^+ l^- \overline{\nu}), \\
\la{eq3.5.6.d}
{\cal A}(\overline{\bf B}^0 \rightarrow h^- l^+ \nu) = 
x_h {\cal A}({\bf B}^0 \rightarrow h^- l^+ \nu), 
\end{gather}
\end{subequations}

\noi
where ${\cal M}_h$ is a common factor. Possible violations of CPT invariance
and of the $\Delta B = \Delta Q$ rule are parametrized by the coefficients 
$x_h$, $y_h$ and $z_h$. In particular, non vanishing $x_h$ and $z_h$ indicate 
violations of $\Delta B = \Delta Q$, because this
rule would forbid the decays ${\bf B}^0 \rightarrow h^+ l^- \overline{\nu}$
and $\overline{\bf B}^0 \rightarrow h^- l^+ \nu$, while a non-vanishing $y_h$
signals violation of CPT invariance, since this symmetry implies equal
probability for the processes ${\bf B}^0 \rightarrow h^- l^+ \nu$
and $\overline{\bf B}^0 \rightarrow h^+ l^- \overline{\nu}$. 

We assume the violations of the discrete symmetries (in mixing or in decay) 
and of the $\Delta B = \Delta Q$ rule to be small. In this way we 
can neglect all terms in wich the coefficients $\theta$, $\xi$, 
$x_h$, $y_h$, $z_h$ are multiplied with the dissipative
parameters $a$, $b$, $c$, $\alpha$, $\beta$ and $\gamma$. With this
assumption, for the semileptonic decays, the parameters introduced 
in~(\ref{eq3.5.4}) assume a simplified form:

\be
\la{eq3.5.7}
\lambda_L^{h^-} = \lambda_H^{h^-} = \sqrt{\sigma^*} x_h \equiv \lambda_h,
\quad
\mu_L^{h^+} = \mu_H^{h^+} = \sqrt{\sigma} z_h \equiv \mu_h. 
\ee

Another class of decays we shall examine are characterized by a final
state $f$ that is an eigenstate of $CP$, with a definite $CP$ 
parity $\zeta_f$:

\be
\la{eq3.5.8}
CP \vert f \rangle = \zeta_f \vert f \rangle, \quad
\zeta_f = \pm 1.
\ee

\noi
At lowest order one finds $\lambda^f_{H,L} = \zeta_f$~\cite{bran}.


\s{Decay rates for a single {\bf B} meson}
\la{s3.6}
\markright{\small{\rm{\thesection. DECAY RATES FOR A SINGLE {\bf B} MESON}}}

In our formalism, the probability rate for the decay of an initial
${\bf B}^0$ or $\overline{\bf B}^0$ meson in a final state $f$ at time
$t$ is given by

\be
\la{eq3.6.1}
{\cal P}_f ({\bf B}^0; t) = \Tr [O_f^d \rho_{{\bf B}^0}^d(t)], \quad
{\cal P}_f (\overline{\bf B}^0; t) = 
\Tr [O_f^d \rho_{\overline{\bf B}^0}^d(t)].
\ee

\noi
Together with these probability rates, it is useful to define integrated 
rates as well:

\be
\la{eq3.6.5}
{\cal P}_{f}({\bf B}^0) = \frac{1}{\Gamma} \int_0^{\infty}
{\cal P}_f({\bf B}^0;\tau) d \tau, \quad
{\cal P}_{f}(\overline{\bf B}^0) = \frac{1}{\Gamma} \int_0^{\infty}
{\cal P}_f(\overline{\bf B}^0;\tau) d \tau.
\ee

\noi
As an example, we will explicitly list the probabilty rates for the 
semileptonic decays. It is convenient to define the parameters

\be
\la{eq3.6.2}
\omega \equiv \frac{\Delta m}{\Gamma}, \quad \quad
\delta \equiv \frac{\Delta \Gamma}{2 \Gamma}
\ee

\noi
where $\Delta m = m_H - m_L$, $\Gamma = (\gamma_H + \gamma_L)/2$ and
$\Delta \Gamma = \gamma_L - \gamma_H$; it is also convenient to rescale the
time as $\tau = t \Gamma$ (that is, we measure the time in units of {\bf B}
lifetime). Note that in the case of ${\bf B}_d$ mesons $\delta$ is 
expected to be small, of the order of $10^{-2}$; this allows to simplify
some expressions in the coming sections.

Explicitely, one finds for the four relevant semileptonic decays 
${\bf B}^0 \rightarrow h^- l^+ \nu$, $h^+ l^- \overline{\nu}$
and $\overline{\bf B}^0 \rightarrow h^- l^+ \nu$, 
$h^+ l^- \overline{\nu}$; for example,

\be
\la{eq3.6.3}
\begin{split}
{\cal P}_{h^-}({\bf B}^0; \tau) = \vert {\cal M}_h \vert ^2 e^{-\tau} 
&\Bigl\{ 
\sinh{\delta \tau} \Bigl[ \frac{\vert r_H \vert ^2 - \vert r_L \vert ^2}
{\vert r_H + r_L \vert ^2} - \re \lambda_h - \frac{\Lambda^{\prime}}{\Delta
\Gamma} \Bigr] + \\
&+ \cosh{\delta \tau} \Bigl[ \frac{\vert r_H \vert ^2 + \vert r_L  
\vert ^2}{\vert r_H + r_L \vert ^2} - \re y_h + \frac{2}{\Gamma}
\frac{\delta}{\delta ^2 + \omega ^2} \re \Delta^{\prime} \Bigr] + \\ 
&+ \sin{\omega \tau} \Bigl[ 2 \frac{\im (r_H r_L^*)}
{\vert r_H + r_L \vert ^2} + \im \lambda_h - \frac{1}{2\Delta m} \re 
\Theta^{\prime} \Bigr] + \\
&+ \cos{\omega \tau} \Bigl[ 2 \frac{\re (r_L r_H^*)}
{\vert r_H + r_L \vert ^2} e^{-(\Omega^{\prime} + \Lambda^{\prime})
\frac{\tau}{\Gamma}}
- \re y_h - \frac{2}{\Gamma} \frac{\delta}
{\delta ^2 + \omega ^2} \re \Delta^{\prime} \Bigr] \Bigr\}; \\
\end{split}
\ee

\noi
and the others are reported in Appendix~\ref{a3}.

In these expressions we can identify the contributions coming from the 
standard evolution, from the hypothetical violations of CPT and $\Delta B
= \Delta Q$ and from the dissipation. The probabilities ${\cal P}_h$
are directly accessible and can be studied in experiments performed at
colliders: a fit of the time dependence of 
these quantities should permit, in line of principle, the evaluation 
of dissipative parameters $\Lambda^{\prime}$, $\Delta^{\prime}$, 
$\Omega^{\prime}$ and $\re \Theta^{\prime}$. In the next section
we shall define some quantities, called asymmetries, more suitable
to extract informations about the dissipative parameters.

Analogously we can compute the 
probability rates for the decay of ${\bf B}^0$ and $\overline{\bf B}^0$
into final states with a definite CP parity, for example 
${\bf D}^+ {\bf D}^-$, $\pi^+ \pi^-$, ${\bf J}/{\bf \Psi} \; {\bf K}$.
Their expressions are more involved and can be found in Appendix~\ref{a3}.


\s{Asymmetries for a single {\bf B} meson}
\la{s3.7}
\markright{\small{\rm{\thesection. ASYMMETRIES FOR A SINGLE {\bf B} MESON}}}

In order to reduce systematic errors, two experimental studies of the
neutral {\bf B} mesons system are usually performed using suitable
combinations of the decay probabilities introduced in the previous section.
They are called asymmetries and give a direct information about interesting 
physical quantities.

Given the process ${\bf B}^0
\rightarrow {\bf B}^0$, its CPT conjugate is $\overline{\bf B}^0 
\rightarrow \overline{\bf B}^0$; given instead the process ${\bf B}^0
\rightarrow \overline{\bf B}^0$ its T conjugate is $\overline{\bf B}^0
\rightarrow {\bf B}^0$. Since, in absence of violation of the
$\Delta B = \Delta Q$ rule, $h^+$ and $h^-$ signal the decay of 
$\overline{\bf B}^0$ and ${\bf B}^0$ respectively, CPT and T violations are 
studied by the asymmetries

\be
\la{eq3.7.1}
A_{CPT}(\tau) \equiv \frac{{\cal P}_{h^+}(\overline{\bf B}^0; \tau) -
{\cal P}_{h^-}({\bf B}^0; \tau)}
{{\cal P}_{h^+}(\overline{\bf B}^0; \tau) +
{\cal P}_{h^-}({\bf B}^0; \tau)},
\ee

\noi
respectively

\be
\la{eq3.7.2}
A_{T}(\tau) \equiv \frac{{\cal P}_{h^-}(\overline{\bf B}^0; \tau) -
{\cal P}_{h^+}({\bf B}^0; \tau)}
{{\cal P}_{h^-}(\overline{\bf B}^0; \tau) +
{\cal P}_{h^+}({\bf B}^0; \tau)}.
\ee

In the standard case (i.e. without dissipative evolution) they
vanish if CPT and T are preserved. In our case, using the 
expressions in Appendix~\ref{a3}, we find

\be
\la{eq3.7.3} 
\begin{split}
A_{CPT}(\tau) =& \frac{(\vert r_L\vert^2 - \vert r_H\vert^2)
\sinh{\delta \tau} + 2\im (r_L r_H^*) \sin{\omega \tau}}
{(\vert r_L\vert^2 + \vert r_H\vert^2)
\cosh{\delta \tau} + 2 \re (r_L r_H^*) \cos{\omega \tau}}
+ 2 \re y_h + \\ 
&+ \frac{1}{\cosh{\delta \tau} + \cos{\omega \tau}}
[\im (\lambda_h - \mu_h) \sin{\omega \tau} + \re (\lambda_h - \mu_h)
\sinh{\delta \tau}] + \\
&- \frac{\cosh{\delta \tau} - \cos{\omega \tau}}
{\cosh{\delta \tau} + \cos{\omega \tau}}\frac{4}{\Gamma}
\frac{\delta}{\delta^2 + \omega^2} \re \Delta^{\prime}
\end{split}
\ee

and

\be
\la{eq3.7.4}
\begin{split}
A_T (\tau) =& \frac{\vert r_L r_H \vert ^2 - 1}{\vert r_L r_H \vert ^2 + 1}
- 2 \re y_h + \\
&+ \frac{1}{\cosh{\delta \tau} - \cos{\omega \tau}} \left\{
\sinh{\delta \tau} \left[\re (\mu_h - \lambda_h) + \frac{4}{\Gamma} 
\frac{\omega}{\delta^2 + \omega^2} \im \Delta^{\prime}\right] + \right. \\
&+ \left. \sin{\omega \tau} \left[\im (\lambda_h - \mu_h) -
\frac{4}{\Gamma} \frac{\delta}{\delta^2 + \omega^2} \im \Delta^{\prime}
\right] \right\}
\end{split}
\ee

We observe that these quantities are affected by the presence of
dissipation, since they contains both the real and imaginary part of
$\Delta^{\prime}$.
However, in their expressions the dissipative parameters 
appear multiplied by the coefficient $\delta$, that, as alredy noticed,
is expected to be small; the dissipative effects are suppressed with 
respect to the standard contributions. As a consequence, these asymmetries 
are not in general suitable for probing the presence of dissipative 
effects in the neutral {\bf B} mesons system.

To isolate the dissipative contribution, more complicated combinations 
of the probability rates are needed. For example, the asimmetry

\be
\la{eq3.7.5}
A_{\Delta m}(\tau) \equiv \frac
{\left[{\cal P}_{h^-}({\bf B}^0;\tau) - {\cal P}_{h^-}
(\overline{\bf B}^0;\tau)\right]
- \left[{\cal P}_{h^+}({\bf B}^0;\tau) - {\cal P}_{h^+}
(\overline{\bf B}^0;\tau)\right]}
{{\cal P}_{h^-}({\bf B}^0;\tau) + {\cal P}_{h^-}(\overline{\bf B}^0;\tau)
+ {\cal P}_{h^+}({\bf B}^0;\tau) + {\cal P}_{h^+}(\overline{\bf B}^0;\tau)}
\ee

\noi
used in the standard case to fit the mass parameter $\Delta m$, in our case
presents a characteristic dependence on the dissipative terms; in the
limit of a vanishing $\delta$, it takes a very simple form:

\be
\la{eq3.7.6}
A_{\Delta m}(\tau) = (\re \theta )^2 (1 - \cos{\omega \tau}) + 
e^{- \Omega^{\prime} \frac{\tau}{\Gamma}} \cos{\omega \tau} - 
\sin{\omega \tau}\left[\frac{1}{\Delta m} \re \Theta^{\prime} + 
\im (\lambda_h + \mu_h)\right] 
\ee

Assuming the validity of the $\Delta B = \Delta Q$ rule (that forbids 
the decays ${\bf B}^0 \rightarrow h^+ l^- \overline{\nu}$
and $\overline{\bf B}^0 \rightarrow h^- l^+ \nu$), and neglecting
the quadratic dependence on the CPT-violating parameter $\theta$, the
different time behaviors in $A_{\Delta m}(\tau)$ should allow a determination
of the non-standard parameters $\Omega^{\prime}$ and the real part
of $\Theta^{\prime}$. The actual accuracy of such a determination
highly depends on the sensitivity of the measure of $A_{\Delta m}(\tau)$.
From the results of available simulations one can expect an accuracy
in the determination of the various terms in $A_{\Delta m}(\tau)$ of
about a few percent~\cite{ball,btev}. This sensitivity is alredy
enough to give interesting bounds on $\Omega^{\prime}$ and 
$\re \Theta^{\prime}$.

The assumption $\delta = 0$ has some important consequences. Indeed, in the
limit $\Delta \Gamma = 0$ the non-Hermitian part 
$\it{\Gamma}$ of the Weisskopf-Wigner
Hamiltonian $H$ becomes proportional to the identity. In this case the
eigenstates of $H$, $\bh$ and $\bl$ becomes orthogonal, so
$r_L r_H^* = 1$, producing $\vert \sigma \vert = 1$ (or $\xi = 0$) 
and, accordingly to~(\ref{eq3.1.7}), $\im \theta = 0$. Then T 
violation in mixing is forbidden and CPT violation in mixing is parametrized
by $\re \theta$. This fact has important consequences for experimental
tests of CPT and T invariance in mixing~\cite{banu1,1banu1,2banu1,banu2,
lavo1,lavo2}.

Another interesting asymmetry is defined using the decay of both ${\bf B}^0$
and $\overline{\bf B}^0$ in an eigenstate $f$ of $CP$:

\be
\la{eq3.7.7}
A_f(\tau)\equiv \frac{{\cal P}_f({\bf B}^0; \tau) - 
{\cal P}_f(\overline{\bf B}^0; \tau)}{{\cal P}_f({\bf B}^0; \tau) +
{\cal P}_f(\overline{\bf B}^0; \tau)};
\ee

\noi
in the standard case, this observable is particularly suitable for studing 
violations of CP invariance. In presence of dissipation, we get

\be
\la{eq3.7.8}
\begin{split}
A_f(\tau) =& \frac{2 \re \lambda_f}{1 + \vert \lambda_f \vert^2} \re \theta
+ \frac{2 \zeta_f}{\omega \Gamma} \im \Delta^{\prime} + \\
&- \sin{\omega \tau} \left[ \frac{2 \im \lambda_f}
{1 + \vert \lambda_f \vert^2} 
+ \frac{2 \zeta_f}{\omega \Gamma} \re \Delta^{\prime} \right] + \\
&+ \cos{\omega \tau} \left[ \frac{1 - \vert \lambda_f \vert^2}
{1 + \vert \lambda_f \vert^2} - \frac{2 \re \lambda_f}
{1 + \vert \lambda_f \vert^2} \re \theta - \frac{2 \zeta_f}{\omega \Gamma} 
\im \Delta^{\prime} \right],
\end{split}
\ee

\noi
where the condition $\delta \approx 0$ has been used; this enables us 
to write $\lambda_H^f$ and $\lambda_L^f$ in terms of a unique amplitude 
ratio $\lambda_f$:

\be
\la{eq3.7.9}
\lambda_f \equiv \sqrt{\sigma ^*}\frac
{{\cal A}(\overline{\bf B}^0 \rightarrow f)}
{{\cal A}({\bf B}^0 \rightarrow f)}
\ee

\noi
and

\be
\la{eq3.7.10}
\lambda^f_L = \lambda_f (1 - \re \theta), \quad
\lambda^f_H = \lambda_f (1 + \re \theta).
\ee

The presence of dissipation in~(\ref{eq3.7.8}) manifests itself in the 
modification of the coefficients of the oscillating terms and in the
presence of a $\tau$-independent piece. Dedicated {\bf B} experiments
at colliders should be able to identify the various $\tau$-dependence
and therfore provide stringent bounds on the parameter 
$\Delta^{\prime}$.

Asymmetries analogous to the ones just discussed can be constructed using
integrated rates, see eq.~(\ref{eq3.6.2}), rather than the probability rates. 
The quantities we obtain in this way are time independent and are denoted 
with a prime index to distinguish them from the previously defined ones. 
Two examples are

\be
\la{eq3.7.11}
\begin{split}
A^{\prime}_{\Delta m} =& \frac{1}{1 + \omega^2}
\left\{ 1 + \omega^2 (\re \theta)^2 - \omega \left[
\frac{1}{\Delta m} \re \Theta^{\prime} + \im (\lambda_h + \mu_h) 
\right] + \right. \\
&\left.+ \frac{1}{\Gamma (1 + \omega^2)} \left[(\omega^2 - 1) 
\Omega^{\prime} + 2 \omega^2 \Lambda^{\prime} \right]
\right\}
\end{split}
\ee

\noi
and

\be
\la{eq3.7.12}
\begin{split}
A_f^{\prime} =& \frac{2 \re \lambda_f}{1 + \vert \lambda_f \vert^2} \re \theta
+ \frac{2 \zeta_f}{\omega \Gamma} \im \Delta^{\prime} + \\
&- \frac{1}{1 + \omega^2} \left\{ \omega \left[ \frac{2 \im \lambda_f}
{1 + \vert \lambda_f \vert^2} 
+ \frac{2 \zeta_f}{\omega \Gamma} \re \Delta^{\prime} \right] + \right. \\
&\left.- \frac{1 - \vert \lambda_f \vert^2}
{1 + \vert \lambda_f \vert^2} + \frac{2 \re \lambda_f}
{1 + \vert \lambda_f \vert^2} \re \theta + \frac{2 \zeta_f}{\omega \Gamma} 
\im \Delta^{\prime} \right\}.
\end{split}
\ee

The result in equation~(\ref{eq3.7.11}) is particularly interesting: assuming 
the validity of the $\Delta B = \Delta Q$ rule and CPT-invariance,
a measure of $A^{\prime}_{\Delta m} (1 + \omega^2)$ not compatible with $1$
whould signal the presence of non standard effects. Such a test can be 
performed at colliders with high accuracy, in particular using the next
generation dedicated {\bf B} mesons experiments.

The integrated asymmetry~(\ref{eq3.7.12}) is not very useful in probing
the presence of dissipative effects in neutral {\bf B} mesons decays:
since there is no time-dependence, the measure of this asymmetry would
give little information on the magnitude of $\Delta^{\prime}$.


\s{Decay rates for entangled {\bf B} meson}
\la{s3.8}
\markright{\small{\rm{\thesection. DECAY RATES FOR ENTANGLED {\bf B} MESONS}}}

In the analysis of neutral {\bf B} mesons physics, the study of pairs 
of mesons produced at the so-called {\bf B}-factories, that is $e^+ e^-$ 
colliders tuned at the energy of the ${\bf \Upsilon}(4S)$ resonance, is of
particular relevance. The pairs of {\bf B} mesons are produced by the 
decay of this resonance,

\be
\la{eq3.8.1}
{\bf \Upsilon}(4S) \rightarrow {\bf B}^0 \overline{\bf B}^0,
\ee

\noi
in an entangled state. Since ${\bf \Upsilon}(4S)$ has spin $1$ whereas 
the two ${\bf B}$ mesons are spinless, they are produced, in the 
${\bf \Upsilon}(4S)$ rest frame, in an antisymmetric spatial state. In 
this reference frame ${\bf B}^0$ and $\overline{\bf B}^0$ are flying apart 
with opposite momenta $\vec{p}$ and $-\vec{p}$; the state encoding this 
configuration is the singlet state:

\be
\la{eq3.8.2}
\vert \psi \rangle = \frac{1}{\sqrt{2}}
(\vert {\bf B}^0, -\vec{p} \rangle \otimes 
\vert \overline{\bf B}^0, \vec{p}\rangle -
\vert \overline{\bf B}^0, -\vec{p} \rangle \otimes 
\vert {\bf B}^0, \vec{p}\rangle),
\ee

\noi
or, in the density matrix formalism, the operator $\rho = \vert \psi \rangle
\langle \psi \vert$, in this case a $4 \times 4$ matrix. 
Its irreversible time evolution is assumed to have a factorized form:

\be
\la{eq3.8.3}
\rho(t_1, t_2) = (\Lambda_{t_1} \otimes \Lambda_{t_2})[\rho(0)],
\ee

\noi
where we consider two possibly different proper times $t_1$ and $t_2$ for 
the two mesons since in general they won't decay at the same time. 

\begin{Rem}
\la{rem3.8.1}
{\rm In the standard treatement of the neutral {\bf B} mesons system
the time-evolution is assumed to have the factorized form~(\ref{eq3.8.3})
since there is not interaction between the two particles in the pair;
in the open system approach an interaction can be induced by the common
thermal bath, but it disappears at the lowest order in the dissipative
effects; whence the hypothesis of factorization.}
\end{Rem}

As in the previous sections, we prefer to express the statistical 
operator in the basis diagonalizing the Hamiltonian contribution, then

\be
\la{eq3.8.4}
\rho^d(t_1, t_2) = (R^{-1} \otimes R^{-1}) \rho(t_1, t_2)
(R^{-1 \dagger} \otimes R^{-1 \dagger}),
\ee

\noi
which in terms of the basis
\{$P_i , i = 1,\ldots 4$\} introduced in Section~\ref{s2.1}, reads

\be
\la{eq3.8.5}
\rho^d(t_1, t_2) = \frac{1}{2 \vert r_L + r_H \vert ^2}
(P_1(t_1)\otimes P_2(t_2) + P_2(t_1)\otimes P_1(t_2) - 
P_3(t_1)\otimes P_4(t_2) - P_4(t_1)\otimes P_3(t_2)),
\ee

\noi
where $P_i(t) = \Lambda_t[P_i]$.

The probability that a meson decays into a final state $f_1$ at proper time
$t_1$ while the other meson decays into the final state $f_2$ at proper
time $t_2$ is given by

\be
\la{eq3.8.6}
{\cal P}(f_1, t_1; f_2, t_2) = \Tr [(O_{f_1}^d\otimes O_{f_2}^d) 
\rho^d(t_1, t_2)]
\ee

\noi
where $O_{f_1}^d$ and $O_{f_2}^d$ are the $2 \times 2$ matrix 
representations, in the basis of eigenstates of $H$, of the Hermitian
operators $O_{f_1}$ and $O_{f_2}$ described in Section~\ref{s3.5}.

Since the neutral {\bf B} lifetime is very short, it is difficult
to identify the two proper times $t_1$ and $t_2$: the 
probability~(\ref{eq3.8.6}) is experimentally unattractive. It is thus 
customary to define integrated rates, as in the case of the decays of a 
single meson. In particular it is possible to consider a fixed time delay 
$t = t_1 - t_2$ between the two decays, obtaining a one-time probability 
rate:

\be
\la{eq3.8.7}
{\cal P}(f_1, f_2; t) \equiv \int_0^{\infty}
{\cal P}(f_1, t^{\prime} + t; f_2, t^{\prime}) dt^{\prime}
\ee

\noi
when $t\geqslant 0$, that is when the first meson decays after the second,
while, for $t < 0$, one defines

\be
\la{eq3.8.7.bis}
{\cal P}(f_1, f_2; t) \equiv \int_0^{\infty}
{\cal P}(f_1, t^{\prime} + t; f_2, t^{\prime}) \theta (t^{\prime} + t) 
dt^{\prime}
\ee

\noi
where the $\theta$ of Heavyside is required since the evolution is of 
semigroup type, with forward in time propagation. This definition
implies

\be
\la{eq3.8.8}
{\cal P}(f_1, f_2; -t) = {\cal P}(f_2, f_1; t).
\ee

In the following we shall always assume $t \geqslant 0$.
It is also possible to define a time independent integrated rate as

\be
\la{eq3.8.9}
{\cal P}(f_1, f_2) = \int_0^{\infty} {\cal P}(f_1, f_2; t) dt
\ee

\noi
in which we sum over all possible delays between the two decays.
This quantity is relevant in the experimental analysis at the so-called
symmetric {\bf B} factories where the ${\bf \Upsilon}(4S)$ decays
pratically at rest in the laboratory frame, and then the time information
is hardly attaineable. On the other side, at the so-called asymmetric
{\bf B}  factories the decay of the ${\bf \Upsilon}(4S)$ resonance takes 
place in a boosted frame respect to the laboratory; in this case the
temporal information can be extracted and the one-time 
probability~(\ref{eq3.8.7}) is suitable to the data analysis.


The explicit expressions for these joint probabilities are rather involved;
some particularly interesting examples are collected in Appendix~\ref{a3}.


One can nevertheless make a general remark. Consider ${\cal P}(f,f;0)$, 
i.e. the probability rate for the simultaneous decay of the two ${\bf B}$ 
mesons in the same final state $f$. Since the initial state is 
antisymmetric, the usual dynamics produces ${\cal P}(f,f;0) = 0$ 
whereas the modified dynamics in general gives ${\cal P}(f,f;0) \ne 0$. 
This quantity could then unambiguously show the presence of non 
standard effects.


\s{Asymmetries for entangled {\bf B} meson}
\la{s3.9}
\markright{\small{\rm{\thesection. ASYMMETRIES FOR ENTANGLED {\bf B} MESONS}}}

As in the study of the decays of single {\bf B} mesons, the experimental 
analysis of our system is not usually performed directly with the 
probability rates introduced before; rather, some suitable asymmetries
are constructed.
We use the assumption $\delta = 0$ to simplify our expressions and,
as before, we refer to a rescaled time scale $\tau = t \Gamma$.

\noi
An interesting quantity involving semileptonic decays in both mesons is

\be
\la{eq3.9.1}
{\cal R}(\tau) \equiv \frac
{{\cal P}(h^+, h^+; \tau) + {\cal P}(h^-, h^-; \tau)}
{{\cal P}(h^+, h^-; \tau) + {\cal P}(h^-, h^+; \tau)}.
\ee

\noi
In the given approximation, one finds

\be
\la{eq3.9.2}
\begin{split}
{\cal R}(\tau) =& \frac{(1 + \vert r_H r_L \vert^2)
(\cosh{\delta \tau})}{(\vert r_L \vert^2 + \vert r_H \vert^2)
\cosh{\delta \tau} + 2 \re (r_L r_H^*) \cos{\omega \tau}} + \\
&+ \frac{2}{(1 + \cos{\omega \tau})^2}\left[
(1 + \tau) \frac{\Omega^{\prime}}{\Gamma} \cos{\omega \tau}  +
\frac{1}{\Delta m}(\sin{\omega \tau} + \omega \cos{\omega \tau})
\re \left( \frac{\Theta^{\prime}}{1 - i \omega} \right)
\right],
\end{split}
\ee

\noi
and its evaluation for simultaneous decays is particularly meaningful:

\be
\la{eq3.9.3}
{\cal R}(0) = \frac{1}{2}\left[
\frac{\Omega^{\prime}}{\Gamma} - \frac{1}{\Delta m}
\frac{\omega}{1 + \omega^2}(\omega \im \Theta^{\prime} - \re \Theta^{\prime})
\right],
\ee

\noi
since in the standard case ${\cal R}(0) = 0$. Therefore a fit of 
${\cal R}(\tau)$ on actual experimental data that produces a non-zero
intercept would clearly signal the presence of non-standard effects.

Another quantity built with the double semileptonic decays and relevant
in the study of dissipative effects is the asymmetry ${\cal A}_{\Delta m}
(\tau)$, analogous to the one defined in~(\ref{eq3.7.5}):

\be
\la{eq3.9.4}
{\cal A}_{\Delta m}(\tau) \equiv \frac
{\left[{\cal P}(h^+,h^-;\tau) + {\cal P}(h^-,h^+;\tau)\right]
- \left[{\cal P}(h^+,h^+;\tau) + {\cal P}(h^-,h^-;\tau) \right]}
{{\cal P}(h^+,h^+;\tau) + {\cal P}(h^-,h^-;\tau) +
{\cal P}(h^+,h^-;\tau) + {\cal P}(h^-,h^+;\tau)}.
\ee

\noi
Its explicit expression is particularly simple and meaningful, even in the 
case of vanishing $\delta$:

\be
\la{eq3.9.5}
\begin{split}
{\cal A}_{\Delta m}(\tau) =& \frac{\Gamma}{\Gamma + \Omega^{\prime}}
e^{-\Omega^{\prime} \frac{\tau}{\Gamma}} \cos{\omega \tau}  +
(\re \theta)^2 (1 - \cos{\omega \tau}) + \\
&- (\sin{\omega \tau}
+ \frac{1}{\Delta m} \omega \cos{\omega \tau}) \re \left( 
\frac{\Theta^{\prime}}{1 - i \omega}\right).
\end{split}
\ee

Thanks to the different time behavior, an experimental study of
this observable should allow to extract precise information on
$\Omega^{\prime}$ and the combination $\omega \im \Theta^{\prime} - 
\re \Theta^{\prime}$. The asymmetry ${\cal A}_{\Delta m}$ is intensively
studied at {\bf B}-factories~\cite{baba,dedo,leon} 
since it allows a precise determination of the 
mass difference $\Delta m$. Indeed, 
${\cal A}_{\Delta m}(\tau) = \cos{\omega \tau}$
in absence of dissipation.

Another asymmetry interesting from a dissipative point of view is
built with one-time integrated rates and it involves a semileptonic
channel and an hadronic channel (in particular a state with definite 
$CP$-parity):

\be
\la{eq3.9.6}
{\cal A}_f (\tau) \equiv
 \frac
{{\cal P}(h^+,f;\tau) - {\cal P}(h^-,f;\tau)}
{{\cal P}(h^+,f;\tau) + {\cal P}(h^-,f;\tau)}.
\ee

If we neglect possible violations of the $\Delta B = \Delta Q$ rule,
the semileptonic decay identify the flavor of the decaying meson. In 
this case this asymmetry is equivalent to the $A_f(\tau)$ defined
in~(\ref{eq3.7.7}). If the final state is $f = {\bf J}/ {\bf \Psi}\; 
{\bf K}$ these observables are directly expressed in terms of one of the 
angles that parametrize the CP violation in the Cabibbo Kobayashi Maskawa
matrix; for this reason they are extensively studied. 
The expression of ${\cal A}_f(\tau)$ in the case of irreversible dynamics
is given by:

\be
\la{eq3.9.7}
\begin{split}
{\cal A}_f (\tau) =& \frac{2 \re \lambda_f}{1 + \vert \lambda_f \vert^2}
\re \theta + \zeta_f \re (\lambda_h - \mu_h + 2\zeta_f y_h) -
\frac{2 \zeta_f}{\omega \Gamma} \im \Delta^{\prime} + \\
&+ \cos{\omega \tau} \left[ 
\frac{1 - \vert \lambda_f \vert^2}{1 + \vert \lambda_f \vert^2}
- \frac{2 \re \lambda_f}{1 + \vert \lambda_f \vert^2} \re \theta 
- \frac{2 \zeta_f}{\omega \Gamma} \re \left( 
\frac{\omega + 2i}{2 + i \omega} \Delta^{\prime}
\right) \right] + \\
&+ \sin{\omega \tau} \left[
\frac{2 \im \lambda_f}{1 + \vert \lambda_f \vert^2}
- \frac{2 \zeta_f}{\omega \Gamma} \im \left( 
\frac{\omega + 2i}{2 + i \omega} \Delta^{\prime}
\right) \right]
\end{split}
\ee

\noi
where $\lambda_f$ is defined in~(\ref{eq3.7.9}) and $\zeta _f$ 
in~(\ref{eq3.5.8}); for simplicity in this formula only first order terms
in CPT violating parameters are kept. 
Equation~(\ref{eq3.9.7}) holds for $\tau \geqslant 0$; for $\tau < 0$
the expression is

\be
\la{eq3.9.7.bis}
\begin{split}
{\cal A}_f (\tau) =& \frac{2 \re \lambda_f}{1 + \vert \lambda_f \vert^2}
\re \theta + \zeta_f \re (\lambda_h - \mu_h + 2\zeta_f y_h) -
\frac{2 \zeta_f}{\omega \Gamma} \re \left( 
\frac{\omega + 2i}{2 + i \omega} \Delta^{\prime}
\right) + \\
&+ \cos{\omega \tau} \left[ 
\frac{1 - \vert \lambda_f \vert^2}{1 + \vert \lambda_f \vert^2}
- \frac{2 \re \lambda_f}{1 + \vert \lambda_f \vert^2} \re \theta 
- \frac{2 \zeta_f}{\omega \Gamma} \im \Delta^{\prime} \right] + \\
&+ \sin{\omega \tau} \left[
\frac{2 \im \lambda_f}{1 + \vert \lambda_f \vert^2}
+ \frac{2 \zeta_f}{\omega \Gamma} \re \Delta^{\prime}
\right];
\end{split}
\ee

\noi
the two expressions coincide for $\tau = 0$ and, in absence of dissipation,
one can be obtained from the other by letting $\tau \rightarrow -\tau$.
The fact that this is not longer true when $\Delta^{\prime} \ne 0$ is a 
signal of irreversibility.

Assuming the validity of the $\Delta B = \Delta Q$ rule and neglecting
CPT-violating effects in mixing ($\theta = 0$) as well as direct
CP violations (i.e. $\vert \lambda_f \vert = 1$, see~\cite{bran}),
a fit of~(\ref{eq3.9.7}) with experimental data allows the determination
of both $\re \Delta^{\prime}$ and $\im \Delta^{\prime}$. The study of
${\cal A}_f$ could result in one of the best tests on the presence
of dissipative effects in {\bf B} physics, in particular for the 
final state ${\bf J}/{\bf \Psi} \; {\bf K}$~\cite{baba}.

The quantities involving the fully integrated probability rates, 
eq.~(\ref{eq3.8.9}), in general are not suitable to study dissipative
effects, since the dissipative parameters are multiplied by $\delta$;
neverthless it is possible to define two ratios unaffected by this problem:

\be
\la{eq3.9.8}
\chi_B \equiv \frac
{{\cal P}(h^+,h^+) + {\cal P}(h^-,h^-)}
{{\cal P}(h^+,h^+) + {\cal P}(h^-,h^-) + {\cal P}(h^+,h^-) + {\cal P}(h^-,h^+)}
\ee

\noi
and

\be
\la{eq3.9.9}
R_B \equiv \frac
{{\cal P}(h^+,h^+) + {\cal P}(h^-,h^-)}
{{\cal P}(h^+,h^-) + {\cal P}(h^-,h^+)}.
\ee

Assuming the $\Delta B = \Delta Q$ rule, $\chi_B$ represents the total 
${\bf B}^0$ - $\overline{\bf B}^0$ mixing probability. In our framework,
one explicitely finds:

\be
\la{eq3.9.10}
\chi_B = \frac{\omega^2}{2(1 + \omega^2)}\left[
1 - (\re \theta)^2 + \frac{2}{\omega^2 (1 + \omega^2)} X \right],
\ee

\be
\la{eq3.9.11}
R_B = \frac{\omega^2 \left[1 - (\re \theta)^2 \right]}
{2 + \omega^2\left[1 + (\re \theta)^2 \right]} +
\frac{4}{(2 + \omega^2)^2} X;
\ee

\noi
the dissipative contribution is the same for the two observables and it
is given by

\be
X \equiv \frac{1}{\Gamma} (\Omega^{\prime} - \omega^2 
\Lambda^{\prime}) - \frac{\omega}{\Delta m} (\omega \im \Theta^{\prime} - 
\re \Theta^{\prime}).
\ee

Independent measures of these two quantities would provide a way
to estimate both $\re \theta$ and $X$, and therefore give limits
on both dissipative and CPT-violating effects.


\s{Discussion and conclusions}
\la{s3.10}
\markright{\small{\rm{\thesection. DISCUSSION AND CONCLUSIONS}}}

In this chapter we have treated the neutral ${\bf B}$ mesons system
in the framework of quantum dynamical semigroups. The formalism is very
general and can be applied to describe the time evolution of this system 
when it is subject to a weak interaction with an external
environment. As described in 
Section~\ref{s3.2}, quantum gravity could give a motivation for such an
approach, since in this case the space-time itself would act as an
effective environment for any physical system. 

As discussed before, the open system framework allows a rough evaluation 
of magnitude of the dissipative parameters; they scale at most as 
$m_{\bf B}^2/M_F$, where $m_{\bf B}$ is the meson mass while $M_F$ is
a large fundamental mass scale. Tipically, $M_F$ coincides with the Planck 
scale so that $m_{\bf B}^2/M_F \sim 10^{-18}~{\rm GeV}$. This value is very 
small; however, the sofistication of the dedicated {\bf B}-experiments,
both at colliders(CDF-II, HERA-B, BTeV, LHC-$b$) and {\bf B}-factories
(BaBar, Belle, CLEO-III), is so high that the sensitivity
required to probe the presence of non-vanishing dissipative parameters
can be reached in just a few years of data 
taking~\cite{ball,btev,baba,dedo,leon}.

\begin{Rem}
\la{rem3.10.1}
{\rm An evaluation of the dissipative parameters for a gravitational bath have 
been alredy performed in the context of the neutral {\bf K} mesons
physics~\cite{benat,benap}. 
Upper bounds for the dissipative parameters have been obtained;
these are in the range of what suggested by gravitational or stringy effects.
More data are expected from ongoing experiments at the so-called 
${\bf \Phi}$-factories, the analogous of {\bf B}-factories for the {\bf K}
mesons.}
\end{Rem}

To conclude, we observe that, besides the study of irreversible
dynamics, these experiments, involving correlated mesons, could give
the possibility to investigate the mathematical notion of complete
positivity from a phenomenological point of view. Let us consider
a simplified situation with a trivial Hamiltonian sector,
${\cal H} = 0$, and a dissipative part with vanishing $a$, $b$ and $c$.
A dissipative contribution of this kind has been used to describe
non-standard effects in the {\bf K} meson system, using a positive
but not completely positive dynamics~\cite{elli2,huet,elli3}.
The eigenvalues of the evolved singlet state in~(\ref{eq3.8.3}), for
$t_1 = t_2$, are

\be
\la{eq3.10.1}
\begin{split}
\lambda_{1,2} & = 2 \pm \{ [E_+(t)]^2 + [E_-(t)]^2 + 2[F(t)]^2 \}, \\
\lambda_{3,4} & = \pm [E_+(t) + E_-(t)]\{[E_+(t) - E_-(t)]^2 
+ 4[F(t)]^2 \}^{1/2},
\end{split}
\ee

\noi
where

\be
\la{eq3.10.2}
\begin{split}
E_{\pm}(t) &= \frac{1}{\nu_+ - \nu_-}[(\nu_+ + 2\alpha)e^{\nu_{\pm}t} -
(\nu_- + 2\alpha)e^{\nu_{\mp}t}], \\
F(t) &= \frac{2\beta}{\nu_+ - \nu_-}[e^{\nu_- t} - e^{\nu_+ t}],
\end{split}
\ee

\noi
with $\nu_{\pm} = - (\alpha + \gamma) \pm \sqrt{(\alpha - \gamma)^2
+ 4\beta^2}$.

The condition of positivity of this dynamics reads $\alpha \gamma \geqslant
\beta^2$ whereas complete positivity requires $\alpha = \gamma$ and
$\beta = 0$. Consequently, in the simply positivity case $\nu_{\pm}$ are both
negative and thus, for $t \ne 0$ two of the eigenvalues~(\ref{eq3.10.1})
are always negative and the statistical interpretation of the density matrix
$\rho(t,t)$ is compromised. In the completely positive case we find
$\nu_+ = \nu_-$ and all the considered eigenvalues remain non negative for
any time.

These results offer a further motivation for studying generalized
dynamics of the form~(\ref{eq3.3.1}) at mesons factories. These set-ups
are in fact high-performance quantum interferometers: at least in principle,
they can verify from the experimental point of view the role of
the condition of complete positivity in the time evolution of correlated
mesons~\cite{benak,1benak,2benak,3benak,4benak,benab}. 
As shown by the simple example above, this is not just a mere
technical question: it is crucial for our physical understanding of the
quantum dynamics of open systems.

%% file: chapter4.tex
\chapter{Neutrinos}
\la{ch4}
\markboth{\small{\rm{\thechapter. NEUTRINOS}}}{}

Neutrino physics is presently a very active area of reserch. 
Recent experiments suggest neutrinos to be massive and the three
flavors, electronic ($\nu_e$), muonic ($\nu_{\mu}$) and
tauonic ($\nu_{\tau}$) neutrinos, to mix among themselves.
Consequently the Standard
Model should be modified by the introduction of a mixing matrix in
the leptonic sector (analogous to the Cabibbo Kobayashi Maskawa matrix
in the hadronic sector). The first hint to the possibility of neutrino
mixing came from the observation of a reduced $\nu_e$ flux from the Sun,
incompatible with the usual model of solar evolution. 
The most favoured explanation of this phenomenon is the so-called
MSW effect which is deduced from the hypothesis that electronic neutrinos
interact with solar matter~\cite{wolf,mikh,1mikh}. 
In this context, flying neutrinos behave as
open systems in interaction with an environment (the Sun), but the MSW 
effect is only the Lamb-shift term (a redefinition of the neutrino effective
Hamiltonian) and no dissipative effects appear.

In other approaches, dissipation has been taken into 
account~\cite{halp,stod,sawy,samu,burg1,burg2,bala1,
bala2,lore1,torr,nuno,raff},
though paying scarce attention to internal consistency. 
In this chapter, following the lines introduced in 
Chapters~\ref{ch1} and \ref{ch2}, we will apply the rigorous formalism 
of open quantum systems to describe neutrino time-evolutions
in matter, with particular care to the derivation of a
Markovian reduced dynamics. In this way, we will get results unaffected 
from physical inconsistencies that otherwise can appear~\cite{burg2}.

We shall consider an environment with constant mean density
and derive the general form of the generator of the time-evolution
in the Markovian approximation. Since, in the standard description of the
MSW effect, it is the variation of this mean density that produces the most 
relevant aspects in the treatment of the flavors 
oscillations~\cite{mikh,1mikh}, we shall describe how to generatize our 
formalism to account for the case of environments with a varying density 
profile.

The transitions probabilities are explicitly written out in two
different situations: a flavor-changing interaction and a flavor-preserving
one (the case of solar matter).


\s{The physics of neutrinos in vacuum}
\la{s4.1}
\markright{\small{\rm{\thesection. THE PHYSICS OF NEUTRINOS IN VACUUM}}}

In this section we shortly outline the principal facts of
neutrino physics in vacuum. For sake of simplicity, we consider 
the case of two neutrino's flavors, $\nu_e$ and $\nu_{\mu}$, 
and we describe the neutrino system by means of the bidimensional Hilbert 
space of flavor, $\mathfrak{H}_2$~\cite{kim,moh,bil}; indeed, in vacuum
the neutrino momentum can be kept fixed and the related degrees of freedom
does not appear in the description. This is the standard phenomenological 
approach to this system and the experimental data are usually analyzed in 
this context. This approach is justified by more rigorous treatements;
see~\cite{grim,giun}.

\noi
As in the case of the neutral {\bf B} mesons system we distinguish 
between two basis for $\mathfrak{H}_2$:

\bi 

\item the basis of physical states (or flavor states),
\{$\nel$, $\nm$\}; these are the states that are actually detected
in experiments;

\item the basis of the eigenstates of the Hamiltonian of the system
$H_S$, \{$\no$, $\nt$\}, characterized by

\be
\la{eq4.1.0}
\left\{
\begin{array}{l}
H_S \no = \epsilon^{\prime}_1 \no, \\ \\
H_S \nt = \epsilon^{\prime}_2 \nt, \\
\end{array}
\right.
\ee

\noi
where the $H_S$ is Hermitian since neutrinos are stable particles and
$\epsilon^{\prime}_{1,2}$ are real eigenvalues. 

\ei

$H_S$ generates the time-evolution of egenstates in vacuum:

\be
\la{eq4.1.0.bis}
\vert \nu_i (t)\rangle = e^{-i H_S t} \vert \nu_i \rangle 
= e^{-i \epsilon^{\prime}_i t} \vert \nu_i \rangle 
\ee

\noi
and the eigenvalues $\epsilon^{\prime}_i$ are the energies of these states,
that depend on the neutrino momentum $p$ and mass $m_i$. Indeed,
in agreement with experimental evidence~\cite{bile}, the states $\nu_1$ and 
$\nu_2$ are assumed to be massive and we denote with $m_1$, respectively 
$m_2$, their masses. In the Standard Model they vanish; for later 
reference we assume they are not degenerate. 

Since $m_1$ and $m_2$ are believed to be very small, $\epsilon^{\prime}_i$ can
be expanded in the ultrarelativistic limit, i.e. neglecting\footnote{These 
terms are supposed smaller than the dissipative effects that we shall 
consider in the cominig sections} the terms of higher order than $m_1^2$ 
and $m_2^2$: 

\be
\la{eq4.1.0.tris}
\epsilon^{\prime}_i = \sqrt{p^2 + m_i^2} \approx p + \frac{m_i^2}{2p} 
\approx p + \frac{m_i^2}{2E},
\ee

\noi
where $E \approx p$ is the neutrino energy.

The two basis are connected, in vacuum,  by an unitary transformation $U_0$
(the so-called mixing matrix) that can be chosen to be orthogonal by a 
suitable redefinition of the states $\no$ and $\nt$. 

\begin{Rem}
\la{rem4.1.1}
{\rm The choice of an orthogonal $U_0$ matrix is a consequence of the 
impossibility to violate the CP symmetry in a two flavors context: in fact 
the CP violation is related to the existence of unavoidable complex phases 
in the mixing matrix and they appear necessarily in a three flavors context.
However, when dealing with dissipative dynamics, in general
the complex phase appearing in $U_0$ is relevant also in the two flavors
context, because
it produces physical consequences~\cite{benan}. In our context, however,
this phase can be consistently reabsorbed; indeed, the particular dissipative 
effects we shall consider in what follows does not produce any
evidence of this phase. Therefore, in the following,
we do not account for possible violations of CP in the leptonic sector.}
\end{Rem}

\noi
We have:

\be
\la{eq4.1.1}
\left\{
\begin{array}{l}
\nel = U_0 \no, \\ \\
\nm = U_0 \nt;
\end{array}
\right.
\quad U_0 = 
\left(
\begin{array}{cc}
\cos{\theta} & \sin{\theta} \\
-\sin{\theta} & \cos{\theta} \\
\end{array}
\right);
\ee

\noi
where the parameter $\theta$ is the mixing angle in vacuum between 
the two families considered.

\noi
The Hamiltonian $H_S$ in the basis above reads:

\be
\la{eq4.1.2}
H_S^d = \left(
\begin{array}{cc}
\epsilon_1 & 0 \\
0 & \epsilon_2 \\
\end{array}
\right), \quad
H_S = \frac{\Delta m^2}{4E} \left(
\begin{array}{cc}
-\cos{2 \theta} & \sin{2 \theta} \\
\sin{2 \theta} & \cos{2 \theta} \\
\end{array}
\right)
\ee

\noi
where\footnote{We use the notation introduced in Chapter~\ref{ch2}
to characterize the operators expressed in this basis}
$H_S = U_0 H_S^d U_0^{-1}$.  
Since we prefer to deal with a traceless matrix $H_S^d$, we have redefined 
the eigenvalues of $H_S$ as $\epsilon_1 = -\Delta m^2/4E$ and 
$\epsilon_2 = \Delta m^2/4E$, where $\Delta m^2 \equiv m_2^2 - m_1^2$
and $E$ is the neutrino energy.

The distinction between the two basis of the Hilbert space
produces the phenomenon of flavor oscillations, i.e. periodical transitions 
between $\nu_e$ and $\nu_{\mu}$. 

Since a physical state is given by a superposition of eigenstates of $H_S$,
an initially pure state $\nel$ will present, for $t > 0$, a non-zero
component along $\nm$. For instance

\be
\la{eq4.1.4.bis}
\vert \nu_e(t) \rangle = \Bigl( e^{i \Delta m^2 t /4E} \cos^2{\theta}
+ e^{-i \Delta m^2 t /4E} \sin^2{\theta} \Bigr) \vert \nu_e \rangle +
\Bigl( i \sin{2 \theta} \sin{\frac{\Delta m^2}{4E}t} \Bigr) \vert 
\nu_{\mu} \rangle.
\ee

\noi
The probability of transition (in vacuum) is 
given by~\cite{kim,moh,bil}:

\be
\la{eq4.1.4}
P_{\nu_e \rightarrow \nu_{\mu}}(t) = \vert \langle \nu_{\mu} \vert
\nu_e(t) \rangle \vert^2 = 
\sin^2{2 \theta} \sin^2{\frac{\Delta m^2}{4E} t}
\ee

\noi
and the surviving probability is

\be
\la{eq4.1.5}
P_{\nu_e\rightarrow \nu_e}(t) = 1 - P_{\nu_e \rightarrow \nu_{\mu}}(t).
\ee

\noi
Vacuum neutrino oscillations were the first explanation of the lack of 
$\nu_e$ in the solar flux.

\begin{Rem}
\la{rem4.1.2}
{\rm This phenomenon should lead to lepton numbers violations, forbidden in
the Standard Model.}
\end{Rem}

\noi

However, the transition probabilities are modified by the interactions 
of neutrinos in matter~\cite{wolf,mich}; in a homogeneous medium a matter 
mixing angle and two redefined 
masses are needed and the probabilities of transition are formally equal
to~(\ref{eq4.1.4}) and (\ref{eq4.1.5}) but with the redefined parameters
$\theta_m$ and $\Delta m_m^2$. These 
transition probabilities are used to fit the parameters $\theta$ and 
$\Delta m^2$ to the experimental data~\cite{kim,moh,bil}.

In what follows,
we shall describe neutrinos in flavor space by means of statistical
operators since we are interested in decohering evolutions. In particular,
$\rho_{\nu_{e}} = \nel \langle \nu_e \vert$ and $\rho_{\nu_{\mu}} = 
\nm \langle \nu_{\mu} \vert$, whence~(\ref{eq4.1.4}) reads

\be
\la{eq4.1.6}
P_{\nu_e \rightarrow \nu_{\mu}}(t) = \Tr[\rho_{\nu_e}(t) \rho_{\nu_{\mu}}].
\ee

\noi
We shall also adopt the vector representation, developed in
Chapter~\ref{ch2}, for the density matrices.


\s{The model of interaction}
\la{s4.2}
\markright{\small{\rm{\thesection. THE MODEL OF INTERACTION}}}

In the previous section we have described the behavior of neutrinos 
in vacuum. Here we consider the case of neutrinos interacting with
a suitable environment, consisting of a slab of matter of volume $V$.
The Hamiltonian describing both neutrinos and environment has the form
$H_T = H_S + H_B + H_I$, where $H_S$ is the neutrino Hamiltonian 
(\ref{eq4.1.2}), $H_B$ is the Hamiltonian of the environment, 
and $H_I$ is the contribution 
containing the interaction between neutrinos and environment. In this 
section we shall derive its form.

\begin{Rem}
\la{rem4.2.0}
{\rm In order to have a general $H_I$, it is not possible to consider 
a two-dimensional $\mathfrak{H}_S$, as in vacuum; momentum dependence
needs also to be taken into account. Nevertheless, we shall see that, 
under certain circumstances, a two-dimensional effective description 
is available also in the open system context.}
\end{Rem}

\noi
The interaction
$H_I$ is determined by a Hamiltonian density $\mathfrak{h}$ by

\be
\la{eq4.2.1}
H_I(t) = \int\limits_V \mathfrak{h}(x) d\vec{x},
\ee

\noi
where $x \equiv (\vec{x}, t)$ (the dependence on time in $H_I$
disappears if we suppose the layer of matter to be uniform).
We assume a linear coupling of the form

\be
\la{eq4.2.2}
\mathfrak{h}(x) = g \sum_{\alpha} V_S^{\alpha}(x) \otimes V_B^{\alpha}(x),
\ee

\noi
with

\be
\la{eq4.2.3}
V_S^{\alpha}(x) \equiv V_S^{lm \mu}(x) =
\overline{\nu}_l(x) \gamma_{\mu}(1 - \gamma_5) \nu_m(x),
\ee

\noi
where $g$ is a suitable coupling constant (dimensionally it is an inverse
energy squared), $V_{S,B}^{\alpha}(x)$ are the neutrino and bath
currents that characterize the interactions and $\alpha$ is a set of flavor 
and Dirac indices. The $\gamma_{\mu}$'s, $\mu = 0, \ldots 3$ are 
the Dirac matrices and $\gamma_5 = i \gamma_0 \gamma_1 \gamma_2 \gamma_3$
while $\nu_l (x)$, $l = e, \mu$, are the fermionic neutrino fields.
To guarantee the Hermiticity of $H_I$ we require

\be
\la{eq4.2.4}
V_B^{lm\mu \dagger}(x) = V_B^{ml \mu}(x),
\ee

\noi
otherwise, $V_B^{\alpha}(x)$ is completely arbitrary.
The indices $S$ and $B$ denote the neutrino system and the environment 
respectively, according to the notation of Chapter~\ref{ch1}.

\begin{Rem}
\la{rem4.2.1}
{\rm The assumption~(\ref{eq4.2.3}) is quite general; for instance,
it includes transitions forbidden in the Standard Model, like neutrinos 
flavor-changing interactions~\cite{mcka,zee,babu,john}. 
These latter are induced by suitable
environment operators $V_{B}^{lm \mu}$ with $l \ne m$.}
\end{Rem}

\begin{Rem}
\la{rem4.2.1.bis}
{\rm To describe neutrino interactions in the Sun, a low-energy Hamiltonian
density is actually introduced; it contains 
neutral currents and charged currents contributions in
the point-interaction approximation (Fermi theory) to the Standard 
Model~\cite{bil}:
\be
\la{eq4.7.1}
\begin{split}
{\mathfrak h}(x) = \frac{G_F}{\sqrt{2}}\sum_l \sum_{\mu}
&\overline{\nu}_l(x) \gamma_{\mu}(1 - \gamma_5)\nu_l(x) \otimes
\Bigl\{ \delta_{le} \overline{e}(x) \gamma^{\mu}(1 - \gamma_5) e(x) + \\
&+ \sum_f \overline{f}(x) \gamma^{\mu} \left[ I_3 (1 - \gamma_5) -
2Q \sin^2{\theta_W} \right] f(x) \Bigr\},
\end{split}
\ee
where $l = e, \mu$ identifies the neutrino fields; $f = e, p, n$ characterizes
the fermionic fields (electrons, protons and neutrons respectively); $I_3$
and $Q$ are the isospin and charge operators; $\theta_W$ is the
weak angle and $G_F$ the Fermi constant. In this case, only flavor-preserving
transitions are allowed.}
\end{Rem}

\noi
In the representation for which $\gamma_5$ is diagonal, the left component 
of neutrino field can be written as

\be
\la{eq4.2.5}
\nu_l(x)_L = \frac{1}{2}(1 - \gamma_5)\nu_l (x) = 
\frac{1}{\sqrt{2}} 
\left(
\begin{array}{c}
\phi_l (x) \\
-\phi_l (x) \\
\end{array} 
\right);
\ee

\noi
correspondingly the neutrino currents become

\be
\la{eq4.2.6}
V_S^{lm \mu}(x) = 2 \phi_l^{\dagger}(x) \sigma_{\mu} \phi_m (x),
\ee

\noi
where \{$\sigma_{\mu}, \; \mu = 0, \ldots ,3$\} is the set of Pauli
matrices plus the $2 \times 2$ identity. The Weyl spinors $\phi_l (x)$ can 
be expanded only in the sector of interest over components of negative 
helicity and positive energy~\cite{mann} so that at $t = 0$

\be
\la{eq4.2.7}
\phi_l (\vec{x},0) = \frac{1}{\sqrt{V}} \sum_{\vec{k}} e^{i \vec{k}
\cdot \vec{x}} \beta (\vec{k}) \phi_l (\vec{k}),
\ee

\noi
where $V$ is the volume of the layer of matter crossed by the neutrino, and
$\beta (\vec{k})$ is the negative elicity eigenstate,

\be
\la{eq4.2.8}
\frac{\sigma \cdot \vec{k}}{k} \beta(\vec{k}) = -\frac{1}{2}\beta(\vec{k}).
\ee

\noi
$\phi_l (\vec{k})$ are the annihilation operators of neutrino states with 
flavor $l$ and momentum $\vec{k}$ and, together with the creation
operators $\phi_l^{\dagger} (\vec{k})$, they satisfy anti-commutation 
relations:

\be
\la{eq4.2.9}
\phi_l^{\dagger} (\vec{k}) \vert 0 \rangle = \vert l, \vec{k} \rangle; \quad
\{ \phi_l (\vec{k}), \phi_l^{\dagger}(\vec{k}^{\prime})\} = 
\delta_{\vec{k} \vec{k}^{\prime}}, \quad
\{ \phi_l (\vec{k}), \phi_l(\vec{k}^{\prime})\} = 
\{ \phi_l^{\dagger} (\vec{k}), \phi_l^{\dagger}(\vec{k}^{\prime})\} = 0.
\ee

\noi
Using~(\ref{eq4.2.7}) the currents~(\ref{eq4.2.3}) read

\be
\la{eq4.2.10}
V_S^{\alpha}(\vec{x},0) = \frac{1}{V} \sum_{\vec{k}, \vec{k}^{\prime}}
e^{i(\vec{k}^{\prime} - \vec{k})\cdot \vec{x}}
S_{\mu}(\vec{k}, \vec{k}^{\prime}) \phi_l^{\dagger}(\vec{k})
\phi_m (\vec{k}^{\prime}),
\ee

\noi
where $\alpha = (l,m,\mu)$ and $S_{\mu}(\vec{k}, \vec{k}^{\prime})
\equiv \beta^{\dagger}(\vec{k}) \sigma_{\mu} \beta (\vec{k}^{\prime})$.
The behavior of this quantity under complex conjugation is

\be
\la{eq4.2.11}
[S_{\mu}(\vec{k}, \vec{k}^{\prime})]^* = S_{\mu}(\vec{k}^{\prime},\vec{k}).
\ee

\noi
These quantities shall be employed in the following sections.

The generic neutrino state $\rho_S$ in the finite volume $V$ can be 
written in terms of flavor and momentum eigenstates:

\be
\la{eq4.2.12}
\rho_S(t) = \sum_{i,j} \sum_{\vec{r}, \vec{s}}
\rho_{ij}(\vec{r},\vec{s};t) \vert i \vec{r} \rangle
\langle j \vec{s} \vert = \sum_{i,j} \sum_{\vec{r}, \vec{s}}
\rho_{ij}(\vec{r},\vec{s};t) \phi_i^{\dagger}(\vec{r}) \vert 0 \rangle
\langle 0 \vert \phi_j (\vec{s}),
\ee

\noi
where the creation and annihilation operators satisfy~(\ref{eq4.2.9}).


\s{The Markovian approximation}
\la{s4.3}
\markright{\small{\rm{\thesection. THE MARKOVIAN APPROXIMATION}}}

Following the standard steps introduced in Chapter~\ref{ch1}, in order
to deal with the environment and to end up with a Markov 
time-evolution for the neutrino state $\rho_S$, we have to perform a
suitable approximation.

In Section~\ref{s1.4} we have described two different procedures to
obtain a Markovian approximation for the evolution of a system 
in a thermal bath. Introduced the typical times $\tau_S$
and $\tau_B$, characterizing system and environment respectively,
a Markovian approximation amounts to $\tau_S / \tau_B \rightarrow +\infty$,
that is either $\tau_S\rightarrow +\infty$ or $\tau_B \rightarrow 0$.
In the physical context we are actually considering, the typical
time of the system is of the order of the period of oscillations,
$\tau_S \sim E/ \Delta m^2$. On the other hand, the characteristic time of 
the environment, $\tau_B$, is related to the decay time of its correlations.
To get a Markovian approximation, we consider the case
$\tau_B \rightarrow 0$; that is, we consider environments whose 
correlation functions decay in sufficiently short times. 
As we have seen in Section~\ref{s2.4}, this procedure
corresponds to the singular coupling limit.

\begin{Rem}
{\rm We are considering a homogeneous thermal bath in a finite volume $V$, 
thus the Markovian approximation, that requires an infinite volume, is to
be performed in the hypothesis of a volume $V$ big enough so that the 
Poincar\'e recurrencies can be neglected~\cite{alic1}.}
\end{Rem}

The singular coupling amounts to considering two-point correlation functions 
behaving like Dirac-deltas in time: 

\be
\la{eq4.3.1}
h_{\alpha \alpha^{\prime}}(x, x^{\prime}) = 
\Tr_B [\rho_B V_B^{\alpha}(\vec{x}, t) V_B^{\alpha^{\prime}}
(\vec{x}^{\prime}, t^{\prime})] \rightarrow c_{\alpha 
\alpha^{\prime}}(\vec{x},\vec{x}^{\prime})\delta(t - t^{\prime}).
\ee

According to the discussion of the previous chapters, the resulting Markovian 
evolution equation is:

\be
\la{eq4.3.2}
\begin{split}
\dot{\rho}_S (t) = -i \left[H_S^{eff}, \rho_S (t)\right] +
\frac{1}{2}\sum_{\alpha, \alpha^{\prime}} \iint\limits_V 
&c_{\alpha \alpha^{\prime}}(\vec{x}, 
\vec{x}^{\prime})([V_S^{\alpha}(\vec{x},0)\rho_S (t), 
V_S^{\alpha^{\prime}}(\vec{x}^{\prime},0)] + \\
&+ [V_S^{\alpha}(\vec{x},0), \rho_S(t) 
V_S^{\alpha^{\prime}}(\vec{x}^{\prime},0)])d\vec{x} 
d\vec{x}^{\prime}
\end{split}
\ee

\noi
and the statistical operator $\rho_S$ lives in the neutrino Hilbert space
$\mathfrak{H}_S$, containing flavor and space degrees of freedom. 
We shall later try to reduce further the description onto the flavor
degrees of freedom.
In~(\ref{eq4.3.2}), $H_S^{eff}$ is the effective neutrino Hamiltonian
obtained from $H_S$ by  
taking into account the mean field term produced by the interaction
with the environment: $H_S^{eff} = H_S + \Tr_B(\rho_B H_I)$, with $H_I$ 
defined in~(\ref{eq4.2.1}) and $\rho_B$ the thermal equilibrium state of 
the bath; see the discussion in Section~\ref{s1.3}.
It is the effective Hamiltonian usually introduced in the 
description of the interactions of neutrinos in matter leading to the
MSW effect~\cite{wolf,mikh,1mikh,bil}. 
In the two-dimensional flavor space, the representation of
$H_S^{eff}$ over the physical states is

\be
\la{eq4.3.5}
H_S^{eff} = \frac{\Delta m^2}{4E}
\left(
\begin{array}{cc}
-\cos{2\theta} & \sin{2\theta} \\
\sin{2\theta} & \cos{2\theta} \\
\end{array}
\right) +
\left(
\begin{array}{cc}
V_{ee} & V_{e\mu} \\
V_{\mu e} & V_{\mu \mu} \\ 
\end{array}
\right)
\ee

\noi
where 

\be
\la{eq4.3.5.bis}
\begin{array}{ll}
V_{ee} \equiv \langle \nu_e \vert \Tr_B(\rho_B H_I) \vert \nu_e \rangle,
\quad
&V_{e\mu} \equiv \langle \nu_e \vert \Tr_B(\rho_B H_I) \vert \nu_{\mu} \rangle,
\quad \\ \\
V_{\mu \mu} \equiv \langle \nu_{\mu} 
\vert \Tr_B(\rho_B H_I) \vert \nu_{\mu} \rangle,
\quad
&V_{\mu e} \equiv \langle \nu_{\mu}\vert \Tr_B(\rho_B H_I) \vert \nu_e \rangle,
\quad
\end{array}
\ee

\noi
are contributions of first order in $H_I$ and 
$V_{ee}, V_{\mu \mu} \in {\bf R}$, $V_{\mu e} = V_{e \mu}^*$. 

\begin{Rem}
\la{rem4.3.1}
{\rm The off-diagonal entries vanish for flavor-preserving interactions.
Indeed, in this case in~(\ref{eq4.2.2}) we have 
\be
\la{eq4.3.6.add}
V_B^{lm \mu}(x)
= \delta_{lm} V_B^{ll \mu}(x),
\ee 
where $l,m = e, \mu$. For instance, for neutrinos in solar matter, we find:
\be
\la{eq4.7.2}
V_{ee} = \sqrt{2} G_F \left( n_e - \frac{1}{2} n_n \right), \quad
V_{\mu \mu} = -\frac{1}{\sqrt{2}} G_F n_n, \quad
V_{e \mu} = V_{\mu e} = 0,
\ee
where $n_e$ and $n_n$ are the density of electrons, neutrons respectively, and
$G_F$ is the Fermi constant.}
\end{Rem}

The effective Hamiltonian is diagonalized by an orthogonal transformation
$U$ that connects the eigenstates of $H_S^{eff}$ to the physical states:

\be
\la{eq4.3.6}
\left\{
\begin{array}{l}
\nel = U \vert \nu_{m1} \rangle, \\ \\
\nm = U \vert \nu_{m2} \rangle; 
\end{array}
\right. \quad
U = 
\left(
\begin{array}{cc}
\cos{\theta_m} & \sin{\theta_m} \\
-\sin{\theta_m} & \cos{\theta_m} \\
\end{array}
\right),
\ee

\noi
where \{$\vert \nu_{m1}\rangle$, $\vert \nu_{m2} \rangle$\} are the 
eigenstates of $H_S^{eff}$ and $\theta_m$ is the mixing angle
in matter. The Hamiltonian eigenvalues in the adopted relativistic 
approximation are $\epsilon_{m1} = -\Delta m_m^2/4E$ and $\epsilon_{m2} =
\Delta m_m^2/4E$ where $\Delta m_m^2 = m_{m2}^2 - m_{m1}^2$ and $m_{m1}$,
$m_{m2}$ are the neutrino masses in the medium. They are related to
the equivalent quantities in vacuum; if $V_{e \mu} = V_{\mu e} = 0$, as
in the case of solar matter, we have~\cite{mikh,1mikh}

\begin{subequations}
\la{eq4.3.7}
\begin{gather}
\la{eq4.3.7.a}
\tan{2\theta_m} = \frac{\Delta m^2 \sin{2\theta}}{\Delta m^2 
\cos{2\theta} - A}, \\
\la{eq4.3.7.b}
\Delta m_m^2 =
\sqrt{A^2 + \left(\Delta m^2 \right)^2 -2A\Delta m^2
\cos{2\theta}},
\end{gather}
\end{subequations}

\noi
where

\be
\la{eqnuova}
A \equiv 2E (V_{ee} - V_{\mu \mu}).
\ee

\begin{Rem}
\la{rem4.3.2}
{\rm The resonance in equation~(\ref{eq4.3.7.a}) produces an enhancement 
of the oscillating behavior that, combined with a suitable variation
of the density of the crossed medium, generates the so-called MSW effect, 
actually the best candidate to the explanation of the problem of solar 
neutrinos~\cite{mikh,1mikh}.}
\end{Rem}

To first order in the interaction (that is neglecting the dissipative 
contributions of equation~(\ref{eq4.3.2})), the transition probability in 
the uniform medium considered reads

\be
\la{eqnuova2}
P_{\nu_e \rightarrow \nu_{\mu}}(t) = 
\sin^2{2 \theta_m} \sin^2{\frac{\Delta m_m^2}{4E} t}.
\ee

The dissipative effects are of second order in $H_I$; they result
in Hamiltonian, damping and noise terms as results from rearranging 
equation~(\ref{eq4.3.2}):

\be
\la{eq4.3.8}
\begin{split}
\dot{\rho}_S (t) = -i \left[H_S^{eff}, \rho_S (t)\right] &-
\frac{1}{2}\sum_{\alpha, \alpha^{\prime}} \iint
\limits_V d\vec{x} d\vec{x}^{\prime}
b_{\alpha \alpha^{\prime}}(\vec{x}, \vec{x}^{\prime})
[V_S^{\alpha}(\vec{x},0) V_S^{\alpha^{\prime}}(\vec{x}^{\prime},0),
\rho_S (t)] + \\
&- \frac{1}{2}\sum_{\alpha, \alpha^{\prime}} \iint
\limits_V d\vec{x} d\vec{x}^{\prime}
a_{\alpha \alpha^{\prime}}(\vec{x}, \vec{x}^{\prime})
\{V_S^{\alpha}(\vec{x},0) V_S^{\alpha^{\prime}}(\vec{x}^{\prime},0),
\rho_S (t)\} + \\
&+ \sum_{\alpha, \alpha^{\prime}} \iint
\limits_V d\vec{x} d\vec{x}^{\prime}
a_{\alpha \alpha^{\prime}}(\vec{x}, \vec{x}^{\prime})
V_S^{\alpha^{\prime}}(\vec{x}^{\prime},0) \rho_S (t) V_S^{\alpha}(\vec{x},0),
\end{split}
\ee

\noi
where the new coefficients are defined as

\begin{subequations}
\la{eq4.3.9}
\begin{gather}
\la{eq4.3.9.a}
a_{\alpha \alpha^{\prime}}(\vec{x}, \vec{x}^{\prime}) \equiv
\int_{-\infty}^{\infty} h_{\alpha \alpha^{\prime}}(\vec{x}, 
\vec{x}^{\prime};u) du; \\
\la{eq4.3.9.b}
b_{\alpha \alpha^{\prime}}(\vec{x}, \vec{x}^{\prime}) \equiv
\int_{0}^{\infty} h_{\alpha \alpha^{\prime}}(\vec{x}, 
\vec{x}^{\prime};u) du - \int_{-\infty}^{0} h_{\alpha \alpha^{\prime}}
(\vec{x}, \vec{x}^{\prime};u) du.
\end{gather}
\end{subequations}

According to Section~\ref{s1.5},
in equation~(\ref{eq4.3.8}) we recognize the typical form of the generator 
of a Markov semigroup. 

\begin{Rem}
\la{rem4.3.3}
{\rm Accordingly to Remark~\ref{rem1.5.3}, in the first row of~(\ref{eq4.3.8})
we note that the vacuum Hamiltonian $H_S$ is redefined by contributions of
first order in the interaction, already accounted for in $H_S^{eff}$, and
of second order, determined by the generalized matrix 
$[b_{\alpha \alpha^{\prime}}(\vec{x},\vec{x}^{\prime})]$.}
\end{Rem}

\begin{Rem}
\la{rem4.3.4}
{\rm Since the time-evolution~(\ref{eq4.3.8}) is obtained from a singular 
coupling, it is completely positive and the generalized matrix
$[a_{\alpha \alpha^{\prime}}(\vec{x},\vec{x}^{\prime})]$ is positive definite.}
\end{Rem}


\s{The generator of the dissipative evolution}
\la{s4.4}
\markright{\small{\rm{\thesection. THE GENERATOR OF THE DISSIPATIVE EVOLUTION}}}

We want to write the form of the generator of the dissipative evolution,
eq.~(\ref{eq4.3.8}) in the flavor space $\mathfrak{H}_2$ only; we thus have
to get rid of the momentum degrees of freedom. We start by considering the
expression~(\ref{eq4.2.12}) for the statistical operator $\rho_S$.
By rearranging the summation indices, 
(\ref{eq4.3.8}) results in the following equation for the entries of
$\rho_S(t)$:

\be
\la{eq4.4.2}
\begin{split}
\dot{\rho}_{ij}(\vec{r},\vec{s};t) = &-i \sum_{k} 
\left[\left(H_S^{eff}\right)_{ik}
\rho_{kj}(\vec{r},\vec{s};t) - \rho_{ik}(\vec{r},\vec{s};t)
(H_S^{eff})_{kj}\right] + \\
&- \frac{g^2}{V^2} \sum_{l,m}\sum_{\vec{k}, \vec{q}}
\left\{ \left[ A_{imlm}(\vec{r},\vec{q},\vec{k},\vec{q})
+ B_{imlm}(\vec{r},\vec{q},\vec{k},\vec{q}) \right] \rho_{lj}
(\vec{k},\vec{s};t) + \right. \\
&\quad\quad + \left[ A_{lmjm}(\vec{q},\vec{k},\vec{s},\vec{k})
- B_{lmjm}(\vec{q},\vec{k},\vec{s},\vec{k}) \right] \rho_{il}
(\vec{r},\vec{q};t) +  \\ 
&\quad\quad - \left. 2A_{ljli}(\vec{k},\vec{s},\vec{q},\vec{r}) \rho_{lm}
(\vec{q},\vec{k};t) \right\},
\end{split}
\ee

\noi
where

\begin{subequations}
\la{eq4.4.3}
\begin{gather}
A_{lml^{\prime}m^{\prime}} 
(\vec{k},\vec{k}^{\prime},\vec{q},\vec{q}^{\prime})
\equiv \sum_{\mu, \mu^{\prime}}
\iint\limits_V d\vec{x} d\vec{x}^{\prime} a_{\alpha \alpha^{\prime}}
(\vec{x},\vec{x}^{\prime})e^{i(\vec{k}^{\prime} - \vec{k})\cdot\vec{x}}
e^{-i(\vec{q}^{\prime} - \vec{q})\cdot\vec{x}^{\prime}}
S_{\mu}(\vec{k},\vec{k}^{\prime}) 
S_{\mu^{\prime}}(\vec{q}^{\prime},\vec{q}), \\
B_{lml^{\prime}m^{\prime}} 
(\vec{k},\vec{k}^{\prime},\vec{q},\vec{q}^{\prime})
\equiv \sum_{\mu, \mu^{\prime}}
\iint\limits_V d\vec{x} d\vec{x}^{\prime} b_{\alpha \alpha^{\prime}}
(\vec{x},\vec{x}^{\prime})e^{i(\vec{k}^{\prime} - \vec{k})\cdot\vec{x}}
e^{-i(\vec{q}^{\prime} - \vec{q})\cdot\vec{x}^{\prime}}
S_{\mu}(\vec{k},\vec{k}^{\prime}) 
S_{\mu^{\prime}}(\vec{q}^{\prime},\vec{q}), 
\end{gather}
\end{subequations}

\noi
and $\alpha = (l,m,\mu)$, $\alpha^{\prime} = 
(l^{\prime},m^{\prime},\mu^{\prime})$.

Equation~(\ref{eq4.4.2}) is rather complex; however, under suitable 
conditions, it is possible to simplify it, reducing the effective Hilbert 
space to a two-dimensional one. The Hilbert space describing
our system is $\mathfrak{H}_S = \mathfrak{H}_2 \otimes \mathfrak{H}_p$,
where $\mathfrak{H}_2$ is the two-dimensional flavor Hilbert space whereas
$\mathfrak{H}_p$ is the momentum Hilbert space. The
reduction to $\mathfrak{H}_2$ is performed by adapting the projector technique
developed in Section~\ref{s1.3}. We shall see that such a reduction
will produce a nicer equation than~(\ref{eq4.4.2}), under conditions that
are not too restrictive.

\begin{Rem}
\la{rem4.4.1}
{\rm Usually the formalism of projectors is applied to
eliminate the environment degrees of freedom, leaving only those 
of the system of interest. In the present case, however, the trace
over the environment has alredy been performed, and an adapted version of
the projection technique allows us to discard the momentum degrees of 
freedom.}
\end{Rem}

Let us start with the time-evolution~(\ref{eq4.4.2}); for later 
convenience we shall adopt the compact notation

\be
\la{eqcompact}
\dot{\rho}_S(t) = L [\rho _S(t)] = 
(L_H + L_D) [\rho_S(t)],
\ee

\noi
in which we have separated the Hamiltonian and the dissipative part 
of the generator~(\ref{eq4.3.8}). In particular, $L_H [\rho_S(t)]$
represents the first row of~(\ref{eq4.3.8}) while $L_D [\rho_S(t)]$
the latter two.

The state $\rho_S$ acts on $\mathfrak{H}_2 
\otimes \mathfrak{H}_p$ and we want to extract the time-evolution of the
reduced statistical operator acting on $\mathfrak{H}_2$ only. We will
denote this latter by $\rho$; it is obtained by tracing $\rho_S$ over 
the momentum degrees of freedom: $\rho = \Tr_p (\rho_S)$.

In order to apply the projector technique, we need a reference
state for the momentum degrees of freedom: we choose $\rho_p = \vert \vec{p} 
\rangle \langle \vec{p} \vert$, where $\vec{p}$ is the momentum of the 
incoming neutrino; accordingly we define the projectors $P[\,\cdot\,] =
\Tr_p (\,\cdot\,) \otimes \rho_p$ and $Q = {\bf I} - P$. Adapting 
relations~(\ref{eq1.3.6}) and (\ref{eq1.3.7}) to this case,
from~(\ref{eqcompact}) we extract the following equation of motion:

\be
\la{eq4.4.4}
\dot{\rho} (t) = P(L_H + L_D) \rho (t) +
\int_0^t dt^{\prime} P(L_H + L_D)
e^{Q L (t - t^{\prime})} Q L_D    \rho (t^{\prime}),
\ee

\noi
where we have used the property $Q L_H P = 0$ and supposed the
initial state to be factorized: $\rho_S(0) = \rho(0) \otimes \rho_p$.

This is not a manageable expression yet because of the convolution term
in the right-hand side. 
In order to get a time-evolution having the structure of semigroup, 
for flavor degrees of freedom only, we have to make further assumptions.

After explicit computation, the dependence of $A(\vec{k},
\vec{k}^{\prime},\vec{q},\vec{q}^{\prime})$ and $B(\vec{k},
\vec{k}^{\prime},\vec{q},\vec{q}^{\prime})$ on the momenta, in the 
operators $P L_D$ and $Q L_D$, is rather
different: in $P L_D$ these coefficients present in any case two
pairs of equal momentum indices, while this never happens for 
$Q L_D$. To understand the physical meaning of this, observe 
that these coefficients are Fourier transforms,
for example

\be
\la{eq4.4.5}
A(\vec{k},\vec{k}^{\prime},\vec{q},\vec{q}^{\prime})
\approx \iint\limits_V d\vec{x} d\vec{x}^{\prime} a(\vec{x}, \vec{x}^{\prime})
e^{i(\vec{k}^{\prime} - \vec{k})\cdot \vec{x}}
e^{-i(\vec{q}^{\prime} - \vec{q})\cdot \vec{x}^{\prime}},
\ee

\noi
where $a(\vec{x}, \vec{x}^{\prime})$ is a spatial correlation function 
of the bath; in general, 
it can be written as the sum of two functions: $f(\vec{x},
\vec{x}^{\prime}) = f(\vert \vec{x} - \vec{x}^{\prime}\vert)$, 
traslationally invariant, and a contribution
$g(\vec{x},\vec{x}^{\prime})$:

\be
\la{eq4.4.6}
a(\vec{x}, \vec{x}^{\prime}) = f(\vert \vec{x} - \vec{x}^{\prime} \vert)
+ g(\vec{x}, \vec{x}^{\prime}).
\ee

\noi
Therefore~(\ref{eq4.4.5}) becomes

\be
\la{eq4.4.7}
A(\vec{k},\vec{k}^{\prime},\vec{q},\vec{q}^{\prime}) \approx
\delta_{(\vec{k} - \vec{k}^{\prime}),(\vec{q} - \vec{q}^{\prime})}
\int\limits_V d\vec{z} f(\vert z \vert) e^{i(\vec{q}^{\prime}- \vec{q})\cdot 
\vec{z}} + \iint\limits_V d \vec{x} d\vec{x}^{\prime} g(\vec{x}, \vec{x}^{\prime})
e^{i(\vec{k}^{\prime} - \vec{k})\cdot \vec{x}}
e^{-i(\vec{q}^{\prime} - \vec{q})\cdot \vec{x}^{\prime}}.
\ee

The contributions to $P L_D$ come from both the terms in the 
right-hand side of this equation whereas the contributions to
$Q L_D$ are produced only from the Fourier transform of $g(\vec{x},
\vec{x}^{\prime})$. Translational
invariance means momentum conservation, therefore the term 
$P L_D$ does not change momentum, $Q L_D$ instead
does. Thus, if the correlation function $a(\vec{x},\vec{x}^{\prime})$ is
traslationally invariant (i.e. $g(\vec{x}, \vec{x}^{\prime}) = 0$) 
automatically $Q L_D = 0$ and the integral term 
in eq.~(\ref{eq4.4.4}) vanishes. In general it could be neglected
if, in $a(\vec{x},\vec{x}^{\prime})$, the traslationally invariant part
$f(\vert \vec{x} - \vec{x}^{\prime}\vert)$ is dominant with respect to 
the non invariant one, $g(\vec{x},\vec{x}^{\prime})$. 

In the following, we work within the hypothesis of almost translational
invariance within $V$, so that~(\ref{eq4.4.4}) becomes:

\be
\la{eq4.4.9}
\dot{\rho} (t) = P(L_H + L_D) \rho (t),
\ee

\noi
and the evolution equation for the entries of the $2 \times 2$ statistical 
operator $\rho$ reads

\be
\la{eq4.4.10}
\begin{split}
\dot{\rho}_{ij}(t) = &-i \sum_k \Bigl[
\left(H_S^{eff}\right)_{ik} \rho_{kj}(t) - 
\rho_{ik}(t) \left(H_S^{eff}\right)_{kj}
\Bigr] + \\
&- \frac{g^2}{V^2} \sum_l \Bigl\{
[ A_{il}(\vec{p}) + B_{il}(\vec{p}) ] \rho_{lj}(t) + 
[ A_{lj}(\vec{p}) - B_{lj}(\vec{p}) ] \rho_{il}(t) + 
2 \sum_m A_{mjli}(\vec{p}) \rho_{lm}(t) \Bigr\},
\end{split}
\ee

\noi
where

\begin{subequations}
\la{eq4.4.11}
\begin{gather}
A_{imjl}(\vec{p}) = \sum_{\vec{k}} A_{imjl}(\vec{p},\vec{k},\vec{p},\vec{k}),
\quad A_{ij}(\vec{p}) = \sum_m A_{imjm}(\vec{p}); \\
B_{ij}(\vec{p}) = \sum_{\vec{k}}\sum_m B_{imjm}
(\vec{p},\vec{k},\vec{p},\vec{k}).
\end{gather}
\end{subequations}

\noi
In vector representation, using the physical basis for 
${\mathfrak H}_2$, this equation assumes the form

\be
\la{eq4.4.12}
\frac{d}{dt}\vert \rho (t) \rangle = {\cal L} \vert \rho(t) \rangle = 
({\cal H} + {\cal D}) \vert \rho(t) \rangle,
\ee

\noi
where

\be
\la{eq4.4.12.bis}
{\cal H} = -2\left(
\begin{array}{cccc}
0 & 0 & 0 & 0 \\
0 & 0 & h_3 & -h_2 \\
0 & -h_3 & 0 & h_1 \\
0 & h_2 & -h_1 & 0 \\
\end{array}
\right), \quad
{\cal D} = 
-2 \left(
\begin{array}{cccc}
0 & 0 & 0 & 0 \\
u & a & b & c \\
v & b & \alpha & \beta \\
w & c & \beta & \gamma \\
\end{array}
\right).
\ee

The matrix ${\cal L} = {\cal H} + {\cal D}$
is the generator of the reduced time-evolution we are looking for.
The parameters characterizing the Hamiltonian part are

\be
\la{eq4.4.13}
\begin{split}
h_1 &= 
\frac{\Delta m^2}{4E} \sin{2\theta} + \re V_{e\mu} + \omega_1, \\
h_2 &= 
- \im V_{e\mu} + \omega_2, \\
h_3 &= 
-\frac{\Delta m^2}{4E} \cos{2\theta} + \frac{1}{2} (V_{ee} - V_{\mu \mu}) + \omega_3,
\end{split}
\ee

\noi
where $\omega_1$, $\omega_2$ and $\omega_3$ are the Hamiltonian contributions
of second order in the interaction (see Remark~\ref{rem4.3.3}), whose 
expressions are reported in Appendix~\ref{a4}.

Following the discussion of Sections~\ref{s1.8} and \ref{s2.3}, 
the coefficients 
$u$, $v$ and $w$ in the dissipative part ${\cal D}$ are related to 
entropy behavior and to the existence of asymptotic states 
different from the maximally mixed state $\vert \rho_2 \rangle =
1/2 (1, 0, 0, 0)$. The remaining parameters, instead, produce decoherence 
and dissipation. The expressions of all of these parameters are listed in
Appendix~\ref{a4}.

These coefficients satisfy the inequalities~(\ref{eq2.2.6}) that 
guarantee the complete positivity of the evolution because they
are obtained via a singular coupling limit~\cite{gori2}.

In the neutrino's case 
the impact of complete positivity cannot be fully appreciated since this 
property is needed for the physical consistency of the time-evolution of 
bipartite entangled systems, and there are not phenomenological contexts 
of this kind in neutrino physics.

\begin{Rem}
\la{rem4.4.2}
{\rm If the interactions are flavor-preserving most of these coefficients 
vanish since we have
\be
\la{eq4.4.16.bis}
A_{imjl} = \delta_{im} \delta_{jl}A_{iijj}, \quad
A_{ij} = \delta_{ij} A_{ii}, \quad B_{ij} = \delta_{ij} B_{ii}.
\ee
Considering the expressions reported in Appendix~\ref{a4}, we find 
$\omega_1 = \omega_2 = 0$, $u = v = w = 0$, $a = \alpha$ and $b = c = 
\gamma = 0$; the dissipative contribution is fully parameterized by only one 
coefficient, $a$, in agreement with~\cite{burg1}.}
\end{Rem}


\s{An example of flavor changing dissipative evolution}
\la{s4.5}
\markright{\small{\rm{\thesection. AN EXAMPLE OF FLAVOR CHANGING DISSIPATIVE EVOLUTION}}}

As already mentioned (see Section~\ref{s2.4}), the exact integration of 
equation~(\ref{eq4.4.12}) is always possible, but the expression of
${\cal G}_t$ is not transparent enough to be amenable to physical
interpretations. It is more convenient to expand ${\cal G}_t$ to first 
order in the dissipative parameters appearing in ${\cal D}$; this computation 
has already been motivated and developed in Section~\ref{s2.4} and
applied to neutrinos in~\cite{benan,benan2}.

In this section, however, we wish to give the exact transition probability 
in a simplified case~\cite{benan}: $h_2 = 0$, $u = v = w = 0$, 
$a = \alpha = \gamma$ and $c = 0$. As explained in Remark~\ref{rem4.4.2},
this latter case amounts to flavor-changing interactions since $b$, 
$\beta$ and $\gamma$ are not vanishing and it is compatible with the 
inequalities expressing the complete positivity of the evolution provided 
that $a^2 \geqslant b^2 + \beta ^2$.

\begin{Rem}
\la{rem4.5.1}
{\rm In our vector representation the transition probability can be expressed
as an inner product:
\be
\la{eq4.5.0}
{\cal P}_{\nu_e \rightarrow \nu_{\mu}}(t) = 
2 \langle \rho_{\nu_{\mu}} \vert \rho_{\nu_e}(t) \rangle = 
2 \langle \rho_{\nu_{\mu}} \vert {\cal G}_t \vert \rho_{\nu_e} \rangle,
\ee
as can be seen by putting in~(\ref{eq4.1.6}) the expansions of both
$\rho_{\nu_e}$ and $\rho_{\nu_{\mu}}$ over the basis of Pauli matrices 
plus identity.}
\end{Rem}

\noi
After exponentiation of the generator ${\cal L}$, the time-evolutor 
${\cal G}_t = e^{{\cal L}t}$ amounts to

\be
\la{eq4.5.1}
{\cal G}_t = \left(
\begin{array}{cccc}
1 & 0 & 0 & 0 \\
0 & & & \\
0 & & \tilde{\cal G}_t & \\
0 & & & \\
\end{array}
\right)
\ee

\noi
where the $3 \times 3$ sub-matrix $\tilde{\cal G}_t$ is given by

\be
\la{eq4.5.2}
\tilde{\cal G}_t = e^{-2at} \left[ 
{\bf I}_3 +
\frac{\sin{2 \Theta_m t}}
{2 \Theta_m} {\cal M} +
\frac{\sin^2{\Theta_m t}}
{4 \Theta_m^2} {\cal M}^2 \right],
\ee

\noi
with

\be
\la{eq4.5.3}
{\cal M} = -2 \left(
\begin{array}{ccc}
0 & b + h_3 & 0 \\
b - h_3 & 0 & \beta + h_1 \\
0 & \beta - h_1 & 0 \\
\end{array}
\right)
\ee

\noi
and ${\bf I}_3$ the $3 \times 3$ identity matrix. We have introduced 
$\Theta_m = \sqrt{h_1^2 + h_3 ^2 - b^2 - \beta^2}$, that can be
either real or complex, depending on the relative size of the coefficients
appearing in it. Since in the present context the dissipative parameters
are small, we consider a real $\Theta_m$; the general case is presented
in~\cite{benan}.

\begin{Rem}
\la{rem4.5.2}
{\rm Note that, that, in absence of dissipation (i.e. $b = \beta = 0$) 
$\Theta_m$ reduces to $\Delta m_m^2 /4E$, and then, in absence of matter,
to $\Delta m^2 /4E$.}
\end{Rem}

\noi
In vectorial notation, the pure $\nu_e$ state (the initial state) and the 
pure $\nu_{\mu}$ state are

\begin{subequations}
\la{eq4.5.4}
\begin{gather}
\rho_{\nu_e} = \vert \nu_e \rangle \langle \nu_e \vert,
\quad \vert \rho_{\nu_e} \rangle = \frac{1}{2}(1, 0, 0, 1), \\
\rho_{\nu_{\mu}} = \vert \nu_{\mu} \rangle \langle \nu_{\mu} \vert,
\quad \vert \rho_{\nu_{\mu}} \rangle = \frac{1}{2}(1, 0, 0, -1), 
\end{gather}
\end{subequations}

\noi
whence

\be
\la{eq4.5.5}
{\cal P}_{\nu_e \rightarrow \nu_{\mu}}(t) = 
\frac{1}{2}\left[
1 - e^{-2at} - 2 \; \frac{\beta^2 - h_1^2}
{\Theta_m^2} \; e^{-2at}
\sin^2{\Theta_m t}
\right].
\ee

Consistently, if there is not dissipation (\ref{eq4.5.5}) gives
the transition probability in matter to first order in the interaction, 
that is equation~(\ref{eqnuova2}). In absence of matter, we get the
vacuum transition probability~(\ref{eq4.1.4}).

The damping exponentials, typical of dissipative phenomena, flatten
the oscillations in the transition probability so that it approaches 
the value $1/2$ for very long times, and this happens regardless of the 
value of the vacuum mixing angle $\theta$.
We remind that in vacuum the transition probability has the upper
limit $\sin^2{2\theta}$ and then tends to vanish for $\theta$ approaching $0$
whereas the asymptotic mixing between the two flavors, in the dissipative 
case, is maximal, regardless of the value of $\theta$.


\s{Non homogeneous matter}
\la{s4.6}
\markright{\small{\rm{\thesection. NON HOMOGENEOUS MATTER}}}

Up to now we have considered a simplified environment with a constant 
mean density. However the physical appeal of neutrinos
interacting with matter arises when the density of the considered medium
is varying along the neutrino path. This is indeed the real situation: in
the solar case the neutrinos are produced near the core of the Sun, the 
region with higher density, then they travel through regions with
gradually decreasing densities. Actually the more plausible explanation of 
the missing neutrino solar flux depends strongly on the varying
profile of the Sun density~\cite{mikh,1mikh}.

Since any slab of matter with a not-constant density profile can be 
well approximated by a sequence of homogeneous layers
with suitable thickness, the time evolutor can be obtained
by composing elementary evolutors computed using the formalism 
described before. That is, if the inhomogeneous medium consists of $n$ 
homogeneous layers, one has

\be
\la{eq4.6.1}
{\cal G}_t = {\cal G}^n_{t_n} \circ
\ldots \circ {\cal G}^2_{t_2} \circ {\cal G}^1_{t_1}, \quad
t = t_1 + t_2 + \ldots + t_n
\ee

\noi
where $t_i$, $i = 1, \ldots n$ are the times needed to traverse the 
successive layers. The composition of maps in~(\ref{eq4.6.1}) is completely 
positive by construction.

Then in our formalism all we need to describe the neutrino evolution 
in the real matter is to divide the traversed medium into suitable slabs, 
possibly of different thickness, depending on the rapidity of the
mean density change.

Traditionally, in the study of neutrino transitions in matter two 
approaches are possible to account for the change of the medium density:
it can vary slowly, and the transitions are called adiabatic, or rapidly.
In the adiabatic approximation, neutrino state evolves in time like
eigenstates of the effective Hamiltonian that, in the neutrino frame,
is varying in time; otherwise neutrinos do not 
necessarily remain in eigenstates of $H_S^{eff}(t)$ but can 
pass from an eigenstate to another. This standard
formalism has been adapted to the case of a dissipative evolution
for statistical operators in~\cite{benan}.

\s{The solar matter case}
\la{s4.7}
\markright{\small{\rm{\thesection. THE SOLAR MATTER CASE}}}

In this section we apply our formalism to describe the dissipative 
time-evolution of a neutrino crossing a layer of solar matter with 
constant mean density and to evaluate the impact of this modified 
evolution on the flavors transition probability. We adopt the
Hamiltonian density~(\ref{eq4.7.1}).


\noi
The autocorrelation functions reflect the decoupling of flavors:

\be
\la{eq4.7.3}
\begin{split}
&a_{lm\mu,l^{\prime}m^{\prime}\mu^{\prime}}(\vec{x},\vec{x}^{\prime}) =
\delta_{lm} \delta_{l^{\prime}m^{\prime}} \delta_{ll^{\prime}} 
a_{lm\mu,l^{\prime}m^{\prime}\mu^{\prime}}(\vec{x},\vec{x}^{\prime}), \\
&b_{lm\mu,l^{\prime}m^{\prime}\mu^{\prime}}(\vec{x},\vec{x}^{\prime}) =
\delta_{lm} \delta_{l^{\prime}m^{\prime}} \delta_{ll^{\prime}} 
b_{lm\mu,l^{\prime}m^{\prime}\mu^{\prime}}(\vec{x},\vec{x}^{\prime}); \\
\end{split}
\ee

\noi
$\delta_{ll^{\prime}}$ means statistical independence of the bath operators
of different flavors. As a consequence of~(\ref{eq4.7.3}) most of the
coefficients in Appendix~\ref{a4} 
vanish and the evolution equation in the flavor basis, eq.~(\ref{eq4.4.12}), 
takes the simple form

\be
\la{eq4.7.4}
\frac{d}{dt}\vert \rho (t) \rangle = -2
\left[ \left(
\begin{array}{cccc}
0 & 0 & 0 & 0 \\
0 & 0 & h_3 & 0 \\
0 & -h_3 & 0 & h_1 \\
0 & 0 & -h_1 & 0 \\
\end{array}
\right) +
\left(
\begin{array}{cccc}
0 & 0 & 0 & 0 \\
0 & a & 0 & 0 \\
0 & 0 & a & 0 \\
0 & 0 & 0 & 0 \\
\end{array}
\right) \right]
\vert \rho (t) \rangle,
\ee

\noi
where, following~(\ref{eq4.4.13}), (\ref{eq4.7.2}) and Appendix~\ref{a4},

\be
\la{eq4.7.5}
\begin{split}
h_1 &= i \frac{\Delta m^2}{4E} \sin{2\theta}, \\
h_3 &= i \left( \frac{G_F}{\sqrt{2}}n_e - \frac{\Delta m^2}{4E}\cos{2\theta} 
\right) + \omega_3, \\
\omega_3 &= \frac{1}{2} \im \left[ B_{\mu \mu}(\vec{p}) - B_{ee}(\vec{p}) 
\right], \\
a &= \frac{1}{2}\left[ A_{ee}(\vec{p}) + A_{\mu \mu}(\vec{p}) \right].
\end{split}
\ee

In order to evaluate the time-evolution of this state, we use the 
formalism developed in Section~\ref{s2.4}, with $r_1 = -\cot{\theta_m}$ 
and $r_2 = -\tan{\theta_m}$. It is convenient to perform the
computation in the basis \{$\vert \nu_{m1}\rangle$, $\vert \nu_{m2}\rangle$\}
that diagonalizes $H_S^{eff}$ (introduced in~(\ref{eq4.3.6})); 
indeed $P_{\nu_e \rightarrow \nu_{\mu}}(t)$ 
is basis independent by definition.

\noi
The resulting expression of the transition probability in the dissipative
regime is thus

\be
\la{eq4.7.8}
\begin{split}
P_{\nu_e \rightarrow \nu_{\mu}}(t) = \frac{1}{2}
&\Bigl\{ 1 - e^{-2a \sin^2{2\theta_m} t} \cos^2{2 \theta_m} + \\
&- \sin^2{2\theta_m} \Bigl[ e^{-a (1 + \cos^2{2\theta_m}) t}
\cos{\frac{\Delta m^2}{2E} t} 
+ a \frac{2E}{\Delta m^2} (1 + 3\cos^2{2 \theta_m}) 
\sin{\frac{\Delta m^2}{2E} t}
\Bigr] \Bigr\},
\end{split}
\ee

\noi
where the parameter $\omega_3$ has been 
absorbed in a redefinition of the parameters appearing in the Hamiltonian
part of the evolution. 
For $a = 0$ we obtain the transition probability at first order
in the interaction for an homogeneous medium, eq.~(\ref{eqnuova2}).
It reduces to the vacuum oscillation probability in absence of matter.

In order to obtain the transition probability in the real, not uniform, solar
matter, the considerations of Section~\ref{s4.6} should be applied;
however, using a magnitude estimate based on dimensional analysis, 
the impact of dissipation for the phenomena that can be
described in our framework is negligeable, as we shall shortly 
explain.


\s{Discussion and conclusions}
\la{s4.8}
\markright{\small{\rm{\thesection. DISCUSSION AND CONCLUSIONS}}}

We have treated the topic of neutrino oscillations in matter from the
rigorous point of view of open quantum systems, using rigorous
Markovian limits that produce a fully consistent dynamics 
for the described system and avoids problems of physical consistency. 

The interaction between neutrinos and environment has been considered to
be as general as possible, including some scenarios forbidden in the 
Standard Model, like flavor changing currents. The formalism of the
singular coupling has been employed, producing a request on the bath
correlation time: it must be much shorter than the characteristic time of
the system.

The conditions to obtain a convolutionless dynamics in the flavor space 
have been discussed, showing the tight connection between the semigroup 
structure of the reduced dynamics and the isotropy of the bath correlation 
functions (i.e. their invariance under translations). No specific models of 
fluctuations have been introduced; we have rather assumed that the 
contributions breaking this isotropy be negligeable. If they
are not so, flavor and momentum degrees of freedom need to be treated 
together and an effective two-dimensional prescription is unavailable.

In this framework, several dissipative phenomena can be described. We 
have evaluated the transition probability between two neutrino flavors
in two different contexts: a flavor-changing and a flavor-preserving
interaction and explicit results for homogeneous  matter
have been obtained. To account for matter inhomogeneities, in
general, numerical computations should be performed.

The typical terms introduced by dissipation are damping exponentials 
whose lifetimes are the reciprocals of the dissipative coefficients;
this means that the effect of the irreversible dynamics should be relevant
on a time scale of order $t_{diss} \sim 1/ (gn)^2 \tau_B$, the inverse of
the typical magnitude of dissipative parameters in non-relativistic
matter, where $n$ is the density of scatterers and $g$ the 
coupling constant in the interaction term.

For example, the thermal fluctuations in the Sun satisfy the limits 
imposed by our formalism; a satisfactory evaluation of the correlation 
time is given by the inverse of the thermal energy at temperature $T$: 
$\tau_B = \hbar / k T$, where $\hbar$ and $k$ are the Planck and Boltzmann 
constants respectively. The explicit evaluation for the Sun core
($T \sim 10^7~{\rm K}$) gives $\tau_B \sim 10^{-21} 
{\rm s}$, much shorter than $\tau_S \sim E/ \Delta m^2 \sim 10^{-4} 
{\rm s}$ (obtained considering the values $E \sim 1~{\rm MeV}$ and
$\Delta m^2 \sim 10^{-5}~{\rm eV}^2$). 
However, neutrinos cross the Sun in a time so short that the
thermal contributions are completely negligeable: $t_{diss} \sim 10^{14}
{\rm s}$ (using the electron density
$n_e \sim 10^{26}~{\rm cm}^{-3}$ and $G_F \sim 10^{-5}~{\rm GeV}^{-2}$), 
versus a crossing time of about $3$ seconds. 

Does it exist a physical instance in which the dissipative effects give 
a strong effect despite the weak interaction with the surroundings? It 
is difficult to imagine a real situation of this kind. However the 
gravitational thermal bath described in Section~\ref{s3.2} could give
such an environment: it should be everywhere present, and plausibly
homogeneous and isotropic. In this case the bath energy is parametrized
by the Plank mass, $M_P \sim 10^{19} {\rm GeV}$, and for a  neutrino 
energy of order $E_S \sim 1 {\rm MeV}$, we can roughly evaluate
$t_{diss} \sim 10^{7} {\rm s}$, that is of order of the time needed for a 
neutrino to cross intergalactic distances. The asymptotical behavior
for the transition probability, with and without dissipation, should be 
very different. In the first case it would approach the $1/2$ value; in the
second one it would oscillate with a maximal amplitude of 
$\sin^2{2\theta}$ (observe that in this scenario it is impossible to 
distinguish between $\theta$ and $\theta_m$). The averaged transition
probability would be certainly increased and its time dependence destroyed.
In general, decoherence in neutrino physics has been studied 
in~\cite{benan,lisi,liu,chan}.

The last remark is about the discrimination of the Dirac or Majorana 
nature of neutrino fields~\cite{sche,li}: although dissipative dynamics
in general allow to distinguish between these two possibilities~\cite{benan},
this is not the case in this chapter, since the microscopic interactions
just presented are diagonal in the flavor neutrino fields. 


%% file: chapter5.tex
\chapter{Neutron interferometry}
\la{ch5}
\markboth{\small{\rm{\thechapter. NEUTRON INTERFEROMETRY}}}{}

In this chapter we apply the formalism of open quantum systems to
study neutron interferometry. 
There are two main motivations for this kind of approach.
First of all, physical setups that are subjected to irreversible 
evolutions are actually available: they give the opportunity 
to study several models of Markovian dynamics, and to test
their physical consistency.
In the second place, since experiments based on quantum interferometry
strongly rely upon the property of quantum correlations
between different degrees of freedom, they can be
suitably employed to investigate the notion of complete positivity (CP) of 
the reduced time-evolution $\Lambda_t$. As we have seen in Chapter~\ref{ch1},
this property, tightly connected to quantum correlations expressed by 
entanglement, has 
a somewhat abstract and technical definition; it could seem at first
sight a mathematical artifact rather than a physically motivated request.
However in this context the relevance of CP becomes apparent since we are 
dealing here with reduced dynamics of the form ${\bf I} \otimes \Lambda_t$,
precisely the one used in its definition (see Section~\ref{s1.2}).
Indeed, as explicitely shown in the models discussed below, lack of CP 
necessarily leads to the appearance of unacceptable negative eigenvalues 
in the time evolving physical state describing entangled degrees of freedom.

After a brief introduction to the topic of neutron interferometry,
and to the mathematical framework usually employed to describe it, we
shall study non standard dissipative evolutions for neutrons inside
the interferometer. These reduced dynamics are the result of the interaction
with an external environment in which the interferometric device is immersed.

We shall be very general, allowing for
evolutions that are not completely positive. Three different 
prototypes will be taken into account, all of them 
experimentally achievable by choosing appropriately the properties of the 
external environment (typically a thermal bath): they cover instances of 
completely positive, simply positive (but not CP) and even not-positive 
time-evolutions. Some of these 
dynamics are commonly used in some branches of physics (for example, 
in chemical physics~\cite{budi,lair,suar}).

Neutron interferometry has been proposed to test the non-contextuality
hypothesis in quantum mechanics~\cite{basu}, that is the possibility 
that all the members of any set of commuting observables posses a 
value independent on the choice of the commuting observables measured
at the same time (that is, independent on the context in which the measure
of that observable is performed). This hypothesisis can be tested by 
studying violations of the CHSH inequality~\cite{clau} (a particular 
Bell-like inequality); in the final part of the chapter, using the 
previously developed results,
we will study how this relation is modified when the interferometer is 
immersed in a stochastic background.


\s{The physical apparatus}
\la{s5.1}
\markright{\small{\rm{\thesection. THE PHYSICAL APPARATUS}}}

For our discussion, we shall refer to an idealized, but standard
model of neutron interferometer\footnote{similar set-ups can be constructed using polarized hotons see~\cite{mich}} \cite{basu}.  
The apparatus, sketched in Figure~\ref{fig5.1.1}, 
consists of three basic components: an initial
beam splitter ($bs_1$), dividing an incoming, polarized, 
neutron beam in two 
components; a spin flipper ($sf$), acting on one of the
two components emerging from the beam splitter; a final beam splitter
($bs_2$) that recombines the two components, equipped with a phase 
shifter ($ps$). The neutrons of the incoming
beam are polarized along a fixed direction, say the $z$-axis; 
the beam splitter $bs_1$ is characterized by a transmission $p$
and reflection $q$ coefficients 
satisfying the condition $\vert p \vert ^2 + \vert q \vert ^2 = 1$
(in the ideal case where no neutrons are absorbed). 
The two components follow 
spatially separated (by distances of the order of centimeters) trajectories, 
denoted by $u$ (for up) and $d$ (for down). Then, the translational degree of
freedom may be described by a two-dimensional Hilbert space $\mathfrak{H}^a_2$ 
generated by the orthogonal states $\vert \psi_u \rangle$ and 
$\vert \psi_d \rangle$, that are associated to the two macroscopically
distinguished paths.

\begin{figure}
\la{fig5.1.1}
\begin{picture}(370,200)(-10,10)
\linethickness{0,5pt}
\put(79,75){\line(1,0){6}}
\put(85,75){\line(0,1){50}}
\put(85,125){\line(-1,0){6}}
\put(79,125){\line(0,-1){50}}
\put(82,128){\makebox(0,0)[b]{$bs_1$}}
\put(82,72){\makebox(0,0)[t]{$p, q$}}
\put(88,100){\line(1,0){3}}
\put(91,100){\line(0,1){12}}
\put(91,112){\line(-1,0){3}}
\put(88,112){\line(0,-1){12}}
\put(94,103){\makebox(0,0)[l]{$sf$}}
\put(263,75){\line(1,0){6}}
\put(269,75){\line(0,1){50}}
\put(269,125){\line(-1,0){6}}
\put(263,125){\line(0,-1){50}}
\put(266,128){\makebox(0,0)[b]{$bs_2 + ps$}}
\put(266,72){\makebox(0,0)[t]{$\theta, \varphi$}}
\put(260,94){\line(1,0){3}}
\put(260,94){\line(0,1){12}}
\put(260,106){\line(1,0){3}}
\put(272,94){\line(-1,0){3}}
\put(272,94){\line(0,1){12}}
\put(272,106){\line(-1,0){3}}
\qbezier(82,100)(174,192)(266,100)
\qbezier(82,100)(174,8)(266,100)
\put(82,100){\line(-1,-1){50}}
\put(266,100){\line(1,1){50}}
\put(266,100){\line(1,-1){50}}
\put(42,47){\makebox(0,0)[t]{$\vert \uparrow_z \rangle$}}
\put(174,47){\makebox(0,0)[t]
{$\vert \psi_d \rangle \otimes\vert \uparrow_z \rangle$}}
\put(174,153){\makebox(0,0)[b]
{$\vert \psi_u \rangle \otimes\vert \downarrow_z \rangle$}}
\put(316,144){\line(1,0){9}}
\put(325,144){\line(0,1){12}}
\put(325,156){\line(-1,0){9}}
\put(316,156){\line(0,-1){12}}
\put(319,144){\line(0,1){12}}
\put(321,159){\makebox(0,0)[b]{$c_1$}}
\put(321,141){\makebox(0,0)[t]{$u, \vec{n}$}}
\put(316,44){\line(1,0){9}}
\put(325,44){\line(0,1){12}}
\put(325,56){\line(-1,0){9}}
\put(316,56){\line(0,-1){12}}
\put(319,44){\line(0,1){12}}
\put(321,59){\makebox(0,0)[b]{$c_2$}}
\put(321,41){\makebox(0,0)[t]{$d, \vec{n}$}}
\end{picture}
\caption{\footnotesize Typical model of neutron interferometer for 
non-contextuality tests. $bs_1$ and $bs_2$ are the two beam splitters, 
$sf$ is the spin flipper, $ps$ the phase shifter, $c_1$ and $c_2$ the 
position plus spin counters.}
\end{figure}
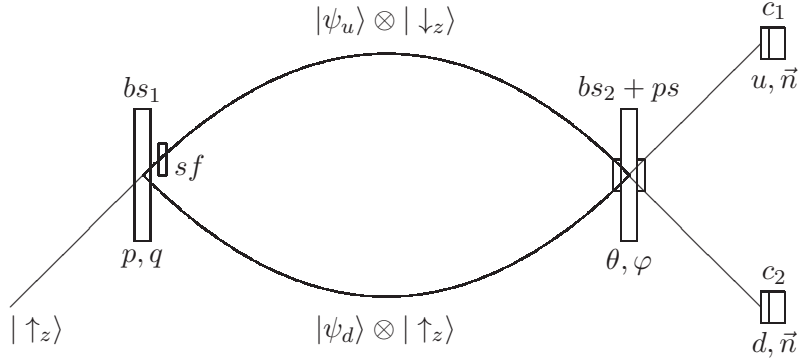

\begin{Rem}
\la{rem5.1.0}
{\rm Since the neutrons are $1/2$-spin particles, their spin degree of 
freedom is also described by a two-dimensional Hilbert space 
$\mathfrak{H}^b_2$, spanned by the states
$\vert \uparrow _z \rangle$, spin-up, and $\vert \downarrow _z \rangle$, 
spin-down along the $z$-axis. On the whole, neutrons in the apparatus
may be appropriately described by the $4$-dimensional tensor Hilbert space 
$\mathfrak{H}^a_2 \otimes \mathfrak{H}^b_2 = \mathfrak{H}_4$.}
\end{Rem}

Inside the interferometer, the component following the $u$ trajectory 
is subjected to a spin-flip,
$\vert \uparrow _z \rangle \rightarrow \vert \downarrow _z \rangle$; as a
result, the state describing a neutron travelling in the inner region of 
the interferometer is entangled 
(where the entanglement relies upon a quantum correlation between internal
and spatial degrees of freedom of the same system):

\be
\la{eq5.1.1}
\vert \psi \rangle = 
p \vert \psi_u \rangle \otimes \vert \downarrow _ z \rangle +
q \vert \psi_d \rangle \otimes \vert \uparrow _ z \rangle.
\ee

\noi
In the density matrix formalism, this state is represented by the 
$4 \times 4$ statistical operator

\be
\la{eq5.1.2}
\rho^{ab} = \vert \psi \rangle \langle \psi \vert = 
\vert p \vert ^2 P_1 \otimes Q_2 + \vert q \vert ^2 P_2 \otimes Q_1 +
p q^* P_3 \otimes Q_4 + p^* q P_4 \otimes Q_3, 
\ee

\noi
where 

\be
\la{eq5.1.3}
\begin{split}
&P_1 \equiv \vert \psi_u \rangle \langle \psi_u \vert, \quad
P_2 \equiv \vert \psi_d \rangle \langle \psi_d \vert, \quad
P_3 \equiv \vert \psi_u \rangle \langle \psi_d \vert, \quad
P_4 \equiv \vert \psi_d \rangle \langle \psi_u \vert; \\
&Q_1 \equiv \vert \uparrow_z \rangle \langle \uparrow_z \vert, \quad
Q_2 \equiv \vert \downarrow_z \rangle \langle \downarrow_z \vert, \quad
Q_3 \equiv \vert \uparrow_z \rangle \langle \downarrow_z \vert, \quad
Q_4 \equiv \vert \downarrow_z \rangle \langle \uparrow_z \vert.
\end{split}
\ee

\begin{Rem}
\la{rem5.1.1}
{\rm We stress that the entanglement in the state~(\ref{eq5.1.1}) is
different from the usual one, for instance the one described in the 
case of neutral ${\bf B}$ mesons, where quantum correlations connect the 
degrees of freedom of two spatially separated physical systems, 
a neutral {\bf B} meson and its antiparticle, both produced 
in the decay of a resonance. In the present case, on the contrary, the 
entanglement is between different degrees of freedom of the same 
physical system: translational and spin degrees of freedom.}
\end{Rem}

By denoting with $\Lambda_t$ the dynamical map describing the 
time-evolution undergone inside the apparatus, at the second beam splitter, 
after a time $t$, the initial state $\rho^{ab}$ in~(\ref{eq5.1.2}) is 
changed according to a law $\rho^{ab} \rightarrow 
\rho^{ab}(t) = \Lambda_t [\rho^{ab}]$.

At the second beam splitter, the $u$-component is further shifted by an
angle $\varphi$, with respect to the $d$-component, by the phase shifter 
$ps$. The effects of the whole operation $bs_2 + ps$ are described by a 
unitary transformation

\be
\la{eq5.1.4}
\begin{split}
\begin{array}{l}
\vert \psi_u \rangle \rightarrow \vert \psi_u(\theta, \varphi) \rangle
= U(\theta, \varphi) \vert \psi_u \rangle, \\
\vert \psi_d \rangle \rightarrow \vert \psi_d(\theta, \varphi) \rangle
= U(\theta, \varphi) \vert \psi_d \rangle, \\
\end{array}
;\quad
U(\theta, \varphi) = \left(
\begin{array}{cc}
e^{-i \varphi} \sin{\theta} & e^{-i \varphi} \cos{\theta} \\
\cos{\theta} & -\sin{\theta} \\
\end{array}
\right)
\end{split}
\ee

\noi
where $\cos^2{\theta}$ and $\sin^2{\theta}$ are the reflection and 
transmission probabilities for $bs_2$ (the analogue of $q$ and $p$);
their values can be modified by changing the angle $\theta$, that is
by changing the physical properties of the beam splitter $bs_2$
(the parameters $p$ and $q$, characterizing the first beam splitter, 
are instead kept fixed). The initial statistical 
operator thus evolves as follows:

\be
\la{eq5.1.5}
\rho^{ab} \rightarrow (U(\theta, \varphi) \otimes {\bf I}_2) \rho^{ab}(t) 
(U^{\dagger}(\theta, \varphi) \otimes {\bf I}_2),
\ee

\noi
where ${\bf I}_2$ is the $2 \times 2$ identity in the translational Hilbert 
space $\mathfrak{H}^a_2$. 

The exiting neutron beam in the state~(\ref{eq5.1.5}) then impinges on
two counters ($\rm c_1$ and $\rm c_2$), 
equipped with spin-analyzers that can be set along any direction
$\vec{n} = (n_x, n_y, n_z)$ in space. These counters are used to measure
the relative frequencies with which neutrons are detected in one of the two 
exiting beams, with spin up or down along $\vec{n}$. These frequencies 
of counts, denoted by ${\cal O}^{j, \vec{n}}_t (\theta, \varphi)$ are 
given by mean values of the projectors $P_j(\theta, \varphi) \otimes 
Q_{\vec{n}}$ in $\mathfrak{H}^a_2 \otimes \mathfrak{H}^b_2$ over 
$\rho^{ab}(t)$:

\be
\la{eq5.1.6}
{\cal O}^{j, \vec{n}}_t (\theta, \varphi) \equiv
\Tr [\rho^{ab}(t) P_j(\theta, \varphi) \otimes Q_{\vec{n}}],
\ee

\noi
where 

\be
\la{eq5.1.6.bis}
P_j(\theta, \varphi) \equiv U^{\dagger}(\theta, \varphi)
P_j U(\theta, \varphi),
\ee

\noi
with $j = 1,2$, are the spatial projectors in~(\ref{eq5.1.3}),
whereas $Q_{\vec{n}} \equiv \vert \uparrow_{\vec{n}}\rangle 
\langle \uparrow_{\vec{n}} \vert$ is the projector over the eigenspace
spanned by the state $\vert \uparrow_{\vec{n}} \rangle$ with spin-up along
the $\vec{n}$-direction.

\begin{Rem}
\la{rem5.1.1.add}
{\rm The last beam splitter together with the counters plus spin
analyzers constitutes the measuring device, since it sets the values 
of the angles $\theta$ and $\varphi$, that can be varied by the
experimenter.}
\end{Rem}

Neutron interferometry has proved to be an extremely powerful tool to
investigate gravitational, inertial and phase shifting effects
occurring inside the interferometer\cite{stau,rauc}: they are
described by a time-evolution for the spatial density matrix. 
Instead, in what follows, we shall assume a dissipative time-evolution 
to affect the spin degrees of freedom, because of the presence of a
stochastic magnetic field inside the apparatus~\cite{gori2,budi}. 
The neutron beam thus behaves like an open system with time-evolution 
given by $\Lambda_t = {\bf I}_2 \otimes \Lambda_t^b$,  
$\Lambda_t^b$ being the dynamical map describing the non-trivial evolution 
in the spin sector. Explicitely, while traversing the interferometer,
the neutron beam state changes according to $\rho^{ab} \rightarrow 
\rho^{ab}(t) = {\bf I}_2 \otimes \Lambda_t^b [\rho^{ab}]$.
Neutron interferometry can then be used to investigate
the property of complete positivity. In particular, the form of
$\Lambda_t$ is the one appearing in the definition of complete
positivity in Section~\ref{s1.2}: 
to guarantee the positivity of $\rho^{ab}(t) = \Lambda_t
[\rho^{ab}(0)]$ for any time $t$ we need necessarily a completely 
positive map $\Lambda_t^b$.


In Section~\ref{s5.4} we shall show explicitely that lack of complete 
positivity generates inconsistencies in the physical interpretation of
the statistical operator $\rho^{ab}(t)$, that is the appeareance of negative
eigenvalues. Noticeably, these eigenvalues can be
directly accessed in experiments by setting appropriately the angles
$\theta$ and $\varphi$ and the direction $\vec{n}$.
Indeed, the entries of $\rho^{ab}(t)$ can be expressed by means 
of the frequencies measured at the counters. Writing $\rho^{ab}(t) = 
\sum_{i,j = 1}^4 \rho_{ij}(t) P_i \otimes Q_j$, with $P_i$ and $Q_j$
as in~(\ref{eq5.1.3}), it turns out that 

\be
\la{eq5.1.7}
\begin{split}
&\rho_{11}(t) = {\cal O}^{1,z}_t (0,0), \quad
\rho_{12}(t) = {\cal O}^{1,-z}_t (0,0), \\ 
&\rho_{21}(t) = {\cal O}^{2,z}_t (0,0), \quad
\rho_{22}(t) = {\cal O}^{2,-z}_t (0,0); 
\end{split}
\ee

\noi
the remaining 
$12$ entries $\rho_{ij}$ cannot be evaluated directly since in these
cases the $P_i \otimes Q_j$ operators are not Hermitian; nevertheless they 
can be written by using suitable combinations of observables
involving the following projectors, see~(\ref{eq5.1.6.bis}):

\be
\la{eq5.1.8}
\begin{split} 
P_{\pm} \equiv P_{1,2}\left(\frac{\pi}{4},0\right), \quad
&P_{\pm i} \equiv P_{1,2}\left(\frac{\pi}{4},-\frac{\pi}{2}\right), \\ 
Q_{\pm x} \equiv \frac{1}{2}
(\vert \uparrow_z \rangle \pm \vert \downarrow_z \rangle)
(\langle \uparrow_z \vert \pm \langle \downarrow_z \vert), \quad
&Q_{\pm y} \equiv \frac{1}{2}
(\vert \uparrow_z \rangle \pm i \vert \downarrow_z \rangle)
(\langle \uparrow_z \vert \mp i \langle \downarrow_z \vert).
\end{split}
\ee

Then, from

\be
\la{eq5.1.9}
P_3 = \frac{1}{2}[(P_+ - P_-) + i(P_{+i} - P_{-i})] = P_4^*, \quad
Q_3 = \frac{1}{2}[(Q_x - Q_{-x}) + i(Q_y - Q_{-y})] = Q_4^*
\ee

\noi
the expressions of the remaining entries can be explicitely evaluated
(see Appendix~\ref{a5}); for example

\be
\la{eq5.1.10}
\begin{split}
\rho_{43}(t) = \frac{1}{4} \Bigl\{
&\Bigl[{\cal O}^{1,x}_{t}\left(\frac{\pi}{4},0\right) - 
{\cal O}^{1,-x}_{t}\left(\frac{\pi}{4},0\right)\Bigr] +i
\Bigl[{\cal O}^{1,y}_{t}\left(\frac{\pi}{4},0\right) - 
{\cal O}^{1,-y}_{t}\left(\frac{\pi}{4},0\right)\Bigr] + \\
- &\Bigl[{\cal O}^{2,x}_{t}\left(\frac{\pi}{4},0\right) - 
{\cal O}^{2,-x}_{t}\left(\frac{\pi}{4},0\right)\Bigr] -i
\Bigl[{\cal O}^{2,y}_{t}\left(\frac{\pi}{4},0\right) - 
{\cal O}^{2,-y}_{t}\left(\frac{\pi}{4},0\right)\Bigr] + \\
- &\Bigl[{\cal O}^{1,y}_{t}\left(\frac{\pi}{4},-\frac{\pi}{2}\right) - 
{\cal O}^{1,-y}_{t}\left(\frac{\pi}{4},-\frac{\pi}{2}\right)\Bigr] +i
\Bigl[{\cal O}^{1,x}_{t}\left(\frac{\pi}{4},-\frac{\pi}{2}\right) - 
{\cal O}^{1,-x}_{t}\left(\frac{\pi}{4},-\frac{\pi}{2}\right)\Bigr] + \\
+ &\Bigl[{\cal O}^{2,y}_{t}\left(\frac{\pi}{4},-\frac{\pi}{2}\right) - 
{\cal O}^{2,-y}_{t}\left(\frac{\pi}{4},-\frac{\pi}{2}\right)\Bigr] -i
\Bigl[{\cal O}^{2,x}_{t}\left(\frac{\pi}{4},-\frac{\pi}{2}\right) - 
{\cal O}^{2,-x}_{t}\left(\frac{\pi}{4},-\frac{\pi}{2}\right)\Bigr] \Bigr\}
\end{split} 
\ee

\noi
Also these entries are thus, in principle, accessible to experimental
measures.


\s{The stochastic thermal bath}
\la{s5.2}
\markright{\small{\rm{\thesection. THE STOCHASTIC THERMAL BATH}}}

We assume that, inside the
interferometer, the neutron beam is affected by a weak stochastic 
magnetic field; as a result, the dynamics is that of a dissipative 
open system. 

We use the formalism introduced in Section~\ref{s1.7}; namely, the spin
degree of freedom is described by a $2 \times 2$ stochastic density
matrix $\Sigma (t)$, satisfying the time-dependent Liouville-Von Neumann
equation

\be
\la{eq5.2.0}
\begin{split}
\frac{d}{dt}\Sigma (t) &= (L_0 + L_I(t))[\Sigma(t)], \\
L_0[\cdot] \equiv -i[H_0, &\cdot]; \quad L_I(t)[\cdot] \equiv -i[H_I(t),\cdot]
\end{split}
\ee

\noi
where $H_0 = \omega_0 \sigma_z /2$; $H_I(t) = \sum_{k = 1}^3 V_k(t) \sigma_k$
is the fluctuating part of the Hamiltonian
and $V_k(t)$, $k = 1, \ldots 3$, are proportional to Gaussian stochastic 
magnetic field components. We assume $\langle V_k(t) \rangle = 0$ 
$\forall k$ and the two-point correlation functions to be real, 
stationary and symmetric, that is

\be
\la{eq5.2.0.add}
h_{ij}(t - s) = \langle V_i(t) V_j(s) \rangle = h_{ij}^* (t - s)
= h_{ji}(s - t).
\ee 

\noi
The covariance matrix, defined as ${\cal W}(t) \equiv [h_{ij}(t)]$, is 
thus real, stationary and symmetric. 

The effective $2\times 2$ spin density matrix of the neutron beam 
when it impinges upon the second beam splitter is given by 
$\rho(t) = \langle \Sigma(t) \rangle$, where an average over the noise
has been performed; its equation of motion has the form

\be
\la{eq5.2.1}
\dot{\rho}(t) = -i[H_0, \rho(t)] + \sum_{k,j = 1}^3 d_{kj}(t)
[\sigma_k,[\sigma_j, \rho(t)]]
\ee

\noi
with

\be
\la{eq5.2.2}
d_{kj}(t) = - \sum_{l = 1}^3 \int_0^t h_{kl}(s) u_{lj}(-s) ds
\ee

\noi
where the $U_{lj}$ are the entries of the orthogonal matrix

\be
\la{eq5.2.3}
{\cal U}(t) = [u_{ij}(t)] = \left(
\begin{array}{ccc}
\cos{\omega_0 t} & -\sin{\omega_0 t} & 0 \\
\sin{\omega_0 t} & \cos{\omega_0 t} & 0 \\
0 & 0 & 1
\end{array}
\right)
\ee

\noi
such that $\sigma_l (-s) = \sum_{j = 1}^3 u_{lj}(-s) \sigma_j$.
Equation~(\ref{eq5.2.1}) is the form assumed by the general 
equation~(\ref{eq1.7.9}) in the two-dimensional case and we refer to 
Section~\ref{s1.7} for details. The only assumptions required to extract 
the reduced dynamics~(\ref{eq5.2.1}) are the Gaussian nature of the stochastic 
magnetic field and the initial decoupling between spin and noise, that is 
$\rho(0) = \langle \Sigma(0) \rangle = \Sigma(0)$.

Our assumptions on the correlation functions make the $3\times 3$ matrix
$D(t) \equiv [d_{ij}(t)]$ real; thus it can be written as the sum of 
a symmetric and a skew-symmetric part. Defining the two $3 \times 3$ 
matrices

\be
\la{eq5.2.4}
D^{S,A}(t) \equiv \frac{1}{2}[D(t) \pm D^T(t)] = [d^{S,A}_{ij}(t)],
\ee

\noi
where $T$ means transposition, the dissipative part in 
equation~(\ref{eq5.2.1}) can be written as the sum of a Hamiltonian
contribution plus a purely dissipative one. Then

\be
\la{eq5.2.5}
\dot{\rho}(t) = -i \Bigr[H_0 - \sum_{i,j,k = 1}^3 d^A_{ij}(t) \epsilon_{ijk}
\sigma_k, \rho(t)\Bigl] + 2 \sum_{i,j = 1}^3 d_{ij}^S(t)
\left(
\frac{1}{2} \{ \sigma_i \sigma_j, \rho(t) \} - \sigma_j \rho(t) \sigma_i
\right).
\ee

In the limit of weak stochastic magnetic field, it is possible to
perform a Markovian approximation for the reduced dynamics~(\ref{eq5.2.5})
as discussed in Chapter~\ref{ch1};
in practice it is usually performed by extending to $+ \infty$ of the upper 
limit of the integral in~(\ref{eq5.2.2}). 
The matrices $D^S$ and $D^A$ become then time-independent and we 
obtain dynamical equations of motion in Lindblad form:

\be
\la{eq5.2.6}
\dot{\rho}(t) = -i \Bigr[H_0 - \sum_{i,j,k = 1}^3 d^A_{ij} 
\epsilon_{ijk}\sigma_k, \rho(t)\Bigl] + 
\frac{1}{2} \sum_{i,j = 1}^3 c_{ij} \left(
\sigma_i \rho(t) \sigma_j - \frac{1}{2}
\{ \sigma_j \sigma_i, \rho(t) \}
\right),
\ee

\noi
where $C \equiv [c_{ij}] = -4 D^S$. 

\begin{Rem}
\la{rem5.2.1}
{\rm Following the result of Theorem~\ref{the1.5.2}, 
the dynamics~(\ref{eq5.2.6}) is completely positive if and only if the 
matrix $C = [c_{ij}]$ is positive definite. In general, the brute force 
Markovian approximation adopted above may provide a Markovian 
equation~(\ref{eq5.2.6}) that generates time-evolutions completely 
positive, simply positive or even not-positive, depending on the 
behavior of the correlation functions, in~(\ref{eq5.2.2}).}
\end{Rem}

Equation~(\ref{eq5.2.6}) can be rewritten using the vector representation
$\vert \rho(t) \rangle$ for density matrices $\rho(t)$ presented
in Section~\ref{s2.1}. The Schr{\" o}dinger-like equation that follows is:

\be
\la{eq5.2.7}
\frac{d}{dt} \vert \rho(t) \rangle = ({\cal H} + {\cal D})
\vert \rho(t) \rangle
\ee

\noi
where ${\cal H}$ and ${\cal D}$ are $4 \times 4$ real matrices representing
the Hamiltonian, respectively dissipative contribution in~(\ref{eq5.2.6}):

\be
\la{eq5.2.8}
{\cal H} = -2 \left(
\begin{array}{cccc}
0 & 0 & 0 & 0 \\
0 & 0 & h_3 & -h_2 \\
0 & -h_3 & 0 & h_1 \\
0 & h_2 & -h_1 & 0 
\end{array}
\right), \quad
{\cal D} = -2 \left(
\begin{array}{cccc}
0 & 0 & 0 & 0 \\
0 & a & b & c \\
0 & b & \alpha & \beta \\
0 & c & \beta & \gamma 
\end{array}
\right)
\ee

\noi
where

\be
\la{eq5.2.9}
h_1 = 2d^A_{32}, \quad
h_2 = 2d^A_{13}, \quad
h_3 = \frac{\omega_0}{2} + 2d^A_{21}
\ee

\noi
and

\be
\la{eq5.2.10}
\begin{array}{ll}
a = -2(d^S_{22} + d^S_{33}), & \quad
b = 2d^S_{12}, \\ \\
\alpha = -2(d^S_{11} + d^S_{33}), & \quad
\beta = 2d^S_{23}, \\ \\
\gamma = -2(d^S_{11} + d^S_{22}), & \quad
c = 2d^S_{13}.
\end{array}
\ee

Notice that, in this parametrization

\be
\la{eq5.2.11}
C = \left(
\begin{array}{ccc}
\alpha + \gamma - a & -2b & -2c \\
-2b & a + \gamma - \alpha & -2\beta \\
-2c & -2\beta & a + \alpha - \gamma
\end{array}
\right).
\ee

From Theorems~\ref{the2.3.2} and \ref{the2.3.3}, it immediately follows that

\begin{enumerate}
\item The reduced dynamics generated by~(\ref{eq5.2.6}) is positive if 
and only if the parameters entering in ${\cal D}$ satisfy the inequalities:
\be
\la{eq.p}
\left\{
\begin{array}{l}
a \geqslant 0 \\
\alpha \geqslant 0 \\
\gamma \geqslant 0 \\
\end{array} 
\right. ; \;
\left\{
\begin{array}{l}
\alpha \gamma \geqslant b^2 \\
a \gamma \geqslant c^2 \\
a \alpha \geqslant \beta^2 \\
\end{array}
\right. ; \;
a \alpha \gamma + 2 b c \beta - \alpha c^2 - \gamma b^2 - a \beta^2 
\geqslant 0;
\ee
\item The reduced dynamics generated by~(\ref{eq5.2.6}) is completely 
positive if and only if the parameters entering in ${\cal D}$ satisfy 
the inequalities:
\be
\la{eq.cp}
\left\{
\begin{array}{l}
2R \equiv \alpha + \gamma - a \geqslant 0 \\
2S \equiv a + \gamma - \alpha \geqslant 0 \\
2T \equiv a + \alpha - \gamma \geqslant 0 \\
\end{array} 
\right. ; \;
\left\{
\begin{array}{l}
RS \geqslant b^2 \\
RT \geqslant c^2 \\
ST \geqslant \beta^2 \\
\end{array}
\right. ; \;
RST - 2bc\beta -
S c^2 - 
R \beta^2 -
T b^2 \geqslant 0
\ee
that is, if $C$ is positive.
\end{enumerate}


\s{Three different stochastic backgrounds}
\la{s5.3}
\markright{\small{\rm{\thesection. THREE DIFFERENT STOCHASTIC BACKGROUNDS}}}

In order to investigate the notion of complete positivity in more detail,
we consider different decaying properties of the covariance matrix 
${\cal W}(t)$. We focus our attention on three particular choices, 
reproducing equations of motion usually employed in the study of 
open systems~\cite{budi,lair,suar}; the aim of this work is to give again 
direct evidence to the fact that the absence of complete positivity in the 
reduced dynamics generates necessarily physical inconsistencies in the 
interpretation of the formalism.
In particular, these reduced dynamics may be experimentally realized 
by suitable stochastic magnetic fields and thus describe concrete physical 
contexts.

\subsection{White noise}
\la{ss5.3.1}

The first case is that with white noise correlations for the 
stochastic magnetic field:

\be
\la{eq5.3.1}
h_{ij}(t) = h_{ij} \delta (t),
\ee

\noi
where $h_{ij}$ ($i,j = 1, \ldots 3$) are time-independent, such that
$h_{ij} = h_{ji} = h_{ij}^*$.

The covariance matrix ${\cal W}(t)$ is thus time-independent, 
positive and symmetric. Consequently $D^A = 0$ and $D^S = -{\cal W} 
\leqslant 0$. Then $C \geqslant 0$ and the evolution is automatically 
completely positive; the dissipative 
coefficients are related to the correlation functions by

\be
\la{eq5.3.3}
\begin{array}{ll}
a \equiv -2 (h_{22} + h_{33}), &
b \equiv 2 h_{12}, \\ \\
\alpha \equiv -2 (h_{11} + h_{33}), &
\beta \equiv 2 h_{23}, \\ \\
\gamma \equiv -2 (h_{11} + h_{22}), &
c \equiv 2 h_{13}.
\end{array}
\ee

\subsection{Diagonal covariance matrix}
\la{ss5.3.2}

In this case, we suppose ${\cal W}(t)$ diagonal with

\be
\la{eq5.3.5}
h_{11}(t) = h_{22}(t) = g^2 B_1^2 e^{-\lambda \vert t \vert},
\quad h_{33}(t) = g^2 B_3^2 e^{- \mu \vert t \vert}
\ee

\noi
where the $B_i$'s are the time-independent components of the magnetic field,
$g$ is a constant proportional to the neutron magnetic moment and $\lambda$,
$\mu$ are decay coefficients.

\noi
We find

\be
\la{eq5.3.6}
D^A = -\frac{g^2 \omega_0 B_1^2}{\lambda ^2 + \omega _0 ^2} \left(
\begin{array}{ccc}
0 & 1 & 0 \\
-1 & 0 & 0 \\
0 & 0 & 0
\end{array}
\right), \quad
C = \left(
\begin{array}{ccc}
\gamma & 0 & 0 \\
0 & \gamma & 0 \\
0 & 0 & 2a - \gamma
\end{array}
\right),
\ee

\noi
where

\be
\la{eq5.3.7}
a \equiv \frac{2g^2 B_3^2}{\mu} + \frac{2 g^2 B_1^2 \lambda}
{\lambda^2 + \omega_0^2}, \quad
\gamma \equiv \frac{4 g^2 B_1^2 \lambda}{\lambda^2 + \omega_0^2}.
\ee

\noi
The $4 \times 4$ matrices ${\cal H}$ and ${\cal D}$, in~(\ref{eq5.2.7}) 
take the form

\be
\la{eq5.3.8}
{\cal H} = \left(
\begin{array}{cccc}
0 & 0 & 0 & 0 \\
0 & 0 & -\omega & 0 \\
0 & \omega & 0 & 0 \\
0 & 0 & 0 & 0  
\end{array}
\right), \quad
{\cal D} = -2 \left(
\begin{array}{cccc}
0 & 0 & 0 & 0 \\
0 & a & 0 & 0 \\
0 & 0 & a & 0 \\
0 & 0 & 0 & \gamma  
\end{array}
\right)
\ee

\noi
where $\omega \equiv \omega_0 + \frac{4g^2 \omega_0 B_1^2}
{\lambda^2 + \omega_0^2}$.

Therefore, the entries of the spin density matrix $\rho(t)$ satisfy the
Redfield-Bloch\cite{slic} equations of motion:

\be
\la{eq5.3.9}
\begin{array}{ll}
\dot{\rho}_1(t) = -\gamma \rho_1(t) + \gamma \rho_2(t), &
\dot{\rho}_2(t) = - \dot{\rho}_1(t), \\ \\
\dot{\rho}_3(t) = -i \omega \rho_3(t) -2a \rho_3(t), &
\dot{\rho}_4(t) = \left[\dot{\rho}_3(t)\right]^*,
\end{array}
\ee

\noi
that can be explicitely solved yielding

\be
\la{eq5.3.11}
\begin{split}
\rho_1(t) &= \frac{1}{2}\left[ \left( 1 + e^{-2 \gamma t} \right)\rho_1(0)
+ \left( 1 - e^{-2 \gamma t} \right)\rho_2(0) \right], \\
\rho_2(t) &= \frac{1}{2}\left[ \left( 1 - e^{-2 \gamma t} \right)\rho_1(0)
+ \left( 1 + e^{-2 \gamma t} \right)\rho_2(0) \right], \\
\rho_3(t) &= e^{-(2a - i\omega)t} \rho_3(0), \\
\rho_4(t) &= e^{-(2a + i\omega)t} \rho_4(0).
\end{split}
\ee

Since $a$ and $\gamma$ are real and positive, the positivity of the dynamics
is assured; complete positivity requires $C\geqslant 0$
and then $2a - \gamma \geqslant 0$. From~(\ref{eq5.3.11}) we observe that
diagonal and off-diagonal elements of $\rho$ decay with different
relaxing times, $T_1$ and $T_2$ respectively; they are the reciprocals
of the coefficients $2 \gamma$ and $2 a$. Then the condition of complete
positivity reads $1/T_2 \geqslant 1/2T_1$~\cite{gori1,budi}. 
In~\cite{budi,lair} it is
shown that this typical order relation can be reversed by setting $B_3 = 0$
and keeping fourth-order terms in the general master 
equation~(\ref{eq1.7.6}) leading to~(\ref{eq5.2.1}). In such a case we
have a positive, but not completely positive time-evolution.

\subsection{Single component field correlation}
\la{ss5.3.3}

In this case the magnetic field is supposed to have only the $x$-component: 
$V_1(t) \ne 0$, $V_2(t) = V_3(t) = 0$ with

\be
\la{eq5.3.12}
h_{11}(t) = g^2 B^2 e^{-\lambda \vert t \vert};
\ee

\noi
consequently 

\be
\la{eq5.3.12.bis}
D^A = 
-\frac{g^2 \omega_0 B^2}{2(\lambda^2 + \omega_0^2)}
\left(
\begin{array}{ccc}
0 & 1 & 0 \\
-1 & 0 & 0 \\
0 & 0 & 0
\end{array}
\right), \quad
C = \left(
\begin{array}{ccc}
2\alpha & -2b & 0 \\
-2b & 0 & 0 \\
0 & 0 & 0
\end{array}
\right)
\ee

\noi
with

\be
\la{eq5.3.13}
\alpha \equiv \frac{2g^2 B^2 \lambda}{\lambda^2 + \omega_0^2}, \quad
b \equiv -\frac{g^2 B^2 \omega_0}{\lambda^2 + \omega_0^2}.
\ee

\noi
In the vector representation~(\ref{eq5.2.7}), the dissipative matrix 
has the form

\be
\la{eq5.3.14}
{\cal D} = -2 \left(
\begin{array}{cccc}
0 & 0 & 0 & 0 \\
0 & 0 & b & 0 \\
0 & b & \alpha & 0 \\
0 & 0 & 0 & \alpha
\end{array}
\right)
\ee

\noi
whereas the Hamiltonian contribution ${\cal H}$, expressed as in 
equation~(\ref{eq5.3.8}), is characterized by $\omega \equiv \omega_0 + 
\frac{2g^2 \omega_0 B^2}{\lambda^2 + \omega_0^2}$.
The corresponding Redfield-Bloch equations are

\be
\la{eq5.3.15}
\begin{array}{ll}
\dot{\rho}_1(t) = - \alpha \rho_1(t) + \alpha \rho_2(t), &
\dot{\rho}_2(t) = - \dot{\rho}_1(t), \\ \\
\dot{\rho}_3(t) = -i \omega \rho_3(t) -\alpha \rho_3(t)
+ \alpha \rho_4(t) + 2ib \rho_4(t),\quad &
\dot{\rho}_4(t) = \left[\dot{\rho}_3(t)\right]^*,
\end{array}
\ee

\noi
with solutions 

\be
\la{eq5.3.17}
\begin{split}
\rho_1(t) &= \frac{1}{2}\left[ \left( 1 + e^{-2 \alpha t} \right)\rho_1(0)
+ \left( 1 - e^{-2 \alpha t} \right)\rho_2(0) \right], \\
\rho_2(t) &= \frac{1}{2}\left[ \left( 1 - e^{-2 \alpha t} \right)\rho_1(0)
+ \left( 1 + e^{-2 \alpha t} \right)\rho_2(0) \right], \\
\rho_3(t) &= e^{-\alpha t} \left[ \rho_4(0) \cosh{\delta t} +
\frac{\alpha - 2i b}{\delta} \rho_4(0) \sinh{\delta t} +
\frac{i \omega}{\delta} \rho_3(0) \sinh{\delta t} \right], \\
\rho_4(t) &= e^{-\alpha t} \left[ \rho_3(0) \cosh{\delta t} +
\frac{\alpha + 2i b}{\delta} \rho_3(0) \sinh{\delta t} -
\frac{i \omega}{\delta} \rho_4(0) \sinh{\delta t} \right],
\end{split}
\ee

\noi
where $\delta \equiv \sqrt{\alpha^2 + 4 b^2 - \omega^2}$.

In this case, positivity and complete positivity asks for $b = 0$ 
(and thus $\omega = \omega_0$), while for $b \ne 0$ the time-evolution 
is not even positive.


\s{Positivity and complete positivity}
\la{s5.4}
\markright{\small{\rm{\thesection. POSITIVITY AND COMPLETE POSITIVITY}}}

The covariance matrices in Subsections~\ref{ss5.3.2}
and \ref{ss5.3.3} do not guarantee the complete positivity of the time 
evolution. In this section we address the problems that arise in
these cases.

In the diagonal covariance matrix approach, let us consider  the 
maximally entangled state, $p = -q = 1/\sqrt{2}$ in~(\ref{eq5.1.1});
it evolves into

\be
\la{eq5.4.1}
\rho^{ab}(t) = \frac{1}{2} [
P_1 \otimes Q_2(t) + P_2 \otimes Q_1(t) -
P_3 \otimes Q_4(t) - P_4 \otimes Q_3(t) ]
\ee

\noi
where $Q_i(t)$ are computed from equations~(\ref{eq5.3.11}). The $4 \times 4$
statistical operator has the form

\be
\la{eq5.4.2}
\rho^{ab}(t) = \left(
\begin{array}{cccc}
E_-(t) & 0 & 0 & 0 \\
0 & E_+(t) & F(t) & 0 \\
0 & F^*(t) & E_+(t) & 0 \\
0 & 0 & 0 & E_-(t)
\end{array}
\right),
\ee

\noi
where

\be
\la{eq5.4.3}
E_{\pm}(t) \equiv \frac{1}{4}\left( 
1 \pm e^{-2 \gamma t} \right), \quad
F(t) \equiv -\frac{1}{2}e^{-(2a - i\omega) t};
\ee

\noi
the eigenvalues of $\rho^{ab}(t)$ at the exit of the interferometer are thus

\be
\la{eq5.4.4}
\begin{split}
\lambda_{1,2}(t) &= \frac{1}{4} \left( 1 - e^{-2 \gamma t} \right), \\
\lambda_{3,4}(t) &= \frac{1}{4} \left( 1 + e^{-2 \gamma t} \pm
2 e^{-2 a t} \right).
\end{split}
\ee

Observe that $\lambda_4(0) = 0$ and $d \lambda_4(0)/dt = (2a - \gamma)/2$;
then, by continuity, if $a < \gamma /2$, there is whole range of $t$ where
$\lambda_4(t) < 0$ and $\rho^{ab}(t)$ loses its physical meaning. On the other
hand, if the complete positivity condition is fulfilled, $\lambda_4(t) 
\geqslant 0$ for any time $t$ and no problem arises.

\begin{figure}
\la{fig5.4.1}
\begin{center}
\includegraphics[angle=0,width=0.5\textwidth]{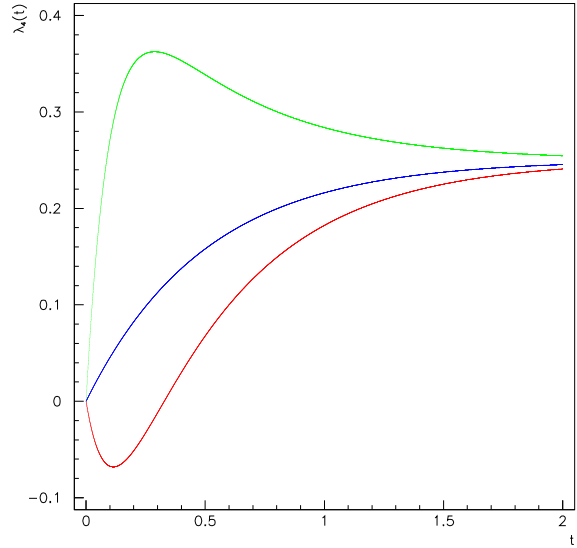}
\end{center}
\caption{\footnotesize Three different behaviors for $\lambda_4(t)$:
for $a > \gamma /2$ (green line), for $a = \gamma /2$ (blue line)
and for $a < \gamma /2$ (red line). We see that in the latter case 
there is a whole range of $t$ where $\lambda_4(t) < 0$ and then the 
density matrix loses its physical meaning; indeed, the time-evolution
is not completely positive.}
\end{figure}

\noi
Similar conclusions hold for the single component field
correlation case. 

\begin{Rem}
\la{rem5.4.1}
{\rm It is necessary an entangled initial state to 
see the problems generated by positive dynamics that are not completely 
positive. Indeed, suppose to have a factorized initial state, i.e. in the form
\begin{equation*}
\rho^{ab} = \sum_{i,j} c_{ij} \rho^a_i \otimes \rho^b_j > 0
\end{equation*}
with $c_{ij}\geqslant 0$, and let it evolve following the factorized
dynamics ${\bf I}_2 \otimes \Lambda_b^s$; then the statistical operator
at time $t$ is positive for any $t$, because of the positivity of 
$\Lambda_b^s$:
\begin{equation*}
\rho^{ab}(t) = ({\bf I}_2 \otimes \Lambda_b^s)[\rho^{ab}] = 
\sum_{i,j} c_{ij} \rho^a_i \otimes \Lambda_b^s[\rho^b_j] \geqslant 0
\end{equation*}}
\end{Rem}

\begin{Rem}
\la{rem5.4.2}
{\rm Besides the appearance of negative eigenvalues, we observe that, in the
single component field correlation case, for $\delta > \alpha$,
$\rho_3(t)$ and $\rho_4(t)$ diverge exponentially, 
see~(\ref{eq5.3.17}); the positivity
(and complete positivity) condition, $b = 0$, avoids this unacceptable
behavior.}
\end{Rem}

\begin{Rem}
\la{rem5.4.3}
{\rm Although apparently formal, these resuls are far from being academic.
Indeed, as alredy stressed at the end of Section~\ref{s5.1}, the entries
of $\rho$ are in principle directly accessible to the experiment. 
For example, in
the diagonal covariance matrix case, it follows from eq.~(\ref{eq5.1.7})
that $E_-(t) = {\cal O}_t^{1,z}(0,0)$ and $E_+(t) = {\cal O}_t^{2,z}(0,0)$
while $F(t)$ coincides with the expression in equation~(\ref{eq5.1.10}).}
\end{Rem}

By modulating a background magnetic field close to the stochastic properties
investigated in the previous three cases, one might reproduce experimentally 
the conditions for three different reduced dynamics and check their 
consequences.

Then, one may conclude that reduced, Markovian time evolutions $\Lambda_b^s$
must be not only positive, but also completely positive, since the lack
of any of these constraints results in experimentally detectable
inconsistencies.

The use of one reduced dynamics instead of another depends on the
Markovian approximation used to derive it and whether, given the properties 
of the stochastic field, it was justified or not. It thus seems appropriate 
to conclude that, whenever a semigroup composition law is expected, the
physically appropriate Markovian approximations are those which lead
to reduced dynamics consisting of completely positive maps~\cite{dumc}.


\s{Noncontextuality: the Clauser-Horne-Shimony-Holt inequality}
\la{s5.5}
\markright{\small{\rm{\thesection. NONCONTEXTUALITY: THE CLAUSER-HORNE-SHIMONY-HOLT INEQUALITY}}}

Quantum Mechanics has been shown to be a non-local and contextual theory,
as it violates the so-called Bell inequalities that have to be 
necessarily satisfyed by local, realistic and noncontextual 
theories~\cite{pere}.

Neutron interferometry is an optimal tool for investigating 
noncontextuality~\cite{basu}; indeed in this case the entanglement 
does not involve spatially separated parts, as in the usual tests 
exhibiting violations of Bell-like inequalities which focus on  
locality.

It is customary to use a particular form of Bell 
inequality, due to Clauser, Horne, Shimony and Holt~\cite{clau}. 
Using the previously defined projectors in configuration and spin space, 
one constucts the observables

\be
\la{eq5.5.1} 
A(\theta, \varphi) \equiv P_1(\theta, \varphi) - P_2(\theta, \varphi), \quad
B(\vec{n}) \equiv Q_{\vec{n}} - Q_{-\vec{n}}. 
\ee

These observables commute as translational and spin observables do; they
are called dichotomic since their eigenvalues are $\pm 1$. Choosing two
different values for both $\theta$ and $\varphi$ and two distinct
directions for the versor $\vec{n}$, we obtain the set of $4$ observables
$A(\theta_i,\varphi_i)$ and $B(\vec{n}_j)$ with $i,j = 1,2$. 

\noi
From the mean values

\be
\la{eq5.5.2}
\begin{split}
C_t(\theta,\varphi;\vec{n}) &\equiv \Tr[\rho^{ab}(t) A(\theta,\varphi)
\otimes B(\vec{n})] = \\
&= {\cal O}_t^{1,\vec{n}}(\theta,\varphi) +
 {\cal O}_t^{2,-\vec{n}}(\theta,\varphi) -
 {\cal O}_t^{1,-\vec{n}}(\theta,\varphi) -
 {\cal O}_t^{2,\vec{n}}(\theta,\varphi),
\end{split}
\ee

\noi
with four possible configurations of the control parameters $\theta$,
$\varphi$ and $\vec{n}$, the Clauser-Horne-Shimony-Holt inequality
is obtained~\cite{clau}:

\be
\la{eq5.5.3}
\vert C_t(\theta_1,\varphi_1;\vec{n}_1) +
C_t(\theta_1,\varphi_1;\vec{n}_2) + C_t(\theta_2,\varphi_2;\vec{n}_1) -
C_t(\theta_2,\varphi_2;\vec{n}_2) \vert \leqslant 2.
\ee

This inequality holds if we suppose the two possible outcomes $\pm 1$ of 
a measurement of the observables $A(\theta_i,\varphi_i)$ and 
$B(\vec{n}_j)$ are predetermined by the state $\rho^{ab}(t)$ independently
of which other commuting observable is simultaneously measured (for 
example, the value of $A(\theta_1,\varphi_1)$ is unchanged if we
measure simultaneously either $B(\vec{n}_1)$ or $B(\vec{n}_2)$). A
violation of this relation implies necessarily the contextuality of
the underlying theory. Equation~(\ref{eq5.5.2}) shows that this analysis
can be directly performed by means of the frequencies of counts
in the interferometric apparatus.

\begin{Rem}
\la{rem5.5.1}
{\rm In the configuration described in~\cite{basu,mich} the state $\rho^{ab}$ 
is not subject to any dynamics inside the 
interferometer, but in our context it does: this is the reason why we 
have considered an explicit time dependence in the $C_t(\theta,
\varphi,\vec{n})$ mean values.}
\end{Rem}

\noi
For the maximally entangled initial state ($p$ = $-q$ = $1/\sqrt{2}$)
these quantities take the form

\be
\la{eq5.5.4}
\begin{split}
C_t(\theta,\varphi;\vec{n}) = \sin^2{\theta} \Tr_b [Q_2(t)B(\vec{n})] &+
\cos^2{\theta} \Tr_b [Q_1(t)B(\vec{n})] + \\
&- \sin{2\theta} \re 
\left(e^{-i \varphi} \Tr_b[Q_4(t)B(\vec{n})]\right)
\end{split}
\ee

\noi
where $\Tr_b$ is the trace over the spin degree of freedom.

In vectorial representation, the time-evolution in the spin sector
reads $\vert \rho(t) \rangle = 
{\cal G}_t \vert \rho(0) \rangle$; in the following we shall consider
the first order approximation of ${\cal G}_t$ in the dissipative parameters,
supposed to be small with respect to the parameter characterizing the standard
Hamiltonian evolution, $\omega_0$ (for the expression of the entries of
${\cal G}_t$, see Section~\ref{s2.4}). 
We introduce two vectors, $\vec{G}(t)$ and $\vec{F}(t)$, whose components
are defined by

\be
\la{eq5.5.5}
G_i(t) \equiv \left({\cal G}_t\right)_{i3},\quad
F_i(t) \equiv \left({\cal G}_t\right)_{i1} -i \left({\cal G}_t\right)_{i2},
\quad i = 1,2,3.
\ee

\noi
Then, the traces in~(\ref{eq5.5.4}) can be written as:

\be
\la{eq5.5.6}
\Tr_b [Q_{1,2}(t)B(\vec{n})] = \pm \vec{G}(t)\cdot \vec{n}, \quad
\Tr_b [Q_{4}(t)B(\vec{n})] = \vec{F}(t)\cdot \vec{n};
\ee

\noi
whence the mean values (5.2) read

\be
\la{eq5.5.7}
C_t(\theta,\varphi;\vec{n})=\vec{n}\cdot\Bigl[
 \vec{G}(t)\cos{2\theta} - 
\re \left(e^{-i\varphi}\vec{F}(t)\right)\sin{2\theta}\Bigr] .
\ee

We will now discuss the explicit behaviour of 
$C_t(\theta,\varphi;\vec{n})$
in the three cases introduced in Section~\ref{s5.3}, to leading order
in the dissipative parameters.

\subsection{White noise}

Though analytic expressions of ${\cal G}_t$ are obtainable in the general
case of a stochastic magnetic field with white noise correlations,
these are rather involved and scarcely illuminating.
More conveniently, we consider the expression of
the time evolutor ${\cal G}_t$ to first
order in the parameters, appearing in the dissipative part $\cal D$ 
of~(\ref{eq5.2.8}), assumed to be small. One calculates

\be
\la{eq5.5.8}
\begin{split}
G_1(t) &= -\frac{4 \vert C \vert}{\omega_0} \sin{\frac{\omega_0 t}{2}}
\cos{\left( \frac{\omega_0 t}{2} + \phi_C \right)}, \\
G_2(t) &= \frac{4 \vert C \vert}{\omega_0} \sin{\frac{\omega_0 t}{2}}
\sin{\left( \frac{\omega_0 t}{2} + \phi_C \right)}, \\
G_3(t) &= e^{-2 \gamma t},
\end{split}
\ee

\noi
where
$\vert C \vert^2=c^2+\beta^2$ and $\tan{\phi_C} = \beta/c$;

\be
\begin{split}
\la{eq5.5.9}
\re \left[ e^{-i\varphi} F_1(t) \right] &=
e^{-(a + \alpha)t} \cos(\omega_0 t - \varphi) +
\frac{\vert B \vert}{\omega_0} \sin{\omega_0 t}
\cos{(\varphi + \phi_B)}, \\
\re \left[ e^{-i\varphi} F_2(t) \right] &=
e^{-(a + \alpha)t} \sin(\omega_0 t - \varphi) -
\frac{\vert B \vert}{\omega_0} \sin{\omega_0 t}
\sin{(\varphi - \phi_B)}, \\
\re \left[ e^{-i\varphi} F_3(t) \right] &=
-\frac{4 \vert C \vert}{\omega_0} \sin{\frac{\omega_0 t}{2}}
\cos{\left( \frac{\omega_0 t}{2} - \varphi - \phi_C \right)},
\end{split}
\ee

\noi
where $\vert B\vert^2=(a - \alpha)^2+4b^2$ and 
$\tan{\phi_B} = 2b/(\alpha - a)$. It thus follows that, in leading
order with respect to $a$, $b$, $c$, $\alpha$, $\beta$ and $\gamma$

\be
\la{eq5.5.10}
\begin{split}
C_t(\theta,\varphi;\vec{n}) = \;&n_x \Bigl\{
-\frac{4 \vert C \vert}{\omega_0} \cos{2 \theta}
\sin{\frac{\omega_0 t}{2}} \cos{\left( \frac{\omega_0 t}{2} + 
\phi_C \right)} + \\
&- \sin{2 \theta} \Bigl[ e^{-(a + \alpha)t} \cos{(\omega_0 t - \varphi)} +
\frac{\vert B \vert}{\omega_0} \sin{\omega_0 t} \cos{(\varphi + \phi_B)}
\Bigr] \Bigr\} + \\
&+ n_y \Bigl\{
\frac{4 \vert C \vert}{\omega_0} \cos{2 \theta}
\sin{\frac{\omega_0 t}{2}} \sin{\left( \frac{\omega_0 t}{2} + 
\phi_C \right)} + \\
&- \sin{2 \theta} \Bigl[ e^{-(a + \alpha)t} \sin{(\omega_0 t - \varphi)} -
\frac{\vert B \vert}{\omega_0} \sin{\omega_0 t} \sin{(\varphi - \phi_B)}
\Bigr] \Bigr\} + \\
&+ n_z \Bigl\{
e^{-2 \gamma t} \cos{2 \theta} + \frac{4 \vert C \vert}{\omega_0}
\sin{2 \theta} \sin{\frac{\omega_0 t}{2}} 
\cos{\left( \frac{\omega_0 t}{2} - \varphi - \phi_C \right)}\Bigr\}.
\end{split}
\ee

\subsection{Diagonal covariance matrix}

For the stochastic magnetic fields with correlation functions
as in~(\ref{eq5.3.5}) we have

\be
\la{eq5.5.11}
\vec{G}(t) = \left(0, 0, e^{-2 \gamma t} \right), \quad
\vec{F}(t) = e^{-(2a - i \omega)t}(1, -i, 0),
\ee

\noi
whence

\be
\la{eq5.5.12}
C_t(\theta,\varphi;\vec{n}) = \;
n_z e^{-2 \gamma t} \cos{2 \theta} -
e^{-2 a t}\sin{2 \theta}[n_x \cos{(\omega t - \varphi)}
+ n_y \sin{(\omega t - \varphi)}].
\ee

\subsection{Single component field correlation}

In the case of magnetic field with correlation
functions as in~(\ref{eq5.3.12}) the computation is similar. One finds:

\be
\begin{split}
\la{eq5.5.13}
\vec{G}(t) &= \left( 0, 0, e^{-2 \alpha t} \right), \quad \\
\vec{F}(t) &= e^{- \alpha t} \Bigl[
\left( \cosh{\delta t} + i \frac{\omega}{\delta} \sinh{\delta t} \right) 
(1, -i, 0) + \frac{\alpha + 2ib}{\delta} \sinh{\delta t} (1, i, 0) \Bigr],
\end{split}
\ee

\noi
whence, assuming $\delta>0$, one obtains:

\be
\la{eq5.5.14}
\begin{split}
C_t(\theta, \varphi; \vec{n}) = \;&n_z e^{-2 \alpha t} \cos{2 \theta} + \\
&- e^{-\alpha t} \sin{2 \theta} \Bigl\{
n_x \Bigl[ \left(
\cosh{\delta t} + \frac{\alpha}{\delta} \sinh{\delta t} \right) 
\cos{\varphi} + \frac{\omega + 2b}{\delta} \sinh{\delta t} \sin{\varphi}
\Bigr] + \\
&+ n_y \Bigl[ \left( -\cosh{\delta t} + \frac{\alpha}{\delta}
\sinh{\delta t} \right) \sin{\varphi} + \frac{\omega - 2b}{\delta}
\sinh{\delta t} \cos{\varphi} \Bigr] \Bigr\}.
\end{split}
\ee

The lack of positivity preservation which characterizes
the time-evolution leading to equation~(\ref{eq5.5.14}) manifests itself 
in that the quantities $C_t(\theta,\varphi;\vec{n})$ diverge in 
$t$ when $\delta>\alpha$.

Expressions~(\ref{eq5.5.10}), (\ref{eq5.5.12}) and (\ref{eq5.5.14})
agree with those used when there is no
dissipation, namely putting $a = b = c = \alpha = \beta = \gamma = 0$:

\be
\la{eq5.5.15}
C_t(\theta,\varphi;\vec{n}) =
-n_x \sin{2 \theta} \cos{(\omega_0 t - \varphi)} - 
n_y \sin{2 \theta} \sin{(\omega_0 t - \varphi)} +
n_z \cos{2 \theta}.
\ee

Notice that the unitary time-evolution generated by the Hamiltonian 
$H_0$ contributes to a time-varying redefinition of the 
angle $\varphi$.

Concerning the issue of complete positivity versus simple positivity, in
expressions~(\ref{eq5.5.10}) and (\ref{eq5.5.12}) the two possibilities 
result in different relaxation properties due to whether 
inequalities~(\ref{eq.cp}) or~(\ref{eq.p}) are fulfilled.
No physical inconsistencies may affect the mean values 
$C_t(\theta,\varphi;\vec{n})$;
indeed, negative probabilities may result in negative mean values
of positive observables only if the latter are entangled.
In the case of the quantities involved in inequality~(\ref{eq5.5.2}),
the observables are factorized, $P_{1,2}(\theta,\varphi)\otimes Q_{\vec{n}}$
and the positivity of their mean values is preserved even when the time 
evolution is only positive and not completely positive.

This can be seen as follows.
By definition of dual map $\Lambda^*_t$, to the Schr{\"o}dinger time-evolution
$\rho^{ab}(t) = ({\bf I}_2\otimes\Lambda_t)\,[\rho^{ab}(0)]$, there
corresponds the Heisenberg time-evolution of observables 
$A(t)=({\bf I}_2 \otimes \Lambda^*_t) [A]$,

\be
\la{eq5.5.16}
\Tr \Bigl[ ({\bf I}_2 \otimes \Lambda_t)[\rho^{ab}(0)] A \Bigr] =
\Tr \Bigl[ \rho^{ab} ({\bf I}_2 \otimes
\Lambda^*_t )[A] \Bigr].
\ee

The maps $\Lambda^*_t$, dual to $\Lambda_t$, form a semigroup
of dynamical maps that transform positive observables into positive
observables, if the $\Lambda_t$'s preserve the positivity of states.
Consequently, even when the initial state $\rho^{ab}$ is entangled and 
the $\Lambda_t$'s positivity preserving, but not completely positive, it 
turns out that

\be
\la{eq5.5.17}
\Tr \Bigl[ \rho^{ab}(t) P_{1,2}(\theta,\varphi)
\otimes Q_{\vec{n}} \Bigr] =
\Tr \Bigl[ \rho^{ab} P_j(\theta,\varphi)
\otimes \Lambda^*_t [Q_{\vec{n}}] \Bigr] \geqslant 0.
\ee

\section{Discussion and conclusions}
\la{s5.6}
\markright{\small{\rm{\thesection. DISCUSSION AND CONCLUSIONS}}}

Complete positivity is a property of quantum time-evolutions
which is enjoyed by the standard dynamics of closed
quantum systems generated by Hamiltonian operators, but not
automatically by the more general reduced dynamics describing time-evolution
of open quantum systems in interaction with suitable environments.
Complete positivity is intimately related to the phenomenon of quantum
entanglement, between two different systems, but also between two
different degrees of freedom of a same physical system.

In this chapter we have considered the two entangled degrees of freedom,
translational and rotational, of a beam of neutrons travelling through
an interferometric apparatus.
We have studied the consequences of placing the
interferometer in a stochastic, gaussian magnetic field weakly
coupled to the spin degree of freedom, providing an experimentally
controllable environment.
As explained
in the three cases of Section~\ref{s5.3},
the same Markov approximation naively yields a
semigroup of dynamical maps ${\bf I}_2 \otimes \Lambda^b_t$, where only
the spin degree of
freedom evolves in time; by varying the decay properties of the
external field correlations, these maps turn out to be alternatively
completely positive, simply positive, not even positive.

The noncontextuality tests proposed in~\cite{basu,mich} are based on the
Clauser-Horne-Shimony-Holt inequality~(\ref{eq5.5.3}) without 
time-dependence.
The presence of a fluctuating magnetic field induces relaxation on the
spin degree of freedom with strength and properties depending on
those of the field.
Typically, the mean values in the inequality are damped and make
it more difficult to be violated.
However, in presence of stochastic fields yielding reduced
dynamics that do not preserve positivity, the inequality might be
dramatically violated because of possible mean values
diverging in time.

This latter possibility is a manifestation of the fact
that any physically consistent
time-evolution $\Lambda^b_t$ must preserve the positivity of spin states
in order that the eigenvalues of the corresponding spin density matrices
might at any time be used as probabilities, in agreement with the
statistical interpretation of quantum mechanics.
If  $\Lambda^b_t$ preserves the trace of spin density matrices, but not
their positivity, spin states
may evolve in time in such a way that some of their eigenvalues
become negative, while others greater than $1$, without upper bounds.
It is this physically unacceptable phenomenon that leads to diverging
mean values.

The request of positivity preservation by the maps $\Lambda^b_t$ with
respect to spin states is thus unexcapable, but
it is not enough to avoid physical inconsistencies when the
time-evolution maps ${\bf I}_2 \otimes \Lambda^b_t$ act on states
$\rho^{ab}$ with correlations between spin and translational degrees
of freedom.

Inequality~(\ref{eq5.5.3}) does reveal the difference 
between completely positive
and simply positivity preserving $\Lambda^b_t$, but only as long as the
relaxation characteristic are concerned, without any further effect
(as the divergence of some contributions to the inequality).
In fact, the positive observables in~(\ref{eq5.5.2})
are factorized, that is they incorporate no entanglement between the
translational and spin degree of freedom.
Even if the initial state does incorporate entanglement,
it nevertheless follows that the mean values of factorized observables
remain positive and bounded.

However, the interferometric apparatus proposed in~\cite{basu,mich}
might also be used to measure the entries of the states of the
neutron beam at the exit of the interferometer.
In this way, one might have access to the spectrum of an initially
entangled state after being subjected to the effects of the stochastic
magnetic field.

In the case of fluctuating fields yielding  reduced dynamics
that preserve positivity, but are not completely positive,
the theoretical predictions indicate the appearance of negative
eigenvalues, that is of negative probabilities, in the spectrum of
the entangled exiting state.
The fact that they are, in line of principle, detectable
experimentally, does not allow to dismiss such an occurrence as
practically negligible.
Rather, it forces to reconsider the Markov approximation used to derive
the time-evolution and to select as physically consistent
only those providing
completely positive reduced dynamics.

%% file: appendixA.tex
\chapter{Dissipative parameters in the eigenstates basis}
\la{a1}
\markboth{\small{\rm{\thechapter. DISSIPATIVE PARAMETERS IN THE EIGENSTATES BASIS}}}{\small{\rm{\thechapter. DISSIPATIVE PARAMETERS IN THE EIGENSTATES BASIS}}}

For an unstable system,
the dissipative parameters entering into~(\ref{eq2.4.16}) are defined by:

\begin{eqnarray*}
\lambda_1 &\equiv& \frac{1}{\vert r_1 + r_2 \vert ^2}
\Bigl\{ -w(1 + \vert r_1 \vert ^2)(1 - \vert r_2 \vert ^2)
-2 (1 + \vert r_1 \vert ^2) \re (U r_2)\Bigr\}, \\
\lambda_2 &\equiv& \frac{1}{\vert r_1 + r_2 \vert ^2}
\Bigl\{ -w(1 + \vert r_2 \vert ^2)(1 - \vert r_1 \vert ^2)
+2 (1 + \vert r_2 \vert ^2) \re (U r_1)\Bigr\}, \\
\sigma &\equiv& \frac{1}{\vert r_1 + r_2 \vert ^2}
\Bigl\{ -w(1 + \vert r_2 \vert ^2)(1 - \vert r_2 \vert ^2) 
- 2 (1 + \vert r_2 \vert ^2) \re (U r_2)\Bigr\}, \\
\xi &\equiv& \frac{1}{\vert r_1 + r_2 \vert ^2}
\Bigl\{ -w(1 + \vert r_1 \vert ^2)(1 - \vert r_1 \vert ^2) 
+ 2 (1 + \vert r_1 \vert ^2) \re (U r_1)\Bigr\}, \\
\delta_1 &\equiv& \frac{1}{\vert r_1 + r_2 \vert ^2}
\Bigl\{ w(1 - r_1 r_2^*)(1 - \vert r_2 \vert ^2) +
2(1 - r_1 r_2^*) \re (U r_2)\Bigr\}, \\
\delta_2 &\equiv& \frac{1}{\vert r_1 + r_2 \vert ^2}
\Bigl\{ -w(1 + \vert r_2 \vert ^2)(1 + r_2 r_1^*) +
(1 + \vert r_2 \vert ^2)(U^* r_1^* - U r_2)\Bigr\}, \\
\phi_1 &\equiv& \frac{1}{\vert r_1 + r_2 \vert ^2}
\Bigl\{ w(1 - r_1 r_2^*)(1 - \vert r_1 \vert ^2) -
2(1 - r_1 r_2^*) \re (U r_1)\Bigr\}, \\
\phi_2 &\equiv& \frac{1}{\vert r_1 + r_2 \vert ^2}
\Bigl\{ -w(1 + \vert r_1 \vert ^2)(1 + r_2 r_1^*) +
(1 + \vert r_1 \vert ^2)(U^* r_1^* - U r_2)\Bigr\}, \\
\omega &\equiv& \frac{1}{\vert r_1 + r_2 \vert ^2}
\Bigl\{ w(1 - r_1 r_2^*)(1 + r_2 r_1^*) - 
(1 - r_1 r_2^*)(U^* r_1^* - U r_2)\Bigr\}, \\
\theta &\equiv& \frac{1}{\vert r_1 + r_2 \vert ^2}
\Bigl\{ w(1 - r_2 r_1^*)(1 + r_2 r_1^*) - 
(1 - r_2 r_1^*)(U^* r_1^* - U r_2)\Bigr\}, \\
\end{eqnarray*}

\begin{eqnarray*}
\Lambda &\equiv& \frac{1}{\vert r_1 + r_2 \vert ^2}
\Bigl\{ \gamma (1 - \vert r_1 \vert ^2)(1 - \vert r_2 
\vert ^2) - 2A \re (r_2 r_1^*) + 2 \re (B r_1^* r_2^*) + \\ 
&&- 2 (1 - \vert r_2 \vert ^2) \re (C r_1^*) 
+ 2 (1 - \vert r_1 \vert ^2) \re (C r_2^*)\Bigr\}, \\
\Xi &\equiv& \frac{1}{\vert r_1 + r_2 \vert ^2}
\Bigl\{ \gamma (1 - \vert r_1 \vert ^2)^2 + 2A \vert r_1 \vert ^2
- 2 \re (B r_1^{*2}) - 4 (1 - \vert r_1 \vert ^2) \re (C r_1^*)
\Bigr\}, \\
\Sigma &\equiv& \frac{1}{\vert r_1 + r_2 \vert ^2}
\Bigl\{ \gamma (1 - \vert r_2 \vert ^2)^2 + 2A \vert r_2 
\vert ^2 - 2 \re (B r_2^{*2}) + 4(1 - \vert r_2 \vert ^2) \re (C r_2^*)
\Bigr\}, \\
\Delta &\equiv& \frac{1}{\vert r_1 + r_2 \vert ^2}
\Bigl\{ -\gamma (1 - \vert r_2 \vert ^2)(1 + r_1 r_2^*) - 
A (\vert r_2 \vert ^2 - r_1 r_2^*) + B r_2^{*2} - B^* r_1 r_2 + \\
&&+ (C^* r_1 - C r_2^*)(1 - \vert r_2 \vert ^2) - 2 (1 + r_1 r_2^*) 
\re (C r_2^*)\Bigr\}, \\
\Phi &\equiv& \frac{1}{\vert r_1 + r_2 \vert ^2}
\Bigl\{ -\gamma (1 - \vert r_1 \vert ^2)(1 + r_1 r_2^*) - 
A (\vert r_1 \vert ^2 - r_1 r_2^*) + B^* r_1^2 - B r_1^* r_2^* + \\
&&+ (C^* r_1 - C r_2^*)(1 - \vert r_1 \vert ^2) + 2 (1 + r_1 r_2^*) 
\re (C r_1^*)\Bigr\}, \\
\Omega &\equiv& \frac{1}{\vert r_1 + r_2 \vert ^2}
\Bigl\{ -\gamma \vert 1 + r_1 r_2^* \vert ^2 - A 
(\vert r_1 \vert ^2 + \vert r_2 \vert ^2) - 2 \re (B r_1^* r_2^*) 
+ \\ 
&&+ 2 \re \left[ C \left( r_1^*(1 - \vert r_2 \vert ^2) 
- r_2^* (1 - \vert r_1 \vert ^2) \right) \right]
\Bigr\}, \\
\Theta &\equiv& \frac{1}{\vert r_1 + r_2 \vert ^2}
\Bigl\{ -\gamma (1 + r_2 r_1^*)^2 + 2A r_2 r_1^* + B r_1^{*2} 
+ B^* r_2^2 - 2 (1 + r_2 r_1^*)(C^* r_2 - C r_1^*)\Bigr\}. \\
\end{eqnarray*}

\noi
For a stable system, they are:

\begin{eqnarray*}
\lambda_1 &=& \Bigl\{ w(1 - \vert r_1 \vert ^2)
-2 \re (U r_1) \Bigr\}, \\
\lambda_2 &=& - \Bigl\{ w(1 - \vert r_1 \vert ^2)
-2 \re (U r_1)\Bigr\}, \\
\sigma &=& \frac{1}{\vert r_1 \vert ^2} \Bigl\{ w(1 - \vert r_1 \vert ^2)
-2 \re (U r_1)\Bigr\}, \\
\xi &=& -\vert r_1 \vert ^2 \Bigl\{ w(1 - \vert r_1 \vert ^2)
-2 \re (U r_1) \Bigr\}, \\
\delta_2 &=& - \Bigl\{ 2w - U^* r_1^* + \frac{U}{r_1^*} \Bigr\}, \\
\phi_2 &=& -\vert r_1 \vert ^2 \Bigl\{ 2w - U^* r_1^* + 
\frac{U}{r_1^*} \Bigr\}, \\
\delta_1 &=& \phi_1 \quad = \quad \omega \quad = \quad \theta 
\quad = \quad 0, \\
\end{eqnarray*}

\begin{eqnarray*}
\Lambda &=& \frac{1}{(1 + \vert r_1 \vert ^2)^2}
\Bigl\{ -\gamma (1 - \vert r_1 \vert ^2)^2
- 2A\vert r_1 \vert ^2 + 2 \re (B r_1^{*2}) + 4 (1 - \vert r_1 \vert ^2) 
\re (C r_1^*) \Bigr\}, \\
\Xi &=& -\frac{\vert r_1 \vert ^2}{(1 + \vert r_1 \vert ^2)^2}
\Bigl\{ -\gamma (1 - \vert r_1 \vert ^2)^2
- 2A\vert r_1 \vert ^2 + 2 \re (B r_1^{*2}) + 4 (1 - \vert r_1 \vert ^2) 
\re (C r_1^*) \Bigr\},\\
\Sigma &=& -\frac{1}{\vert r_1 \vert ^2(1 + \vert r_1 \vert ^2)^2}
\Bigl\{ -\gamma (1 - \vert r_1 \vert ^2)^2
- 2A\vert r_1 \vert ^2 + 2 \re (B r_1^{*2}) + 4 (1 - \vert r_1 \vert ^2) 
\re (C r_1^*) \Bigr\}, \\
\Delta &=& \frac{1}{(1 + \vert r_1 \vert ^2)^2}
\Bigl\{ 2\gamma (1 - \vert r_1 \vert ^2)^2 - 
A (1 - \vert r_1 \vert ^2) + \frac{B}{r_1^2} - B^* r_1^2 + \\ 
&&- \Bigl(C^* r_1 - \frac{C}{r_1}\Bigr)(1 - \vert r_1 \vert ^2) 
- 4 \re (C r_1^*)\Bigr\}, \\
\Phi &=& -\frac{\vert r_1 \vert ^2}{(1 + \vert r_1 \vert ^2)^2}
\Bigl\{ 2\gamma (1 - \vert r_1 \vert ^2)^2 - 
A (1 - \vert r_1 \vert ^2) + \frac{B}{r_1^2} - B^* r_1^2 + \\ 
&&- \Bigl(C^* r_1 - \frac{C}{r_1}\Bigr)(1 - \vert r_1 \vert ^2) 
- 4 \re (C r_1^*)\Bigr\},\\
\Omega &=& \frac{1}{(1 + \vert r_1 \vert ^2)^2}
\Bigl\{ -4\gamma \vert r_1 \vert ^2 - A 
(\vert r_1 \vert ^4 + 1) - 2 \re (B r_1^{*2}) 
- 4 (1 - \vert r_1 \vert ^2) \re (C r_1^*)
\Bigr\}, \\
\Theta &=& \frac{\vert r_1 \vert ^2}{(1 + \vert r_1 \vert ^2)^2}
\Bigl\{ -4\gamma + 2A + B r_1^{*2} + \frac{B^*}{r_1^{*2}}
- 4\Bigl(\frac{C^*}{r_1^*} - Cr_1^*\Bigr)\Bigr\}. \\
\end{eqnarray*}

\noi
Note that, in this case, the following relations hold:

\begin{eqnarray*}
\lambda_2 = - \lambda_1, \quad
\xi = -\vert r_1 \vert ^2 \lambda_1, \quad
\phi_2 = \vert r_1 \vert ^2 \delta_2, \\ \\
\Xi = -\vert r_1 \vert ^2 \Lambda, \quad 
\Sigma = -\frac{1}{\vert r_1 \vert ^2}\Lambda, \quad 
\Phi = -\vert r_1 \vert ^2 \Delta.
\end{eqnarray*}

%% file: appendixB.tex
\chapter{Expressions of the time evolutors}
\la{a2}
\markboth{\small{\rm{\thechapter. EXPRESSIONS OF THE TIME EVOLUTORS}}}{\small{\rm{\thechapter. EXPRESSIONS OF THE TIME EVOLUTORS}}}

The entries of the time evolutor $\hat{\cal G}^d_t$ in~(\ref{eq2.4.2})
for an unstable system are\footnote{By definition, $\lambda \equiv (\lambda_1 + \lambda_2)/2$}:

\begin{eqnarray*}
(\hat{\cal G}_t^{d})_{11} &=& e^{-\gamma_1 t}, \\ 
(\hat{\cal G}_t^{d})_{12} &=& \frac{1}{\Delta \Gamma}(\Sigma + \sigma)
(e^{-\gamma_2 t} - e^{-\gamma_1 t}), \\
(\hat{\cal G}_t^{d})_{13} &=& \frac{2}{\Delta \Gamma_+}(\Delta + \delta_1)
(e^{-\Gamma_- t} - e^{-\gamma_1 t}), \\
(\hat{\cal G}_t^{d})_{14} &=& \frac{2}{\Delta \Gamma_-}(\Delta^* + \delta_1^*) 
(e^{-\Gamma_+ t} - e^{-\gamma_1 t}),  \\
(\hat{\cal G}_t^{d})_{21} &=& \frac{1}{\Delta \Gamma}(\Xi + \xi)
(e^{-\gamma_2 t} - e^{-\gamma_1 t}), \\ 
(\hat{\cal G}_t^{d})_{22} &=& e^{-\gamma_2 t}, \\
(\hat{\cal G}_t^{d})_{23} &=& \frac{2}{\Delta \Gamma_-}(\Phi + \phi_1)
(e^{-\gamma_2 t} - e^{-\Gamma_- t}), \\
(\hat{\cal G}_t^{d})_{24} &=& \frac{2}{\Delta \Gamma_+}(\Phi^* + \phi_1^*)
(e^{-\gamma_2 t} - e^{-\Gamma_+ t}), \\
(\hat{\cal G}_t^{d})_{31} &=& \frac{2}{\Delta \Gamma_+}(\phi_2 - \Phi^*)
(e^{-\Gamma_- t} - e^{-\gamma_1 t}), \\
(\hat{\cal G}_t^{d})_{32} &=& \frac{2}{\Delta \Gamma_-}(\delta_2 - \Delta^*)
(e^{-\gamma_2 t} - e^{-\Gamma_- t}), \\
(\hat{\cal G}_t^{d})_{33} &=& e^{-(\Gamma_- + \Lambda - \Omega 
+ \lambda - \omega)t}, \\
(\hat{\cal G}_t^{d})_{34} &=& \frac{1}{\Delta m}(\Theta + \theta)
e^{-\Gamma t} \sin{\Delta m t}, \\
(\hat{\cal G}_t^{d})_{41} &=& \frac{2}{\Delta \Gamma_-}(\phi_2^* - \Phi)
(e^{-\Gamma_+ t} - e^{-\gamma_1 t}), \\
\end{eqnarray*}

\begin{eqnarray*}
(\hat{\cal G}_t^{d})_{42} &=& \frac{2}{\Delta \Gamma_+}(\delta_2^* - \Delta)
(e^{-\gamma_2 t} - e^{-\Gamma_+ t}), \\
(\hat{\cal G}_t^{d})_{43} &=& \frac{1}{\Delta m}(\Theta^* + \theta^*)
e^{-\Gamma t} \sin{\Delta m t}, \\
(\hat{\cal G}_t^{d})_{44} &=& e^{-(\Gamma_+ + \Lambda - \Omega 
+ \lambda - \omega^*)t}. \\
\end{eqnarray*}

\noi
For a stable system they take the form:

\begin{eqnarray*}
(\hat{\cal G}_t^{d})_{11} &=& \frac{1}{2} (e^{2\Lambda t} + 1) +
\frac{\lambda_1}{2\Lambda}(e^{2\Lambda t} - 1) \\
(\hat{\cal G}_t^{d})_{12} &=& \frac{\Sigma + \sigma}{2 \Lambda}
(e^{2\Lambda t} - 1) \\
(\hat{\cal G}_t^{d})_{13} &=& -\frac{i}{\Delta m} \Delta
(e^{i \Delta m t} - 1) \\
(\hat{\cal G}_t^{d})_{14} &=& \frac{i}{\Delta m} \Delta^* 
(e^{-i \Delta m t} - 1) \\
(\hat{\cal G}_t^{d})_{21} &=& \frac{\Xi + \xi}{2 \Lambda}
(e^{2\Lambda t} - 1) \\ 
(\hat{\cal G}_t^{d})_{22} &=& \frac{1}{2} (e^{2\Lambda t} + 1) -
\frac{\lambda_1}{2\Lambda}(e^{2\Lambda t} - 1)  \\
(\hat{\cal G}_t^{d})_{23} &=& -\frac{i}{\Delta m}\Phi
(e^{i \Delta m t} - 1) \\
(\hat{\cal G}_t^{d})_{24} &=& \frac{i}{\Delta m}\Phi^*
(e^{-i \Delta m t} - 1) \\
(\hat{\cal G}_t^{d})_{31} &=& -\frac{i}{\Delta m}(\phi_2 - \Phi^*)
(e^{i \Delta m t} - 1) \\
(\hat{\cal G}_t^{d})_{32} &=& -\frac{i}{\Delta m}(\delta_2 - \Delta^*)
(e^{i\Delta m t} - 1) \\
(\hat{\cal G}_t^{d})_{33} &=& e^{(\Omega + i \Delta m)t} \\
(\hat{\cal G}_t^{d})_{34} &=& \frac{1}{\Delta m}\Theta \sin{\Delta m t} \\
(\hat{\cal G}_t^{d})_{41} &=& \frac{i}{\Delta m}(\phi_2^* - \Phi)
(e^{-i \Delta m t} - 1) \\
(\hat{\cal G}_t^{d})_{42} &=& \frac{i}{\Delta m}(\delta_2^* - \Delta)
(e^{-i \Delta m t} - 1) \\
(\hat{\cal G}_t^{d})_{43} &=& \frac{1}{\Delta m}\Theta^* \sin{\Delta m t} \\
(\hat{\cal G}_t^{d})_{44} &=& e^{(\Omega - i \Delta m)t}. \\
\end{eqnarray*}

%% file: appendixC.tex
\chapter{Neutral {\bf B} mesons decays}
\la{a3}
\markboth{\small{\rm{\thechapter. NEUTRAL {\bf B} MESONS DECAYS}}}{\small{\rm{\thechapter. NEUTRAL {\bf B} MESONS DECAYS}}}

\section{Semileptonic decays}

\begin{equation*}
\begin{split}
{\cal P}_{h^-}({\bf B}^0; \tau) = \vert {\cal M}_h \vert ^2 e^{-\tau} 
&\Bigl\{ 
\sinh{\delta \tau} \Bigl[ \frac{\vert r_H \vert ^2 - \vert r_L \vert ^2}
{\vert r_H + r_L \vert ^2} - \re \lambda_h - \frac{\Lambda^{\prime}}{\Delta
\Gamma} \Bigr] +  \\
&+ \cosh{\delta \tau} \Bigl[ \frac{\vert r_H \vert ^2 + \vert r_L  
\vert ^2}{\vert r_H + r_L \vert ^2} - \re y_h + \frac{2}{\Gamma}
\frac{\delta}{\delta ^2 + \omega ^2} \re \Delta^{\prime} \Bigr] + \\ 
&+ \sin{\omega \tau} \Bigl[ 2 \frac{\im (r_H r_L^*)}
{\vert r_H + r_L \vert ^2} + \im \lambda_h - \frac{1}{2\Delta m} \re 
\Theta^{\prime} \Bigr] + \\
&+ \cos{\omega \tau} \Bigl[ 2 \frac{\re (r_L r_H^*)}
{\vert r_H + r_L \vert ^2} e^{-(\Omega^{\prime} + \Lambda^{\prime})
\frac{\tau}{\Gamma}}
- \re y_h - \frac{2}{\Gamma} \frac{\delta}
{\delta ^2 + \omega ^2} \re \Delta^{\prime} \Bigr] \Bigr\}; \\
\end{split}
\end{equation*}

\begin{equation*}
\begin{split}
{\cal P}_{h^+}({\bf B}^0; \tau) = \vert {\cal M}_h \vert ^2 e^{-\tau} 
&\Bigl\{ 
\sinh{\delta \tau} \Bigl[ - \re \mu_h - \frac{2}{\Gamma} \frac{\omega}
{\delta ^2 + \omega ^2} \im \Delta^{\prime} - 
\frac{\Lambda^{\prime}}{\Delta \Gamma} \Bigr] + \\
&+ \cosh{\delta \tau} \Bigl[ \frac{2}{\vert r_H + r_L \vert ^2} 
+ \re y_h \Bigr] + \\ 
&+ \sin{\omega \tau} \Bigl[ \im \mu_h + \frac{2}{\Gamma}
\frac{\delta}{\delta ^2 + \omega ^2} \im \Delta^{\prime} + 
\frac{1}{2\Delta m} \re \Theta^{\prime} \Bigr] + \\
&+ \cos{\omega \tau} \Bigl[ - \frac{2}{\vert r_H + r_L \vert ^2}
e^{-(\Omega^{\prime} + \Lambda^{\prime})\frac{\tau}{\Gamma}} - 
\re y_h \Bigr] \Bigr\}; \\
\end{split}
\end{equation*}

\begin{equation*}
\begin{split}
{\cal P}_{h^+}(\overline{\bf B}^0; \tau) = \vert {\cal M}_h \vert ^2 e^{-\tau} 
&\Bigl\{ 
\sinh{\delta \tau} \Bigl[ \frac{\vert r_L \vert ^2 - \vert r_H \vert ^2}
{\vert r_H + r_L \vert ^2} - \re \mu_h - \frac{\Lambda^{\prime}}{\Delta
\Gamma} \Bigr] + \\
&+ \cosh{\delta \tau} \Bigl[ \frac{\vert r_H \vert ^2 + \vert r_L  
\vert ^2}{\vert r_H + r_L \vert ^2} + \re y_h - \frac{2}{\Gamma}
\frac{\delta}{\delta ^2 + \omega ^2} \re \Delta^{\prime} \Bigr] + \\ 
&+ \sin{\omega \tau} \Bigl[ 2 \frac{\im (r_L r_H^*)}
{\vert r_H + r_L \vert ^2} - \im \mu_h - \frac{1}{2\Delta m} \re 
\Theta^{\prime} \Bigr] + \\
&+ \cos{\omega \tau} \Bigl[ 2 \frac{\re (r_H r_L^*)}
{\vert r_H + r_L \vert ^2} e^{-(\Omega^{\prime} + \Lambda^{\prime})
\frac{\tau}{\Gamma}}
+ \re y_h + \frac{2}{\Gamma} \frac{\delta}
{\delta ^2 + \omega ^2} \re \Delta^{\prime} \Bigr] \Bigr\}; \\
\end{split}
\end{equation*}

\begin{equation*}
\begin{split}
{\cal P}_{h^-}(\overline{\bf B}^0; \tau) = \vert {\cal M}_h \vert ^2 
e^{-\tau} 
&\Bigl\{ 
\sinh{\delta \tau} \Bigl[ - \re \lambda_h + \frac{2}{\Gamma} \frac{\omega}
{\delta ^2 + \omega ^2} \im \Delta^{\prime} - 
\frac{\Lambda^{\prime}}{\Delta \Gamma} \Bigr] + \\
&+ \cosh{\delta \tau} \Bigl[ 2 \frac{\vert r_L r_H \vert ^2}
{\vert r_H + r_L \vert ^2} - \re y_h \Bigr] + \\ 
&+ \sin{\omega \tau} \Bigl[ - \im \lambda_h - \frac{2}{\Gamma}
\frac{\delta}{\delta ^2 + \omega ^2} \im \Delta^{\prime} + 
\frac{1}{2\Delta m} \re \Theta^{\prime} \Bigr] + \\
&+ \cos{\omega \tau} \Bigl[ -2 \frac{\vert r_L r_H \vert ^2}
{\vert r_H + r_L \vert ^2} e^{-(\Omega^{\prime} + \Lambda^{\prime})
\frac{\tau}{\Gamma}}
+ \re y_h \Bigr] \Bigr\}. \\
\end{split}
\end{equation*}

\section{Decay in a CP eigenstate $f$}

\begin{equation*}
\begin{split}
{\cal P}_{f}({\bf B}^0; \tau) = \frac{1}{2}\vert 
{\cal M}_h \vert ^2 e^{-\tau} 
&\Bigl\{ 
\sinh{\delta \tau} \Bigl[ 2 \frac{\vert 1 - \lambda^f_H \vert^2 
\vert r_H \vert^2 - \vert 1 + \lambda^f_L \vert^2 \vert r_L \vert^2}
{\vert r_H + r_L \vert ^2} + \\
& \quad - \frac{4}{\Gamma}\zeta_f \frac{\delta}{\delta^2 +
\omega^2} \re \Delta^{\prime} - \frac{4 \Lambda^{\prime}}{\Delta \Gamma} -
\frac{4}{\Gamma} \frac{\omega}{\delta^2 + \omega^2} \im \Delta^{\prime} 
\Bigr] + \\
&+ \cosh{\delta \tau} \Bigl[ 2 \frac{\vert 1 - \lambda^f_H \vert^2 
\vert r_H \vert^2 + \vert 1 + \lambda^f_L \vert^2 \vert r_L \vert^2}
{\vert r_H + r_L \vert ^2} \Bigr] +  \\ 
& \quad + \frac{4}{\Gamma}\zeta_f \frac{\omega}{\delta^2 +
\omega^2} \im \Delta^{\prime} + 
\frac{4}{\Gamma} \frac{\delta}{\delta^2 + \omega^2} \re \Delta^{\prime} 
\Bigr] + \\
&+ \sin{\omega \tau} \Bigl[ -4 \frac{\im(r_L r_H^* (1 + \lambda^f_L)
(1 - \lambda^{f*}_H))}{\vert r_H + r_L \vert ^2} 
e^{-(\Omega^{\prime} + \Lambda^{\prime}) \frac{\tau}{\Gamma}} + \\
& \quad - \frac{4}{\Gamma}\zeta_f \frac{\omega}{\delta^2 +
\omega^2} \re \Delta^{\prime} + 
\frac{4}{\Gamma} \frac{\delta}{\delta^2 + \omega^2} \im \Delta^{\prime} 
\Bigr] +  \\
&+ \cos{\omega \tau} \Bigl[ 4 \frac{\re(r_L r_H^* (1 + \lambda^f_L)
(1 - \lambda^{f*}_H))}{\vert r_H + r_L \vert ^2} 
e^{-(\Omega^{\prime} + \Lambda^{\prime}) \frac{\tau}{\Gamma}} + \\
& \quad - \frac{4}{\Gamma}\zeta_f \frac{\omega}{\delta^2 +
\omega^2} \im \Delta^{\prime} - 
\frac{4}{\Gamma} \frac{\delta}{\delta^2 + \omega^2} \re \Delta^{\prime} 
\Bigr] \Bigr\}; \\
\end{split}
\end{equation*}

\begin{equation*}
\begin{split}
{\cal P}_{f}(\overline{\bf B}^0; \tau) = \frac{1}{2}\vert 
{\cal M}_h \vert ^2 e^{-\tau} 
&\Bigl\{ 
\sinh{\delta \tau} \Bigl[ 2 \vert r_L r_H \vert^2 
\frac{\vert 1 - \lambda^f_H \vert^2 - \vert 1 + \lambda^f_L \vert^2}
{\vert r_H + r_L \vert ^2} + \\
& \quad + \frac{4}{\Gamma}\zeta_f \frac{\delta}{\delta^2 +
\omega^2} \re \Delta^{\prime} - \frac{4 \Lambda^{\prime}}{\Delta \Gamma} +
\frac{4}{\Gamma} \frac{\omega}{\delta^2 + \omega^2} \im \Delta^{\prime} 
\Bigr] + \\
&+ \cosh{\delta \tau} \Bigl[ 2 \vert r_L r_H \vert^2
\frac{\vert 1 - \lambda^f_H \vert^2 + \vert 1 + \lambda^f_L \vert^2}
{\vert r_H + r_L \vert ^2} \Bigr] +  \\ 
& \quad - \frac{4}{\Gamma}\zeta_f \frac{\omega}{\delta^2 +
\omega^2} \im \Delta^{\prime} - 
\frac{4}{\Gamma} \frac{\delta}{\delta^2 + \omega^2} \re \Delta^{\prime} 
\Bigr] + \\
&+ \sin{\omega \tau} \Bigl[ -4 \vert r_L r_H \vert^2
\frac{\im((1 - \lambda^f_H)(1 + \lambda^{f*}_L))}{\vert r_H + r_L \vert ^2} 
e^{-(\Omega^{\prime} + \Lambda^{\prime}) \frac{\tau}{\Gamma}} + \\
& \quad + \frac{4}{\Gamma}\zeta_f \frac{\omega}{\delta^2 +
\omega^2} \re \Delta^{\prime} -
\frac{4}{\Gamma} \frac{\delta}{\delta^2 + \omega^2} \im \Delta^{\prime} 
\Bigr] +  \\
&+ \cos{\omega \tau} \Bigl[ 4 \vert r_L r_H \vert^2
\frac{\re((1 - \lambda^f_H)(1 + \lambda^{f*}_L))}{\vert r_H + r_L \vert ^2}
e^{-(\Omega^{\prime} + \Lambda^{\prime}) \frac{\tau}{\Gamma}} + \\
& \quad + \frac{4}{\Gamma}\zeta_f \frac{\omega}{\delta^2 +
\omega^2} \im \Delta^{\prime} + 
\frac{4}{\Gamma} \frac{\delta}{\delta^2 + \omega^2} \re \Delta^{\prime} 
\Bigr] \Bigr\}. \\
\end{split}
\end{equation*}

%% file: appendixD.tex
\chapter{Dissipative parameters in the neutrino case}
\la{a4}
\markboth{\small{\rm{\thechapter. DISSIPATIVE PARAMETERS IN THE NEUTRINO CASE}}}{\small{\rm{\thechapter. DISSIPATIVE PARAMETERS IN THE NEUTRINO CASE}}}

\begin{eqnarray*}
\omega_1 &=& \frac{g^2}{2V^2} \im \left[
2 B_{e \mu}(\vec{p}) + A_{e \mu \mu \mu}(\vec{p}) -
A_{e e \mu e}(\vec{p}) - A_{e e e \mu}(\vec{p}) +
A_{\mu e \mu \mu}(\vec{p})
\right], \\
\omega_2 &=& -\frac{g^2}{2V^2} \im \left[
2 B_{\mu e}(\vec{p}) + A_{e e \mu e}(\vec{p}) -
A_{e \mu \mu \mu}(\vec{p}) - A_{e e e \mu}(\vec{p}) + 
A_{\mu e \mu \mu}(\vec{p})
\right], \\
\omega_3 &=& -\frac{g^2}{2V^2} \im \left[B_{e e}(\vec{p}) -
B_{\mu \mu}(\vec{p}) - 2A_{e e \mu \mu}(\vec{p}) \right], \\
u &=& -\frac{g^2}{V^2} \re \left[
A_{e \mu e e}(\vec{p}) + A_{\mu \mu \mu e}(\vec{p}) - 
A_{e \mu}(\vec{p})
\right], \\
v &=& \frac{g^2}{V^2} \im \left[
A_{e \mu e e}(\vec{p}) + A_{\mu \mu \mu e}(\vec{p}) - 
A_{e \mu}(\vec{p})
\right], \\
w &=& \frac{g^2}{V^2} \left[
A_{e \mu e \mu}(\vec{p}) - A_{\mu e \mu e}(\vec{p})
\right], \\
a &=& \frac{g^2}{2V^2} \left\{
A_{ee}(\vec{p}) + A_{\mu \mu}(\vec{p}) -
2\re \left[ A_{e e \mu \mu}(\vec{p}) +
A_{e \mu \mu e}(\vec{p}) \right] \right\}, \\
b &=& \frac{g^2}{V^2} \im \left[
A_{e \mu \mu e}(\vec{p}) \right], \\
c &=& -\frac{g^2}{2V^2} \re \left[
A_{e e \mu e}(\vec{p}) - A_{e \mu \mu \mu}(\vec{p}) +
A_{e e e \mu} - A_{\mu e \mu \mu}
\right], \\
\alpha &=& \frac{g^2}{2V^2} \left\{
A_{ee}(\vec{p}) + A_{\mu \mu}(\vec{p}) -
2\re \left[ A_{e e \mu \mu}(\vec{p}) -
A_{e \mu \mu e}(\vec{p}) \right] \right\}, \\
\beta &=& -\frac{g^2}{2V^2} \im \left[
A_{e e e \mu}(\vec{p}) - A_{\mu e \mu \mu}(\vec{p}) -
A_{e e \mu e}(\vec{p}) + A_{e \mu \mu \mu}(\vec{p})
\right], \\
\gamma &=& \frac{g^2}{V^2} \left[
A_{e \mu e \mu}(\vec{p}) + A_{\mu e \mu e}(\vec{p})
\right].
\end{eqnarray*}

%% file: appendixE.tex
\chapter{Frequencies of counts}
\la{a5}
\markboth{\small{\rm{\thechapter. FREQUENCIES OF COUNTS}}}{\small{\rm{\thechapter. FREQUENCIES OF COUNTS}}}

\begin{eqnarray*}
\rho_{13}(t) &=& \frac{1}{2} \Bigl\{
\Bigl[{\cal O}^{1,x}_{t}\left(0,0\right) - 
{\cal O}^{1,-x}_{t}\left(0,0\right)\Bigr] +i
\Bigl[{\cal O}^{1,-y}_{t}\left(0,0\right) - 
{\cal O}^{1,y}_{t}\left(0,0\right)\Bigr]\Bigr\} \\ \\
\rho_{14}(t) &=& \frac{1}{2} \Bigl\{
\Bigl[{\cal O}^{1,x}_{t}\left(0,0\right) - 
{\cal O}^{1,-x}_{t}\left(0,0\right)\Bigr] -i
\Bigl[{\cal O}^{1,-y}_{t}\left(0,0\right) - 
{\cal O}^{1,y}_{t}\left(0,0\right)\Bigr]\Bigr\} \\ \\
\rho_{23}(t) &=& \frac{1}{2} \Bigl\{
\Bigl[{\cal O}^{2,x}_{t}\left(0,0\right) - 
{\cal O}^{2,-x}_{t}\left(0,0\right)\Bigr] +i
\Bigl[{\cal O}^{2,-y}_{t}\left(0,0\right) - 
{\cal O}^{2,y}_{t}\left(0,0\right)\Bigr]\Bigr\} \\ \\
\rho_{24}(t) &=& \frac{1}{2} \Bigl\{
\Bigl[{\cal O}^{2,x}_{t}\left(0,0\right) - 
{\cal O}^{2,-x}_{t}\left(0,0\right)\Bigr] -i
\Bigl[{\cal O}^{2,-y}_{t}\left(0,0\right) - 
{\cal O}^{2,y}_{t}\left(0,0\right)\Bigr]\Bigr\} \\ \\
\rho_{31}(t) &=& \frac{1}{2} \Bigl\{
\Bigl[{\cal O}^{1,z}_{t}\left(\frac{\pi}{4},0\right) - 
{\cal O}^{2,z}_{t}\left(\frac{\pi}{4},0\right)\Bigr] -i
\Bigl[{\cal O}^{1,z}_{t}\left(\frac{\pi}{4},-\frac{\pi}{2}\right) - 
{\cal O}^{2,z}_{t}\left(\frac{\pi}{4},\frac{\pi}{2}\right)\Bigr]\Bigr\} \\ \\
\rho_{32}(t) &=& \frac{1}{2} \Bigl\{
\Bigl[{\cal O}^{1,-z}_{t}\left(\frac{\pi}{4},0\right) - 
{\cal O}^{2,-z}_{t}\left(\frac{\pi}{4},0\right)\Bigr] -i
\Bigl[{\cal O}^{1,-z}_{t}\left(\frac{\pi}{4},-\frac{\pi}{2}\right) - 
{\cal O}^{2,-z}_{t}\left(\frac{\pi}{4},\frac{\pi}{2}\right)\Bigr]\Bigr\} \\ \\
\rho_{41}(t) &=& \frac{1}{2} \Bigl\{
\Bigl[{\cal O}^{1,z}_{t}\left(\frac{\pi}{4},0\right) - 
{\cal O}^{2,z}_{t}\left(\frac{\pi}{4},0\right)\Bigr] +i
\Bigl[{\cal O}^{1,z}_{t}\left(\frac{\pi}{4},-\frac{\pi}{2}\right) - 
{\cal O}^{2,z}_{t}\left(\frac{\pi}{4},\frac{\pi}{2}\right)\Bigr]\Bigr\} \\ \\
\rho_{42}(t) &=& \frac{1}{2} \Bigl\{
\Bigl[{\cal O}^{1,-z}_{t}\left(\frac{\pi}{4},0\right) - 
{\cal O}^{2,-z}_{t}\left(\frac{\pi}{4},0\right)\Bigr] +i
\Bigl[{\cal O}^{1,-z}_{t}\left(\frac{\pi}{4},-\frac{\pi}{2}\right) - 
{\cal O}^{2,-z}_{t}\left(\frac{\pi}{4},\frac{\pi}{2}\right)\Bigr]\Bigr\}
\end{eqnarray*}

\begin{eqnarray*}
\rho_{33}(t) &=& \frac{1}{4} \Bigl\{
\Bigl[{\cal O}^{1,x}_{t}\left(\frac{\pi}{4},0\right) - 
{\cal O}^{1,-x}_{t}\left(\frac{\pi}{4},0\right)\Bigr] -i
\Bigl[{\cal O}^{1,y}_{t}\left(\frac{\pi}{4},0\right) - 
{\cal O}^{1,-y}_{t}\left(\frac{\pi}{4},0\right)\Bigr] + \\
&&\quad- \Bigl[{\cal O}^{2,x}_{t}\left(\frac{\pi}{4},0\right) - 
{\cal O}^{2,-x}_{t}\left(\frac{\pi}{4},0\right)\Bigr] +i
\Bigl[{\cal O}^{2,y}_{t}\left(\frac{\pi}{4},0\right) - 
{\cal O}^{2,-y}_{t}\left(\frac{\pi}{4},0\right)\Bigr] + \\
&&\quad- \Bigl[{\cal O}^{1,y}_{t}\left(\frac{\pi}{4},-\frac{\pi}{2}\right) - 
{\cal O}^{1,-y}_{t}\left(\frac{\pi}{4},-\frac{\pi}{2}\right)\Bigr] -i
\Bigl[{\cal O}^{1,x}_{t}\left(\frac{\pi}{4},-\frac{\pi}{2}\right) - 
{\cal O}^{1,-x}_{t}\left(\frac{\pi}{4},-\frac{\pi}{2}\right)\Bigr] + \\
&&\quad+ \Bigl[{\cal O}^{2,y}_{t}\left(\frac{\pi}{4},-\frac{\pi}{2}\right) - 
{\cal O}^{2,-y}_{t}\left(\frac{\pi}{4},-\frac{\pi}{2}\right)\Bigr] +i
\Bigl[{\cal O}^{2,x}_{t}\left(\frac{\pi}{4},-\frac{\pi}{2}\right) - 
{\cal O}^{2,-x}_{t}\left(\frac{\pi}{4},-\frac{\pi}{2}\right)\Bigr] \Bigr\}
\end{eqnarray*}

\begin{eqnarray*}
\rho_{34}(t) &=& \frac{1}{4} \Bigl\{
\Bigl[{\cal O}^{1,x}_{t}\left(\frac{\pi}{4},0\right) - 
{\cal O}^{1,-x}_{t}\left(\frac{\pi}{4},0\right)\Bigr] +i
\Bigl[{\cal O}^{1,y}_{t}\left(\frac{\pi}{4},0\right) - 
{\cal O}^{1,-y}_{t}\left(\frac{\pi}{4},0\right)\Bigr] + \\
&&\quad- \Bigl[{\cal O}^{2,x}_{t}\left(\frac{\pi}{4},0\right) - 
{\cal O}^{2,-x}_{t}\left(\frac{\pi}{4},0\right)\Bigr] -i
\Bigl[{\cal O}^{2,y}_{t}\left(\frac{\pi}{4},0\right) - 
{\cal O}^{2,-y}_{t}\left(\frac{\pi}{4},0\right)\Bigr] + \\
&&\quad+ \Bigl[{\cal O}^{1,y}_{t}\left(\frac{\pi}{4},-\frac{\pi}{2}\right) - 
{\cal O}^{1,-y}_{t}\left(\frac{\pi}{4},-\frac{\pi}{2}\right)\Bigr] -i
\Bigl[{\cal O}^{1,x}_{t}\left(\frac{\pi}{4},-\frac{\pi}{2}\right) - 
{\cal O}^{1,-x}_{t}\left(\frac{\pi}{4},-\frac{\pi}{2}\right)\Bigr] + \\
&&\quad- \Bigl[{\cal O}^{2,y}_{t}\left(\frac{\pi}{4},-\frac{\pi}{2}\right) - 
{\cal O}^{2,-y}_{t}\left(\frac{\pi}{4},-\frac{\pi}{2}\right)\Bigr] +i
\Bigl[{\cal O}^{2,x}_{t}\left(\frac{\pi}{4},-\frac{\pi}{2}\right) - 
{\cal O}^{2,-x}_{t}\left(\frac{\pi}{4},-\frac{\pi}{2}\right)\Bigr] \Bigr\}
\end{eqnarray*}

\begin{eqnarray*}
\rho_{43}(t) &=& \frac{1}{4} \Bigl\{
\Bigl[{\cal O}^{1,x}_{t}\left(\frac{\pi}{4},0\right) - 
{\cal O}^{1,-x}_{t}\left(\frac{\pi}{4},0\right)\Bigr] +i
\Bigl[{\cal O}^{1,y}_{t}\left(\frac{\pi}{4},0\right) - 
{\cal O}^{1,-y}_{t}\left(\frac{\pi}{4},0\right)\Bigr] + \\
&&\quad- \Bigl[{\cal O}^{2,x}_{t}\left(\frac{\pi}{4},0\right) - 
{\cal O}^{2,-x}_{t}\left(\frac{\pi}{4},0\right)\Bigr] -i
\Bigl[{\cal O}^{2,y}_{t}\left(\frac{\pi}{4},0\right) - 
{\cal O}^{2,-y}_{t}\left(\frac{\pi}{4},0\right)\Bigr] + \\
&&\quad- \Bigl[{\cal O}^{1,y}_{t}\left(\frac{\pi}{4},-\frac{\pi}{2}\right) - 
{\cal O}^{1,-y}_{t}\left(\frac{\pi}{4},-\frac{\pi}{2}\right)\Bigr] +i
\Bigl[{\cal O}^{1,x}_{t}\left(\frac{\pi}{4},-\frac{\pi}{2}\right) - 
{\cal O}^{1,-x}_{t}\left(\frac{\pi}{4},-\frac{\pi}{2}\right)\Bigr] + \\
&&\quad+ \Bigl[{\cal O}^{2,y}_{t}\left(\frac{\pi}{4},-\frac{\pi}{2}\right) - 
{\cal O}^{2,-y}_{t}\left(\frac{\pi}{4},-\frac{\pi}{2}\right)\Bigr] -i
\Bigl[{\cal O}^{2,x}_{t}\left(\frac{\pi}{4},-\frac{\pi}{2}\right) - 
{\cal O}^{2,-x}_{t}\left(\frac{\pi}{4},-\frac{\pi}{2}\right)\Bigr] \Bigr\}
\end{eqnarray*}

\begin{eqnarray*}
\rho_{44}(t) &=& \frac{1}{4} \Bigl\{
\Bigl[{\cal O}^{1,x}_{t}\left(\frac{\pi}{4},0\right) - 
{\cal O}^{1,-x}_{t}\left(\frac{\pi}{4},0\right)\Bigr] -i
\Bigl[{\cal O}^{1,y}_{t}\left(\frac{\pi}{4},0\right) - 
{\cal O}^{1,-y}_{t}\left(\frac{\pi}{4},0\right)\Bigr] + \\
&&\quad- \Bigl[{\cal O}^{2,x}_{t}\left(\frac{\pi}{4},0\right) - 
{\cal O}^{2,-x}_{t}\left(\frac{\pi}{4},0\right)\Bigr] +i
\Bigl[{\cal O}^{2,y}_{t}\left(\frac{\pi}{4},0\right) - 
{\cal O}^{2,-y}_{t}\left(\frac{\pi}{4},0\right)\Bigr] + \\
&&\quad+ \Bigl[{\cal O}^{1,y}_{t}\left(\frac{\pi}{4},-\frac{\pi}{2}\right) - 
{\cal O}^{1,-y}_{t}\left(\frac{\pi}{4},-\frac{\pi}{2}\right)\Bigr] +i
\Bigl[{\cal O}^{1,x}_{t}\left(\frac{\pi}{4},-\frac{\pi}{2}\right) - 
{\cal O}^{1,-x}_{t}\left(\frac{\pi}{4},-\frac{\pi}{2}\right)\Bigr] + \\
&&\quad- \Bigl[{\cal O}^{2,y}_{t}\left(\frac{\pi}{4},-\frac{\pi}{2}\right) - 
{\cal O}^{2,-y}_{t}\left(\frac{\pi}{4},-\frac{\pi}{2}\right)\Bigr] -i
\Bigl[{\cal O}^{2,x}_{t}\left(\frac{\pi}{4},-\frac{\pi}{2}\right) - 
{\cal O}^{2,-x}_{t}\left(\frac{\pi}{4},-\frac{\pi}{2}\right)\Bigr] \Bigr\}
\end{eqnarray*}

%% file: tesi.bbl
\begin{thebibliography}{999}
\addcontentsline{toc}{chapter}{Bibliography}
\markboth{\small{\rm{BIBLIOGRAPHY}}}{\small{\rm{BIBLIOGRAPHY}}}

\bibitem{alic1}
R. Alicki and K. Lendi, {\it Quantum Dynamical Semigroups and Applications},
Lecture Notes in Physics, vol. 286, Springer-Verlag, 1987

\bibitem{breu}
H.-P. Breuer and F. Petruccione, {\it The Theory of Open Quantum
Systems}, Oxford University Press, 2002

\bibitem{davi1}
E.B. Davies, {\it Quantum Theory of Open Systems}, Academic Press, 1976

\bibitem{slic}
C.P. Slichter, {\it Principles of magnetic resonance, with examples from 
solid state physics}, Springer-Verlag, 1990

\bibitem{loui}
W.H. Louisell, {\it Quantum Statistical Properties of Radiation},
Wiley, 1973

\bibitem{puri}
R.R. Puri, {\it Mathematical Methods of Quantum Optics}, Springer, 2001

\bibitem{gard}
C.W. Gardiner and P. Zoller, {\it Quantum Noise}, Springer, 2000

\bibitem{scul}
M.O. Scully and M.S. Zubairy, {\it Quantum Optics}, Cambridge University
Press, 1997

\bibitem{fond}
L. Fonda, G.C. Ghirardi and A. Rimini, Rep. Prog. Phys. {\bf 41} (1978),
587

\bibitem{benai1}
F. Benatti and R. Floreanini, Phys. Rev. {\bf D 62} (2000), 125009
({\tt hep-ph/0009283})

\bibitem{benai2}
F. Benatti and R. Floreanini, Phys. Lett. {\bf B 451} (1999), 422
({\tt quant-ph/9902026})

\bibitem{benai3}
F. Benatti, R. Floreanini and R. Romano, J. Phys. {\bf A 35} (2002), 4955
({\tt quant-ph/0205007})

\bibitem{nakaz}
H. Nakazato, M. Namiki and S. Pascazio, Int. J. Mod. Phys. {\bf B 10}
(1996), 247


\bibitem{take}
M. Takesaki, {\it Theory of Operator Algebras}, vol.~1, Springer, 1979

\bibitem{horo}
M. Horodecki, P. Horodecki and R. Horodecki, Phys. Lett. {\bf A 223} 
(1996), 1 ({\tt quant-ph/9605038})

\bibitem{krau}
K. Kraus, Ann. Phys. {\bf 64} (1971), 311

\bibitem{stin}
W. F. Stinespring, Proc. Amer. Math. Soc. {\bf 6} (1955), 211

\bibitem{choi}
M. D. Choi, Canad. J. Math. {\bf 24} (1972), 520

\bibitem{choi2}
M. D. Choi, Linear Alg. Appl. {\bf 10} (1975), 285

\bibitem{nach}
O. Nachtmann, {\it Elementary particle physics: concepts and phenomena},
Springer-Verlag, 1990

\bibitem{naka}
S. Nakajima, Progr. Theor. Phys. {\bf 20} (1958), 948

\bibitem{zwan}
R. Zwanzig, Lect. Theor. Phys. {\bf 3} (1960), 106

\bibitem{haak}
F. Haake, {\it Statistical tratement of open systems by generalized master
equations}, Springer Tracts in Modern Physics, vol. 66, 1973

\bibitem{pech}
P. Pechukas, Phys. Rev. Lett. {\bf 73} (1994), 1060

\bibitem{roye}
A. Royer, Phys. Rev. Lett. {\bf 77} (1996), 3272

\bibitem{rome}
V. Romero-Rochin and I. Oppenheim, Physica {\bf 155 A} (1989), 52

\bibitem{suar}
A. Suarez, R. Silbey and I. Oppenheim, J. Chem. Phys. {\bf 97} (1992),
5101

\bibitem{gnut}
S. Gnutzmann and F. Haake, Z. Phys. {\bf B 101} (1996), 263

\bibitem{yu}
T. Yu, L. Diosi, N. Gisin and W.T. Strunz, Phys. Rev. {\bf B 60} (1999), 91

\bibitem{gasp}
P. Gaspard and M. Nagaoka, J. Chem. Phys. {\bf 111} (1999), 5668 

\bibitem{1gasp}
P. Gaspard and M. Nagaoka, J. Chem. Phys. {\bf 111} (1999), 5676

\bibitem{alic2}
R. Alicki, Phys. Rev. Lett. {\bf 75} (1995), 3020

\bibitem{spoh}
H. Spohn, Rev. Mod. Phys. {\bf 52} (1980), 569

\bibitem{gori1}
V. Gorini, A. Frigerio, M. Verri, A. Kossakowski and E.C.G. Sudarshan,
Rep. Math. Phys. {\bf 13} (1978), 149

\bibitem{dumc}
R. D{\" u}mcke and H. Spohn, Z. Phys. {\bf B 34} (1979), 419

\bibitem{davi2}
E. B. Davies, Math. Ann. {\bf 219} (1976), 147

\bibitem{hepp}
K. Hepp and E.H. Lieb, Helv. Phys. Acta {\bf 46} (1973), 573

\bibitem{palm}
P. F. Palmer, J. Math. Phys. {\bf 18} (1977), 527

\bibitem{gori2}
V. Gorini and A. Kossakowski, J. Math. Phys. {\bf 17} (1976), 1298

\bibitem{frig}
A. Frigerio, C. Novellone and M. Verri, Rep. Math. Phys. {\bf 12} (1977), 
279

\bibitem{lind}
G. Lindblad, Commun. Math. Phys. {\bf 48} (1976), 119

\bibitem{gori3}
V. Gorini, A. Kossakowski and E.C.G. Sudarshan, J. Math. Phys. {\bf 17} 
(1976), 821

\bibitem{benfac}
F. Benatti, R. Floreanini and R. Romano, J. Phys. {\bf A}: 
Math Gen. {\bf 35} (2002), L551 ({\tt quant-ph/0205007})

\bibitem{benak}
F. Benatti and R. Floreanini, Phys. Lett. {\bf B 389} (1996), 100
({\tt hep-th/9607059}) 

\bibitem{1benak}
F. Benatti and R. Floreanini, Phys. Lett. {\bf B 401} (1997), 337
({\tt hep-ph/9704283}) 

\bibitem{2benak}
F. Benatti and R. Floreanini, Nucl. Phys. {\bf B 488} (1997), 335 

\bibitem{3benak}
F. Benatti and R. Floreanini, Nucl. Phys. {\bf B 511} (1998), 550
({\tt hep-ph/9711240}) 

\bibitem{4benak}
F. Benatti and R. Floreanini, Phys. Lett. {\bf B 468} (1999), 287
({\tt hep-ph/9910508})

\bibitem{benab}
F. Benatti, R. Floreanini and R. Romano, Nucl. Phys. {\bf B 602}
(2001), 541 ({\tt hep-ph/0103239})

\bibitem{gelf}
I.M. Gel'fand {\it Lectures on Linear Algebra}, Interscience Tracts in 
Pure and Applied Mathematics, vol. 9, 1961

\bibitem{pian}
F. Benatti, R. Floreanini and M. Piani, ``Curing non-positive evolutions
in open system dynamics'', submitted to EPJ {\bf B}

\bibitem{lend}
K. Lendi, J. Phys. {\bf A 20} (1987), 15

\bibitem{spo}
H. Spohn, Rep. Math. Phys. {\bf 10} (1976), 189

\bibitem{fri}
A. Frigerio, Comm. Math. Phys. {\bf 63} (1978), 269


\bibitem{benat}
F. Benatti and R. Floreanini, Banach Center Publications {\bf 43}
(1998), 71

\bibitem{wehr}
A. Wehrl, Rev. Mod. Phys. {\bf 50} (1978), 221


\bibitem{bran}
G.C. Branco, L. Lavoura and J.P. Silva, {\it CP Violation}, Clarendon Press,
1999

\bibitem{bigi}
I.I. Bigi and A.I. Silva, {\it CP Violation}, Cambridge University Press,
2000

\bibitem{hawk}
S. Hawking, Commun. Math. Phys. {\bf 87} (1983), 395 

\bibitem{1hawk}
S. Hawking, Phys. Rev. {\bf D 37} (1988), 904 

\bibitem{2hawk}
S. Hawking, Phys. Rev. {\bf D 53} (1996), 3099 

\bibitem{3hawk}
S. Hawking and C. Hunter, Phys. Rev. {\bf D 59} (1999), 044025

\bibitem{elli1}
J. Ellis, N.E. Mavromatos and D.V. Nanopoulos, Phys. Lett. {\bf B 293}
(1992), 37 

\bibitem{1elli1}
J. Ellis, N.E. Mavromatos and D.V. Nanopoulos, Int. J. Mod. Phys. {\bf A 11}
(1996), 1489

\bibitem{bena1}
F. Benatti and R. Floreanini, Ann. Phys. {\bf 273} (1999), 58

\bibitem{mari}
M.S. Marinov, JETP Lett. {\bf 15} (1972), 479 

\bibitem{1mari}
M.S. Marinov, Sov. J. Nucl. Phys. {\bf 19} (1974), 173 

\bibitem{2mari}
M.S. Marinov, Nucl. Phys. {\bf B 253} (1985), 609

\bibitem{elli2}
J. Ellis, J.S. Hagelin, D.V. Nanopoulos and M. Srednicki, Nucl. Phys. 
{\bf B 241} (1984), 381

\bibitem{bert}
R.A. Bertlmann and W. Grimus, Phys. Rev. {\bf D 64} (2001), 056004
({\tt hep-ph/0101160}) 

\bibitem{cole}
S. Coleman, Nucl. Phys. {\bf B 307} (1988), 867

\bibitem{gidd}
S.B. Giddings and A. Strominger, Nucl. Phys. {\bf B 307} (1988), 854

\bibitem{sred}
M. Srednicki, Nucl. Phys. {\bf B 410} (1993), 143

\bibitem{unru}
W.G. Unruh and R.M. Wald, Phys. Rev. {\bf D 52} (1995), 2176

\bibitem{gara}
L.J. Garay, Phys. Rev. Lett. {\bf 80} (1998), 2508 

\bibitem{1gara}
L.J. Garay, Phys. Rev. {\bf D 58} (1998), 124015 


\bibitem{huet}
P. Huet and M.E. Peskin, Nucl. Phys. {\bf B 434} (1995), 3 

\bibitem{elli3}
J. Ellis, J.L. Lopez, N.E. Mavromatos and D.V. Nanopoulos, 
Phys. Rev. {\bf D 53} (1996), 3846

\bibitem{ball}
P. Ball {\it et al.}, {\bf B} decays at the LHC, CERN-TH-2000-101
({\tt hep-ph/0003238})

\bibitem{btev}
BTeV proposal, 2000, http://www-btev.fnal.gov

\bibitem{banu1}
M.C. Banuls and J. Bernabeu, Phys. Lett. {\bf B 423} (1998), 151
({\tt hep-ph/9710348}) 

\bibitem{1banu1}
M.C. Banuls and J. Bernabeu, Phys. Lett. {\bf B 464} (1999), 117
({\tt hep-ph/9908353}) 

\bibitem{2banu1}
M.C. Banuls and J. Bernabeu, Nucl. Phys. {\bf B 590} (2000), 19
({\tt hep-ph/0005323}) 

\bibitem{banu2}
M.C. Banuls, Nucl. Phys. Proc. Suppl. {\bf 93} (2001), 268
({\tt hep-ph/0009317})

\bibitem{lavo1}
L. Lavoura, Phys. Lett. {\bf B} (1998), 390 ({\tt hep-ph/9806470})

\bibitem{lavo2}
L. Lavoura and J.P. Silva, Phys. Rev. {\bf D 60} (1999), 066003
({\tt hep-ph/9902348})

\bibitem{baba}
BaBar Collaboration, The Babar Physics Book, SLAC-R-504, 1998

\bibitem{dedo}
G. De Domenico and C. Y{\` e}che, BaBar-Note-409, 1998

\bibitem{leon}
C. Leonidopoulos, Ph.D. Thesis, Princeton University, 2000

\bibitem{benap}
F. Benatti and R. Floreanini, Mod. Phys. Lett. {\bf A 14} (1999), 1519


\bibitem{wolf}
L. Wolfenstein, Phys. Rev {\bf D 17} (1978), 2369

\bibitem{mikh}
S.P. Mikheyev and A.Y. Smirnov, Sov. J. Nucl. Phys. {\bf 42} (1985), 913 

\bibitem{1mikh}
S.P. Mikheyev and A.Y. Smirnov, Nuovo Cimento {\bf 9C} (1986), 17

\bibitem{halp}
A. Halprin, Phys. Rev. {\bf D 34} (1986), 3462

\bibitem{stod}
L. Stodolsky, Phys. Rev. {\bf D 36} (1987), 2273

\bibitem{sawy}
R.F. Sawyer, Phys. Rev. {\bf D 42} (1990), 3908

\bibitem{samu}
S. Samuel, Phys. Rev. {\bf D 48} (1993), 1462

\bibitem{burg1}
C.P. Burgess and D. Michaud, Annals Phys. {\bf 256} (1997), 1
({\tt hep-ph/9606295})

\bibitem{burg2}
P. Bamert, C.P. Burgess and D. Michaud, Nucl. Phys. {\bf B 513} 
(1998), 319 ({\tt hep-ph/9707542})

\bibitem{bala1}
A.B. Balantekin, Phys. Rept. {\bf 315} (1999), 123
({\tt hep-ph/9808281})

\bibitem{bala2}
A.B. Balantekin, J.M. Fetter and F.N. Loreti, Phys. Rev. {\bf D 54} 
(1996), 3941 ({\tt astro-ph/9604061})

\bibitem{lore1}
F.N. Loreti and A.B. Balantekin, Phys. Rev. {\bf D 50} (1994), 4762
({\tt nucl-th/9406003})

\bibitem{torr}
E. Torrente-Lujan, Phys. Rev. {\bf D 59} (1999), 073001
({\tt hep-ph/9807361})

\bibitem{nuno}
H. Nunokawa, A. Rossi, V.B. Semikoz and J.W.F. Valle, Nucl. Phys. 
{\bf B 472} (1996), 495 ({\tt hep-ph/9602307})

\bibitem{raff}
G. Raffelt, G. Sigl and L. Stodolsky,  Phys. Rev. Lett. {\bf 70} (1993), 2363
({\tt hep-ph/9209276})



\bibitem{kim}
C.W. Kim and A. Pevsner, {\it Neutrinos in physics and astrophysics},
Contemporary Concepts in Physics, vol. 8

\bibitem{moh}
R.N. Mohapatra and P.B. Pal, {\it Massive Neutrinos in Physics and 
Astrophysics}, $2^{nd}$ ed., World Scientific, 1999

\bibitem{bil}
S.M. Bilenky, C. Giunti and W. Grimus, Prog. Part. Nucl. Phys. {\bf 43} 
(1999), 1 ({\tt hep-ph/9812360})

\bibitem{bile}
S.M.Bilenky, ``On the status of neutrino mixing and oscillations'',
Report at the XVI Recontres de Physique de La Vallee d'Aoste, La Thuile, 
Aosta Valley (Italy) March 3-9, 2002
({\tt hep-ph/0205047})

\bibitem{grim}
W. Grimus, P. Stockinger and S. Mohanty, Phys.Rev. {\bf D 59} (1999), 013011
({\tt hep-ph/9807442})

\bibitem{giun}
C. Giunti, JHEP {\bf 0211} (2002), 017 ({\tt hep-ph/0205014})

\bibitem{mcka}
D. McKay and H. Munczek, Phys. Rev. {\bf D 19} (1979), 985

\bibitem{zee}
A. Zee, Phys. Lett. {\bf B 93} (1980), 389

\bibitem{babu}
K.S. Babu, Phys. Lett. {\bf B 203} (1988), 132

\bibitem{john}
L.M. Johnson and D.W. McKay, Phys. Lett. {\bf B 508} (2001), 290
({\tt hep-ph/9909355})

\bibitem{mann}
P.D. Mannheim, Phys. Rev. {\bf D 37} (1988), 1935

\bibitem{benan}
F. Benatti and R. Floreanini, Phys. Rev. {\bf D 64} (2001), 085015
({\tt hep-ph/0105303})

\bibitem{benan2}
F. Benatti and R. Floreanini, JHEP {\bf 0002} (2000), 032,
({\tt hep-ph/0002221})

\bibitem{lisi}
E. Lisi, A. Marrone and D. Montanino, Phys. Rev. Lett. {\bf 85} (2000), 1166

\bibitem{liu}
Y. Liu, L. Hu and M.-L. Ge, Phys. Rev. {\bf D 56} (1997), 6648

\bibitem{chan}
C.-H. Chang, W.-S. Dai, X.-Q. Li, Y. Liu, F.-C. Ma and Z. Tao,
Phys. Rev. {\bf D 60} (1999), 033006

\bibitem{sche}
J. Schechter and J.W.F. Valle, Phys. Rev. {\bf D 22} (1980), 2227

\bibitem{li}
L.F. Li and F. Wilczek, Phys. Rev. {\bf D 25} (1982), 143

\bibitem{koh}
J.K. Koh, C. Giunti and C.W. Kim, J. Korean Phys. Soc. {\bf 24} (1991), 275


\bibitem{budi}
J. Budimir and J.L. Skinner, J. Stat. Phys. {\bf 49} (1987), 1029

\bibitem{lair}
B.B. Laird, J.L. Skinner, J. Chem. Phys. {\bf 94} (1991), 4405

\bibitem{pere}
A. Peres, {\it Quantum Theory: Concepts and Methods}, Fundamental theories 
of physics, vol.57, Kluwer, 1993.

\bibitem{clau}
J.F. Clauser, M.A. Horne, A. Shimony and R.A. Holt, Phys. Rev. Lett.
{\bf 23} (1969), 880

\bibitem{basu}
S. Basu, S. Bandyopadhyay, G. Kar and D. Home, Phys. Lett. {\bf A 279}
(2001), 281 ({\tt quant-ph/9907030})

\bibitem{mich}
M. Michler, H. Weinfurter and M. Zukowski, Phys. Rev. Lett. {\bf 84}
(2000), 5457 ({\tt quant-ph/0009061})

\bibitem{stau}
J.L. Staudenmann, S.A. Werner, R. Colella and A.W. Overhauser, 
Phys. Rev. {\bf A 21} (1980), 1419

\bibitem{wern}
S.A. Werner and A.G. Klein, Meth. Exp. Phys. {\bf A 23} (1986), 259

\bibitem{sear}
V.F. Sears, {\it Neutron Optics: an introduction to the theory of neutron 
optical phenomena and their applications}, Oxford University Press, 1989

\bibitem{utsu}
M. Utsuro, S. Kawano, T. Kawai and A. Kawaguchi, {\it Advance in neutron 
optics and related research facilities}, The Physical Society of Japan, 1996

\bibitem{litt}
K.C. Littrell, B.E. Allmann and S.A. Werner, Phys. Rev. {\bf A56} (1997), 
1767

\bibitem{allm}
B.E. Allmann, H. Kaiser, S.A. Werner, A.G. Wagh, V.C. Rakhecha and J.
Summhammer, Phys. Rev. {\bf A 56} (1997), 4420

\bibitem{rauc}
H. Rauch and S.A. Werner, {\it Neutron interferometry: lessons in 
experimental quantum mechanics}, Oxford University Press, 2000


\end{thebibliography}
